\documentclass[fleqn,usenatbib]{mnras}

\usepackage{newtxtext,newtxmath}

\usepackage[T1]{fontenc}

\DeclareRobustCommand{\VAN}[3]{#2}
\let\VANthebibliography\thebibliography
\def\thebibliography{\DeclareRobustCommand{\VAN}[3]{##3}\VANthebibliography}

\usepackage{physics}
\usepackage{hyperref}
\usepackage{multirow,tabularx}
\usepackage{array}
\usepackage[table,svgnames,dvipsnames]{xcolor}
\usepackage{makecell,cellspace,caption}
\usepackage{mathtools}
\usepackage{orcidlink}
\usepackage{graphicx}	                
\usepackage{amsmath}	                
\usepackage{bm}                         
\usepackage{enumitem}
\usepackage{url}
\usepackage{braket}
\usepackage{threeparttable}
\usepackage{placeins}

\usepackage[final]{zebra-goodies}     
\zebranewnote{question}{orange}
\zebranewnote{revised}{red}

\newcommand\dex{\,{\rm dex}}
\newcommand\Kelvin{\,{\rm K}}
\newcommand\Mpc{\,{\rm Mpc}}
\newcommand\au{\,{\rm au}}
\newcommand\pc{\,{\rm pc}}
\newcommand\mpc{\, h^{-1}{\rm {Mpc}}}

\newcommand\Kpc{\, {\rm {kpc}}}
\newcommand\kpc{\, h^{-1}{\rm {kpc}}}

\newcommand\Kyr{\,{\rm kyr}}
\newcommand\Myr{\,{\rm Myr}}
\newcommand\Gyr{\,{\rm Gyr}}

\newcommand\kms{\,{\rm {km\, s^{-1}}}}
\newcommand\Msun{\,{\rm M_\odot}}
\newcommand\msun{\, h^{-1}{\rm M_\odot}}
\newcommand\msunperyr{\, {\rm M_\odot}{\rm yr}^{-1}}

\newcommand\Lsun{\, {\rm L_\odot}}
\newcommand\perccm{\, {\rm cm}^{-3}}
\newcommand\gperccm{\, {\rm g}\,{\rm cm}^{-3}}
\newcommand\Msunperpcsq{\,{\rm M_\odot}{\rm pc}^{-2}}
\newcommand\Msunpercpc{\,{\rm M_\odot}{\rm pc}^{-3}}
\newcommand\Hatom{{\rm H}}
\newcommand\Hmol{{{\rm H}_2}}
\newcommand\Zsun{\, {\rm Z_\odot}}

\newcommand\figem{\bf}                  
\newcommand\textem{\em}                 
\newcommand\texteem{\bf\em}           
\def\softwarenamestyle[#1]{\textsc{#1}}

\newcommand\ssc{\rm\scriptscriptstyle}  
\newcommand\sscBondi{{\ssc B}}
\newcommand\sscKep{{\ssc K}}
\newcommand\sscBH{{\ssc BH}}
\newcommand\sscAGN{{\ssc AGN}}
\newcommand\sscSGC{{\ssc SGC}}
\newcommand\sscLW{{\ssc LW}}

\defcitealias{moTwophaseModelGalaxy2024}{Paper-I}
\defcitealias{chenTwophaseModelGalaxy2024}{Paper-II}
\defcitealias{chenTwophaseModelGalaxy2025}{Paper-III}
\defcitealias{hopkinsAnalyticModelMagneticallyDominated2024}{H24}
\defcitealias{luEmpiricalModelStar2014}{L14}

\title[Seeding and growing SMBHs]{A two-phase model of galaxy formation: 
IV. Seeding and growing supermassive black holes in dark matter halos}

\author[Yangyao Chen et al.]{
Yangyao Chen\orcidlink{0000-0002-4597-5798},$^{1,2,3,4}$\thanks{E-mail: yangyaochen.astro@foxmail.com}
Houjun Mo\orcidlink{0000-0001-5356-2419}$^{5}$
and
Huiyuan Wang\orcidlink{0000-0002-4911-6990}$^{3,4}$
\\
$^{1}$School of Astronomy and Space Science, Nanjing University, Nanjing, Jiangsu 210093, China\\
$^{2}$Key Laboratory of Modern Astronomy and Astrophysics, Nanjing University, Ministry of Education, Nanjing, Jiangsu 210093, China\\ 
$^{3}$Department of Astronomy, University of Science and Technology of China, Hefei, Anhui 230026, China\\
$^{4}$School of Astronomy and Space Science, University of Science and Technology of China, Hefei, Anhui 230026, China\\
$^{5}$Department of Astronomy, University of Massachusetts, Amherst, MA 01003, USA
}

\date{Accepted XXX. Received YYY; in original form ZZZ}

\pubyear{\the\year{}}

\begin{document}
\label{firstpage}
\pagerange{\pageref{firstpage}--\pageref{lastpage}}

\maketitle


\begin{abstract}
We present a theoretical framework for seeding and growing supermassive black holes (SMBHs)
in dark matter halos along their assembly histories. Seeds are bred out of Pop-III 
stars formed during the first collapse of pristine gas in mini-halos that have reached 
the $\Hmol$-cooling limit, modulated by UV radiation from star formation and 
dynamical heating from fast halo assembly. Such breeding persists until the 
enrichment of the intergalactic medium (IGM) enables Pop-II stars to form.
Post-seeding growth of black holes (BHs) is driven by distinct channels, starting with episodic
super-Eddington accretion associated with nuclear bursts induced by 
global disturbances of galaxies, followed by sustained sub-Eddington accretion
via capturing sub-clouds formed in self-gravitating gas 
clouds (SGCs) in halos of fast assembly, and ending with merger-dominated, 
quiescent growth. We implement the model in subhalo merger trees to build a coherent
framework to follow SMBH-galaxy-halo co-evolution across the whole history
and structural hierarchy. BH seeds are bred with a broad mass 
spectrum of $M_\sscBH = 10$--$10^5\Msun$ at $z \approx 20$--$30$ in mini-halos with 
masses of $10^5$--$10^8\Msun$. Nuclear bursts provide the key condition for
seeds to grow into SMBHs.
The $M_\sscBH$-$M_*$ relation is a multi-piece, 
redshift-dependent function shaped by the interplay among different growth channels.
Our model predictions are broadly consistent with existing observations;
in particular, a population of BHs reminiscent of `little red dots' (LRDs) discovered 
by JWST naturally results from the seeding and growing processes. Potential future
tests of the model are discussed.
\end{abstract}





\section{Introduction}
\label{sec:intro}


The observable Universe exhibits a structure hierarchy composed of populations 
of objects across multiple scales
\citep[e.g.][]{moGalaxyFormationEvolution2010}. 
Galaxies, luminous stellar islands characterized by distinct luminosities and morphologies, 
constitute the building blocks of this hierarchy
\citep{peacockLargescaleSurveysCosmic2003,abazajianFirstDataRelease2003,
wrightGalaxyMassAssembly2016}.
On larger scales, galaxies are gravitationally clustered in groups and clusters of galaxies
\citep{abellCatalogRichClusters1989,lumsdenEdinburghDurhamSouthernGalaxy1992,
daltonAPMGalaxySurvey1997,yangConstrainingGalaxyFormation2003,
vandenboschLinkingEarlyLatetype2003,yangGalaxyGroupsSDSS2007,
tinkerSelfCalibratingHaloBasedGroup2021,wangIdentifyingGalaxyGroups2020},
which themselves are connected through complex structures such as filaments 
and sheets to form the cosmic web
\citep{bondHowFilamentsGalaxies1996,hahnPropertiesDarkMatter2007,
wangELUCIDEXPLORINGLOCAL2016,zhangUnexpectedClusteringPattern2025}.
Within individual galaxies, the microscopic architecture, including individuals and clusters of 
stars and black holes (BHs), exhibits diverse properties and interacts with 
the host galaxy through environmental
\citep{springelFormationSpiralGalaxy2005,dekelColdStreamsEarly2009,
pimbbletBacksplashGalaxiesIsolated2011,wangELUCIDIVGalaxy2018,
wiseFormationMassiveBlack2019,latifTurbulentColdFlows2022,
wangDissectTwohaloGalactic2023,
mengGalaxyPopulationsGroups2023,puskasConstrainingMajorMerger2025} 
and feedback processes
\citep{crotonManyLivesActive2006,bowerBreakingHierarchyGalaxy2006,
somervilleSemianalyticModelCoevolution2008,boothCosmologicalSimulationsGrowth2009,
vogelsbergerModelCosmologicalSimulations2013,
chenCircumgalacticMediumTraced2025,lyuFirstStatisticalDetection2026,yuStellarFeedbackDrives2026}.
According to modern cosmology, galaxies, halos and BHs do not exist statically as they 
are observed. Instead, they are all seeded from primordial density perturbations
in the early Universe, amplified and shaped by various physical processes over 
the long history of the Universe.
With the advent of powerful 
telescopes, such as the 
James Webb Space Telescope (JWST) and Atacama Large Millimeter Array (ALMA),
it is now possible to observe galaxies and BHs from their infancy all the way
to the present day
\citep{finkelsteinCEERSKeyPaper2023,
bezansonJWSTUNCOVERTreasury2024,
parlantiALMAHintsPresence2023,
rowlandREBELS25DiscoveryDynamically2024,
fujimotoPrimordialRotatingDisk2025,
suessMediumBandsMega2024,
naiduAllLittleThings2024,
wangJWSTMIRIReveals2025,
jiaPotentialdrivenMetalCycling2025,
scholtzTentativeRotationGalaxy2025,
songTransitionOutsideinInsideout2026}. 
This allows us to gather discrete evolutionary 
snapshots in observations to form a unified picture of structure formation.
Theoretically, great efforts have been made in understanding the origin and evolution 
of galaxies and BHs in the $\Lambda$ cold dark matter ($\Lambda$CDM) paradigm
\citep[e.g.][]{peeblesLargescaleStructureUniverse1980,
davisEvolutionLargescaleStructure1985,
osheaPopulationIIIStar2008,
schayeEAGLEProjectSimulating2015,
stacyBuildingPopulationIII2016,
wangNIHAOProjectReproducing2015,
pillepichFirstResultsIllustrisTNG2018,
daveSimbaCosmologicalSimulations2019,
wangElevationSuppressionStar2019,
latifRadiationHydrodynamicalSimulations2021,
wangOriginExponentialStarforming2022,
hopkinsFIRE3UpdatedStellar2023}. 
This paradigm provides not only a unified cosmological background for interpreting observational
data, but also initial conditions, matter/energy contents and physical principles 
to model the evolutionary processes and to predict the structure hierarchy 
over the entire history of the Universe. Substantial progress has been made in modeling 
structure formation within this paradigm using different approaches
\citep[e.g.][]{kauffmannFormationEvolutionGalaxies1993,peacockHaloOccupationNumbers2000,springelCosmologicalSmoothedParticle2003},
but many problems remain unresolved \citep[e.g.][]{naabTheoreticalChallengesGalaxy2017,wangTestingGalaxyFormation2025}. 
The main challenge comes from the many components
that make up the structure hierarchy, the wide range of scales involved in the processes that 
drive the evolution, and the complex interplay among these processes. 

A key step in tackling the challenge is, therefore, to build a coherent model that can cover 
the whole structure hierarchy and the full evolution history within 
$\Lambda$CDM, so that the connection and comparison between model and 
observation can be made across the whole structure hierarchy and over the 
entire history of the Universe.
In a recent paper \citep[][hereafter \citetalias{moTwophaseModelGalaxy2024}]{moTwophaseModelGalaxy2024},
we proposed a `two-phase' scenario for galaxy formation in the context of current cosmology. 
The key hypothesis of this two-phase scenario is that the growth of each central 
galaxy can be divided into two phases according to the growth rate of its host halo: a fast phase, in which 
the halo accretion is fast and there is a rapid change in the gravitational potential,   
and a slow phase, in which the halo accretion is slow and the gravitational potential 
remains steady \citep{zhaoGrowthStructureDark2003,luOriginColdDark2006,
zhaoAccurateUniversalModels2009,klypinMultiDarkSimulationsStory2016}. 
The distinct physical processes and the results expected from them  
in the two phases are discussed extensively and modeled in detail
in the papers of this series. The model addresses:
(i) a reversed dynamical evolution in which a galaxy first builds its bulge in the 
fast phase and then grows its disk in the slow phase, 
and efficient growth of the central supermassive black hole (SMBH) via capturing dense sub-clouds 
formed in the turbulent, self-gravitating gas cloud (SGC) 
associated with bulge formation \citepalias{moTwophaseModelGalaxy2024};
(ii) a homology relation between bulge size and halo size (\citealt{chenTwophaseModelGalaxy2024}, hereafter \citetalias{chenTwophaseModelGalaxy2024});
(iii) the two-channel formation of globular clusters (GCs; \citealt{chenTwophaseModelGalaxy2025}, hereafter \citetalias{chenTwophaseModelGalaxy2025}).

One important subject in understanding galaxy formation is to identify the 
co-evolution pattern of BHs and their host galaxies and halos
\citep[e.g.][]{kormendyCoevolutionNotSupermassive2013,bowerDarkNemesisGalaxy2017,habouzitSupermassiveBlackHoles2021,
zhuangEvolutionaryPathsActive2023,grahamAppreciatingMergersUnderstanding2023,jiangDissectingMassQuenching2025,liPhysicalProcessesCoevolution2025}. 
In observations, such evolutionary information is often not accessible 
for individual objects. Instead, properties of objects can be measured for 
representative samples within discrete intervals of redshift ($z$), 
and relations between the properties can be constructed to form a statistical 
description for galaxies at each snapshot of the Universe. 
Since the imprints of physical processes driving galaxy 
formation are encoded in these relations, albeit in a degenerate way, 
retrieving physical information from the relations has been a major
endeavor in linking observations to formation processes
\citep[e.g.][]{wangRelatingGalaxiesDifferent2023}.
One of the most important relations is the $M_{\sscBH}$-$M_*$ relation,
which has been measured observationally in the local Universe
\citep[e.g.][]{kormendyCoevolutionNotSupermassive2013,greeneIntermediateMassBlackHoles2020,
zhuangEvolutionaryPathsActive2023,grahamAppreciatingMergersUnderstanding2023}
and has been recently extended to high $z$ by observations with e.g. JWST and 
ALMA \citep{izumiSubaruHighzExploration2021,
ublerGANIFSMassiveBlack2023,harikaneJWSTNIRSpecFirst2023,
maiolinoJADESDiversePopulation2024,onouePoststarburstPathwayFormation2025,
chenHostGalaxyIf2025}. 
Theoretical efforts have been made to recover the observed relation, 
to predict its extension to higher $z$, and to link the relation 
to underlying processes driving the co-evolution
\citep[e.g.][]{bowerDarkNemesisGalaxy2017,
habouzitSupermassiveBlackHoles2021,liPhysicalProcessesCoevolution2025,
hongDynamicalHotnessStar2023,voitBlackHoleGrowth2024,
kidoBlackHoleEnvelopes2025,jiangFormationLittleRed2025,
dekelFeedbackfreeStarClusters2026,
bellovaryLittleRedDots2025}.

A critical condition for the build-up of the $M_{\sscBH}$-$M_{\rm *}$ relation
is effective regulation by the feedback from 
the accreting SMBH (e.g. \citealt{weinbergerSupermassiveBlackHoles2018}; 
\citealt{hongDynamicalHotnessStar2023}; \citealt{habouzitSupermassiveBlackHoles2021};
\citetalias{moTwophaseModelGalaxy2024}; 
\citealt{liPhysicalProcessesCoevolution2025}). However, this can happen 
only when the SMBH has grown sufficiently massive so that the radiative 
efficiency of the associated AGN, which is $\propto M_{\sscBH}$, 
is sufficient to affect gas cooling and collapse. One way to achieve this 
in modeling is to grow BHs from massive seeds, usually with $M_{\sscBH} \gtrsim 10^5\Msun$, 
which are on or above the observed $M_{\sscBH}$-$M_{\rm *}$ relation, 
and to rely on regulations by the stellar and AGN feedback to drive them 
down onto, and to keep them moving along, the relation. This approach 
is commonly adopted in cosmological hydrodynamic simulations
\citep[e.g.][]{schayeEAGLEProjectSimulating2015,bowerDarkNemesisGalaxy2017,
weinbergerSupermassiveBlackHoles2018,daveSimbaCosmologicalSimulations2019,
habouzitSupermassiveBlackHoles2021,
liPhysicalProcessesCoevolution2025}.
However, it is unclear whether and how such seeds can be bred. 
Alternatively, one can start with seeds that are much less massive, and 
boost their growth via various mechanisms, such as mergers with star 
clusters brought in by dynamical friction, as suggested by 
\citet{dekelGrowthMassiveBlack2025}, or the capture of the 
low angular-momentum sub-clouds that are regenerated continuously by injecting  
disturbance into the turbulent gas, as suggested by
\citetalias{moTwophaseModelGalaxy2024} (see their \S4.3.3 and fig.~7;
see also \citealt{shiHyperEddingtonBlackHole2023}). 
Thus, a realistic model for the formation of BH seeds is required to break this 
seeding-growing degeneracy.

Considerable effort has been devoted to modeling the conditions for seed formation 
and predicting the properties of the resulting BH seeds.
Environmental factors, such as 
enrichment in intergalactic medium \citep[IGM][]{brommFormationFirstSupermassive2003,valianteOriginDustHighredshift2011,costaMassiveBinaryBlack2023,spinosoMultiflavourSMBHSeeding2023}, 
Lyman-Werner (LW) radiation 
\citep{agarwalFirstBillionYears2014,latifUVRegulatedStar2019,wiseFormationMassiveBlack2019,agarwalOptimalNeighbourhoodNurture2019,incatasciatoModellingCosmologicalLymanWerner2023}, 
streaming velocity between baryonic and dark matter
\citep{greifDelayPopulationIII2011,maioImpactPrimordialSupersonic2011,
stacyEffectStreamingMotion2011}, 
dynamical heating or turbulence support acquired through fast accretion
\citep{wiseFormationMassiveBlack2019,latifTurbulentColdFlows2022,wuImpactSubhaloDynamical2026},
and heating from dark matter self-annihilation \citep{natarajanDarkMatterAnnihilation2009,freeseSupermassiveDarkStars2010}, 
are proposed to modify the conditions of seeding.
Depending on the types and properties of the progenitor objects, 
the resulting seeds are proposed to have diverse `flavors', such as
descendants of ordinary Pop-III stars \citep{
schaererPropertiesMassivePopulation2002,
greifTwoPopulationsMetalfree2006,costaMassiveBinaryBlack2023} or 
more massive Pop-III.1 stars \citep{banikFormationSupermassiveBlack2019,singhFormationSupermassiveBlack2023,tanOriginSupermassiveBlack2024,cammelliFormationSupermassiveBlack2025,sanatiEmergenceIonizingFeedback2025,petkovaFormationSupermassiveBlack2026},
direct collapse black holes \citep[DCBHs; e.g.][]{omukaiCanSupermassiveBlack2008,choiSupermassiveBlackHole2013,hosokawaFormationPrimordialSupermassive2013,
chonRadiationHydrodynamicsSimulations2018,mayerDirectFormationMassive2024},
or merger remnants in dense star clusters \citep{antoniniBlackHoleGrowth2019,bocoGrowthMassiveBlack2021,fujiiSimulationsPredictIntermediatemass2024,dekelGrowthMassiveBlack2025}.
As demonstrated by semi-analytic models \citep{ahnInhomogeneousBackgroundH2Dissociating2009,trincaSeekingGrowthFirst2023,spinosoMultiflavourSMBHSeeding2023,pianaSuperEddingtonAccretionHighredshift2024,
venturaSemianalyticModellingPop2024,cammelliFormationSupermassiveBlack2025},
the conditions for seeding and the properties of the resulting seeds can be 
derived self-consistently within the cosmological context,
so that realistic initial conditions for the post-seeding growth
of the BHs can be obtained.

In this paper, we extend the model presented previously in the papers of 
this series toward high $z$, and we implement the extended model 
semi-analytically in halo merger trees to make predictions. 
The main extensions are:
(i) a realistic model of BH seeding that predicts when, where and how BH seeds 
are bred in individual halos, as well as the properties of these seeds;
(ii) a new channel of BH growth that boosts the low-mass seeds to SMBHs.
With the two extensions, the growth history of each BH population at low $z$ 
can be fully and self-consistently recovered up to the redshifts at which 
the seeds are bred. By examining the seeds and growth trajectories that lead 
different populations of BHs onto the observed $M_\sscBH$-$M_*$ relation,
we show that the aforementioned seeding-growing degeneracy can be 
disentangled. More importantly, the properties of both populations and individual 
halos, galaxies, star clusters and BHs, and the relations among
these properties, can be obtained by the model up to redshifts that are far 
beyond the current observational limit ($z \sim 30$), providing testable 
predictions for future observations.

The paper is organized as follows.
In \S\ref{sec:modeling}, we start with a review of the cosmological context 
relevant to the modeling, and then introduce the two components of the model, 
the seeding procedure and the growing procedure, with key extension 
in each procedure emphasized.
The main text focuses on the motivation, the key concepts and the outline
of the model, and refers to the appendices for the detailed prescriptions.
Analytical estimates are provided as much as possible in the appendices 
so that intuition and order-of-magnitude quantification can be obtained 
without diving into the details.
In \S\ref{sec:results}, we synthesize the physical prescriptions 
semi-analytically by implementing them into halo merger trees
in a cosmological volume, present the results predicted by the model, 
and make comparisons with observations. In \S\ref{sec:summary}, 
we summarize our main results, and discuss their implications and potential 
follow-up investigations prompted by them.
Throughout this paper, we adopt a flat $\Lambda$CDM cosmology given by
\citet{planckcollaborationPlanck2015Results2016}: 
$h=0.6774$, $\Omega_{\rm m,0}=0.3089$, $\Omega_{\Lambda,0}=0.6911$, 
$\Omega_{\rm b,0}=0.0486$, and a Gaussian initial power
spectrum specified by $n_{\rm s}=0.9667$ and $\sigma_8=0.8159$.
We use `BH' to refer to the central BH of a host galaxy, 
unless specified otherwise.

\section{Modeling black-hole seeds and their growth in dark matter halos}
\label{sec:modeling}

\begin{figure} \centering
    \includegraphics[width=0.99\columnwidth]{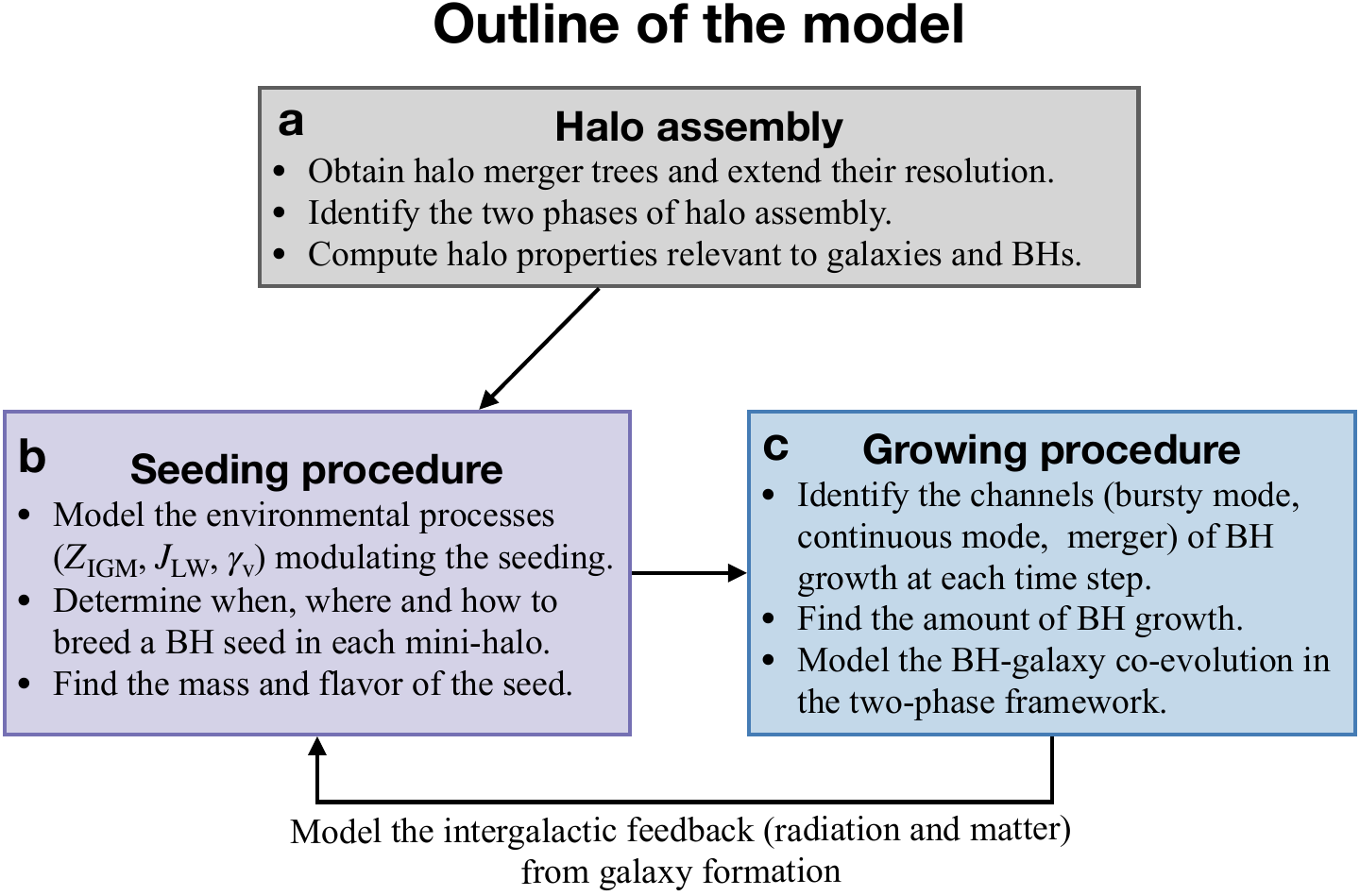}
    \caption{
    {\figem A brief outline of the model.} The model takes halo merger trees 
    in a cosmological volume as input, uses the `seeding procedure' to set the 
    initial condition for every mini-halo, 
    and follows a `growing procedure' to predict the 
    post-seeding evolution over cosmic time. 
    See \S\ref{sec:modeling} for an introduction of the model.
    A more detailed version of this outline is available in 
    Fig.~\ref{fig:flowchart}, and the details
    of numerical implementation are provided in 
    Appendix~\ref{app:sec:impl-details}.
    }
    \label{fig:flowchart-brief}
\end{figure}

Fig.~\ref{fig:flowchart-brief} outlines the key components of our model and the 
pipeline that links them together.
Briefly, the model starts with the merger trees of dark matter halos (\S\ref{ssec:halo-assembly})
and processes of galaxy formation expected in such trees (\S\ref{ssec:two-phase-galaxy}). 
A seeding procedure is then applied to the merger trees to determine when, where and how BH 
seeds are bred in mini-halos (\S\ref{ssec:seeding-main}). The breeding of the seed marks the beginning 
of the formation of a galaxy in the halo and initiates the growing procedure that follows the subsequent 
evolution of the galaxy and the BH (\S\ref{ssec:growing-main}). The output of the model thus 
covers the full lifetimes of BHs and their host galaxies in a cosmological volume. 

\begin{figure*} \centering
    \includegraphics[width=0.85\textwidth]{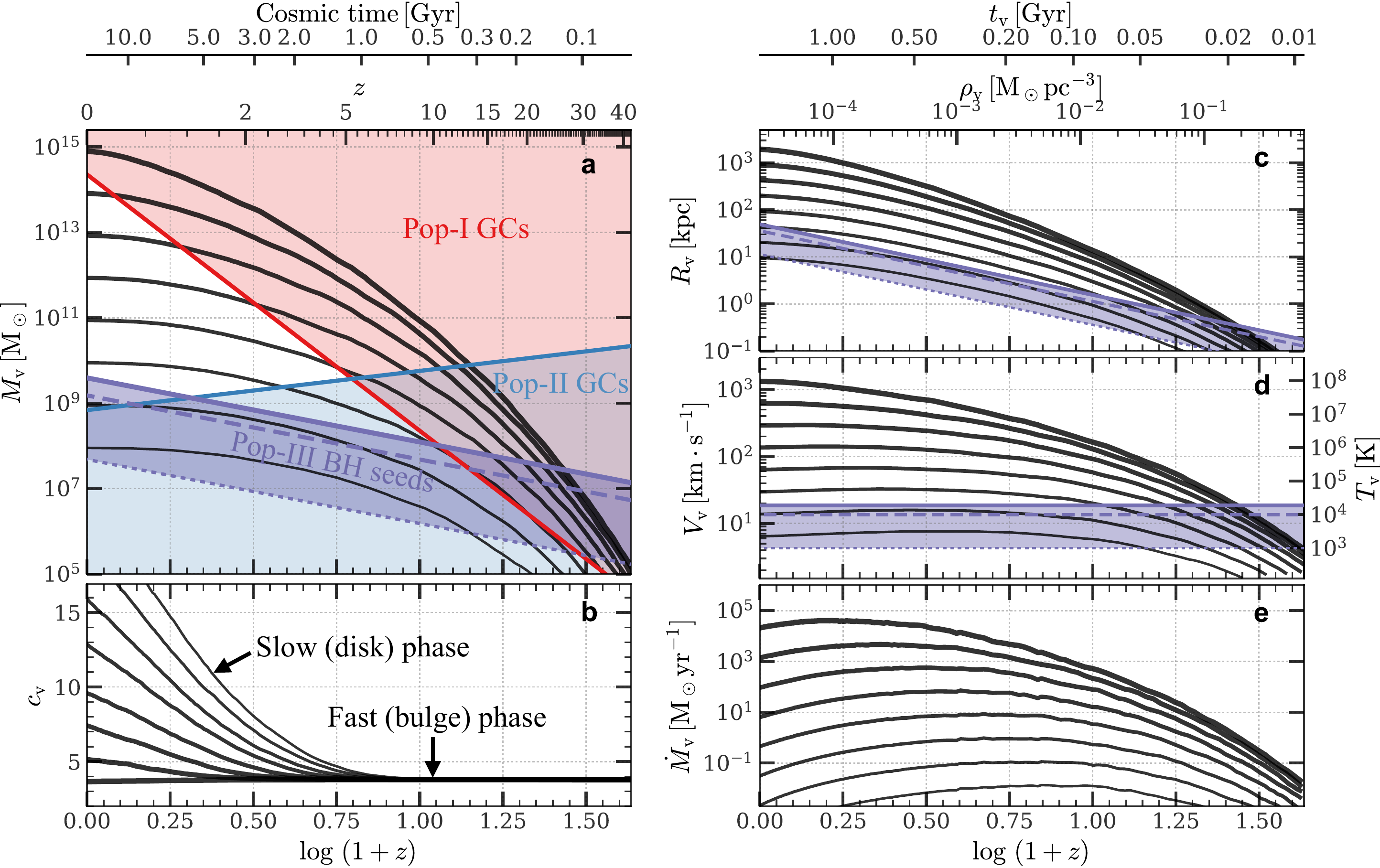}
    \caption{
        {\figem Assembly histories of dark matter halos.}
        Here we show the evolution of different virial properties:
        {\figem a}, mass ($M_{\rm v}$);
        {\figem b}, concentration ($c_{\rm v}$);
        {\figem c}, radius ($R_{\rm v}$, in physical scale);
        {\figem d}, velocity and temperature ($V_{\rm v}$ and $T_{\rm v}$);
        {\figem e}, accretion rate ($\dot{M}_{\rm v}$).
        The assembly histories are computed along the main branches of halos at $z=0$
        sampled by a Monte Carlo algorithm, 
        and a median is taken over the halos with each given
        $M_{{\rm v}, z=0}$ from $10^8\Msun$ to $10^{15}\Msun$
        ({\figem black curves} from thin to thick).
        {\figem Purple bands} in {\figem a}, {\figem c} and {\figem d}
        indicate the regime where the first collapse of an 
        SGC is expected to be triggered so that Pop-III 
        stars can form and a BH seed can be bred, with the associated
        purple {\figem dotted}, {\figem dashed} and {\figem solid} lines 
        representing 
        $M_{\rm v,\Hmol}$ (the mass for $\Hmol$-cooling to be effective; Eq.~\ref{eq:m-v-cool-H2-main}), 
        $M_{\rm v,\Hatom}$ (the mass for $\Hatom$-cooling to be effective; Eq.~\ref{eq:m-v-cool-H-main}),
        and $f_{\rm b}^{-1} M_{\rm J,max}$ 
        (the mass for the SGC with strong turbulence support
        to be Jeans unstable; Eq.~\ref{eq:m-jeans-sgc-max-main}), respectively.
        The exact location for a halo to be seeded within this band
        depends on its assembly history and radiation environment.
        For reference, {\figem red} and {\figem blue} regions in {\figem a} 
        show the regimes 
        where Pop-I and Pop-II GCs can actively form, respectively, according
        to \citetalias{chenTwophaseModelGalaxy2025},
        highlighting the role of IGM enrichment in modulating the 
        stellar population and seed-formation pathway.
        This figure summarizes the properties of halos over wide ranges of 
        mass and redshift (see \S\ref{ssec:halo-assembly}), 
        which will be used as input to our model
        to determine when, where and how the first generation of stars and 
        BH seeds form (see \S\ref{ssec:seeding-main}), as well as their 
        post-seeding growth (see \S\ref{ssec:growing-main}).
    }
    \label{fig:halo-mahs}
\end{figure*}



\subsection{Assembly of dark matter halos} 
\label{ssec:halo-assembly}

In the current $\Lambda$CDM paradigm of galaxy formation, galaxies form and evolve
in dark matter halos. The early assembly of halos at high $z$ is thus crucial for 
determining when, where and how the formation of the first generation of stars and 
the seeds of BHs occurs, as well as the subsequent growth of 
these seeds to SMBHs. To model the full lifetimes of BHs, the assembly of halos 
has to be tracked up to very high $z$, at which the halos are in their infancy (hereafter `mini-halos').
We thus develop a method to extend the assembly history of each
simulated halo towards high $z$ by joining an assembly history sampled from a 
Monte-Carlo algorithm to the simulated history at the mass right above the
resolution of the simulation, with the constraint that the joined history 
matches the value and first-order derivative of the simulated history
at the joining point (see Appendix~\ref{app:ssec:extension-merger-tree}
for details). Appendix~\ref{app:sec:halo-assembly} presents a detailed review 
of the halo population and methods used to obtain their assembly histories. The results most 
relevant to our modeling are summarized below.

Throughout this paper, a halo is defined by a spherical volume of virial 
radius $R_{\rm v}$, within which the mean mass density is $\Delta_{\rm v} = 200$ 
times the critical density of the Universe (\citealt{navarroUniversalDensityProfile1997};
see e.g. \citealt{hanManyBoundariesStratified2026} for alternative definitions). 
The assembly history of a halo can then be described by the evolution 
of various virial properties, such as virial mass ($M_{\rm v}$), virial density ($\rho_{\rm v}$), virial velocity
($V_{\rm v}$), virial temperature ($T_{\rm v}$), 
dynamical timescale ($t_{\rm v} \equiv R_{\rm v}/V_{\rm v}$),
and accretion rate ($\dot{M}_{\rm v}$). Halo density profiles are modeled by the 
NFW form \citep{navarroUniversalDensityProfile1997} specified by a scale radius
$R_{\rm s}$ and a concentration parameter $c_{\rm v} = R_{\rm v}/R_{\rm s}$ that
is related to the assembly history \citep{zhaoAccurateUniversalModels2009}.

Fig.~\ref{fig:halo-mahs} shows the median assembly histories of halos 
of given masses at $z=0$, obtained from large ensembles of halo merger trees 
sampled by a Monte Carlo algorithm (see Appendix~\ref{app:ssec:extension-merger-tree}). 
Typically, more massive halos assemble at higher rates, have higher 
$M_{\rm v}$, $R_{\rm v}$, $V_{\rm v}$ and $T_{\rm v}$ and 
smaller $c_{\rm v}$ across the cosmic history. However, halos with 
$M_{{\rm v},z=0} > 10^{12}\Msun$ have similar progenitors at $z \gtrsim 15$, with mass 
$M_{\rm v} \sim 10^8\Msun$, $R_{\rm v}\sim 1 \Kpc$, 
$V_{\rm v}\sim 20 \kms$ and $T_{\rm v}\sim 10^{4.5}\Kelvin$. 
These `mini-halos', with a typical density $\rho_{\rm v}\sim 0.1 \Msun {\rm pc}^{-3}$, 
are places where BH seeds are bred, as we will see below.

A prominent property of CDM halos is their `two-phase' assembly
(\citealt{zhaoGrowthStructureDark2003}; see also \citealt{luOriginColdDark2006,zhaoAccurateUniversalModels2009,klypinMultiDarkSimulationsStory2016}):
an earlier, fast-accretion phase (hereafter {\texteem fast phase}) where the gravitational potential of a halo 
rapidly deepens with a roughly constant concentration parameter $c_{\rm v} = c_{\rm f} \approx 4$, 
followed by a slow-accretion regime (hereafter {\texteem slow phase}) where the halo grows  
in the outskirts while the inner radius $R_{\rm f} \approx c_{\rm f} R_{\rm s}$ established in the 
{\texteem fast phase} changes little so that $c_{\rm v}$ increases with time. 
These two phases are evident in Fig.~\ref{fig:halo-mahs}, where the growth of $M_{\rm v}$ 
slows over time and $V_{\rm v}$ even decreases over time in later epochs.
The epoch separating the two phases is visible in panel b 
as the deviation of $c_{\rm v}$ from the constant value of $c_{\rm f}$. 
In addition to the median assembly history, variations in individual halos can cause temporary 
excursions between the fast and slow phases. Our model includes such excursions by using 
simulated assembly histories, as described in the following.


To obtain a realistic sample of halos, we use halos from a cosmological N-body 
simulation, TNG100-1-Dark \citep{pillepichFirstResultsIllustrisTNG2018, 
nelsonIllustrisTNGSimulationsPublic2019}. 
This simulation uses $1820^3$ dark matter particles, each with 
a mass of $6.0\times 10^6 \msun$, to sample the density field
in a periodic box with a side length of $L_{\rm box} = 75\mpc$.
Halos, subhalos and their merger trees were obtained from the simulation
using the friends-of-friends (\softwarenamestyle[FoF]) algorithm
with a scaled linking length of $0.2$ \citep{davisEvolutionLargescaleStructure1985},
the \softwarenamestyle[Subfind] algorithm \citep{springelPopulatingClusterGalaxies2001}, 
and the \softwarenamestyle[SubLink] algorithm \citep{springelCosmologicalSimulationCode2005}, 
respectively. 
The lower limit for the FoF halo mass is about $2\times 10^8 \msun$,
insufficient to resolve the mini-halos needed to breed the seeds.
To overcome this, we apply a Monte Carlo algorithm to the simulated 
merger trees to extend them down to the mini-halo regime
(see Appendix~\ref{app:ssec:extension-merger-tree}).
The merger trees in our sample are those rooted in subhalos with 
host halo mass $M_{\rm v} \geqslant 10^{9.5}\Msun$, evaluated
at $z=0$ for a central subhalo and at the infall time for a satellite subhalo.
Subhalos above this mass limit are expected to grow above the hydrogen atomic-cooling limit 
at some epochs, and the choice of this limit thus ensures that 
galaxies with substantial star formation after the cosmic 
reionization are included in our modeling \citep{luEmpiricalModelStar2014}. 
Halos that do not grow above this limit 
may also have some star formation when, for example, the IGM 
has a sufficiently high neutral fraction to shield UV radiation 
\citep[e.g.][]{dijkstraPhotoionizationFeedbackLowMass2004,bensonEpochReionization2006,
maConstraintsGalaxyFormation2025}, which may produce extremely low-mass 
galaxies at low $z$ that are missed by our model. Some of the halo mass limits 
relevant for the formation of early galaxies and BH seeds are 
plotted in Fig.~\ref{fig:halo-mahs} and will be discussed later.

\subsection{The two-phase formation of galaxies} 
\label{ssec:two-phase-galaxy}

As a halo forms, gas within it is heated by virialization processes
and must first cool down before collapsing further to fragment and form stars. 
The competition between heating and cooling processes thus determines galaxy formation 
in the halo. A critical consequence of the two-phase assembly of halos, as proposed in 
\citetalias{moTwophaseModelGalaxy2024} and examined in 
\citet{maTwophaseFormationGalaxies2026}, is the two-phase
formation of galaxies. During the fast phase, gas cools effectively   
before the halo reaches $M_{\rm v} \approx M_{\rm cool} \sim 10^{12} \Msun$, 
collapses rapidly into the halo center, and forms
a self-gravitating gas cloud (SGC) before rotation-support becomes important.   
Density perturbations are seeded and amplified by turbulence in the SGC, 
causing the SGC to fragment into sub-clouds that are dense enough to be free 
of drag force from ambient gas and cloud-cloud collision. The mixing of angular momentum among the sub-clouds 
is inefficient, so that the SGC sustains a dynamically-hot 
structure with a wide distribution of specific angular momentum ($j$).
Stars formed in the sub-clouds inherit the hot dynamics, 
forming a dynamically hot, bulge-like stellar system
with a homology relation between the sizes of the bulge and halo, 
$r_{\rm bulge} \approx (1/100)R_{\rm f}$, that is independent of halo mass 
and redshift \citep{chenTwophaseModelGalaxy2024}.
Stable cold gaseous/stellar disks form in the slow phase as 
the formation of sub-clouds stalls due to the stabilized
inner halo potential and the reduced gas fraction 
by feedback. To highlight the distinct galactic dynamics during the two phases,
we refer to the two phases interchangeably as the `{\texteem bulge phase}' and 
the `{\texteem disk phase}', respectively. 

The formation of sub-clouds in fast-accreting halos provides the 
condition for forming and growing different types of compact sub-structures.
As proposed in \citetalias{chenTwophaseModelGalaxy2025}, the density of 
sub-clouds can be elevated by two distinct channels,
the compression by supersonic turbulence and the slow cooling due to low metallicity,
before they fragment to form stars.
Some sub-clouds can reach a `{\texteem supernova-free}' density, $n_{\rm snf}
\approx 10^{3.5}\perccm$, 
at which the free-fall timescale ($t_{\rm ff}$) of the gas falls below 
$t_{\rm snf} \approx 1\Myr$, so that the gas is converted into stars before the onset 
of supernova feedback. 
The two channels of density elevation thus lead to the formation 
of two populations of globular clusters (GCs), Pop-I and Pop-II, 
that preferentially occupy halos in different evolutionary stages.
Fig.~\ref{fig:halo-mahs}a shows the `atlas' of GC formation, where
the red and blue regions indicate the regimes dominated by 
the two channels, respectively. In the extreme case where the halo is pristine, 
Pop-III stars are expected to form and seed a BH in the halo 
(see \S\ref{ssec:seeding-main} for details). Some of the sub-clouds 
formed in the halo may sink to the halo center to feed the central BH, 
providing a channel for BH growth (see \S\ref{ssec:growing-main} 
for details). 

\subsection{Producing black-hole seeds in mini-halos}
\label{ssec:seeding-main}


\begin{figure} \centering
    \includegraphics[width=0.99\columnwidth]{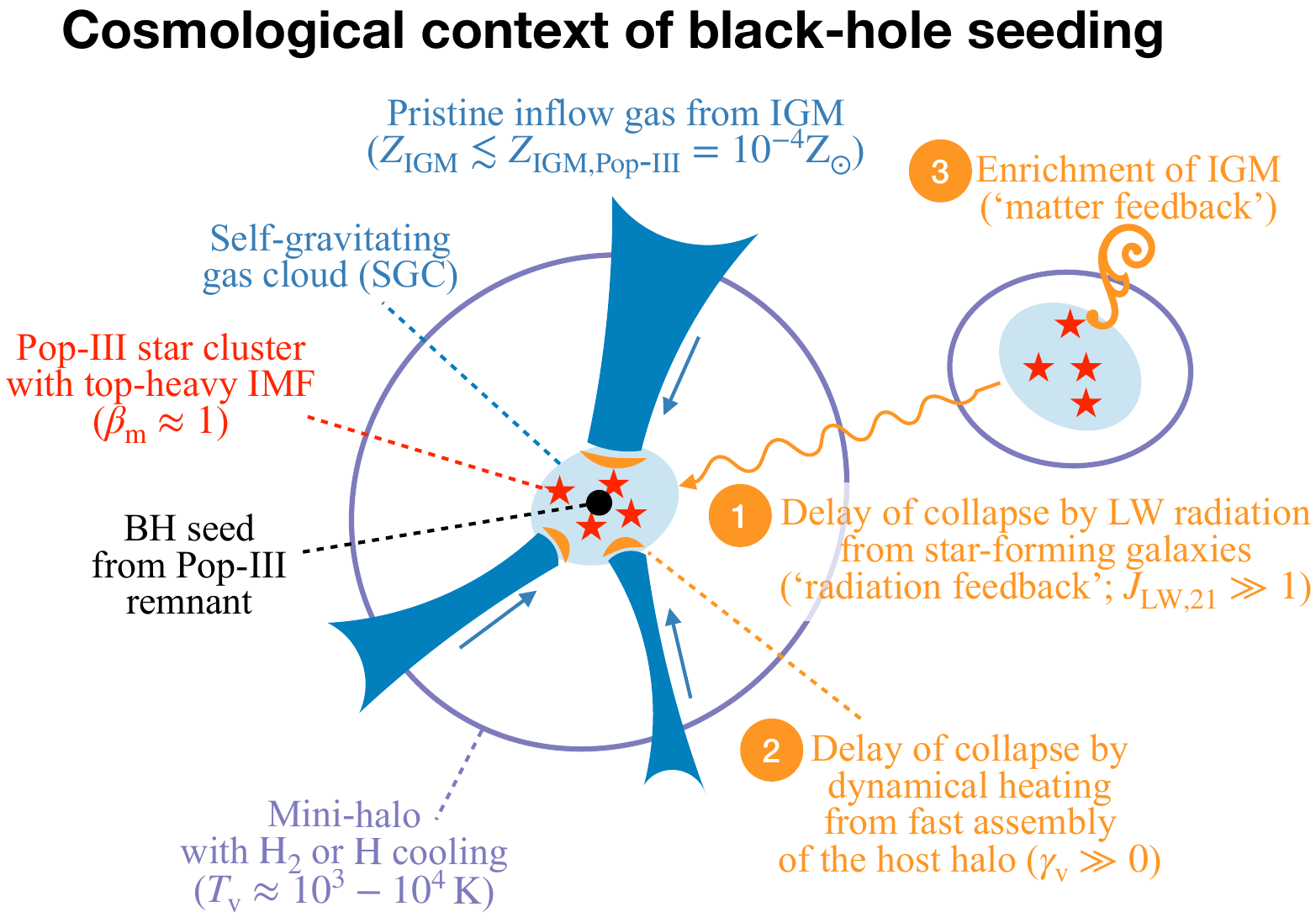}
    \caption{
        {\figem Schematic diagram summarizing the cosmological context of
        BH seeding.} Pristine gas in a mini-halo that has reached the 
        $\Hmol$-cooling limit ($T_{\rm v}\approx 10^3\Kelvin$; 
        Eq.~\ref{eq:m-v-cool-H2-main})
        can start to collapse to form an SGC at the halo center,
        contract further to form a Jeans core, and fragment to form a 
        Pop-III star cluster.
        A light-weight BH seed can be bred when the most massive star 
        (dominant star) in the cluster ends its life by CCSN. Three environmental 
        effects can modulate the seeding process
        (orange texts labeled 1--3): 
        UV radiation and dynamical heating can delay 
        the collapse and raise the 
        mass of the dominant star so that PISN or DCBH can appear and leave
        a more massive seed; enrichment in IGM can promote the cooling,
        cause Pop-II stars to form and leave a less massive seed.
        A radiation-regulated scenario is proposed here to model the IMF
        of Pop-III stars, and the mass of BH seed is determined accordingly 
        by following the stellar evolution of these stars. 
        Our scenario suggests that the environmental effects 
        -- concepts that have been built for galaxy formation at low $z$,
        have already been ubiquitous in the early Universe at $z \gtrsim 20$.
        See \S\ref{ssec:seeding-main} for details.
    }
    \label{fig:cosm-context-seeding}
\end{figure}

To seed a BH in a mini-halo, we identify key processes of cooling and heating
so that conditions for the halo gas to collapse and fragment can be determined.
Once these conditions are met, we consider the feedback from proto-stars
to determine the properties of the first generation stars that form in 
the halo. We then follow the evolution tracks of these stars to predict their 
fates and the properties of the BHs left as their remnants. The output of the 
seeding procedure is the flavor (representing the formation pathway) and mass of the BH seed.
Fig.~\ref{fig:cosm-context-seeding} sketches the key factors in the 
cosmological context to be considered in the seeding procedure.
A detailed description of the seeding procedure is presented in Appendix~\ref{app:sec:seeding}.
The following is a brief summary of the key points.

At high $z$ where the intergalactic environment is pristine, the first important 
coolant during the growth of a mini-halo is molecular hydrogen ($\Hmol$), which 
can keep the gas temperature at a level of $T_\Hmol= 10^3\Kelvin$
\citep{latifUVRegulatedStar2019}. The first collapse of halo gas is thus 
expected when the halo grows to $T_{\rm v} \approx T_\Hmol$, corresponding 
to a halo mass of
\begin{equation} 
    M_{\rm v, \Hmol} \approx \left[ 3.87\times 10^5 \Msun \right] 
    (1+z)_{25}^{-3/2} T_{\Hmol,3}^{3/2} 
    \,, \label{eq:m-v-cool-H2-main}
\end{equation}
where  $(1+z)_{25} \equiv (1+z)/25$ and
$T_{\Hmol,3} \equiv T_\Hmol/(10^3\Kelvin)$. 
At $1+z=25$, the collapse proceeds on a timescale of $t_{\rm v} \approx 20.7\Myr$, 
leading to the formation of a compact SGC with a 
mass $M_\sscSGC \approx 6.08\times 10^4\Msun$,
a radius $R_\sscSGC = 3.63\pc$, 
a density $n_\sscSGC = 1.03\times 10^4\perccm$, 
and a free-fall timescale of $t_{\rm ff, \sscSGC} = 0.463 \Myr$
(see Appendix~\ref{app:sssec:gas-collapse-in-mini-halos}).
The thermal sound speed at $T_{\Hmol}$ is $c_{\rm s} \approx 3.4 \kms$,
implying a Jeans mass of
\begin{equation} 
    M_{\rm J,\Hmol} = \left[ 2.30 \times 10^4 \Msun \right]
        (1+z)_{25}^{-3/2} T_{\Hmol,3}^{3/2}
    \,. \label{eq:m-jeans-sgc-H2-main}
\end{equation}
This Jeans mass is comparable to the SGC mass, indicating that further contraction of the SGC 
proceeds globally without fragmentation until other cooling channels are opened
at a higher density. 

With a primordial composition, the contraction leads to 
a core in the SGC with a density of $n = n_{\rm core} \approx 10^6 \perccm$, 
at which three-body reactions rapidly cool the gas, 
increasing the production of $\Hmol$ and making the core unstable 
\citep[see \S3.3 of][]{latifUVRegulatedStar2019}. 
The Jeans mass in the core is $M_{\rm J,core} \approx 1.88 \times 10^3 \Msun \ll M_{\rm v,\Hmol}$,
corresponding to a size of $R_{\rm core} \approx 0.25 \pc$ for the density in question.
Within such a core, conditions for star formation are satisfied, 
and fragmentation starts from the center and leads to the formation of a central 
(proto-)star followed by the formation of other stars.
Subsequent rapid inflow and frequent mergers are expected to increase the mass 
of the central star, producing a top-heavy IMF for the resulting Pop-III star cluster
whose mass is dominated by the central star
\citep[see e.g.][]{stacyBuildingPopulationIII2016,latifBirthMassFunction2022}. 
The central star dies on a ${\rm Myr}$ timescale in a 
core-collapse supernova (CCSN) explosion and leaves behind a `light-weight'
BH seed with a mass of a few $10\Msun$
\citep[see e.g. \S3.1 of][]{costaMassiveBinaryBlack2023}. 

We consider three environmental effects that can modulate the collapse and fragmentation of the SGC described above, promote or delay the seeding in a mini-halo, and modify the mass of the resulting seed:
\begin{enumerate}[topsep=0pt,parsep=0pt,itemsep=0pt]
    \item 
    {\texteem UV radiation} from galaxies surrounding the mini-halo;
    \item 
    {\texteem dynamical heating} associated with fast assembly of the mini-halo; 
    \item
    {\texteem metal enrichment} in the inter-galactic environment.
\end{enumerate}
The first two can delay the collapse of the halo gas, leading to a more massive 
SGC and thus to a more massive seed.
The last one leads to earlier fragmentation of the SGC, prevents the formation of 
Pop-III stars, and produces a less massive seed. These effects are summarized as 
orange texts in Fig.~\ref{fig:cosm-context-seeding}. 
In the following, we briefly describe our modeling of these effects and 
their consequences for seeding, referring details to 
Appendix~\ref{app:ssec:seed-in-minihalo}.

UV radiation in the Lyman-Werner band $11.2$--$13.6\,{\rm eV}$ (hereafter LW radiation)
has been proposed (e.g. \S2.1 of \citealt{haimanRadiativeFeedbackFirst2000}; \citealt{incatasciatoModellingCosmologicalLymanWerner2023})
as a mechanism to dissociate $\Hmol$, suppress gas cooling, and delay the first collapse
of gas in mini-halos. Radiation in this band, mainly produced
by young stars in nearby galaxies around a mini-halo, can escape the attenuation from 
neutral hydrogen in the IGM and propagate over long distances to
provide external feedback to the mini-halo.
To model the LW radiation at any spacetime location, we consider both the 
radiation background and the contribution from individual nearby galaxies, 
with spherical approximation for the radiation from individuals 
and an analytical fitting function for the IGM attenuation 
adopted to accelerate the computation (see Appendix~\ref{app:ssec:LW-radiation}
for details).
The intensity of LW radiation irradiating a mini-halo,
$J_{\sscLW}$, is defined as the energy in the LW band received at a location 
per unit time per unit area per unit frequency and per unit solid angle, and is 
conveniently expressed as $J_{\sscLW} = J_{\rm \sscLW,21}\,10^{-21} {\rm erg\,s^{-1}cm^{-2}Hz^{-1}sr^{-1}}$.
The threshold mass for gas collapse in a $\Hmol$-cooling mini-halo 
is expected to increase with increasing $J_{\sscLW}$.
Under a strong radiation of $J_{\rm \sscLW,21} \gg 1$, the collapse of the 
halo gas can be delayed until atomic cooling starts to operate, i.e., when 
$T_{\rm v} \approx T_{\Hatom} \approx 10^4\Kelvin$. This corresponds 
to a halo mass of
\begin{equation}
    M_{\rm v, \Hatom} = \left[ 1.22\times 10^7 \Msun \right] 
    (1+z)_{25}^{-3/2} T_{\Hatom, 4}^{3/2}
    \,,\label{eq:m-v-cool-H-main}
\end{equation}
where $T_{\Hatom,4}\equiv T_{\Hatom}/(10^4\Kelvin)$.
In other cases with moderate $J_{\sscLW}$, the collapse is expected to be 
delayed until $M_{\rm v}$ reaches a value between $M_{\rm v, \Hmol}$ and $M_{\rm v, \Hatom}$, 
depending on the intensity of the radiation.
We therefore parameterize the collapse threshold, $M_{\rm v, \sscLW}$, 
as a function of $J_{\sscLW}$, so that it smoothly transits from 
$M_{\rm v, \Hmol}$ (at $J_{\sscLW}=0$) to $M_{\rm v, \Hatom}$ 
(at $J_{\sscLW} \gg 1$), and calibrate the parameterization with 
results from zoom-in hydrodynamic simulations (see Appendix~\ref{app:sssec:delay-by-uv-radiation} for details).

The SGC in a fast-accreting halo is expected to be a turbulence-supported system
(see \citetalias{moTwophaseModelGalaxy2024}). 
The situation is the same here because mini-halos hosting seeds are almost certainly in 
the fast-accreting regime, and because the lack of coolant in the primordial 
gas makes the turbulence hard to dissipate 
\citep[e.g.][]{wiseFormationMassiveBlack2019,wuImpactSubhaloDynamical2026}. 
We refer to the energy injection associated with fast accretion as 
`dynamical heating', regardless of the details of how the energy 
subsequently cascades in the SGC.
To model the effect of dynamical heating in delaying the collapse of the SGC, 
we first consider an extreme case where the heating is so efficient that 
radiative cooling is inefficient to dissipate it.
Taking a turbulence velocity of $v_{\rm t} = 20 \kms$ reported by
\citet{latifTurbulentColdFlows2022}, the SGC has a Jeans mass of 
\begin{equation} 
    M_{\rm J,max} = \left[4.94\times 10^6 \Msun\right] 
    (1+z)_{25}^{-3/2}
    \,, \label{eq:m-jeans-sgc-max-main}
\end{equation}
above which it can collapse globally. The corresponding host halo mass, 
$M_{{\rm v},\gamma} = M_{\rm J,max}/f_{\rm b} = 3.14\times 10^7\Msun$, 
is slightly larger than $M_{\rm v,\Hatom}$ given by Eq.~\eqref{eq:m-v-cool-H-main}, indicating 
that dynamical heating can delay the formation of the first generation of compact 
objects by a timescale similar to that implied by a strong UV radiation field.
In other cases of weaker dynamical heating, we parameterize the threshold 
mass, $M_{\rm v, \gamma}$, for the SGC to collapse as a function of the 
halo specific assembly rate, 
$\gamma_{\rm v} \equiv \dot{V}_{\rm v} / \left[V_{\rm v}H(z)\right]$ 
(Eq.~9 in \citetalias{moTwophaseModelGalaxy2024}),
so that it changes smoothly from $f_{\rm b}^{-1} M_{\rm J, \Hmol}$ 
(at $\gamma_{\rm v} = 0$) to $f_{\rm b}^{-1} M_{\rm J,max}$ 
(at $\gamma_{\rm v} \gg$ some critical value, $\gamma_{\rm crit}$).
See Appendix~\ref{app:sssec:delay-by-fast-accretion} for details.

In the presence of both UV radiation and dynamical heating, the collapse 
threshold is computed as the higher of the two:
\begin{equation}
    M_{\rm v, th} = \max\left[M_{\rm v, \sscLW}, M_{\rm v, \gamma}\right]
    \,.\label{eq:m-v-collapse}
\end{equation}
When either effect is sufficiently strong, the central proto-star 
formed in the Jeans core is expected to be fueled by a higher gas accretion 
rate of $\lesssim 10.7 \msunperyr$. The accretion can be sustained at an 
Eddington-level, causing the formation of a supermassive star (SMS), until the 
feedback from the star itself becomes strong enough to quench the 
accretion (Appendix~\ref{app:sssec:seed-in-atomic-cooling-halo}). At sufficiently
high mass, the central star ends its life by a pair-instability supernova 
(PISN) explosion, leaving no remnant, or by a direct collapse to a BH, 
leaving a `heavy-weight' seed.
We thus use a {\texteem feedback-regulated model} to obtain the 
initial mass function (IMF) of Pop-III stars in the star cluster 
formed during the first collapse in a mini-halo and determine the 
mass of the BH seed by following the stellar evolution of these stars
(see Appendix~\ref{app:ssec:mass-func-pop3} for details).

Star formation in the early universe produces metals that enrich the IGM 
through galactic outflows and elevate the cooling of gas in nearby 
mini-halos that have not yet formed stars. To model the IGM enrichment,
we evaluate the metal production by star formation in galaxies, 
follow the expansion of SN-driven bubbles around individual galaxies 
using spherical approximation for bubble shape and an 
analytical model for expansion rate, and
determine the zones of enrichment and the metallicity within the zones
(see Appendix~\ref{app:ssec:IGM-enrichment} for details).
We adopt a threshold value of $Z_{\ssc IGM,Pop\text{-}III} = 10^{-4}\Zsun$
for the formation of Pop-III stars \citep{brommFormationFirstSupermassive2003,valianteOriginDustHighredshift2011,costaMassiveBinaryBlack2023,spinosoMultiflavourSMBHSeeding2023}. 
We also assume that halos embedded in the IGM with metallicity 
$Z_{\ssc IGM} > Z_{\ssc IGM,Pop\text{-}III}$ form their 
first generation of stars in the form of a Pop-II star cluster,
leaving a light-weight BH seed with $M_{\sscBH} \sim 10 \Msun$
(see Appendix~\ref{app:ssec:mass-func-pop3} for details).

Fig.~\ref{fig:flowchart-brief}b outlines the steps of the seeding procedure.
The procedure takes halo merger trees as input and models the above three 
environmental effects that modulate the seeding.
A list of key conditions regarding the competition between cooling and heating 
is examined for every mini-halo to determine whether the gas can collapse 
to form the first generation of stars and the BH seed.
The output of the procedure is the mass of the seed, as well as the `{\texteem flavor}'
that indicates the formation pathway of the seed:
being modulated by strong UV radiation or fast assembly before the collapse; 
yielding `Pop-III' or `Pop-II' stars at formation; experiencing radiation regulation 
at sub-Eddington or Eddington level; and triggering CCSN, PISN or DCBH in the production of the seed.
Once seeded, the galaxy is assumed to be self-enriched and evolves according to 
the growing procedure to be described in \S\ref{ssec:growing-main}.

Fig.~\ref{fig:halo-mahs}a shows an `{\texteem atlas}' of BH seeds. Here, the purple band 
covers the region in the $M_{\rm v}$-$z$ plane to breed Pop-III seeds.
The intersection of the band with the median assembly history of halos of
a given $M_{{\rm v}, z=0}$ (black curve) provides a reference for 
when (at which $z$), where (at which $M_{\rm v}$) and how (modulated by which 
factor) BHs can be seeded. However, the exact location for a mini-halo 
to be seeded in the $M_{\rm v}$-$z$ plane within the band 
depends on the detailed assembly history and radiation environment, 
which varies from halo to halo. The regime of Pop-III seeds is almost fully 
(partly) covered by those of Pop-II (Pop-I) GC formation, highlighting the critical 
role of IGM enrichment in determining the first population of stars and BH seed
that form in a mini-halo. In \S\ref{sssec:seeding-atlas}, we will come 
back to the seeding atlas with all these effects self-consistently modeled
in a cosmological volume.

\subsection{The growth from seeds to supermassive black holes}
\label{ssec:growing-main}

\begin{figure*} \centering
    \includegraphics[width=0.99\textwidth]{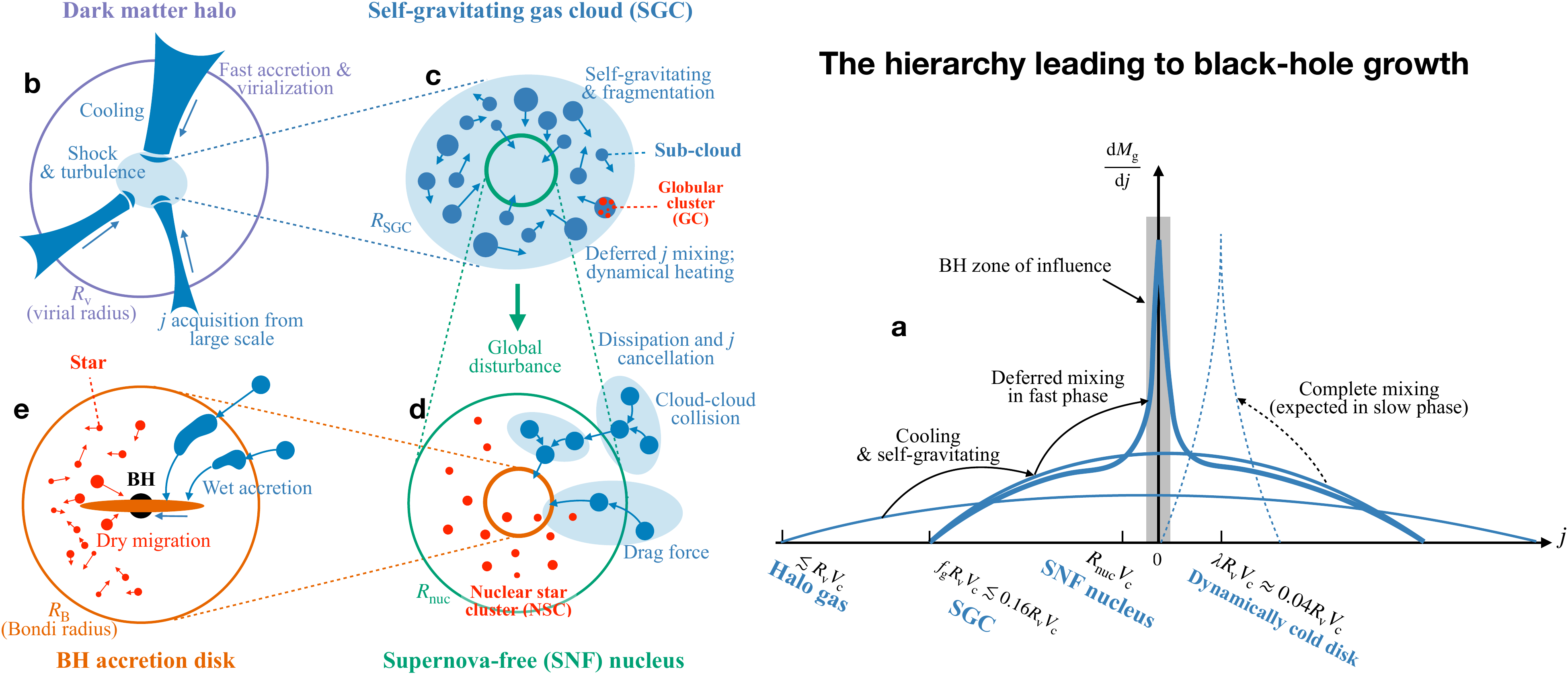}
    \caption{
    {\figem Schematic diagram showing the structure hierarchy that leads to 
    black-hole growth.}
    {\figem a}, the distributions of gas specific angular momentum ($j$) 
    shaped by different processes in the hierarchy.
    {\figem b}--{\figem d}, each level of the hierarchy and processes therein.
    Representations are not to scale, but to show their relative sizes. 
    In a fast-accreting halo, gas rapidly cools and flows into 
    halo center ({\figem b}). $j$ distribution of inflow gas is initially 
    broad, and becomes narrower due to radiative cooling. 
    Inflow gas collides to generate strong shocks, yields supersonic 
    turbulence, becomes self-gravitating before $j$ starts to support
    the gas structure, and leads to the formation of a turbulent 
    SGC ({\figem c}).
    Density perturbations are seeded and amplified by turbulence in the SGC, 
    causing the SGC to fragment into sub-clouds that are dense enough to be 
    free of drag force from ambient gas and cloud-cloud collision.
    The densest sub-clouds can form GCs.
    Dissipation that mixes $j$ of sub-clouds is thus deferred, 
    preventing the formation of a dynamically cold disk and leaving the 
    $j$ distribution broad. 
    A global disturbance on the SGC causes gas to replenish galactic nucleus,
    forming a dense, SNF nucleus ({\figem d}). $j$ mixing is effective 
    in the nucleus, restructuring $j$ distribution to be peaked and centralized 
    and routing gas towards the central BH.
    Within the zone of influence of the BH ({\figem e}), gas surviving 
    feedback joins into the turbulent and magnetized accretion disk 
    and feeds the BH through a `wet channel'.
    Some stars formed in the nucleus migrate inwards via dynamical 
    relaxation, and feed the BH through a `dry channel'.
    In the slow phase of halo assembly, angular-momentum mixing 
    at galactic scale is efficient,
    causing $j$ distribution to be peaked but off-center, 
    and building up a barrier for gas to reach the BH 
    (dashed distribution in {\figem a}). 
    See \S\ref{ssec:growing-main} for the details.
    }
    \label{fig:am-redistribute}
\end{figure*}


The formation of a BH seed is expected to be associated with gas depletion 
in the host mini-halo due to radiative and potentially SN feedback,
causing a transient phase of quiescence in the star formation and AGN activity, 
until replenished gas reactivates the growth.
As detailed in \citetalias{moTwophaseModelGalaxy2024}, the critical condition 
for growing a BH is to break the barrier of angular momentum (denoted by $J$, 
or by $j$ for the specific value per unit mass) so that gas with low $j$ can 
reach the zone of influence of the BH within which the gravity is 
dominated by the BH. Such a condition is naturally achieved in an SGC 
owing to the early fragmentation that leads to the ballistic motion of 
sub-clouds \citep[see also][]{hobbsFeedingSupermassiveBlack2011,gaspariChaoticColdAccretion2015}, 
making the continuous capturing of sub-clouds the main channel of 
BH growth (referred to as the {\texteem continuous mode} thereafter), in contrast to 
{\texteem galaxy-galaxy mergers} that emerge as the main channel of BH growth 
at later epochs \citep[e.g.][]{liPhysicalProcessesCoevolution2025}.

Toward the galactic nucleus, however, the situation differs as sub-clouds are 
affected by three additional processes operating at small scales: 
(i) BH feedback, which regulates the amount of gas that can be 
sustained in the nucleus to power the growth of the BH itself; 
(ii) regeneration of low-$j$ sub-clouds that would otherwise be absent after 
the BH has captured them all or destroyed them via feedback
\citep[see their fig.~5]{sivasankaranAGNFeedbackIsolated2025}; 
(iii) mixing of $j$, as the gas density in the nucleus can be high enough 
to impart a significant drag force and frequent cloud-cloud collisions. 
These processes may cause the gas structure and dynamics to deviate from those 
expected in the outer part of the SGC, thereby modifying BH accretion.

We extend the scenario of \citetalias{moTwophaseModelGalaxy2024} by considering these nuclear 
processes, building the numerical model as a refinement engine that operates with much 
finer time steps on top of the two-phase scenario.
Fig.~\ref{fig:am-redistribute} outlines the four levels of the hierarchy
from the dark matter halo to the BH accretion disk
for the BHs experiencing super-Eddington accretion in an MW-size galaxy and 
its progenitors experiencing fast assembly. 
The fast assembly of the halo and the formation of the SGC make up the first two levels 
of the hierarchy (panels b and c), and their effects on sustaining a broad 
distribution of $j$ are indicated in panel a. These provide a boundary condition for the
activation of nuclear processes that establish the last two levels of 
the hierarchy (panels d and e) and further redistribute $j$.
A detailed description of the growing procedure and the related analytic
estimation is presented in Appendix~\ref{app:sec:growing}, and the remainder of 
this section is a brief summary of the key points.

To avoid ambiguity, we define the nuclear region of an SGC as the inner part 
where the gas density is above a `supernova-free' (SNF) threshold,
$n_{\rm snf} \equiv 10^{3.5}\perccm$ \citep{chenTwophaseModelGalaxy2025}.
This corresponds to a free-fall timescale of $t_{\rm ff, g} \lesssim 1\Myr$
within the nucleus, implying that gas dynamics there proceeds faster than the lifetime 
of massive stars born therein and is thus largely free of SNe.
In the following, we refer to this volume as the {\texteem SNF nucleus},
and denote its radius and gas mass by $R_{\rm nuc}$ and $M_{\rm nuc}$, respectively.
As estimated by \citet{dekelEfficientFormationMassive2023},
other feedback channels from stars, such as radiation and wind from 
both inside and outside the nucleus, are also ineffective in dispersing the gas 
at such a high density. Thus, the only ways remaining to quench an SNF nucleus are
(i) gas consumption by star formation/BH;
(ii) gas ejection by BH feedback.
The consequence of the two is either intense star formation to grow an extremely 
compact nuclear star cluster (NSC) with a 
density $\rho_{\rm *} \sim \rho_{\rm snf} \equiv \mu m_{\rm p} n_{\rm snf}$,
or rapid BH growth, or both.
The growth of both NSC and BH is expected to be a runaway process
due to the short $t_{\rm ff,g}$. We therefore refer to such a bursty 
event as a {\texteem nuclear burst}, and to the channel opened 
by such bursts to grow BH as the {\texteem bursty mode}.


As an approximation, we divide the processes in an SNF nucleus into two stages,
a formation stage where gas inflow is boosted to replenish the nucleus,
and a dispersion stage where the nuclear burst consumes and disperses the gas in the nucleus.
The trigger of the formation of an SNF nucleus in a galaxy is modeled as 
the time when the host halo experiences a temporary excursion to a high accretion 
rate (see Appendix~\ref{app:ssec:snf-nucleus}).
Such an excursion is expected to be able to impart a global disturbance
on the galaxy, generating a certain level of dynamical hotness, removing the barrier 
of $j$ and driving a strong gas inflow towards the nucleus.
The boosted inflow then overpowers (at least over a short period of time) 
the gas consumption by star formation, the accretion of the BH and the ejection by the 
associated feedback, causing gas to pile up and establishing the SNF nucleus.
The structure of the SNF nucleus is expected to be supported by the turbulence generated by 
the rapid inflow, and the gas profile within the nucleus ($r \leqslant R_{\rm nuc}$) is 
therefore modeled as a smooth extrapolation from the outer SGC ($r > R_{\rm nuc}$),
with one parameter to characterize the turbulence structure.

The turbulent and high-density environment in the SNF nucleus provides ideal 
conditions to channel gas through the nucleus towards the BH, which 
we take into account in our model for the dispersion stage.
Sub-clouds in the nucleus are expected to experience frequent collisions 
with other sub-clouds as well as strong drag forces from the ambient gas.
Efficient cooling implied by the high density makes such interactions
highly dissipative, causing further condensation and loss of $j$ of the 
sub-clouds. A net inflow toward the BH is thus generated as the sub-clouds 
sink their orbits (Fig.~\ref{fig:am-redistribute}d).
The essential role of turbulence is thus a `deferred mixing':
in the outer SGC, turbulence seeds the formation of sub-clouds 
and amplifies their density. Inefficient mixing among sub-clouds
preserves the broad distribution of $j$, allowing sub-clouds with low $j$ to 
approach the nucleus. As the low-$j$ sub-clouds approach the inner SGC, 
efficient mixing drives a gas inflow to feed the BH.
The effect of deferred mixing on the distribution of $j$ is schematically 
illustrated in Fig.~\ref{fig:am-redistribute}a, where a sharply peaked
and centralized distribution forms and powers efficient BH accretion. 
This is different from the case where a dynamically-cold disk
forms, where the distribution of $j$ is peaked but off-centered,
leaving little fuel for the BH (dashed distribution in Fig.~\ref{fig:am-redistribute}a). 

The modeling of the dispersion stage for an SNF nucleus follows the continuity
equation of the gas density within the nucleus, taking a spherical approximation 
for the relevant profiles. The evolution of gas in this equation is contributed 
by gas inflow from the outer part of the SGC into the nucleus, star formation in the nucleus, 
accretion of gas into the zone of influence of the BH, 
gas ejection by BH feedback, and a hard time window for gas depletion set by 
the delay time of SNe. Timescale analysis suggests significant competition 
among the sink terms:
BH feedback is almost always the dominant contributor to gas depletion;
star formation outpaces (is outpaced by) BH accretion if 
$M_{\sscBH} \ll M_{\rm nuc}$ ($M_{\sscBH} \gg M_{\rm nuc}$).
A {\texteem turbulence-modified Bondi accretion model} \citep{bondiMechanismAccretionStars1944} is
used to account for the role of turbulence in driving efficient gas inflow,
as discussed above, and to determine the exact accretion rate 
($\dot{M}_\sscBondi$, the Bondi rate) of gas into 
the zone of influence ($r \leqslant R_\sscBondi$, the Bondi radius) of the 
BH. Given the high Bondi rate, a {\texteem turbulent and magnetized accretion disk model} 
\citep[thereafter \citetalias{hopkinsAnalyticModelMagneticallyDominated2024}, 
for the details]{hopkinsAnalyticModelMagneticallyDominated2024} is used 
to solve the structure of the accretion disk, the star 
formation within $R_\sscBondi$, and the resulting 
BH growth rate in such a `{\texteem wet channel}';
stars formed in the nucleus are also expected to experience frequent dynamical 
interactions and contribute a `{\texteem dry channel}' to the BH growth
(Fig.~\ref{fig:am-redistribute}e).
Numerical treatment is applied to evolve the BH mass and the gas and stellar densities 
within the nucleus, taking the structure established in the formation stage as the initial 
condition. The nuclear region is assumed to enter a quiescent phase 
after a burst, waiting for the next excursion event of halo assembly to 
trigger the formation of the next SNF nucleus.
Super-Eddington accretion is found to be achieved in most nuclear bursts
owing to the inefficiency of all feedback channels (see Appendix~\ref{app:ssec:modified-bondi} 
and Fig.~\ref{fig:bh-engine-example}).
The growth history of a BH thus consists of flash episodes of 
super-Eddington accretion and intermittent quiescent stages where other channels of BH growth dominate.
The co-growth of NSC and BH is also a natural consequence of the model.

To summarize, the following channels of BH growth are considered in our model after seeding:
\begin{enumerate}[topsep=0pt,parsep=0pt,itemsep=0pt]
    \item {\texteem Bursty mode}, which appears as episodic 
    super-Eddington accretion of the BH and intense starburst in the SNF nucleus formed 
    due to global disturbances caused by excursions of halo assembly.
    Two distinct channels, the wet channel and the dry channel, are expected to 
    contribute to the BH growth in this mode. 
    \item {\texteem Continuous mode}, in which the BH continuously captures 
    sub-clouds formed in the SGC during the fast phase of a halo.
    The associated star formation in the SGC is expected to build a 
    dynamically hot stellar component (bulge) of the host galaxy.
    In contrast, during the slow phase of a halo, the BH stops growing, and 
    star formation builds up the dynamically cold stellar component (disk).
    \item {\texteem Merger}, in which two galaxies and their central BHs 
    merge to form a new galaxy and a new central BH.
\end{enumerate}
As we will show, these channels dominate BH growth at different epochs 
in the history of a halo, thereby shaping a redshift-dependent, 
multi-piece $M_{\sscBH}$-$M_*$ relation.
Specifically, the bursty mode, a new channel of BH growth introduced
in this paper, plays a key role in boosting the growth of low-mass
seeds to the regime of SMBHs.

An interesting conclusion can be reached by linking the evolution
of structures at different levels of the cosmic hierarchy.
The structure of SNF nuclei formed at different epochs in the history of a 
halo is related to the assembly history of the halo (see \S\ref{app:sssec:snf-structure}). For example,
the derivative of the gas surface density of the SNF nucleus, $\Sigma_{\rm nuc}$, 
can be derived analytically as
\begin{equation} 
    \dv{\ln \Sigma_{\rm nuc}}{\ln a} 
    \approx \frac{2 \Gamma_{\rm v} - 3}{6} 
    \,,\label{eq:diff-sigma-nuc-main}
\end{equation}
where $a = 1/(1+z)$, and $\Gamma_{\rm v} \equiv \dot{M}_{\rm v} / (M_{\rm v} H)$
is the specific growth rate of the halo mass. 
The derivative changes its sign at a critical value of $\Gamma_{\rm v} = 1.5$, 
coinciding with the epoch found by \citet{bocoTwOParametersSemi2023}
that separates the fast- and slow-accretion phases 
of a halo. The same argument can also be applied to $R_{\rm nuc}$ and $M_{\rm nuc}$,
yielding the same critical value for $\Gamma_{\rm v}$.
Moving deeper into the SNF nucleus, the Bondi radius,
the Bondi-accretion rate and the structure of BH accretion disk
are all determined by $\Sigma_{\rm nuc}$.
{\textem Hence, the cosmological information can propagate through 
galaxy ecosystems to the innermost regions of galaxies, and the 
transitions of halo assembly also lead to transitions in BH accretion.}
This is a highly non-trivial result of our model,
as it links the cosmological assembly of structures over many orders of 
magnitude in spatial scale through physical processes. 

The details of analytical estimations and the evolution of the structure hierarchy 
are presented in Appendix~\ref{app:sec:evolution-of-hierarchy}.
The numerical implementation of the model in a cosmological 
N-body simulation is described in Appendix~\ref{app:sec:impl-details}.
In the following, we show the numerical results predicted by the 
{\textem Default} model and discuss their physical and observational implications. 
Several variants of the model are defined and studied in 
Appendix~\ref{app:sec:model-variants}
to show the degrees of freedom allowed by the physical prescriptions 
and to pinpoint the model space to be explored in the future.

\section{Model predictions and comparison with observational data}
\label{sec:results}


\begin{figure*} \centering
    \includegraphics[width=0.925\textwidth]{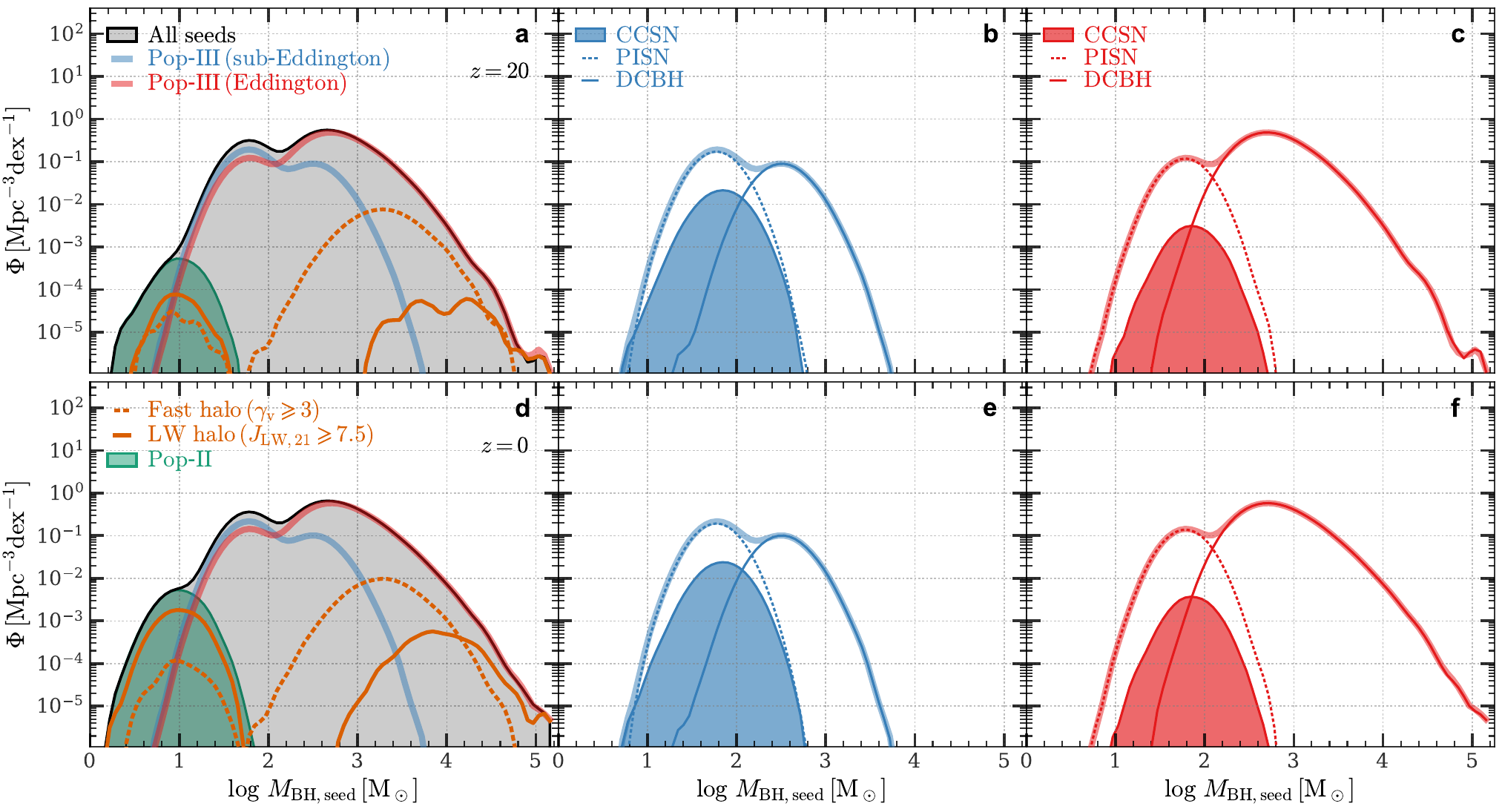}
    \caption{{\figem Mass function of black-hole seeds.}
    {\figem a}, $\Phi(M_{\rm \sscBH,seed},z=20)$, mass functions of all BH seeds 
    that have been bred until $z = 20$ in the simulation volume 
    ({\figem black} curve enclosing gray area)
    and of seeds with different `flavors' (formation pathways): 
    descendants of Pop-III stars that experienced sub-Eddington ({\figem blue})
    or Eddington ({\figem red}) accretion during their formation, 
    those born from dynamically heated SGC in `fast halos' 
    (i.e. fast-accreting halos with $\gamma_{\rm v} \geqslant 3$; {\figem orange dashed})
    or from `LW halos'
    (i.e. LW-irradiated halos with $J_{\sscLW,21} \geqslant 7.5$; {\figem orange solid}),
    and descendants of Pop-II stars fed by enriched IGM
    ({\figem green}).
    {\figem b}, {\figem c}, mass functions of sub-categories of 
    Pop-III seeds.
    Here each category of Pop-III seeds (sub-Eddington or Eddington) 
    is further divided into three 
    sub-categories according to the stellar evolution (CCSN, PISN or DCBH)
    that ends the life of the dominant star in the host 
    Pop-III star cluster.
    {\figem d}--{\figem f}, similar but including all seeds 
    bred until $z = 0$.
    The mass spectrum of BH seeds is quite broad and exhibits two main 
    modes, suggesting that galaxy bimodality has already appeared at
    $z \approx 20$.
    Environmental processes have a divergent impact on the masses 
    of BH seeds, producing both high-mass and low-mass bumps in the 
    mass function, with DCBH and Pop-II flavors, respectively.
    See \S\ref{sssec:seed-mass-func} for details.
    }
    \label{fig:seed-mass-func}
\end{figure*}

\subsection{The expected seed populations}
\label{ssec:result-seed}

\subsubsection{Seed mass functions}
\label{sssec:seed-mass-func}

We define the seed mass function as the number of BH seeds per unit 
logarithmic mass interval per unit comoving volume formed before
a given redshift $z$:
\begin{equation}
    \Phi(M_{\rm {\sscBH}, seed}, z) = 
    \frac{1}{V_{\rm u}} \dv{N_{\rm seed}(> z)}{\log M_{\rm {\sscBH}, seed}} 
    \,, \label{eq:def-seed-mass-func}
\end{equation}
where $V_{\rm u}$ is the comoving volume of the simulation, 
and $\dd{N_{\rm seed}(> z)}$ is the number of BH seeds that have been bred 
by $z$ within the mass interval $\dd{M_{\rm {\sscBH}, seed}}$ in the volume.

Fig.~\ref{fig:seed-mass-func}a--c show the total seed mass function (black)
and the contributions of different flavors (colored)
at $z=20$. The peak of $\Phi$ is at 
$M_{\rm \sscBH, seed} \approx 500\Msun$, recovering the 
typical mass assumed in theoretical and observational 
studies of Pop-III stars
\citep{schaererPropertiesMassivePopulation2002,wangStrongHeII2024}.
The shape of $\Phi$ is quite broad, 
demonstrating the diversity of the environment that modulates the seeding process 
(see below).
PISN is preferentially generated at high mass, which destroys the central star, 
leaves a less massive BH seed (the remnant of another star surviving PISN), 
and produces a prominent
bimodality with a valley at $M_{\rm \sscBH, seed} \approx 100\Msun$, 
in $\Phi$.
DCBH is preferentially generated at even higher mass and produces the 
seeds that dominate the high-mass end of $\Phi$.
In particular, a non-negligible fraction of seeds, with an expected number density of
$\Phi \lesssim 10^{-3} \Mpc^{-3}\dex^{-1}$, falls in the mass range of the SMS 
($M_{\rm \sscBH, seed} \gtrsim 10^4\Msun$). SMSs are believed 
to be promising candidates for the progenitors of bright quasars observed at 
$z \sim 6$ (e.g. \citealt{wiseFormationMassiveBlack2019,latifTurbulentColdFlows2022};
see also our discussion of archaeology in \S\ref{ssec:arch-and-forecast}),
and our model implies that SMSs are indeed sufficiently abundant for such a 
link, provided that some of the seeds can grow further via 
different channels (see Figs.~\ref{fig:growth-effect-seeding} and \ref{fig:scaling-evol}).

A galaxy is expected to experience a short quiescent phase immediately 
after a seed is bred, because the associated AGN feedback strongly suppresses
star formation and BH growth (see \S\ref{ssec:result-history} for details). 
Since BHs with higher mass are more efficient 
in feedback (in terms of the energy released per unit of accreted mass), 
seeds with the DCBH flavor may quench the galaxy for a longer period of time,
while seeds with the CCSN and PISN flavors may allow the galaxy to 
rejuvenate earlier. 
The bimodal distribution of the seed mass is thus expected to produce a bimodal 
distribution in the age of stars after the seeding.
Observations targeting the first generation of quiescent galaxies 
may thus be able to reveal the properties of the infant galaxies 
that have just completed the seeding.

\begin{figure*} \centering
    \includegraphics[width=\textwidth]{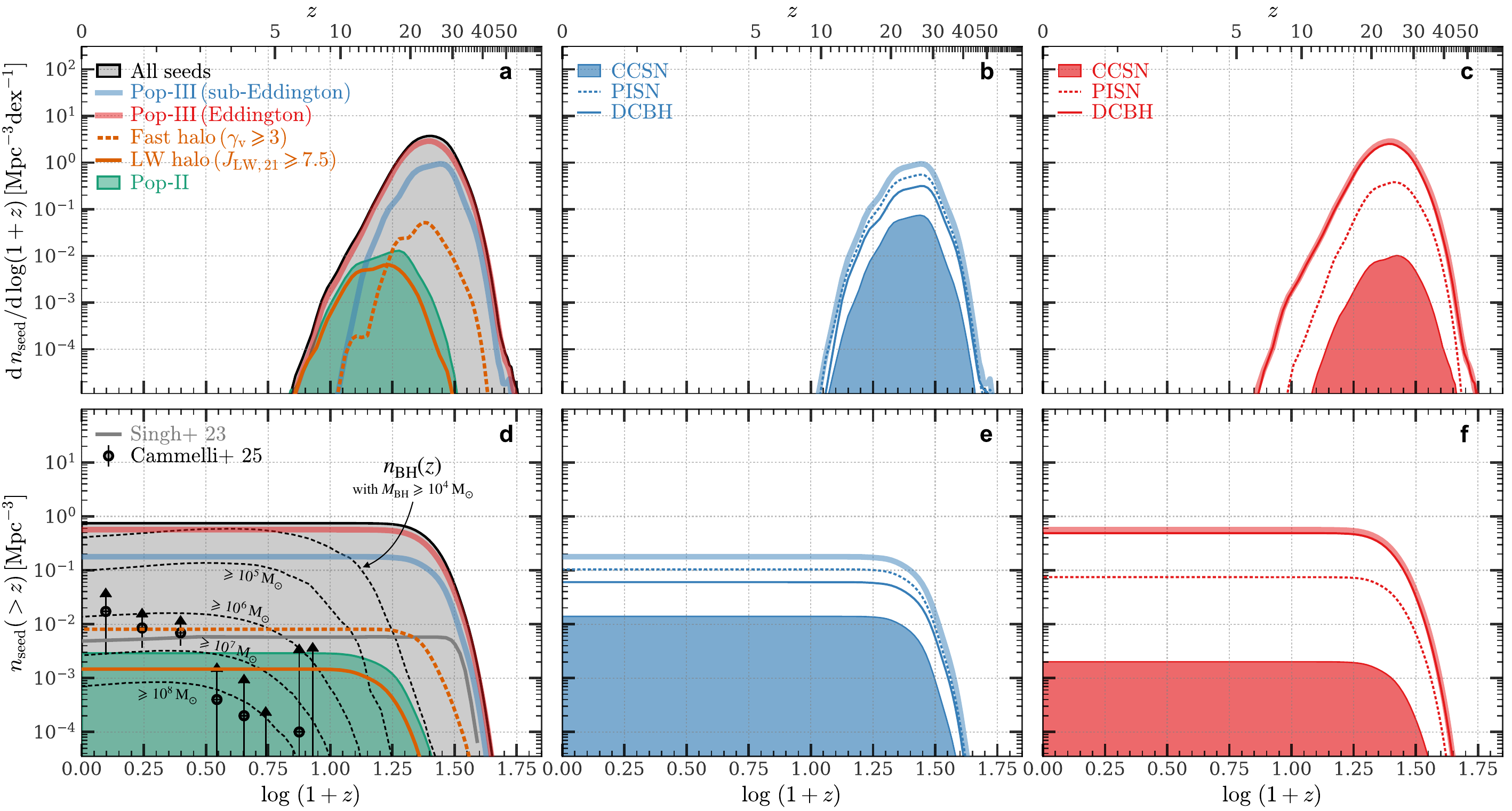}
    \caption{{\figem Cosmic black-hole seeding history.}
        {\figem a}--{\figem c}, cosmic seeding rate density 
        (i.e. number of BH seeds bred per unit logarithmic scale factor
        per unit comoving volume) as a function of
        $z$.
        {\figem d}--{\figem f}, cosmic cumulative seeding density 
        (i.e. number of BH seeds bred before $z$) as a function of $z$.
        The results are shown for all seeds ({\figem black} curve)
        and for seeds with different `flavors' defined in the 
        same way as in Fig.~\ref{fig:seed-mass-func}.
        At $z \approx 20$, most seeds have already been bred. 
        Pop-II seeds form at systematically
        lower redshift than Pop-III seeds,
        and account for a significant fraction of the seed population 
        at $z \lesssim 10$.
        To illustrate the effects of post-seeding evolution,
        each {\figem black dashed} curve in {\figem d} shows $n_\sscBH (z)$, 
        the number density of BHs above a given mass threshold 
        at $z$, predicted after the growing procedure (including the treatment 
        for mergers) is applied to the seeds.
        For comparison, we include the semi-analytical predictions by \citet{singhFormationSupermassiveBlack2023}
        and observational estimates by \citet{cammelliGlimmersCosmicDawn2025}.
        See \S\ref{sssec:cosmic-seeding-history} for details.
    }
    \label{fig:seed-z-func}
\end{figure*}

At $z=20$, the peak value of $\Phi$ is above $0.1\Mpc^{-3}\dex^{-1}$, 
higher than the number density of dwarf galaxies with 
$M_*=10^8\Msun$ at $z=0$ \citep[see e.g. fig.~15 of][]{chenELUCIDVICosmic2019},
suggesting that most such galaxies may have their lifetime starting 
at $z \gtrsim 20$ (see also Fig.~\ref{fig:seed-z-func}).
These stars can either remain in the inner galaxy, be dynamically heated to  
higher orbits, be ejected as halo stars, or merge into other galaxies.
The search for fossils of these stars has been carried out in and around the
Milky Way (MW) and has yielded interesting cases with
Pop-III-like metallicity \citep[see e.g.][for a review]{bonifacioMostMetalpoorStars2025}, or with a peculiar PISN-compatible
chemical pattern \citep[e.g.][]{aguadoPISNexplorerHuntingDescendants2023,
xingMetalpoorStarAbundances2023}. 
Our model suggests that these stars may be
used to reveal the formation of the MW back to $z \approx 20$, the epoch when 
the Universe is younger than $200\Myr$.

Both dynamical heating and UV radiation
delay the collapse of an SGC, elevate the gas mass available for accretion,
bring the accretion of the central star to the Eddington level,
and grow it to higher mass. However, the relative importance of these two environmental factors 
differs between halos. Dynamical heating is stronger in halos with faster assembly
and is thus more important at higher $M_{\rm v}$ and higher 
$z$ (e.g. Fig.~\ref{fig:halo-mahs}).
LW radiation is stronger if galaxies with higher SFRs are present in 
the neighborhood, and is thus more important
at lower $z$ when a larger number of massive, star-forming galaxies 
have formed in over-density regions.
To illustrate the difference between these two factors, we select seeds bred in
(i) `{\texteem fast halos}', defined as those with
$\gamma_{\rm v} \geqslant 3$ at the seeding epoch; 
(ii) `{\texteem LW halos}', defined as those with 
$J_{\rm \sscLW, 21} \geqslant 7.5$.
The thresholds are chosen so that the expected collapse threshold
is raised to $M_{\rm v} \approx 10^7\Msun$ and massive seeds can be 
bred thereafter (see Fig.~\ref{fig:m-v-cool-fit}).
As shown in Fig.~\ref{fig:seed-mass-func},
seeds in both types of halos tend to produce accretion at the Eddington level,
highlighting the importance of the large-scale environment in 
modulating the seeding process.
The masses of seeds in LW halos are higher than those in halos of fast assembly,
consistent with the expectation that the LW radiation is more efficient
at lower $z$ when halos that have not been seeded can reach 
higher masses.

For comparison, we show $\Phi(M_{\rm \sscBH, seed}, z=0)$, the seed mass function
at $z=0$ in Fig.~\ref{fig:seed-mass-func}d--f.
This is similar to that at $z=20$ in terms of normalization, 
bimodal shape and contributions of different flavors. 
A noticeable difference is that $\Phi$ at $z=0$ is
broadened towards both the low- and high-mass ends.
Both tails are dominated by LW halos and fast halos, implying that both
environmental factors are important in delaying the seeding.
The environmental factors now have a {\texteem divergent effect} on the 
masses of BH seeds:
on the one hand, they delay the collapse of an SGC, so that more massive 
DCBHs are bred owing to the higher gas mass available for 
the formation of Pop-III stars (red thin curve in panel f);
on the other hand, it gives more time for SNe from nearby galaxies 
to enrich the  mini-halo, so that Pop-II seeds are more likely to form 
via efficient cooling provided by the metals (green shade in panel d). 
In fact, most Pop-II seeds are bred at $z \lesssim 20$, 
in contrast to the early formation of Pop-III seeds (mostly bred at 
$z \approx 20$--$30$; see also Fig.~\ref{fig:seed-z-func}a).
The interplay between dynamical heating and the
two forms of feedback (matter and radiation) from galaxy formation to the 
large-scale environment may thus leave interesting imprints on 
the seed populations.

\subsubsection{The cosmic seeding history}
\label{sssec:cosmic-seeding-history}

To quantify the redshift distribution of seed formation, we define the 
cosmic seeding rate density as the number of BH seeds bred per unit
logarithmic cosmic scale factor per unit comoving volume:
\begin{equation}
    \dv{n_{\rm seed}}{\log (1+z)} =
    \frac{1}{V_{\rm u}} \dv{N_{\rm seed}
    (>z)}{\log(1+z)} 
    \,.\label{eq:def-seed-dn-dz}
\end{equation}
The cumulative value of the cosmic seeding rate density,
\begin{equation}
    n_{\rm seed}( > z) = N_{\rm seed}(>z) / V_{\rm u} 
    \,,\label{eq:def-seed-n-z}
\end{equation}
is referred to as the cosmic cumulative seeding density.
This mimics the definition of cosmic star formation history 
\citep{madauCosmicStarFormation2014}, but is specific 
to BH seeding.

Fig.~\ref{fig:seed-z-func}a--c show the evolution of seeding rate density
for all seeds and by flavor, and panels d--f show the corresponding cumulative
distributions. 
The total seeding rate density increases steeply with time, reaches 
a peak at $z \gtrsim 25$, and then decreases rapidly.
Most seeds have already been bred by $z \approx 20$,
consistent with the expectation of the seed mass functions 
(see Fig.~\ref{fig:seed-mass-func}).
The fraction of Pop-II seeds is less than $1/100$ at $z = 20$, but increases 
significantly with time,
again reflecting the divergent effect of heating sources 
on seed mass (see \S\ref{sssec:seed-mass-func}).

For Pop-III seeds, those formed with Eddington accretion of the progenitor stars
dominate the seeding rate at all $z$, and become 
increasingly so at lower $z$;
those formed with Eddington (sub-Eddington) accretion are more likely to have 
DCBH (PISN) flavor.
The seeding rates of fast halos and LW halos cross over at 
$z \approx 15$--$20$, with the former (latter) dominating at higher 
(lower) $z$, reinforcing the conclusion drawn
in \S\ref{sssec:seed-mass-func} that the two environmental factors 
work in different regimes.
Most seeds bred at $z \lesssim 15$ are delayed by the LW radiation, 
implying that the optimal strategy for searching for 
ongoing formation of Pop-III stars and BH seeds, as well as their 
immediate descendants, is to target the surroundings of over-dense regions.
Recent JWST projects
\citep[e.g.][]{mattheeEIGERIIFirst2023,naiduAllLittleThings2024,
fujimotoGLIMPSEUltrafaint1052025,hsiaoSAPPHIRESExtremelyMetalPoor2025}
have already been able to cover such high redshifts,
and our model provides a theoretical framework to optimize observational strategies
and interpret their findings.

As to be shown in \S\S\ref{ssec:result-history} and \S\ref{ssec:mbh-ms-relation},
post-seeding evolution can significantly alter the masses of BH seeds, 
which is essential for them to grow to the regime of SMBHs. BH populations observed
at low $z$, and summary statistics made for them, are thus a mixed outcome of 
seeding and growing. To illustrate this, in Fig.~\ref{fig:seed-z-func}d, we 
show the number density of BHs ($n_\sscBH $) with different masses 
as a function of $z$ after our growing procedure has been applied.
At $z \lesssim 5$, BHs with $M_\sscBH \geqslant 10^5\Msun$ only account for 
$\approx 1/10$ of all seeds bred in the simulation, indicating that
most seeds have not grown to a mass high enough to produce observable signatures 
and that a vast population of low-mass BHs is yet to be discovered.
BH-BH mergers slightly reduce the number of BHs by a factor of $\lesssim 2$,
as seen from the reduction in $n_\sscBH $. At $z \lesssim 5$, we 
predict $n_\sscBH \approx 0.1\,{\rm cMpc}^{-3}$ for BHs 
with $M_\sscBH \geqslant 10^5\Msun$, a number density 
more than $1\dex$ higher than that predicted by the semi-analytic model of 
\citet{singhFormationSupermassiveBlack2023}, who use Pop-III.1 stars to 
seed BHs (grey curve; their model variant with $d_{\rm iso}=100\Kpc$),
and than that observed by \citet[black markers; combining their variability-incompleteness-corrected estimate at $z < 4$ 
and luminosity-incompleteness-corrected estimate at $z \geqslant 4$;
see also \citealt{banikFormationSupermassiveBlack2019}]{cammelliGlimmersCosmicDawn2025}. Since almost all seeds in our 
model have $M_\sscBH < 10^5\Msun$ (see Fig.~\ref{fig:seed-mass-func}), 
the predicted high $n_\sscBH$ lies in the post-seeding growth,
mainly the bursty mode, which is efficient in growing low-mass 
BHs and makes their $M_\sscBH$-$M_*$ ratio higher than that expected from
extrapolating the $M_\sscBH$-$M_*$ relation at higher masses
(see \S\ref{ssec:result-history} and Fig.~\ref{fig:growth-examples}
for details). Note that we did not calibrate the bursty mode to match 
observations, and the predictions here serve as a blind test. 
The difference between our model and observations thus
indicates that either the bursty mode in the real Universe is less 
efficient than assumed, or the observations underestimate the abundance 
of low-mass BHs. The factors likely responsible for the uncertainties 
in observations include cosmic variance \citep{somervilleCosmicVarianceGreat2004,mosterCOSMICVARIANCECOOKBOOK2011,
chenELUCIDVICosmic2019,chenMassiveDarkMatter2023,jespersenSignificanceRareObjects2025}, 
diverse selection criteria of AGNs \citep{lyuAGNSelectionDemographics2022,lyuActiveGalacticNuclei2024},
and incomplete calibration of BH-mass estimators due to the lack of 
low-mass BHs with dynamical mass measurements (e.g. \citealt{greeneIntermediateMassBlackHoles2020}; see also Fig.~\ref{fig:scaling-evol}),
all of which warrant caution when extending the observations 
to low $M_\sscBH$. 

\subsubsection{The seeding atlas: when, where and how seeds are bred}
\label{sssec:seeding-atlas}


\begin{figure*} \centering
    \includegraphics[width=0.85\textwidth]{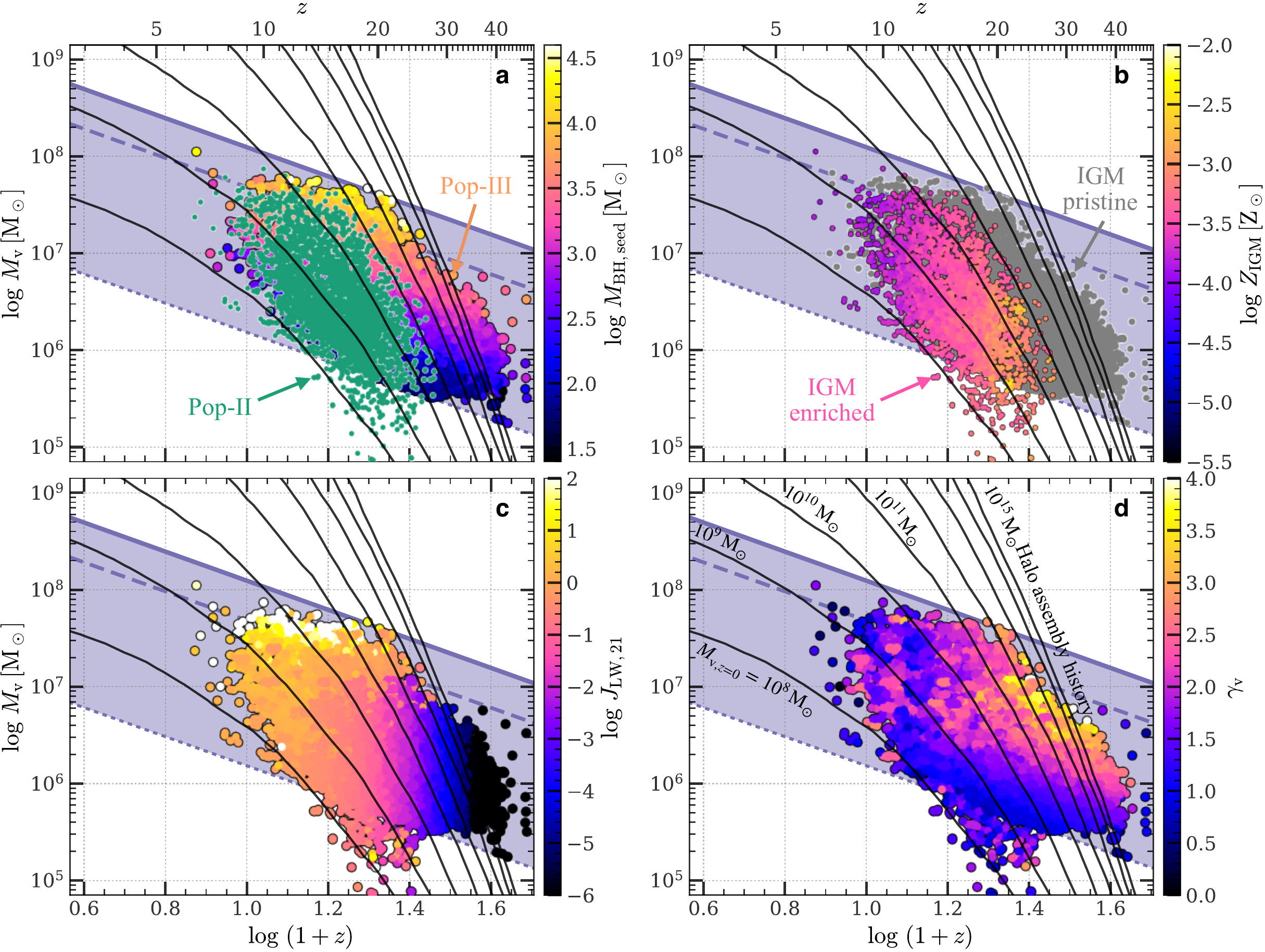}
    \caption{{\figem The atlas of black-hole seeds.}
    Here we show the distribution of BH seeds in the $M_{\rm v}$-$z$ plane 
    (host halo mass versus redshift at the seeding epoch), 
    color-coded by the mass of the BH seed ($M_{\rm {\sscBH}, seed}$, {\figem a}), 
    the metallicity of the IGM surrounding the subhalo ($Z_{\ssc IGM}$, {\figem b}), 
    the intensity of LW radiation irradiating the subhalo ($J_{\rm \sscLW, 21}$, {\figem c}), 
    or the specific halo growth rate ($\gamma_{\rm v}$, {\figem d}), 
    all evaluated at the seeding epoch.
    In {\figem a}, green dots represent Pop-II seeds, while the others represent
    Pop-III seeds. 
    In {\figem b}, grey dots represent those embedded in pristine IGM 
    ($Z_{\ssc IGM} = 0$), while the others are color-coded by $Z_{\ssc IGM}$.
    All seeds bred in the simulation volume are included in this plot.
    Kernel smoothing is applied to the colored properties to clarify the trends.
    Halo assembly histories ({\figem black curves}) and the regime of Pop-III 
    seeds ({\figem purple band}) are overlaid for reference
    (the same as those in Fig.~\ref{fig:halo-mahs}).
    This figure illustrates when, where and how BH seeds are bred, highlighting 
    the role of environmental processes (quantified by 
    $Z_{\ssc IGM}$, $J_{\sscLW}$ and $\gamma_{\rm v}$) 
    in modulating the seeding process.
    See \S\ref{sssec:seeding-atlas} for details.
    }
    \label{fig:seeding-atlas}
\end{figure*}

The purple band in the $M_{\rm v}$-$z$ plane (Fig.~\ref{fig:halo-mahs}a) 
provides an atlas to identify when and where seeds are expected to be bred.
Implementing the model into the cosmological N-body simulation, 
we are now able to track the diverse environments of individual mini-halos and use
them to dissect the atlas into regimes where seeds with distinct flavors 
preferentially occupy.
Fig.~\ref{fig:seeding-atlas} shows such a refined atlas, where individual seeds
are color-coded by the mass ($M_{\rm {\sscBH},seed}$) or by one 
of the three environmental properties ($Z_{\rm IGM}$, 
$J_{\rm \sscLW}$ and $\gamma_{\rm v}$).

The first generation of stars and BH seeds emerges at $z \gtrsim 40$ when the 
progenitors of extremely massive halos ($M_{{\rm v}, z=0} \gtrsim 10^{14}\Msun$) 
reach $M_{\rm v, \Hmol}$ (Eq.~\ref{eq:m-v-cool-H2-main}).
In this early era of galaxy formation,
IGM has not been enriched (panel b), LW field has not been built
(panel c), and dynamical heating is the main environmental factor that 
modulates seeding (panel d).
A mini-halo with a temporary excursion to slow accretion gets
a chance to escape the dynamical heating and form a low-mass, 
Pop-III seed, while a mini-halo sustaining fast accretion breeds a more 
massive seed later on.
A direct prediction of our model is that cosmic star formation 
and seeding activity start to prevail at the locations of extremely 
massive halos that reach $M_{\rm v, \Hmol}$ in pristine regions
and have temporary excursions to low $\gamma_{\rm v}$ at high $z$. 

Given the extremely compact sizes of SGCs at such high $z$ (a few pc; see the 
texts below Eq.~\ref{eq:m-v-cool-H-main}) and
the high density threshold for the pristine gas to fragment,
these earliest Pop-III star clusters are expected to be more 
compact than the Pop-I/II GCs formed later.
The low masses of their host mini-halos also imply low masses 
of the BH seeds (mostly $\lesssim$ a few hundred ${\rm M}_\odot$),
and thus the dominance of CCSNe and PISNe in their flavors.
There are also cases where mini-halos form very early and 
remain at high $\gamma_{\rm v}$ at $z \gtrsim 40$, and our model 
predicts that DCBHs with masses $\gtrsim$ a few hundred ${\rm M}_\odot$ 
can be bred in these mini-halos (panel a).

If a massive halo can sustain a high $\gamma_{\rm v}$ over a long period, 
seeding can be further delayed and the seed mass can be elevated 
(yellow dots in panel d).
For most halos with $M_{{\rm v}, z=0} \geqslant 10^{13}\Msun$, the seeding epoch 
is $z \gtrsim 20$ and the seed mass spans from a few tens to about
$10^4\Msun$ (panel a). This wide range of seed mass also 
implies a diverse flavor (formation pathway) of seeds.
The LW field has not been established and is thus unimportant for the seeding 
in these halos (panel c). Individual halos may violate the general trend 
due to the diverse assembly histories and environments of massive halos
\cite[see e.g. \S4.2 of][]{chenMassiveDarkMatter2023}.

Less massive halos with $M_{{\rm v}, z=0} \lesssim 10^{13}\Msun$ follow
a similar trend driven by dynamical heating, albeit with some upper-left 
shift in the $M_{\rm v}$-$z$ plane. In the atlas,
dynamical heating thus defines a fusiform-shaped regime
filled with tilted, stripe-like equal-$M_{\rm {\sscBH},seed}$ 
or equal-$\gamma_{\rm v}$ contours (panels a and d).
LW radiation starts to modulate seeding in these halos due to 
the higher cosmic SFR density (see Fig.~\ref{fig:growth-cosmic-history} 
and \citealt{madauCosmicStarFormation2014}) and
less effective IGM attenuation (see Appendix~\ref{app:ssec:LW-radiation})
at lower $z$. 
These LW-illuminated halos have the most significant delay of seeding,
reach the highest seed masses, and populate a tip-like regime in the 
atlas (panel c).

The divergent effect of dynamical heating and LW radiation on seed mass
(see \S\ref{sssec:seed-mass-func}) is also reflected in the atlas.
Many low-mass halos ($M_{{\rm v},z=0} \lesssim 10^{12}\Msun$, particularly
those with $M_{\rm v}\lesssim 10^{10}\Msun$) have 
enriched seeding environments, and Pop-II seeds with 
low masses are therefore bred (green cloud in panel a).
However, metal diffusion is much slower than radiation propagation.
The enriched IGM thus only appears for halos already affected 
by the LW radiation. Hence, the distribution of Pop-II seeds in the atlas 
appears as an elongated stripe pulled off by the two sources of heating.

In summary, the seeding atlas can be dissected into three regimes 
according to the environmental processes that modulate the seeding:
\begin{enumerate}[topsep=0pt,parsep=0pt,itemsep=0pt]
    \item {\texteem The dynamical-heating regime}, which spans a fusiform area 
    in the $M_{\rm v}$-$z$ plane. 
    \item {\texteem The LW-radiation regime}, 
    a tip-shaped structure that extends the low-$z$, high-$M_{\rm v}$ end 
    of the dynamical-heating regime.
    \item {\texteem The regime of enriched IGM}, which is a
    two-component stripe
    on top of the low-$z$ parts covered by the above two regimes.
\end{enumerate}

Seeds modulated by different environmental effects or bred with 
different masses differ in spatial distribution, 
since processes establishing the environments operate on different spatial 
scales and have different temporal persistence.
The clustering pattern of seeds may thus encode information about 
how they were bred, and this pattern may persist for a long time, allowing 
the seed properties to be retrieved from low-$z$ observations
targeting the descendants. This can be investigated in the 
future through clustering analyses.

\subsection{The co-evolution of galaxies and black holes}
\label{ssec:result-history}

\begin{figure*} \centering
    \includegraphics[width=0.675\textwidth]{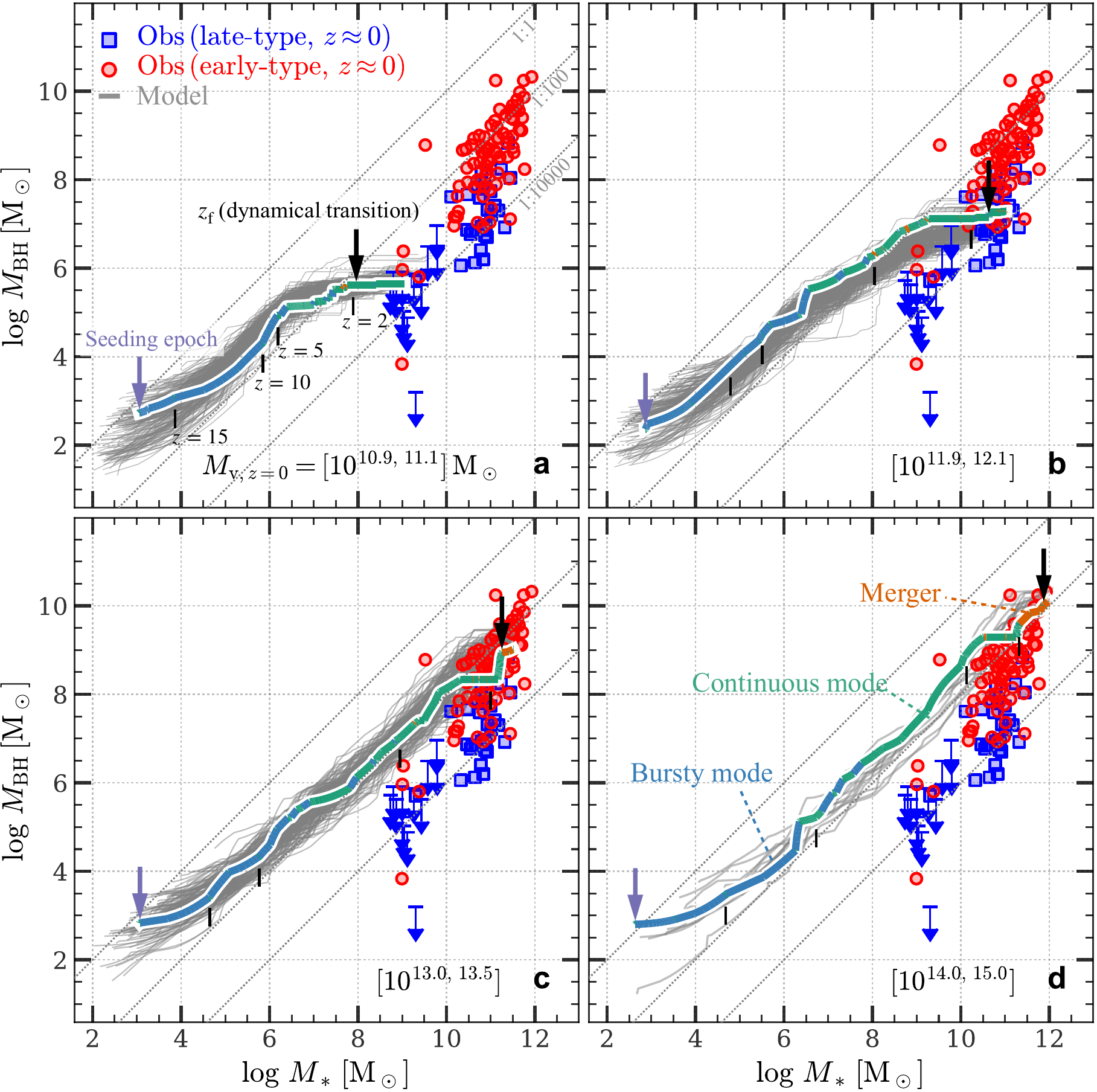}
    \caption{{\figem Predicted growth paths of individual galaxies in the 
    $M_\sscBH$-$M_*$ plane.}
    {\figem a}--{\figem d}, prediction for central galaxies
    with different halo masses at $z=0$, as labeled.
    In each panel, {\figem grey} curves show individual modeled galaxies,
    and a {\figem colored} curve shows one example galaxy
    with the channel dominating the growth of the BH at each time step 
    indicated by the color of the path segment
    ({\figem blue}, bursty mode, wet channel; {\figem green}, continuous mode; {\figem brown}, merger).
    Note that the dry channel in the bursty mode is always 
    subdominant compared with the wet channel, and is thus totally absent.
    Critical epochs for the example galaxy are marked 
    ({\figem purple arrow}: seeding epoch; 
    {\figem black arrow}: $z_{\rm f}$; 
    {\figem black ticks}, $z = 15$, $10$, $5$ and $2$). 
    For reference,
    {\figem blue} and {\figem red} scatters represent
    individual late-type and early-type galaxies, respectively,
    at $z \approx 0$, 
    compiled by \citet{greeneIntermediateMassBlackHoles2020}.
    Tilted grey lines indicate the loci of 
    $M_{\sscBH}$:$M_*=$1:1, 1:100 and 1:10000, respectively.
    In {\figem a},{\figem b}, a random subsample of 256 galaxies is shown 
    for clarity.
    The channels that dominate the growth of a BH emerge sequentially 
    in time, and are expected to produce different observational
    signatures as the galaxy ecosystem evolves from early to late stages.
    See \S\ref{sssec:growth-channels} for details.}
    \label{fig:growth-examples}
\end{figure*}

\begin{figure*} \centering
    \includegraphics[width=0.975\textwidth]{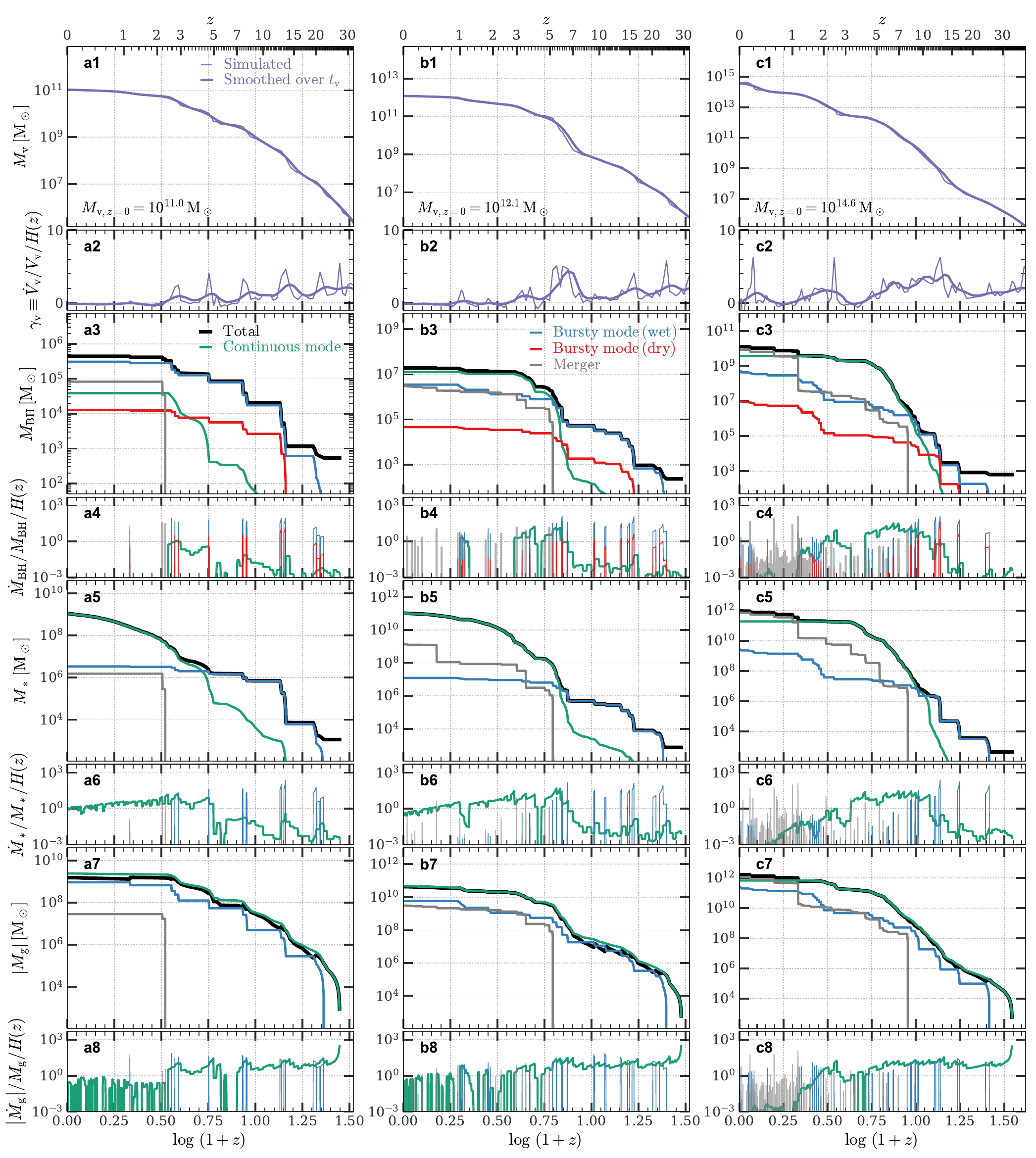}
    \caption{{\figem Evolution of different mass components and
    contributions from different channels for three example galaxies.}
    Columns {\figem a}--{\figem c}, 
    three example central galaxies (the same galaxies as 
    the colored ones in Fig.~\ref{fig:growth-examples}a,b,d), 
    respectively, predicted by our model,
    with their $M_{{\rm v}, z=0}$ indicated in the top row.
    For each component of mass (halo mass, $M_{\rm v}$;    
    BH mass, $M_{\sscBH}$; 
    stellar mass, $M_*$; 
    gas mass, $M_{\rm g}$), we show the cumulative growth and specific 
    growth rate in two rows, respectively.
    For $M_{\rm v}$, the simulated growth history and that smoothed over 
    $t_{\rm v}$ are both shown.
    Curves with different colors represent the growth contributed by different
    channels (continuous mode, wet and dry channels in bursty mode, and 
    merger). 
    The total growth, obtained by summing the contributions
    from all the channels (including the seeds), 
    is shown in {\figem black}.
    The contribution of nuclear burst to $M_{\rm g}$ is negative due to the 
    consumption of gas, so we show the absolute value.
    The growth rates by nuclear burst (both wet and dry) and merger are 
    evaluated by averaging over $10\Myr$, given their bursty nature.
    See \S\ref{sssec:growth-channels} for the details.}
    \label{fig:growth-channels}
\end{figure*}

Our model is built on the seeding and the multiple channels of BH growth  
and star formation (see \S\ref{ssec:growing-main}). 
Here we show the contribution of these channels to the growth of individual galaxies and 
to the cosmic history of the galaxy population. As we will show, the evolution pathways of galaxies 
also provide a tool for galaxy archaeology and futurology that links observed galaxies across cosmic time.

\subsubsection{Growth channels}
\label{sssec:growth-channels}


Fig.~\ref{fig:growth-examples} shows the evolution of individual central 
galaxies in the $M_{\sscBH}$-$M_*$ plane. 
In each range of $M_{{\rm v}, z=0}$, a representative galaxy is highlighted,
with each colored segment indicating the channel that contributes the most 
to the growth of $M_\sscBH$ during a time step.
In more detail, Fig.~\ref{fig:growth-channels} shows the growth of each 
mass component, $M_{\rm X}$ ($X =$ halo, BH, stars or gas),
for three of the highlighted galaxies in Fig.~\ref{fig:growth-examples}.
The specific growth rate, $\dot{M}_{\rm X}/M_{\rm X}/H(z)$, for BH, stars and gas, 
and $\gamma_{\rm v}$ for halo is shown for reference.
The contribution of growth from each channel is shown separately 
and in total.

Most dwarf-size galaxies ($M_{{\rm v}, z=0} \approx 10^{11}\Msun$) are 
seeded between $z=20$ and $30$, roughly the duration when
the cosmic seeding activity peaks (Fig.~\ref{fig:seed-z-func}a).
The highlighted galaxy (Fig.~\ref{fig:growth-examples}a) has
a seed mass of $M_{\sscBH,{\rm seed}} \lesssim 10^{3}\Msun$ and a DCBH 
flavor. Dwarfs are born in the $M_{\sscBH}$-$M_*$ plane
close to the line of $M_{\sscBH}$:$M_* = 1$:$1$, due to the 
top-heavy IMF assumed for the Pop-III stars.
After seeding, gas accretion into the SGC is efficient due to the fast 
halo assembly (Fig.~\ref{fig:growth-channels}a7,a8).
However, the over-massive BH produces efficient AGN feedback that, 
together with the efficient stellar feedback in a low-mass halo, 
prevents the accumulation of star-forming gas and quenches star formation 
and BH growth at $z\gtrsim 20$
(Fig.~\ref{fig:growth-channels}a3--a6).
The continuous mode contributes negligibly to the growth of 
$M_\sscBH$ and $M_*$ at $z \gtrsim 10$ (with a specific growth rate of 
$\lesssim 0.1$). 
Frequent excursions of the halo assembly (Fig.~\ref{fig:growth-channels}a2)
globally disturb the galaxy, cause episodic formation of SNF nuclei,
trigger nuclear bursts, and drive rapid rises of $M_{\sscBH}$ and 
$M_*$ at $z \approx 10$--$20$.
The wet channel in the bursty mode dominates the total growth at $z \gtrsim 5$
and keeps $M_\sscBH / M_* \approx 1/10$.
At $z \approx 5$, the galaxy has reached $M_* \approx 10^{6}\Msun$ 
and $M_{\sscBH} \approx 10^{5}\Msun$.
Thus, the evolution during the bursty stage has two distinct features: 
\begin{enumerate}[topsep=0pt,parsep=0pt,itemsep=0pt]
    \item the back-and-forth switch between star-forming and quiescent 
    states;
    \item the formation of a compact, globular-like object expected from 
    the dominance of the nuclear star formation within the SGC.
\end{enumerate}
Our model thus predicts that the formation of {\texteem the first generation of quiescent 
galaxies} is as early as $z\approx20$--$30$ soon after the seeds are bred, 
that the population of quiescent galaxies is ubiquitous at $z \gtrsim 10$ 
as they fall into gaps between nuclear bursts, 
and that galaxies with {\texteem compact morphology and prominent NSCs} 
are common in that era.

At $z \lesssim 5$, the bursty mode diminishes in the dwarf due to the 
less frequent halo excursion and the large ratio of $M_\sscBH/M_{\rm nuc}$.
A few nuclear bursts, each growing $M_\sscBH$ ($M_*$) by about $10^5\Msun$ 
($10^6\Msun$), are seen at $z = 2$--$3$ when halo excursions
are triggered by halo-halo mergers. These mergers also bring satellites 
that later merge into the central, contributing small amounts of 
$M_\sscBH$ and $M_*$ through the merger channel.
The continuous mode dominates the growth of $M_*$ at $z \lesssim 5$ and keeps the 
galaxy in the star-forming main sequence (with the specific SFR $\gtrsim 1$).
The specific growth rate of $M_\sscBH$ by the continuous mode 
never exceeds $1$ due to the regulation of the BH and stellar 
feedback, and $M_{\sscBH} / M_*$ is reduced to $\sim 1/100$ at 
$z \approx 2$, a value comparable to local massive ETGs (Fig.~\ref{fig:growth-examples}a). 
At $z \lesssim 2$, the halo assembly transits to a slow phase,
causing the galaxy to transit to a dynamically cold disk.
BH growth is quenched due to inefficient viscosity
to drive inward gas migration, and a horizontal path of the galaxy is seen in 
the $M_{\sscBH}$-$M_*$ plane (Fig.~\ref{fig:growth-examples}a).
The paths of dwarfs end at $M_{\sscBH}/M_* \lesssim 1/1000$ at $z=0$, intersecting 
with the upper bounds of the observational data.
Since nuclear bursts dominate the overall growth of $M_\sscBH$ in dwarfs,
{\textem precise measurements of the masses of intermediate-mass black holes (IMBHs) 
in dwarfs may put stringent constraints on the physical processes 
involved in the bursts}.

MW-sized galaxies ($M_{{\rm v}, z=0} \approx 10^{12}\Msun$) follow a growth 
pattern similar to that of dwarfs, but with systematic shifts in the values of 
physical properties.
The seeding in an MW-sized galaxy is earlier due to the earlier formation of 
the host halo, and the chance to breed a Pop-II seed is lower due to the reduced
intersection with the Pop-II regime in the seeding atlas 
(Fig.~\ref{fig:seeding-atlas}a).
The post-seeding growth is still dominated by the wet channel 
in the bursty mode (Fig.~\ref{fig:growth-examples}b; 
Fig.~\ref{fig:growth-channels}b3,b5), but
the continuous mode starts to dominate the growth of $M_*$ at 
an earlier epoch ($z \approx 5$--$10$).
The late-time growth of $M_\sscBH$ at $z \lesssim 5$
is mainly contributed by the continuous mode
(Fig.~\ref{fig:growth-channels}b3), indicating
a stronger resistance to feedback owing to the deeper potential well of the host 
halo -- a conclusion also reached 
by hydro simulations \citep{habouzitSupermassiveBlackHoles2021,
liPhysicalProcessesCoevolution2025,bowerDarkNemesisGalaxy2017} 
and analytical estimations (\citealt{bowerDarkNemesisGalaxy2017};
\citealt{hongDynamicalHotnessStar2023}; \citetalias{moTwophaseModelGalaxy2024}).
The transition of the host halo to slow assembly is earlier 
(at $z \approx 3$; Fig.~\ref{fig:growth-channels}b2), and
the high star formation efficiency (SFE) in MW-mass disks
\citep[e.g.][]{yangConstrainingGalaxyFormation2003,
yangEvolutionGalaxyDarkMatter2012,behrooziAVERAGESTARFORMATION2013},
together with the cessation of BH growth, drives a horizontal migration 
in the $M_{\sscBH}$-$M_*$ plane (Fig.~\ref{fig:growth-examples}b).
Such a migration is significant among all galaxies,
results in a minimal ratio of $M_{\sscBH} / M_* \approx 1/10000$,
and causes the $M_{\sscBH}$-$M_*$ relation of LTGs to systematically 
shift to the right compared to that for ETGs, 
as seen also in observations (Fig.~\ref{fig:growth-examples}b). 

The brightest cluster galaxies (BCGs; $M_{{\rm v}, z=0} \gtrsim 10^{13}\Msun$)
follow the general growth pattern found above, but again with systematic 
shifts in the values of physical properties. 
However, a distinct, merger-driven evolution in $M_\sscBH$ and $M_*$
appears in their late stages (Fig.~\ref{fig:growth-examples}d). 
The host halo of the example galaxy remains in the fast phase
throughout the history (Fig.~\ref{fig:growth-channels}c3,c5), 
drives effective growth of $M_{\sscBH}$ in the continuous mode, 
causes early quenching by BH feedback at $z \approx 1.5$,
and leaves mergers as the only channel to grow $M_\sscBH$ and $M_*$ 
at later times.
A similar conclusion for the dominance of mergers at low $z$ was reached by 
\citet[see their figs.~2 and 10]{liPhysicalProcessesCoevolution2025} 
using a cosmological hydro simulation \citep[see also][]{weinbergerSupermassiveBlackHoles2018,
habouzitSupermassiveBlackHoles2021,maInvestigatingResidualsRelation2025}.

One interesting conclusion is that BCGs are unique among central galaxies in that all 
the channels of BH growth are relevant sequentially in the history and 
become dominant in some stage (Fig.~\ref{fig:growth-examples}d). 
{\texteem Dynamical hotness} in each of the 
stages plays a fundamental role in maintaining the growth of the BH: 
the fast assembly of halo delays seeding and elevates seed mass;
temporary excursions of halo assembly in the post-seeding era induce
global disturbances that trigger episodic nuclear bursts and super-Eddington 
accretion; accretion in the fast phase sustains 
a dynamically hot SGC and a continuous mode of BH growth;
once the galaxy is quenched, the large number of satellites
that have been brought in by the fast assembly of the halo 
sustain the growth of the BH through dry mergers.
The sequential dominance of these channels results in a high but constant ratio of 
$M_{\sscBH} / M_* \approx 1/100$ throughout the post-seeding history
of a BCG, and in a tight $M_{\sscBH}$-$M_*$ relation 
with a slope of $\approx 1$ (Fig.~\ref{fig:growth-examples}c and d). 

In summary, Figs.~\ref{fig:growth-examples} and \ref{fig:growth-channels}
show that the dominant channels sequentially emerging in the growth history of a 
central galaxy are the nuclear burst (wet and dry channels), the continuous  
growth within the self-gravitating gas cloud/disk (with a bursty-to-continuous 
transition mostly at $z \approx 5$--$15$), 
and the ex-situ growth via merger (most significant in BCGs after quenching). 
This sequence is universal for all galaxies, although some of the later channels 
may be absent due to, e.g. dynamical 
transition to a cold disk and infrequent merger in a low-mass halo.

\subsubsection{The cosmic histories of black-hole accretion and star formation}
\label{sssec:cosmic-bh-sf-history}

\begin{figure*} \centering
    \includegraphics[width=0.95\textwidth]{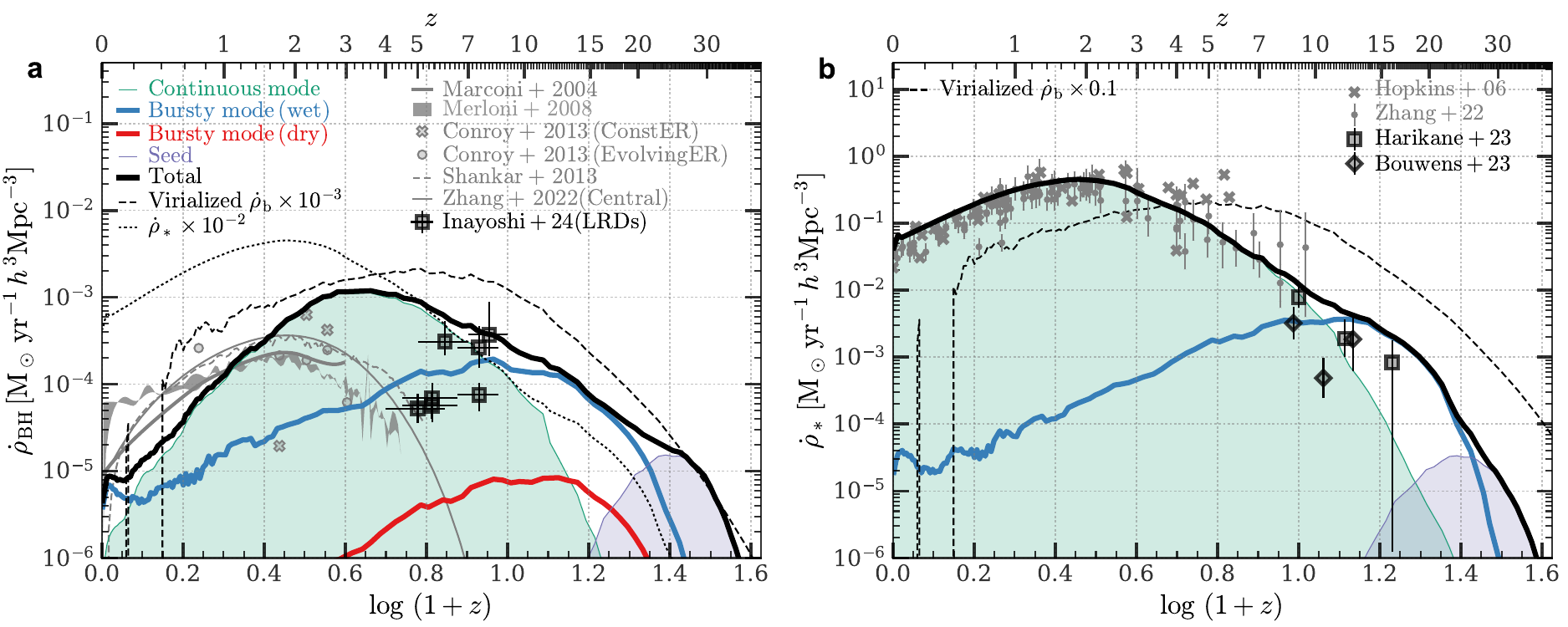}
    \caption{{\figem Evolution of cosmic black hole accretion rate (BHAR)  
    density and star formation rate (SFR) density}.
    {\figem a}, cosmic BHAR density ($\dot{\rho}_{\sscBH}$).
    {\figem b}, cosmic SFR density ($\dot{\rho}_*$).
    Both densities are defined as the mass formed per unit comoving volume
    per unit cosmic time.
    In each panel, {\figem black thick} curve represents the total rate
    predicted by our model,
    while colored curves show the contributions from
    different channels ({\figem green}, continuous mode; 
    {\figem blue}, wet channel in bursty mode; 
    {\figem red}, dry channel in bursty mode). 
    The contribution from the seeding procedure is also 
    included and shown by {\figem purple} curve.
    For comparison, we show a fixed fraction of the growth rate of 
    the virialized baryon mass density (virialized $\dot{\rho}_{\rm b}$; 
    see Eq.~\ref{eq:virialized-rho-b-dot} for definition).
    We also scale $\dot{\rho}_*$ by a factor of $10^{-2}$ and show it 
    in {\figem a}. For $\dot{\rho}_{\sscBH}$, 
    we show the empirical results calibrated by observations obtained by a number 
    of studies, including \citet{marconiLocalSupermassiveBlack2004}, 
    \citet{merloniSynthesisModelAGN2008}, 
    \citet[with either constant or redshift-dependent Eddington ratio 
        distribution assumed]{conroySimpleModelQuasar2013},
    \citet{shankarAccretiondrivenEvolutionBlack2013} and
    \citet[including only central BHs]{zhangTrinitySelfConsistentlyModeling2022},
    and the results compiled by \citet{inayoshiBirthRapidlySpinning2024} 
    based on recent JWST observations for the LRDs
    \citep{akinsCOSMOSWebOverabundancePhysical2025,kokorevCensusPhotometricallySelected2024,mattheeLittleRedDots2024,greeneUNCOVERSpectroscopyConfirms2024}.
    For $\dot{\rho}_*$, we show the observational data points 
    compiled by \citet{hopkinsNormalizationCosmicStar2006} 
    and \citet{zhangTrinitySelfConsistentlyModeling2022}, and 
    recent observational results obtained by
    \citet{harikaneComprehensiveStudyGalaxies2023}
    and \citet{bouwensEvolutionUVLF2023} based on JWST.
    This figure shows that the cosmic histories of BH accretion and 
    star formation can be divided into three sequential eras:
    the seeding era ($z \gtrsim 20$), the bursty era ($5 \lesssim z \lesssim 20$)
    and the continuous era ($z \lesssim 5$), with each era 
    dominated by a distinct channel of growth.
    See \S\ref{sssec:cosmic-bh-sf-history} for details.
    }
    \label{fig:growth-cosmic-history}
\end{figure*}

The universal sequence of the dominant channels in the histories of 
individual galaxies implies that the cosmic history of galaxy formation 
(BH accretion or star formation) is also sequentially dominated 
by these channels.
This conclusion can be reached by considering the cosmic history as a 
convolution of the halo mass function at $z = 0$ and the histories of 
individuals with different $M_{{\rm v}, z=0}$, plus some contributions 
from satellite galaxies.
The transition epoch of the cosmic history from one channel to the next
is thus a weighted average of the transition epochs of individuals.
This leads to the expectation that the cosmic history can be
divided into the following eras based on the driving channel:
\begin{enumerate}[topsep=0pt,parsep=0pt,itemsep=0pt]
    \item {\texteem The seeding era} 
    ($z \gtrsim 20$, during which 
    most seeds have been bred; see \S\ref{sssec:cosmic-seeding-history} 
    and Fig.~\ref{fig:seed-z-func}), 
    in which the cosmic history is dominated by the seeding process 
    in mini-halos (formation of Pop-III stars and BH seeds).
    Over-massive BHs with $M_{\sscBH}/M_* \approx 1$ are expected.
    The latest seeding events in our sample
    are at $z \approx 5$, in the progenitors of dwarfs
    (see Fig.~\ref{fig:seeding-atlas}). 
    The products of galaxy formation in this era constitute only a small 
    fraction of the total $M_\sscBH$ and $M_*$ 
    to be observed at $z \approx 0$, given the low masses of  
    Pop-III star clusters. If star formation stops in a mini-halo
    after seeding, the remnant would be an ultra-faint dwarf.
    The search for such a population in the low-$z$ universe is thus challenging.
    \item {\texteem The bursty era} 
    ($5\lesssim z \lesssim 20$, before the 
    transition of dwarf-size galaxies to the continuous mode; see 
    \S\ref{sssec:growth-channels} and Figs.~\ref{fig:growth-examples} and 
    \ref{fig:growth-channels}), 
    in which the BH accretion and star formation take place mostly as 
    episodic nuclear bursts.
    Compact and concentrated morphology, dominance of NSCs,
    and over-massive BHs with $M_{\sscBH}/M_* \approx 1/10$
    are expected for galaxies in this era.
    Quiescent galaxies are expected to be ubiquitous  
    between the gaps of nuclear bursts.
    \item {\texteem The continuous era} ($z \lesssim 5$), in which both the 
    BH accretion and the star formation are dominated by the continuous mode.
    Galaxies in this era first appear as bulges and may later transit to form 
    extended disk components. 
    Super-Eddington accretion is much less frequent, and
    the ratio of $M_{\sscBH}/M_*$ is reduced to about $1/100$--$1/10000$,
    depending on the dynamical hotness of the galaxy.
\end{enumerate} 

Fig.~\ref{fig:growth-cosmic-history} shows the evolution of the cosmic BH accretion 
rate (BHAR) density, $\dot{\rho}_\sscBH$, and the SFR density, 
$\dot{\rho}_*$, defined respectively as
\begin{align}
    \dot{\rho}_\sscBH &= \frac{1}{V_{\rm u}} \sum_{i} \dot{M}_{\sscBH,i}
    \,,\\
    \dot{\rho}_* &= \frac{1}{V_{\rm u}} \sum_{i} \dot{M}_{*,i} / (1-R)
    \,,
\end{align}
where $R = 0.4$ is the returned mass fraction given by the Chabrier IMF 
\citep{bruzualStellarPopulationSynthesis2003} 
assumed in our model for Pop-I/II stars, 
and the summations are over all galaxies predicted by 
our model in the simulated volume, $V_{\rm u}$, in each given snapshot.
Contributions from different channels are shown in total
and separately. For completeness, the contribution from seeding is 
included in the total densities.
For reference, we also show the growth rate of the `virialized' baryon mass 
density, defined as 
\begin{equation}
    \text{Virialized } \dot{\rho}_{\rm b} \equiv \frac{f_{\rm b}}{V_{\rm u}} \dv{}{t}
    \sum_i M_{{\rm v}, i}
    \,,\label{eq:virialized-rho-b-dot}
\end{equation} 
scaled down by factors of $10^{-3}$ and $10^{-1}$ in the two panels 
of Fig.~\ref{fig:growth-cosmic-history}, respectively,
and the summation is over all halos in our sample at each given snapshot. 
The cosmic history and the transition between the eras follow the 
expectation above, with the following details worth noting.

At $z \gtrsim 10$, the contribution of the bursty mode
to $\dot{\rho}_\sscBH$ rises rapidly with time, due to
the fact that halos accrete rapidly (with $\gamma_{\rm v} \geq 1$
for $M_{{\rm v},z=0} \gtrsim 10^{9.5}\Msun$; see Fig.~\ref{fig:m-v-cool-fit})
and undergo frequent excursions, and densities of entire SGCs are close to the
supernova-free threshold (Eq.~\ref{eq:n-sgc}).
About $1/10,000$--$1/1000$ of the total virialized baryon mass
is accreted into BHs (Fig.~\ref{fig:growth-cosmic-history}a).
All the above factors promoting the bursty growth
diminish with time. The contribution of the bursty mode to 
$\dot{\rho}_\sscBH$ reaches a plateau of 
$\approx 10^{-4}\msunperyr h^3{\rm Mpc}^{-3}$ 
at $z\approx 10$--$5$, is surpassed by that of the continuous mode at 
$z \approx 8$, and constitutes 
$\lesssim 1/10$ of the total $\dot{\rho}_\sscBH$
at $z \lesssim 5$. 

In the continuous era,
$\dot{\rho}_\sscBH$ reaches a peak 
of $\approx 10^{-3}\msunperyr h^3{\rm Mpc}^{-3}$
at $z \approx 3$--$4$, and rapidly
declines at $z \lesssim 2$ as most halos have now transited to the 
slow phase (see Fig.~\ref{fig:halo-mahs}b) in which the BH growth 
is quenched.
$\dot{\rho}_\sscBH$ is less than $1\%$ of $\dot{\rho}_*$
in this era, as expected from the low
$M_{\sscBH}/M_*$ of the LTGs (Fig.~\ref{fig:growth-examples}).
At $z \approx 0$, the bursty mode appears outpacing the continuous mode
again, but the rate of nuclear bursts and the growth of BHs during each burst 
are small, as seen from the large fluctuations in, and the small values of 
$\dot{\rho}_\sscBH$.

The evolution of $\dot{\rho}_*$ (Fig.~\ref{fig:growth-cosmic-history}b) 
has an evolution pattern similar to that of $\dot{\rho}_\sscBH$,
albeit with different transition epochs between the eras 
and different relative contributions from different channels.
The bursty mode dominates $\dot{\rho}_*$ at $z \gtrsim 10$, converting 
$\lesssim 1\%$ of the virialized baryon mass into stars.
Such an efficiency is comparable to those of dwarfs, but is much 
lower than that in a MW-sized galaxy at $z \approx 0$,
indicating a strong depletion of gas by AGN feedback in nuclear bursts 
(see Appendix~\ref{app:sssec:nuclear-burst-basic-eq}).
The dominance of the continuous mode in $\dot{\rho}_*$ starts at $z \approx 10$, 
slightly earlier than that in $\dot{\rho}_\sscBH$, 
due to the competition between BH accretion and star formation 
in nuclear bursts and the difference of feedback in regulating 
BH growth and star formation in the continuous mode.
The subsequent evolution of $\dot{\rho}_*$ follows the observations,
with a single peak at $z\approx 2$, the `cosmic noon'.
The virialized $\dot{\rho}_{\rm b}$ shows a decline with time due to the 
slowdown of the halo assembly. The values of $\dot{\rho}_{\rm b}$ and $\dot{\rho}_*$ 
are comparable at $z<2$, indicating that some of the baryons ejected by previous feedback 
processes can be recycled back to galaxies to fuel star formation.

Recent JWST observations have extended the constraints on 
$\dot{\rho}_\sscBH$ and $\dot{\rho}_*$ to redshifts higher than previously possible. 
Remarkably, a population of `little red dots' (LRDs), featured by their 
peculiar V-shaped spectral energy distributions (SEDs), extremely compact 
morphologies and large population density (about $100\times$ that of UV-selected 
quasars), has been observed at $z \gtrsim 4$ \citep{akinsCOSMOSWebOverabundancePhysical2025,barroComprehensivePhotometricSelection2026,greeneUNCOVERSpectroscopyConfirms2024,kokorevCensusPhotometricallySelected2024,mattheeLittleRedDots2024,pacucciMildlySuperEddingtonAccretion2024,yueStackingXRayObservations2024,chenHostGalaxyIf2025,labbeUNCOVERCandidateRed2025,liLittleRedDots2025,pergerDeepSilenceRadio2025}. 
The compact morphologies and the presence of broad emission lines 
suggest that they may harbor over-massive BHs with 
$M_{\sscBH}/M_* \gtrsim$ a few percent \citep{chenHostGalaxyIf2025}.
This, if confirmed, will increase the cosmic $\dot{\rho}_\sscBH$
to be significantly higher than extrapolations from low-$z$ results,
as seen by comparing the black scatters and grey curves/scatters 
in Fig.~\ref{fig:growth-cosmic-history}a.
Interestingly, the predicted $\dot{\rho}_\sscBH$ of the bursty mode
is consistent with the observational results based on the LRDs, thus offering
a potential explanation of LRDs as the mixture of NSCs and over-massive BHs 
formed in nuclear bursts.

The JWST observations of the cosmic $\dot{\rho}_*$ also suggest a moderate 
enhancement of the SFE at high $z$
\citep{naiduTwoRemarkablyLuminous2022,chenMassiveDarkMatter2023,
yungAreUltrahighredshiftGalaxies2024,harikaneComprehensiveStudyGalaxies2023,
xiaoAcceleratedFormationUltramassive2024,sabtiInsightsHSTUltramassive2024,
wangJWSTMIRIReveals2025}. 
The predictions of our model at $z \gtrsim 10$ match the JWST results 
(Fig.~\ref{fig:growth-cosmic-history}b), even without a calibration 
using these observations.
Given the dominance of the bursty mode at $z \gtrsim 10$,
{\textem precise measurements of the cosmic $\dot{\rho}_\sscBH$ and 
$\dot{\rho}_*$ in the bursty era may thus offer 
clean constraints on processes that affect the properties of nuclear bursts.
}

\subsubsection{Effects of seeding methods}
\label{sssec:growth-effect-seeding}

\begin{figure*} \centering
    \includegraphics[width=0.785\textwidth]{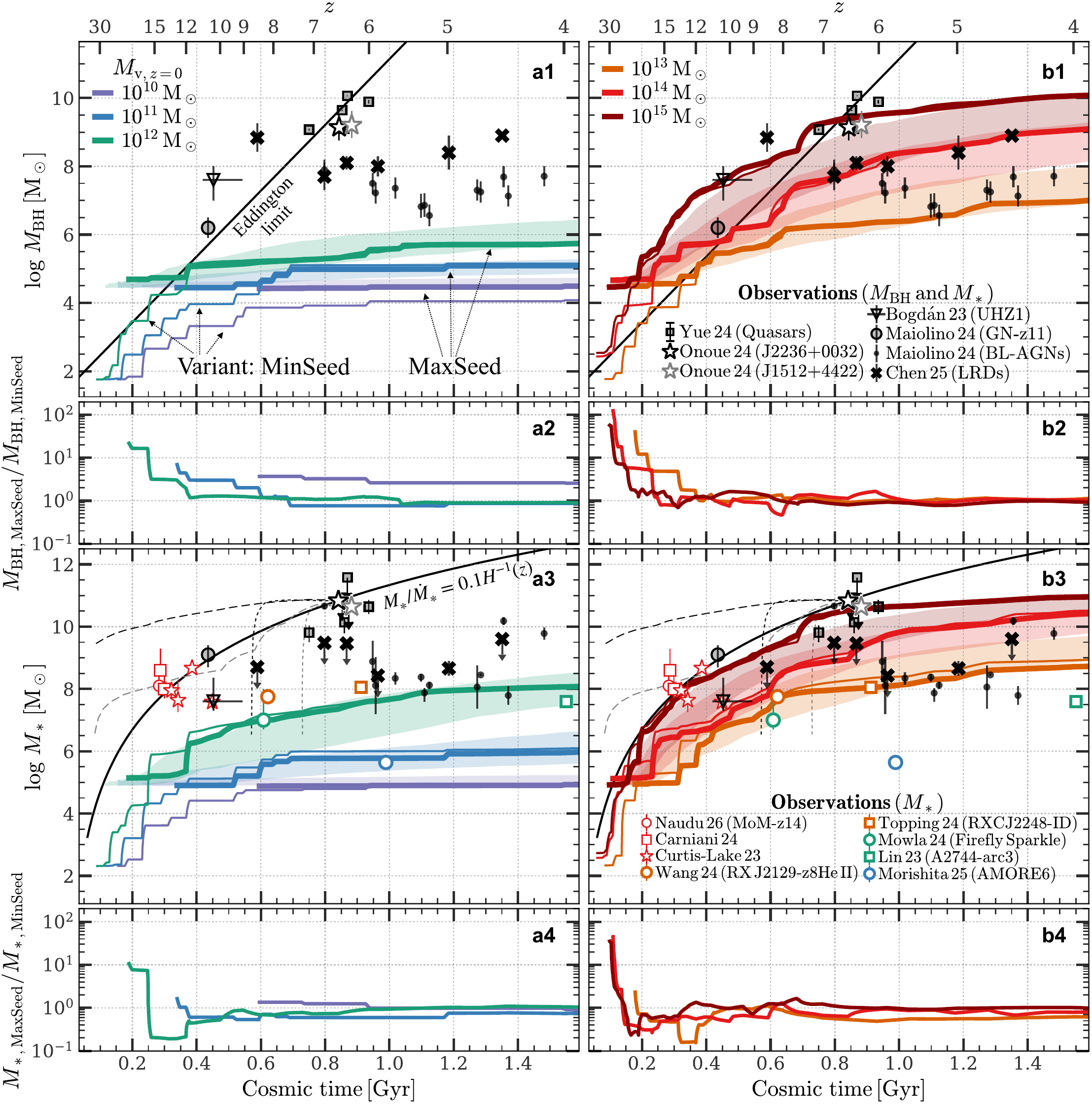}
    \caption{{\figem Idealized and controlled experiments 
    showing effects of seeding methods 
    and demonstrating the use of model predictions for archaeology and futurology}.
    Each {\figem colored} curve shows the growth history predicted
    by our model for the central galaxy in a halo
    with a given mass at $z = 0$. The results are separated into two columns 
    for clarity: {\figem a},
    $M_{{\rm v}, z=0}=10^{10}$, $10^{11}$ and $10^{12}\Msun$; 
    {\figem b}, $M_{{\rm v}, z=0}=10^{13}$, $10^{14}$ and $10^{15}\Msun$.
    For each halo, the {\textem MinSeed} and {\textem MaxSeed} variants of our model are applied, 
    and the results are shown by {\figem thin} and {\figem thick} curves,
    respectively.
    These two variants differ only in the seeding method, and produce the
    minimal and maximal masses of BH seeds, respectively, among 
    all the variants in the entire parameter space.
    All the other ingredients of the model are strictly 
    the same as those in the {\textem Default} variant, thus isolating the 
    effects of seeding methods on the growth of galaxies.
    Each row shows a property as a function of cosmic time: 
    {\figem 1}, BH mass ($M_{\rm BH}$); 
    {\figem 2}, the ratio of $M_{\rm BH}$ between {\textem MaxSeed} and {\textem MinSeed};
    {\figem 3}, stellar mass ($M_*$); 
    {\figem 4}, the ratio of $M_*$ between {\textem MaxSeed} and {\textem MinSeed}.
    {\figem Shaded} region covers the 
    $1$-$\sigma$ ($16^{\rm th}$--$84^{\rm th}$) percentiles, at each $z$, 
    of the masses predicted by {\textem MaxSeed} for
    a sample of $500$ halos with $M_{{\rm v}, z=0}$ the 
    same as the example halo.
    In {\figem a1} and {\figem b1}, 
    a tilted {\figem black line} shows the path of $M_\sscBH$
    expected for a BH growing at the Eddington rate
    (with a radiative efficiency of $\epsilon_{\rm r} = 0.1$, 
    i.e. $\dot{M}_\sscBH = 10 M_\sscBH / t_{\ssc Sal} $ in our 
    notation; see Appendix~\ref{app:sssec:Bondi-accretion-rate}).
    In {\figem a3} and {\figem b3},
    a {\figem black solid curve} shows the path of $M_*$ 
    expected for a galaxy with a specific star formation rate 
    (${\rm sSFR} \equiv \dot{M}_*/M_*$) 
    equal to $t_{\rm v}^{-1} = 10H(z)$.
    For comparison, we show the observations of $M_{\sscBH}$ and $M_*$
    for high-$z$ objects
    \citep{yueEIGERCharacterizingHost2024,
    onouePoststarburstPathwayFormation2025,
    bogdanEvidenceHeavyseedOrigin2023,maiolinoSmallVigorousBlack2024,
    maiolinoJADESDiversePopulation2024,chenHostGalaxyIf2025}.
    In {\figem a3} and {\figem b3},
    the long- and short-dashed curves connecting each post-starburst (PSB) galaxy
    represent the star formation histories, assuming non-parametric and
    delayed-$\tau$ forms, respectively, obtained by their 
    spectrophotometric fitting method.
    Systems with only $M_*$ measurements are also shown 
    \citep{naiduCosmicMiracleRemarkably2026,
    carnianiSpectroscopicConfirmationTwo2024,
    curtis-lakeSpectroscopicConfirmationFour2023,wangStrongHeII2024,
    toppingMetalpoorStarFormation2024,
    mowlaFormationLowmassGalaxy2024,
    linMetalenrichedNeutralGas2023,
    morishitaPristineMassiveStar2025}.
    Monte Carlo trees are used as input of the model for this figure to 
    ensure a large sample size.
    This figure shows that the BHs produced by different seeding methods 
    converge quickly to a similar $M_\sscBH$ owing to the self-regulated
    post-seeding growth, thus ensuring the robustness of the 
    predictions (see \S\ref{sssec:growth-effect-seeding}).
    Matching the observed galaxies with predicted 
    evolutionary tracks can be performed to reconstruct the formation 
    histories of the galaxies, and to forecast their future growth
    (see \S\ref{ssec:arch-and-forecast}).
    }
    \label{fig:growth-effect-seeding}
\end{figure*}

The small contributions of BH seeds and Pop-III stars to 
$\dot{\rho}_\sscBH$ and $\dot{\rho}_*$ (Fig.~\ref{fig:growth-cosmic-history})
imply that these earliest products of galaxy formation are soon mixed into,
and outshone by the $M_\sscBH$ and $M_*$ formed after seeding.
Given the difficulty in targeting the seeds directly, 
we examine whether seeds bred by different methods can preserve some 
observable properties, and whether specific predictions of the model 
are sensitive to the seeding method.

Here we perform an {\texteem idealized and controlled experiment} for each target halo 
using two extreme variants of our model, {\textem MinSeed} and {\textem MaxSeed}, to bracket the seed masses 
expected from variants covering the entire parameter space. 
Specifically, {\textem MinSeed} closes all the `heating' channels that may delay seeding, 
breeds seeds once halos reach the $\Hmol$-cooling 
threshold (Eq.~\ref{eq:m-v-cool-H2-main}), and yields minimal seed masses.
{\textem MaxSeed} forces the `heating' to be maximally effective, thus maximizing the delay of 
seeding and the seed masses (see Appendix~\ref{app:sec:model-variants} for details). 
All other ingredients are controlled to be the same
as the {\textem Default} variant, so that the effects of seeding method can be isolated.
To have a large halo sample covering a large range of $M_{\rm v}$,  here we use halo merger trees 
generated by the Monte-Carlo method described in Appendix~\ref{app:ssec:extension-merger-tree}. 
For any target $M_{{\rm v},z=0}$, we randomly generate a merger tree that is rooted in the halo.
Since we focus on galaxies at high $z$ when the difference
between the variants is significant (see below), mergers do not play 
a significant role in the growth of $M_{\sscBH}$ and $M_*$ (see \S\ref{sssec:growth-channels} and 
Fig.~\ref{fig:growth-examples}). Thus, we only retain the main branch of each tree
to save computational cost. This removes the `dry' contribution to $M_\sscBH$ 
and $M_*$ from side branches, i.e. masses contained in side branches and added to the 
main branch in mergers, but retains the `wet' contribution, i.e. masses 
added by the continuous and bursty modes due to mergers. The two extreme variants 
are then applied independently to the main branch of the same target halo.

Fig.~\ref{fig:growth-effect-seeding} shows the growth of $M_\sscBH$ and $M_*$
in halos of several $M_{{\rm v},z=0}$. The predictions of {\textem MaxSeed} and {\textem MinSeed}  
for an example halo in each sample of $M_{{\rm v},z=0}$ are shown by the thick and thin lines 
of the same color, respectively. To show the variance due to the diversity of
halo assembly, we apply the {\textem MaxSeed} variant to a sample of $500$ halos with the 
same $M_{{\rm v},z=0}$ and plot the 1-$\sigma$ range as the shaded region.   
In the {\textem MinSeed} variant, seeds are bred with $M_{\sscBH} \approx 10^2\Msun$,
mostly at $z \gtrsim 20$. In the {\textem MaxSeed} variant, seeding is delayed and 
seeds (mainly DCBHs) are bred with $M_{\sscBH}\gtrsim 10^4 \Msun$.
For the example of $M_{{\rm v},z=0} = 10^{10}\Msun$,
the delay is the most significant, to $z \approx 8$--$9$.
This again highlights the critical role of environments in 
modulating seeding, as seen in the seeding atlas (Fig.~\ref{fig:seeding-atlas}).

For halos with $M_{{\rm v},z=0} > 10^{10}\Msun$, post-seeding growth histories of $M_\sscBH$ 
predicted by the two variants are very similar. In {\textem MinSeed}, the early growth of each 
example BH, regardless of $M_{{\rm v}, z=0}$, shows repeated jumps with large amplitudes,
with each jump powered by super-Eddington accretion during a short period of $\lesssim 1\Myr$ 
in a nuclear burst. In contrast, the jumps in the early growth predicted by {\textem MaxSeed} have much smaller 
amplitudes so that the overall growth is much slower. Consequently,  
the original differences in $M_\sscBH$ between the two variants are almost completely 
erased shortly after seeding. However, the convergence in the predicted $M_\sscBH$
occurs at a redshift that depends on halo mass, because a halo with 
higher $M_{{\rm v}, z=0}$ is seeded earlier and has a faster post-seeding growth. 
In fact, the growth rates predicted by {\textem MinSeed} are around or above the Eddington limit
for halos with $M_{{\rm v},z=0} \gtrsim 10^{13}\Msun$, but below the Eddington limit 
for halos of lower masses. The growth of $M_*$ in the controlled experiments also follows 
the same convergent pattern as that of $M_\sscBH$, and the epoch of convergence 
is again earlier for higher $M_{{\rm v},z=0}$.

The convergence of the predicted post-seeding growth between the two model variants  
originates from the self-regulation of BH growth by the associated feedback.
In a nuclear burst, a larger $M_\sscBH$ implies a higher Eddington luminosity 
($L_{\ssc Edd}$), a faster depletion of the nuclear gas and a less efficient 
growth of $M_\sscBH$ and $M_*$ (see Appendix~\ref{app:ssec:nuclear-burst} and Fig.~\ref{fig:bh-engine-example}). 
In the continuous mode, a larger $M_\sscBH$ also causes a stronger
reduction in the accumulation rate of cold gas, thus limiting the growth 
of both $M_\sscBH$ and $M_*$ (see Eq.~\ref{eq:f-AGN}). Thus, post-seeding growth in both 
channels is expected to be more suppressed by a higher $M_{\rm {\sscBH},seed}$, 
which causes the difference in seeded mass to be erased by the growth.
Such self-regulated growth of BH has also been found in some  
hydro simulations \citep{liPhysicalProcessesCoevolution2025,
suSelfregulationHighredshiftBlack2025,sivasankaranAGNFeedbackIsolated2025},
albeit with a focus on growth in later epochs.

The fact that the predicted post-seeding evolution of $M_\sscBH$ and $M_*$
is insensitive to the adopted seeding method has important implications for 
using the model to interpret observations. The small difference in 
$M_\sscBH$ and $M_*$ between the two extreme variants after convergence implies that the 
predictions at low $z$ are robust against the seeding method. On the other hand, 
this also implies that it is difficult to use low-$z$ observations 
\citep[e.g.][]{dantonaGNz11WitnessingFormation2023,pacucciJWSTCEERSJADES2023,
onouePoststarburstPathwayFormation2025,jeonPhysicalPathwaysJWSTobserved2025}
to constrain the seeding method due to the degeneracy between seeding and growing.
The only exception is the case of $M_{{\rm v},z=0} \approx 10^{10}\Msun$, 
in which the convergence of $M_\sscBH$ is never reached. This suggests that a potential way to 
distinguish different seeding scenarios is to observe IMBHs hosted by (the progenitors of) 
dwarf galaxies. The main challenge here is that IMBHs may be too faint to be captured by 
current observations, as expected from their low masses 
($M_{\sscBH} \lesssim 10^5\Msun$) and low accretion rates (implied 
by the flat growth paths in Fig.~\ref{fig:growth-effect-seeding}a1 
at $z\lesssim 8$). Additional data, such as abundance patterns of 
relic stars, may provide useful information
\citep[e.g.][]{xingMetalpoorStarAbundances2023,skuladottirPairinstabilitySupernovaOrigin2024}. 

\subsubsection{Archaeology and futurology: linking black holes and galaxies across cosmic time}
\label{ssec:arch-and-forecast}

The controlled experiments can provide some prior constraints 
for model inferences from observations.
This idea is demonstrated below by recovering the growth histories of 
observed galaxies through galactic archaeology and model forecasting (referred 
to as futurology).
In Fig.~\ref{fig:growth-effect-seeding}, we overplot 
JWST-based measurements of $z$, $M_\sscBH$ and $M_*$
for individual galaxies.
By comparing the locations of the observed galaxies with the growth histories 
predicted by our model,
useful conclusions for the general population of observed galaxies 
can be drawn. All observational data fall into the converged regions predicted by
the {\textem MinSeed} and {\textem MaxSeed} variants, suggesting that these observations cannot 
be used to reconstruct the early growth history before the epoch of convergence 
and to distinguish seeding methods, as already mentioned in \S\ref{sssec:growth-effect-seeding}.
The forward forecast of the growth histories, although not sensitive to 
seeding methods, can take into account the diversity and fluctuations in the growth histories
of dark matter halos. For halos with $M_{{\rm v},z=0} = 10^{15}\Msun$, 
the $1$-$\sigma$ range of $M_\sscBH$ ($M_*$) is about 
$2\dex$ ($1\dex$) at $z \gtrsim 4$. The growth history of an individual 
galaxy can also fluctuate significantly, often exhibiting rapid jumps 
followed by flat periods due to the back-and-forth switch between star-forming 
and quiescent states (see \S\ref{sssec:growth-channels}).
Such large intrinsic diversity and fluctuation 
imply a high probability of catching extreme objects in observations,
such as those with $M_{\sscBH}/M_*$ much larger than the expectation from the 
mean $M_{\sscBH}$-$M_*$ relation
\citep{chenMassiveDarkMatter2023,boylan-kolchinStressTesting$Lambda$CDM2023,
maiolinoSmallVigorousBlack2024,pacucciRedshiftEvolutionRelation2024}.

Fig.~\ref{fig:growth-effect-seeding} demonstrates that
archaeological and forecasting information of individual galaxies can be
obtained, and objects observed at different redshifts may be linked together
to form a coherent picture of evolution.
The two massive, post-starburst (PSB) quasar-host galaxies observed at $z \approx 6$--$7$ 
\citep{onouePoststarburstPathwayFormation2025} appear 
to overlap with the predicted evolution of $M_\sscBH$ and $M_*$ of 
the example BCG with $M_{{\rm v},z=0} = 10^{15}\Msun$.
The observed $M_\sscBH$ and $M_*$ are slightly above the upper 
$1$-$\sigma$ ($84^{\rm th}$) percentile predicted for the 
BCG sample (shaded region), indicating 
that the two galaxies are massive outliers even among the 
BCG progenitors of Coma-size halos 
($M_{\rm v} = 10^{14.96}\Msun$; \citealt{okabeSubaruWeaklensingSurvey2014}).
The quasar-host galaxies observed at $z \gtrsim 6$
\citep{yueEIGERCharacterizingHost2024} 
occupy the ranges of $M_\sscBH$ and $M_*$ similar to those of the two PSB galaxies, 
suggesting that they have similar origins and formation pathways.
Among the sample of LRDs \citep{chenHostGalaxyIf2025}, three (at $z=4.46$, $6.34$ and $6.76$) 
are within the $1$-$\sigma$ ranges of $M_\sscBH$ and $M_*$ predicted for BCG 
progenitors with $M_{{\rm v},z=0} \approx 10^{14}\Msun$; two (at $z=4.96$ and $5.84$)
have $M_*$ (or upper bound) close to the example BCG progenitor with 
$M_{{\rm v},z=0} \approx 10^{13}\Msun$, but their $M_\sscBH$ are over-massive,
close to that for $M_{{\rm v},z=0} \approx 10^{14}\Msun$;
one (at $z=8.50$) has an obviously over-massive $M_\sscBH$, even compared with the $1$-$\sigma$ range of the BCG progenitors with
$M_{{\rm v},z=0} \approx 10^{15}\Msun$. The broad-line AGNs
\citep{maiolinoJADESDiversePopulation2024} are consistent with
BCG progenitors with $M_{{\rm v},z=0} \approx 10^{13}\Msun$, 
but some have under-massive $M_*$ that fall into the 1-$\sigma$ range of the
modeled galaxies with $M_{{\rm v},z=0} \approx 10^{12}\Msun$.
The tension between our predictions and the observed 
over-massive $M_\sscBH$ (under-massive $M_*$) for LRDs and 
faint AGNs thus indicates a bias in the broad-line based estimators, 
and lower $M_\sscBH$ may be expected for these objects 
(see e.g. \citealt{chenSupereddingtonAccretionBlack2026,lambridesCaseSuperEddingtonAccretion2026,greeneWhatYouSee2026}).

Towards earlier epochs, the galaxy GN-z11 at $z=10.60$ (\citealt{maiolinoSmallVigorousBlack2024};
see also \citealt{tacchellaJADESImagingGNz112023,jiJADESSmallBlue2025})
has an $M_*$ close to the example BCG with $M_{{\rm v},z=0} \approx 10^{15}\Msun$
(Fig.~\ref{fig:growth-effect-seeding}, column b), making it a viable progenitor for 
galaxies like the two PSBs and other quasar hosts.
Its $M_\sscBH$ is significantly under-massive, but still consistent with the
$1$-$\sigma$ range of BCG progenitors with $M_{{\rm v},z=0} \approx 10^{15}\Msun$.
A galaxy with such high $M_{\rm v}$ is predicted to have experienced 
repeated nuclear bursts by this epoch, and thus to have built a compact and 
massive NSC (see \S\ref{sssec:growth-channels} and Fig.~\ref{fig:growth-channels}).
This is consistent with the observational inference by e.g. \citet{dantonaGNz11WitnessingFormation2023}
that GN-z11 was formed out of pristine gas mixed with ejecta from asymptotic 
giant branch (AGB) stars in a massive GC or a NSC.
A similar conclusion can be reached for the galaxy UHZ1 at $z \approx 10.3$,
which hosts a heavily obscured but X-ray luminous quasar 
\citep{bogdanEvidenceHeavyseedOrigin2023},
and for the samples of Lyman-break galaxies at $z \gtrsim 10$--$14$ 
\citep{naiduCosmicMiracleRemarkably2026,carnianiSpectroscopicConfirmationTwo2024,
curtis-lakeSpectroscopicConfirmationFour2023}
whose $M_*$ align with the evolution predicted for BCG progenitors with 
$M_{{\rm v},z=0} \approx 10^{15}\Msun$.
Thus, their fates should be bright quasars at $z \gtrsim 6$ and BCGs of 
Coma-like halos at $z \approx 0$.

Observations for low-mass galaxies in the early Universe are more challenging, 
but have been achieved by JWST with strong lensing. 
In the third row of Fig.~\ref{fig:growth-effect-seeding}, we show three examples: 
`Firefly Sparkle', `A2744-arc3' and `AMORE6'.
The former two are occupied by compact star clusters,
and the latter is a galaxy with a compact morphology.
All three have a peculiar enrichment pattern:
Firefly Sparkle has ISM and stellar metallicities of about $1\%$ of 
solar values; A2744-arc3 has ISM metallicity that appears to be 
lower than that of the CGM; AMORE6 has an extremely low 
ISM metallicity of $\lesssim 0.1\%$ of the solar value.
The peculiarity in both morphology and metallicity of these galaxies
suggests that they are in the early stages of evolution, where 
an extended, smooth stellar body and well-enriched, well-mixed gas environment 
have not yet been established.
The observed stellar masses of these galaxies are consistent with the 
evolutionary paths of modeled galaxies 
with $M_{{\rm v},z=0} \approx 10^{11}$--$10^{12}\Msun$ 
(Fig.~\ref{fig:growth-effect-seeding}a3).
The prediction is that a large fraction of their stellar mass is formed
by nuclear bursts or in the form of globular clusters 
\citep[see their figs. 17 and 18]{chenTwophaseModelGalaxy2025},
which provides a plausible explanation for their peculiar properties.

The mismatch of individual galaxies between model predictions and 
observations is not unexpected, given that our model has not been calibrated 
to match the observations and that the observations themselves are subject to systematics, 
such as uncertainties in sample selection and SED fitting. 
Examples of such uncertainties are demonstrated in row 3 of
Fig.~\ref{fig:growth-effect-seeding} by the star formation histories 
of the two PSB galaxies shown in dashed curves. 
Their spectrophotometric fitting with two different prior forms
for the star formation histories (non-parametric and delayed-$\tau$) 
yields unconstrained results at high $z$, highlighting the challenges 
in observational data modeling.
Observational data on galaxy environments may provide additional information 
about the formation history, as demonstrated by using 2PCCFs
for galaxies with different structure parameters reported in 
a recent observational study \citep{zhangUnexpectedClusteringPattern2025}.
Similarly, our model can be implemented into a constrained N-body
simulation \citep[e.g.][]{wangELUCIDEXPLORINGLOCAL2016,stiskalekVelocityFieldOlympics2026}
in which large-scale dark-matter environments around individual halos 
are constrained, leaving the modeling of baryonic ($Z_{\ssc IGM}$) and 
radiative ($J_{\rm \sscLW,21}$) environments for the seeding less uncertain.

\subsection{The build-up of the black hole mass-stellar mass relation}
\label{ssec:mbh-ms-relation}

\begin{figure*} \centering
    \includegraphics[width=0.775\textwidth]{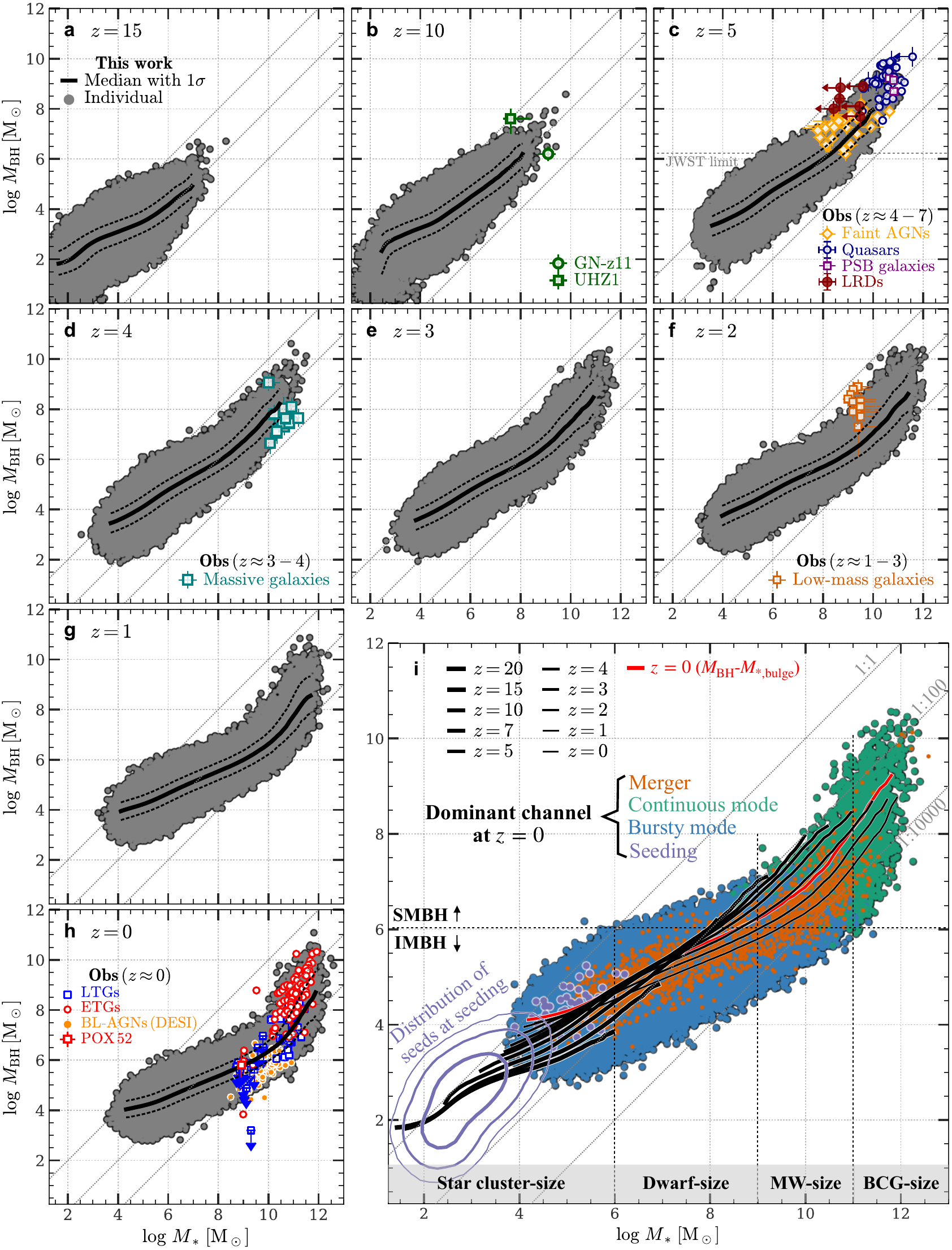}
    \caption{{\figem $M_{\sscBH}$-$M_*$ relations.}
    {\figem a}--{\figem h}, each showing the relation at a given redshift, 
    as labeled. 
    {\figem Grey dots} represent individual galaxies predicted by the 
    model, and
    {\figem black curves} are the running median and 
    1-$\sigma$ ($16^{\rm th}$--$84^{\rm th}$) percentiles of $M_\sscBH$ at given 
    $M_*$. 
    {\figem i}, a summary of the predicted median $M_{\sscBH}$-$M_*$ relations 
    at different redshifts ({\figem black curves}).
    For comparison, we also show the predicted median black hole 
    mass-stellar bulge mass ($M_\sscBH$-$M_{\rm *,bulge}$) relation 
    at $z=0$ ({\figem red curve}),
    and the distribution of BH seeds with their $M_{\rm BH}$ and $M_*$ evaluated at the seeding epochs 
    ({\figem purple contours}; from inner to outer, enclosing 
    $68\%$, $95\%$ and $99.7\%$ of the seeds, respectively). 
    Dots represent galaxies at $z = 0$, with color of each indicating the channel 
    that contributes the most to the $M_{\rm BH}$
    ({\figem purple}, seeding; 
    {\figem blue}, bursty mode; {\figem green}, continuous mode; 
    {\figem orange}, merger).
    {\figem Colored markers} in {\figem a}--{\figem h} show observations
    obtained/compiled by different studies \citep{maiolinoSmallVigorousBlack2024,
    bogdanEvidenceHeavyseedOrigin2023,pacucciJWSTCEERSJADES2023,
    harikaneJWSTNIRSpecFirst2023,ublerGANIFSMassiveBlack2023,
    maiolinoJADESDiversePopulation2024,chenHostGalaxyIf2025,
    onouePoststarburstPathwayFormation2025,carnallMassiveQuiescentGalaxy2023,
    izumiSubaruHighzExploration2021,
    stoneDetectionLowstellarmassHost2023,
    stoneUndermassiveHostGalaxies2024,
    yueEIGERCharacterizingHost2024,liPrevalentPopulationNormalmass2025,
    mezcuaOvermassiveBlackHoles2024,greeneIntermediateMassBlackHoles2020,
    puchaTriplingCensusDwarf2025,
    kawamuroCoevolutionNuclearStructure2024,
    sunIntermediatemassBlackHole2025}.
    A horizontal dashed line in {\figem c} indicates the 
    detection limit of JWST at $z \approx 5$.
    In all panels, the tilted grey lines indicate constant ratios of
    $M_{\sscBH}/M_*$. 
    A log-normal noise with a standard deviation of $0.3\dex$
    is added to the $M_{\sscBH}$ and $M_*$ of each modeled galaxy to mimic the
    observational uncertainties
    \citep[e.g.][]{behrooziAVERAGESTARFORMATION2013,kormendyCoevolutionNotSupermassive2013}.
    This figure demonstrates that the $M_{\sscBH}$-$M_*$ relation is 
    a redshift-dependent, multi-piece relation, with each piece driven 
    by a distinct channel of galaxy formation.
    See \S\ref{ssec:mbh-ms-relation} for the details.
    }
    \label{fig:scaling-evol}
\end{figure*}

The $M_\sscBH$-$M_*$ relation is observationally accessible and 
encodes the co-evolution of BHs and their host galaxies. It is thus interesting 
to see how multi-channel growth in our model shapes the $M_\sscBH$-$M_*$ 
relation over cosmic time. 
Fig.~\ref{fig:scaling-evol} shows the $M_\sscBH$-$M_*$ relations 
at different $z$, as well as $M_\sscBH$ and $M_*$ of individual galaxies
predicted by the {\textem Default} variant.
In the last panel, we summarize the build-up of the $M_\sscBH$-$M_*$ relation
predicted by our model by showing
(i) the distribution of seeds at their seeding epoch;
(ii) a collection of modeled relations at different $z$;
(iii) the distribution of individual galaxies at $z=0$, 
color-coded according to the channel that contributes the most to $M_{\sscBH}$.
In Appendix~\ref{app:sec:model-variants}, we explore the
$M_\sscBH$-$M_*$ relations in other model variants.

At $z = 15$, the $M_\sscBH$-$M_*$ relation shows two branches, 
with $M_{\sscBH}:M_* \approx 1:1$ and $M_{\sscBH}:M_* \approx 1:100$,
respectively, connected by a smooth transition at 
$M_* \approx 10^4\Msun$. 
All galaxies fall into the mass ranges of star clusters and 
dwarfs ($< 10^9\Msun$),
with a few extremely massive outliers reaching $M_*\approx 10^9$ and $M_{\sscBH} \gtrsim 10^7\Msun$.
At $z \gtrsim 10$, the two branches persist, but more galaxies have evolved 
to the upper limit of dwarf mass ($M_* = 10^9\Msun$). 
The galaxy GN-z11 ($z=10.60$; \citealt{tacchellaJADESImagingGNz112023,
maiolinoSmallVigorousBlack2024,jiJADESSmallBlue2025})
falls into the most massive tip of the galaxy distribution, 
with an under-massive BH touching the lower boundary of $M_\sscBH$ 
of the modeled galaxies with similar $M_*$.
The galaxy UHZ1 ($z \approx 10.32$; \citealt{bogdanEvidenceHeavyseedOrigin2023}) 
falls onto the other side of the same tip, with an over-massive BH 
touching the upper boundary of $M_\sscBH$ of the modeled galaxies with similar $M_*$.
At $z = 5$ and $4$, the branch with $M_{\sscBH}:M_* \approx 1:1$
fades away, leaving only a tail that joins the low-mass end of the other branch.
Observed galaxies at these redshifts exhibit significant diversity 
in the $M_\sscBH$-$M_*$ plane: at a given $M_*$, the observed $M_\sscBH$ can vary 
by about $3\dex$. The most massive galaxies in our sample have $M_* \gtrsim 10^{10}\Msun$ 
and $M_{\sscBH} \gtrsim 10^8\Msun$, overlapping with the range covered by bright quasars 
\citep{izumiSubaruHighzExploration2021,
stoneDetectionLowstellarmassHost2023,
stoneUndermassiveHostGalaxies2024,
yueEIGERCharacterizingHost2024} 
and massive PSB galaxies \citep{carnallMassiveQuiescentGalaxy2023,onouePoststarburstPathwayFormation2025} 
found in observations.
Less massive galaxies with $10^8 \lesssim M_*/{\rm M}_\odot \lesssim 10^{10}$
partially overlap with the reported $M_\sscBH$ and $M_*$
for faint AGNs \citep{pacucciJWSTCEERSJADES2023,harikaneJWSTNIRSpecFirst2023,ublerGANIFSMassiveBlack2023,maiolinoJADESDiversePopulation2024}, 
LRDs \citep{chenHostGalaxyIf2025} and massive galaxies 
hosting normal-mass BHs 
\citep{liPrevalentPopulationNormalmass2025},
indicating that each sample only represents a biased sub-population of galaxies.
At $z \lesssim 3$, the surviving branch starts to deviate from 
the line of $M_{\sscBH}:M_* \approx 1:100$. This deviation is
mass-dependent, with galaxies around the MW-mass
($M_* \approx 10^{10}\Msun$) shifted most significantly to the right.
The observed low-mass galaxies with $M_* \lesssim 10^{10}\Msun$
at $z \approx 1$--$3$ \citep{mezcuaOvermassiveBlackHoles2024} 
appear to overlap with the upper boundary 
of the $M_\sscBH$-$M_*$ scatter. At $z \lesssim 1$, the $M_\sscBH$-$M_*$ relation stabilizes 
as a down-bending, bow-shaped curve. The observed ETGs and LTGs 
compiled by \citet{greeneIntermediateMassBlackHoles2020}
occupy the high-$M_*$ end and the knee of the bow,
with $M_\sscBH$ above and below the median relation at the corresponding $M_*$, 
respectively (see also \citealt{grahamAppreciatingMergersUnderstanding2023}).
The broad-line AGN candidates in low-mass galaxies 
\citep{puchaTriplingCensusDwarf2025}
from the Dark Energy Spectroscopic Instrument (DESI) survey,
and POX 52, a local early-type dwarf showing evidence for IMBH 
\citep{kawamuroCoevolutionNuclearStructure2024,sunIntermediatemassBlackHole2025},
also occupy the knee of the relation.

The broad consistency between model predictions and observations 
indicates the plausibility of the physical processes implemented.
The predicted $M_\sscBH$-$M_*$ relation (Fig.~\ref{fig:scaling-evol}i) 
at any $z$ is not a single power-law but consists of separable pieces.
Thus, it is interesting to dissect the relation and reveal the underlying 
physical mechanisms that shape individual pieces.
As shown in Figs.~\ref{fig:growth-examples} and \ref{fig:growth-channels}, 
the halo mass ($M_{{\rm v},z=0}$) appears to be an important
factor that determines the growth stage reached by a galaxy at a given $z$, and 
the dominant growth channels (seeding, bursty or continuous mode, or merger)
emerge sequentially as a galaxy evolves from one stage to another,
establishing a distinct $M_\sscBH$-$M_*$ relation in each of the stages
(see \S\ref{sssec:growth-channels}).
Thus, the overall $M_\sscBH$-$M_*$ relation at a given $z$ 
is a convolution between the halo mass function at $z=0$ and the 
relations of the sub-populations in different stages 
at the redshift $z$ in question.  

During the seeding era ($z \approx 20$--$30$; \S\ref{sssec:cosmic-bh-sf-history}), 
BH seeds are bred with $M_{\sscBH} \sim M_*$ due to the top-heavy IMF 
of Pop-III stars adopted by our model.
This sets the starting loci of the $M_\sscBH$-$M_*$ relation
and produces the highest ratio, $M_{\sscBH}/M_* \approx 1$,
that can ever be achieved in the growth history of galaxies.
Most seeds are IMBHs, waiting for the post-seeding channels to boost them to SMBHs.
The post-seeding growth of each galaxy starts with the bursty mode,
consisting of episodic nuclear bursts triggered by global disturbances on the galaxy. 
The $M_{\sscBH}/M_*$ ratio is $\sim 1/10$ during this era  (Fig.~\ref{fig:growth-examples}),
and a few nuclear bursts are sufficient to boost a BH seed to $\gtrsim 10^6\Msun$ (Fig.~\ref{fig:growth-channels}, 
rows 3 and 4). {\textem The bursty mode turns out to be the key condition for
raising the masses of all BHs, especially low-mass seeds, to the regime of SMBHs}, 
so that the subsequent growth of BHs in the continuous mode is initialized with sufficiently 
high $M_\sscBH$ and galaxies are well-regulated by the AGN feedback 
to sustain an $M_\sscBH$-$M_*$ relation similar to that observed at low $z$. 
The bursty mode thus provides a physical explanation for the 
origin of the `enhancement factor', inferred phenomenologically 
in \citetalias{moTwophaseModelGalaxy2024} (see their fig.~7 and 
\S4.3.3), needed to boost the BHs in low-mass halos to the regime of SMBHs.

Following the bursty era, the growth of BHs via nuclear bursts 
is self-limited by the associated feedback and the continuous-mode growth gradually 
takes over and keeps $M_{\sscBH}/M_*$ around $1/100$ (Fig.~\ref{fig:growth-examples}).
Towards lower $z$, the efficiency of AGN feedback at a given $M_{\rm v}$
increases due to the shallower gravitational potential (quantified by $V_{\rm max}$ of 
the halo) and the stronger coupling between feedback energy and the SGC (see Appendix~\ref{app:ssec:growing}). 
The growth of BHs in dynamically hot SGCs at lower $z$ is thus expected to have stronger 
self-regulation, leading to a $M_\sscBH$-$M_*$ relation with a lower 
amplitude, $M_\sscBH/M_* \approx 1/1000$, as shown by the
red curve in Fig.~\ref{fig:scaling-evol}i at $z=0$. 
A logarithmic slope $\geqslant 1$ is seen at the high-mass end,
as expected from the energy argument of AGN feedback \citep[][\S4.3 
therein]{moTwophaseModelGalaxy2024}. 
The net effect is that the $M_\sscBH$-$M_*$ relation around 
$M_*\sim 10^{10}\Msun$ bends downward at $z \lesssim 5$.

At $z \approx 2$, halos hosting progenitors of MW-size galaxies
(with $M_{{\rm v}, z=0} \approx 10^{12}\Msun$ and 
$M_{*,z=0} \approx 10^{10.5}\Msun$) start to transit from the fast phase to 
the slow phase (see Fig.~\ref{fig:halo-mahs}b). 
Galaxies then start to grow their dynamically cold (disk) 
components effectively in the continuous mode, but the BH accretion is 
terminated almost completely. These MW progenitors thus migrate horizontally in 
the $M_\sscBH$-$M_*$ plane (Fig.~\ref{fig:growth-examples}), 
and eventually reach the knee of the bow-shaped $M_\sscBH$-$M_*$ relation at $z=0$.
Halos hosting less massive galaxies (e.g. dwarfs) transit to the slow 
phase earlier (Fig.~\ref{fig:halo-mahs}b), but
the post-transition star formation is less efficient than that in
progenitors of MW-size galaxies due to their shallower gravitational potentials to 
resist the feedback, so that the horizontal shift in the $M_\sscBH$-$M_*$ 
plane is smaller.
Halos hosting more massive galaxies (e.g. BCGs) transit to the slow phase 
later, and both the SFE and the duration for star formation 
in the dynamically cold phase in these halos are smaller than 
those in MW-size galaxies, which reduces the amount of horizontal migration 
in the $M_\sscBH$-$M_*$ plane.
The mass-dependence of halo transition time and the post-transition SFE
thus gradually `draw the bow' at $z \lesssim 3$, and eventually build up
the bow-shaped $M_\sscBH$-$M_*$ relation at $z = 0$ (Fig.~\ref{fig:scaling-evol}i).
Dry mergers of galaxies are not important in most parts of the 
$M_\sscBH$-$M_*$ relation, since the mass added by each merger only takes a 
temporary dominance in a short period after the merger and is soon surpassed 
by the growth via other channels (see the brown segments in Fig.~\ref{fig:growth-examples}).
The exception is in the high-mass end ($M_*\gtrsim 10^{11}\Msun$) of the relation 
at low $z$ ($z \lesssim 2$), where repeated mergers drive galaxies towards a relation 
of $M_\sscBH \propto M_*$ and produce the scatter in the high-mass tip in 
Fig.~\ref{fig:scaling-evol}i (see also Fig.~\ref{fig:growth-examples}d).

As a summary, our model predicts a {\texteem redshift-dependent, multi-piece 
$M_\sscBH$-$M_*$ relation}, and the growth of galaxies within each piece 
is driven by a distinct channel.
The snapshot views provided by observations, although subject to various uncertainties, 
appear to support the model predictions.
Observations of dwarfs ($M_* \leqslant 10^9\Msun$), which are still missing,
are expected to be the key to testing the model. 
In particular, galaxies of star-cluster size 
($M_{*, z=0} \lesssim 10^6 \Msun$) are predicted to be close to the seeding locations 
in the $M_\sscBH$-$M_*$ plane, and a small fraction of them have $M_\sscBH$
dominated by the seeded mass (Fig.~\ref{fig:scaling-evol}i, purple dots). 
The post-seeding growth of these galaxies is the least important among all galaxies, 
and the search for IMBHs in this population may provide crucial 
constraints on the seeding model.

\section{Summary and discussion}
\label{sec:summary}

In this paper, we have developed a model for the formation and co-evolution of
(SM)BHs with their host galaxies and halos (see Fig.~\ref{fig:flowchart-brief}
for an outline). 
The model is built upon the two-phase framework of galaxy formation established in 
\citetalias{moTwophaseModelGalaxy2024} (see \S\ref{ssec:two-phase-galaxy} 
for a summary), with extensions specific to the seeding and the early growth of BHs.
The model covers wide ranges of spatial, temporal and mass scales
($r \approx 10^{-3}$--$10^6\pc$, $z \approx 40$--$0$, $M_\sscBH \approx 10$--$10^{10}\Msun$),
so that the entire lives of SMBHs and their host galaxies
can be self-consistently initialized and followed within a 
cosmological context. 
We have detailed physical processes relevant to the modeling,
derived equations describing each process and presented analytical approximations 
to obtain order-of-magnitude estimates.
The main features of the model are summarized as follows.

\begin{enumerate}[topsep=0pt,parsep=0pt,itemsep=0pt]
    \item The model takes halo merger trees (\S\ref{ssec:halo-assembly}; Fig.~\ref{fig:halo-mahs}) 
    in a cosmological volume as input to determine when, where and how BH seeds 
    are bred: the seeding epoch of a mini-halo is set by key coolants in pristine 
    gas ($\Hatom$ and $\Hmol$) and by environmental effects that modulate cooling 
    (LW radiation, dynamical heating and IGM enrichment); the seed is the remnant 
    of the most massive star in the first Pop-III cluster formed from the 
    collapsed gas, with a feedback-regulated model giving the mass of this star 
    and, together with an assumed top-heavy IMF, the mass spectrum of the cluster.
    This procedure thus yields both the mass ($M_{\rm {\sscBH}, seed}$) and the 
    `flavor' (formation pathway) of each seed (\S\ref{ssec:seeding-main};
    Fig.~\ref{fig:cosm-context-seeding}).

    \item 
    The post-seeding growth of BHs is modeled as a multi-channel process, with 
    each channel emerging once sufficient dynamical hotness removes the 
    angular-momentum ($j$) barrier to accretion.
    A new channel proposed in this paper is the bursty mode: episodic nuclear
    bursts, each triggered by a global disturbance that forms a dense, turbulent, 
    supernova-free nucleus, in which super-Eddington accretion grows the BH while 
    intense nuclear star formation grows the NSC. Two further channels are 
    included: a continuous mode, in which the BH captures low-$j$ sub-clouds 
    from the turbulent SGC during the fast phase of the host halo, and a merger 
    channel, in which BHs merge following a galaxy-galaxy merger
    (\S\ref{ssec:growing-main}; Fig.~\ref{fig:am-redistribute}).
    
    \item
    To resolve the wide range of scales involved in post-seeding BH growth, 
    we characterize the density around each BH with a ladder-like profile of 
    five steps, each describing one level of the cosmic hierarchy: the Universe, 
    halos, galaxies, galactic nuclei, and BH accretion disks.
    Equations linking adjacent levels let one `climb' from the outer boundary 
    down to the gas structure near a BH
    (see Appendix~\ref{app:ssec:hierarchy} and Fig.~\ref{fig:profile-ladder}), 
    propagating cosmological information from halo scales down to the central 
    BH and thus linking halo assembly to BH growth
    (\S\ref{ssec:growing-main}).    
\end{enumerate}

We have implemented the model into the merger trees constructed from a 
N-body simulation (see Fig.~\ref{fig:flowchart-brief} for an outline),
presented our model predictions, and demonstrated the potential 
applications to be explored in the future. Our main results and conclusions 
are summarized as follows.
\begin{enumerate}[topsep=0pt,parsep=0pt,itemsep=0pt]
    \item The BH seed mass function is broad ($\sim 10$--$10^5\Msun$) and 
    bimodal, with a PISN-carved valley at $M_{\sscBH}\approx10^2\Msun$ 
    separating a DCBH-dominated high-mass peak ($\approx500\Msun$) from a 
    CCSN/PISN-dominated low-mass peak ($\approx50\Msun$). LW radiation and 
    dynamical heating have a divergent effect, producing high-mass (DCBH) and 
    low-mass (Pop-II) bumps in the mass function, 
    respectively (\S\ref{sssec:seed-mass-func}; 
    Fig.~\ref{fig:seed-mass-func}).

    \item Most BH seeds are bred at $z = 20$--$30$ as Pop-III remnants,
    and most Pop-II seeds are bred at $z = 10$--$20$. The seeding rate 
    drops at $z \lesssim 10$ and becomes negligible at $z \lesssim 5$. 
    LW radiation and dynamical heating act mainly on mini-halos at 
    $z \lesssim 20$ and $z \gtrsim 10$, respectively
    (\S\ref{sssec:cosmic-seeding-history}; Fig.~\ref{fig:seed-z-func}).

    \item An environment-modulated `seeding atlas' dissects seeds in the 
    $M_{\rm v}$-$z$ plane into three regimes: a fusiform area shaped by 
    dynamical heating, a tip-like structure shaped by LW radiation, and a 
    two-component stripe shaped by IGM enrichment (\S\ref{sssec:seeding-atlas}; 
    Fig.~\ref{fig:seeding-atlas}).

    \item Post-seeding channels of BH growth emerge sequentially in the history of 
    any given galaxy: the bursty mode dominates soon after seeding 
    ($z\approx20$--$30$), and the continuous mode takes over at 
    $z \approx 5$--$15$ once $M_\sscBH \sim 10^6\Msun$, earlier in more 
    massive halos. In the continuous stage, the dynamically hot (bulge) 
    component forms before the cold (disk) component. Mergers become 
    important only in BCG-size galaxies at $z \lesssim 2$ 
    (\S\ref{sssec:growth-channels}; Figs.~\ref{fig:growth-examples} and 
    \ref{fig:growth-channels}).

    \item The bursty stage features back-and-forth switching between 
    star-forming and quiescent states and the formation of a compact, 
    globular-like NSC, predicting that quiescent galaxies appear as early as 
    $z \approx 20$--$30$ and that galaxies with extremely compact morphology 
    are ubiquitous at $z \gtrsim 10$ 
    (\S\ref{sssec:growth-channels}).

    \item The cosmic histories of BH accretion and star formation, as 
    compositions of individual galaxies, are likewise driven sequentially by 
    these channels and divide into a seeding era ($z \approx 20$--$30$), a 
    bursty era ($z \approx 5$--$20$) and a continuous era ($z \lesssim 5$), 
    with predicted $\dot{\rho}_{\sscBH}$ and $\dot{\rho}_*$ broadly consistent 
    with JWST observations (\S\ref{sssec:cosmic-bh-sf-history}; 
    Fig.~\ref{fig:growth-cosmic-history}).

    \item Controlled experiments with two extreme model variants, 
    {\textem MinSeed} and {\textem MaxSeed} (minimizing/maximizing seed mass), show 
    that self-regulated BH accretion erases the memory of the seeding method: 
    $M_*$ and $M_\sscBH$ at low $z$ become indistinguishable between the 
    variants, except in halos with $M_{{\rm v},z=0} \lesssim 10^{11}\Msun$, 
    where the seeding-driven difference in $M_\sscBH$ persists to 
    $z \lesssim 8$ (\S\ref{sssec:growth-effect-seeding}; 
    Fig.~\ref{fig:growth-effect-seeding}).

    \item Comparing predicted growth histories with JWST-observed BHs and 
    galaxies enables archaeological and futurological inference:
    the bright galaxies/quasars at $z \gtrsim 6$--$10$ evolve in 
    massive halos ($M_{{\rm v},z=0} \approx 10^{15}\Msun$) into Coma-like BCGs; 
    faint AGNs (including LRDs) at $z \gtrsim 4$, assumed to have high $M_{\rm BH}$ as
    suggested by observations, are hosted by halos with 
    $M_{{\rm v},z=0} \gtrsim 10^{13}\Msun$; a small population of low-mass, 
    morphologically and chemically peculiar galaxies at $z \gtrsim 4$ can also 
    be linked to specific formation pathways (\S\ref{ssec:arch-and-forecast}; 
    Fig.~\ref{fig:growth-effect-seeding}).

    \item Our model predicts a redshift-dependent, multi-piece 
    $M_\sscBH$-$M_*$ relation spanning the full lifetime ($z \gtrsim 20$ to 
    present) and mass range (star-cluster-size to BCG-size galaxies) of 
    galaxies, with each piece shaped by a growth channel. 
    The bursty mode turns out to be the key condition
    for growing the seeds to the regime of SMBHs.
    The model recovers the observed $M_\sscBH$-$M_*$ relation 
    from $z\approx10$ to $0$ (\S\ref{ssec:mbh-ms-relation}; Fig.~\ref{fig:scaling-evol}).
\end{enumerate}

\bigskip
The model developed in this paper provides a framework to examine the co-evolution
pattern of BHs and their host galaxies and halos across cosmic time, to 
distinguish the physical processes driving the evolution in different regimes
(redshift, mass and assembly of halo, and environment), and to interpret 
the origin and predict the fate of observed BHs and galaxies. Clearly, the model is 
not definitive; it leaves many interesting questions to be addressed in the future 
using both theoretical and observational approaches. Here we discuss a few of them.

The environment provided by dark matter halos sets the fundamental context of 
galaxy formation in our model, and in most other empirical and 
semi-analytical models, but the dynamical nature of halos and its effects 
have not been fully explored. We assume the baryon content of a halo to be 
$f_{\rm b}$ (the cosmic baryon fraction) times its mass 
(Eqs.~\ref{eq:r-sgc}--\ref{eq:t-ff-sgc}). This is not valid for halos that 
interact strongly with other halos or dense environments, as illustrated by 
the bullet cluster \citep[e.g.][]{cloweWeakLensingMass2004,markevitchDirectConstraintsDark2004} 
and dark matter-deficient galaxies \citep{guoFurtherEvidencePopulation2020,
leeDarkMatterDeficient2021,vandokkumTrailDarkmatterfreeGalaxies2022,
ogiyaTidalFormationDark2022,morenoGalaxiesLackingDark2022,contreras-santosThreeHundredExistence2024}, 
and is expected to be more common at high $z$, when halo mergers are 
frequent \citep{chenMassiveDarkMatter2023} and the dark matter-baryon stream 
velocity is significant relative to the small virial velocities of 
mini-halos \citep{stacyEffectStreamingMotion2011,maioImpactPrimordialSupersonic2011,greifDelayPopulationIII2011}.

Our seeding procedure adopts simplified environmental treatments: 
LW emission is assumed isotropic and to propagate depending only on 
redshift and source-target distance, whereas the ISM, CGM and IGM in 
reality have complex structures that can make the escape and propagation of 
LW radiation 
highly anisotropic and time-dependent 
\citep[e.g.][]{pallottiniSurveyHighzGalaxies2022,latifUVRegulatedStar2019}, 
and dynamical heating is sourced not only by halo assembly but also by the 
dark matter-baryon stream velocity noted above.

The structure and evolution of Pop-III stars, especially SMSs, remain uncertain: their structure can 
fluctuate 
strongly, complicating their evolutionary tracks and radiative feedback 
\citep{hosokawaFormationPrimordialSupermassive2013}; 
gas accretion onto them, considered critical in shaping their mass spectrum 
\citep{omukaiCanSupermassiveBlack2008,schleicherMassiveBlackHole2013}, is 
usually anisotropic, with disk-like structure, clumps 
\citep{chonSupermassiveStarFormation2020,latifBirthMassFunction2022} and 
turbulence-amplified magnetic fields \citep{shardaImportanceMagneticFields2020};
alternative models of dark matter, such as those invoking 
Weakly Interacting Massive Particles (WIMPs), are proposed to exert
additional heating that bloats the stellar structure, lowers the surface 
temperature, elevates gas accretion and yields more massive 
Pop-III.1 stars \citep{natarajanDarkMatterAnnihilation2009,freeseSupermassiveDarkStars2010,banikFormationSupermassiveBlack2019,singhFormationSupermassiveBlack2023,tanOriginSupermassiveBlack2024,cammelliFormationSupermassiveBlack2025,sanatiEmergenceIonizingFeedback2025,petkovaFormationSupermassiveBlack2026}.

Resolving these issues regarding BH seeding requires hydro simulations, with the 
relevant processes implemented accurately and with dynamical ranges beyond current 
computational power, that can quantify the fraction of halos with elevated 
baryon content, test the environmental treatments above, and clarify the 
structure and evolution of Pop-III (proto-)stars. 
On the other hand, if environmental information is unavailable,
an empirical seeding strategy for other empirical/semi-analytical models 
and hydro simulations may be constructed from our seeding atlas 
(\S\ref{sssec:seeding-atlas}) to account for the statistical expectation and 
variation of the seed population at given $M_{\rm v}$ and $z$.

A key assumption of our model is that the formation of an SNF nucleus is 
triggered by a global disturbance on the SGC, associated with a temporary 
excursion of the halo assembly to high $\gamma_{\rm v}$. In reality, such 
disturbances may arise from diverse sources, e.g. a close flyby of another 
galaxy \citep{liEffectGalaxyInteractions2025}, galaxy-galaxy mergers 
\citep{luoConnectionsGalaxyMergers2014,puskasConstrainingMajorMerger2025}, 
counter-rotation of cold streams \citep{danovichFourPhasesAngularmomentum2015}, 
violent disk instability \citep{agertzLargescaleGalacticTurbulence2009,mengStructureStabilityHighredshift2019,renaudGiantClumpsClouds2021}, 
or misalignment-induced tidal forces \citep{chisariGalaxyhaloAlignmentsHorizonAGN2017}. 
High-resolution simulations with large volumes and controlled experiments 
are needed to verify whether nuclei denser than $n_{\rm snf}$ indeed form 
from such disturbances and to resolve their density profiles.
Super-Eddington accretion is a natural outcome once SNF nuclei form, but 
whether such high rates can be sustained depends on the structure 
(anisotropy, clumpiness) of the nucleus and the properties (geometry, 
magnetic field, radiation and jet feedback) of the accretion disk. The dry 
channel (BH-BH mergers and star capture) may also contribute to BH growth, 
as explored by simulations 
\citep{tremmelBeatenPathNew2015,gaeteSupermassiveBlackHole2024,
fujiiSimulationsPredictIntermediatemass2024} and (semi-)analytical models 
\citep{partmannDifficultPathCoalescence2024,liuGravitationalWavesMergers2024,
dekelGrowthMassiveBlack2025}, and can produce observable TDEs and GWs. Both 
channels remain poorly understood and require further model refinement 
\citep[e.g.][]{angles-alcazarCosmologicalSimulationsQuasar2021,latifBirthMassFunction2022,
hopkinsFORGEdFIREResolving2024}, which observational constraints can also 
help to tighten.

NSCs form as by-products of nuclear bursts, and the competition between 
star formation and BH accretion naturally produces an anti-correlation 
between the NSC-BH mass ratio ($M_{\rm NSC}/M_\sscBH$) and $M_\sscBH$, as 
suggested by observations \citep[e.g.][]{georgievMassesScalingRelations2016,
nguyenNearbyEarlytypeGalactic2018,
neumayerNuclearStarClusters2020}. However, dry evolution of an NSC, e.g. 
via mergers with other star clusters 
\citep{tremaineFormationNucleiGalaxies1975,
capuzzo-dolcettaSelfconsistentSimulationsNuclear2008,
neumayerTwodimensionalHaKinematics2011,
arca-seddaGlobularClusterMigratory2014,
guillardNewInsightsFormation2016,
tsatsiRotationNuclearStar2017,
fahrionNuclearStarCluster2022} 
or interaction with the central BH \citep{antoniniCoevolutionNuclearStar2015}, can introduce significant contamination and needs careful modeling.

The bursty mode provides a key condition for BH seeds to grow to SMBHs, but 
as shown in \S\ref{sssec:growth-effect-seeding}, the required mass boost 
degenerates with the seeding method, and current observations cannot break 
this degeneracy. Key targets that can isolate the effects of the bursty 
mode are galaxies at $z \gtrsim 10$, when BH growth is dominated by nuclear 
bursts (\S\ref{sssec:cosmic-bh-sf-history}, Fig.~\ref{fig:growth-cosmic-history}), 
and BHs in low-mass galaxies (dwarfs with $M_* < 10^9\Msun$, especially 
star cluster-size galaxies with $M_* < 10^6\Msun$), where the bursty mode 
contributes a large fraction of the total $M_\sscBH$ 
(\S\ref{sssec:growth-channels}, Figs.~\ref{fig:growth-examples}, 
\ref{fig:growth-channels} and \ref{fig:scaling-evol}). Low-mass halos 
($M_{\rm v} \lesssim 10^{9.5}\Msun$ at infall) are around or below the 
atomic-cooling limit (Fig.~\ref{fig:halo-mahs}) and thus expected to host 
little or no star formation after reionization. Detecting galaxies in such 
halos may provide fossil information of galaxy formation at $z \gtrsim 6$ 
and help calibrate the seeding method and bursty-mode growth.

\section*{Acknowledgements}


YC is supported by Double First-Class Discipline Construction-Astronomy Discipline at Nanjing University,
the National Natural Science Foundation of China (Grant no. 12503014),
and the Fundamental Research Funds for the Central Universities (Grant no. KG202502).
HYW is supported by the National Natural Science Foundation of China (Nos. 12192224) 
and CAS Project for Young Scientists in Basic Research (Grant no. YSBR-062).
YC thanks Tao Wang, Kai Wang, Fangzhou Jiang, Ziwen Zhang, Yunjing Wu, Enci Wang, Long Wang,
Qi Guo, Baitian Tang, Zhaozhou Li, Houzun Chen and Xi Kang for their valuable insights and discussions, 
and thanks Yi Mao and Chen Chen for their technical support.
The authors would like to express their gratitude to the Tsinghua Astrophysics 
High-Performance Computing platform at Tsinghua University and the 
Supercomputer Center of the University of Science and Technology of China for 
providing the necessary computational and data storage resources that have 
significantly contributed to the research results presented in this paper.

The computations and presentations in this paper are supported by various software 
tools, including the HPC toolkits 
\softwarenamestyle[Hipp] \citep{chenHIPPHIghPerformancePackage2023}
\footnote{\url{https://github.com/ChenYangyao/hipp}}
and 
\softwarenamestyle[PyHipp]\footnote{\url{https://github.com/ChenYangyao/pyhipp}},
interactive computation environment 
\softwarenamestyle[IPython] \citep{perezIPythonSystemInteractive2007},
numerical libraries \softwarenamestyle[NumPy] \citep{harrisArrayProgrammingNumPy2020}, 
\softwarenamestyle[Astropy] \citep{
robitailleAstropyCommunityPython2013,
astropycollaborationAstropyProjectBuilding2018,
astropycollaborationAstropyProjectSustaining2022}
and \softwarenamestyle[SciPy] \citep{virtanenSciPy10Fundamental2020},
graphical library 
\softwarenamestyle[Matplotlib] \citep{hunterMatplotlib2DGraphics2007},
and correlation function calculator
\citep{sinhaCorrfuncBlazingFast2019,sinhaCORRFUNCSuiteBlazing2020}\footnote{\url{https://github.com/manodeep/Corrfunc}}.
The implementation of the Monte Carlo trees
is revised from the public code 
\softwarenamestyle[SatGen]\footnote{\url{https://github.com/shergreen/SatGen}} \citep{jiangSatGenSemianalyticalSatellite2021}.
This research has made extensive use of the arXiv and NASA’s Astrophysics Data System.
Data compilations used in this paper have been made much more accurate and 
efficient by the software \softwarenamestyle[WebPlotDigitizer]. 

\section*{Data Availability}
\label{sec:data-availability}


The code repository \softwarenamestyle[TwoPhaseGalaxyModel]\footnote{\url{https://github.com/ChenYangyao/two-phase-galaxy-model}}
implements the model described in this series of papers.
All data used in this paper, including data points displayed in figures
and observational results used for comparison, 
will be distributed along with the repository.



\bibliographystyle{mnras}
\bibliography{ref} 




\appendix


\section{Dark matter halos and their assembly}
\label{app:sec:halo-assembly}

\begin{figure*} \centering
    \includegraphics[width=\textwidth]{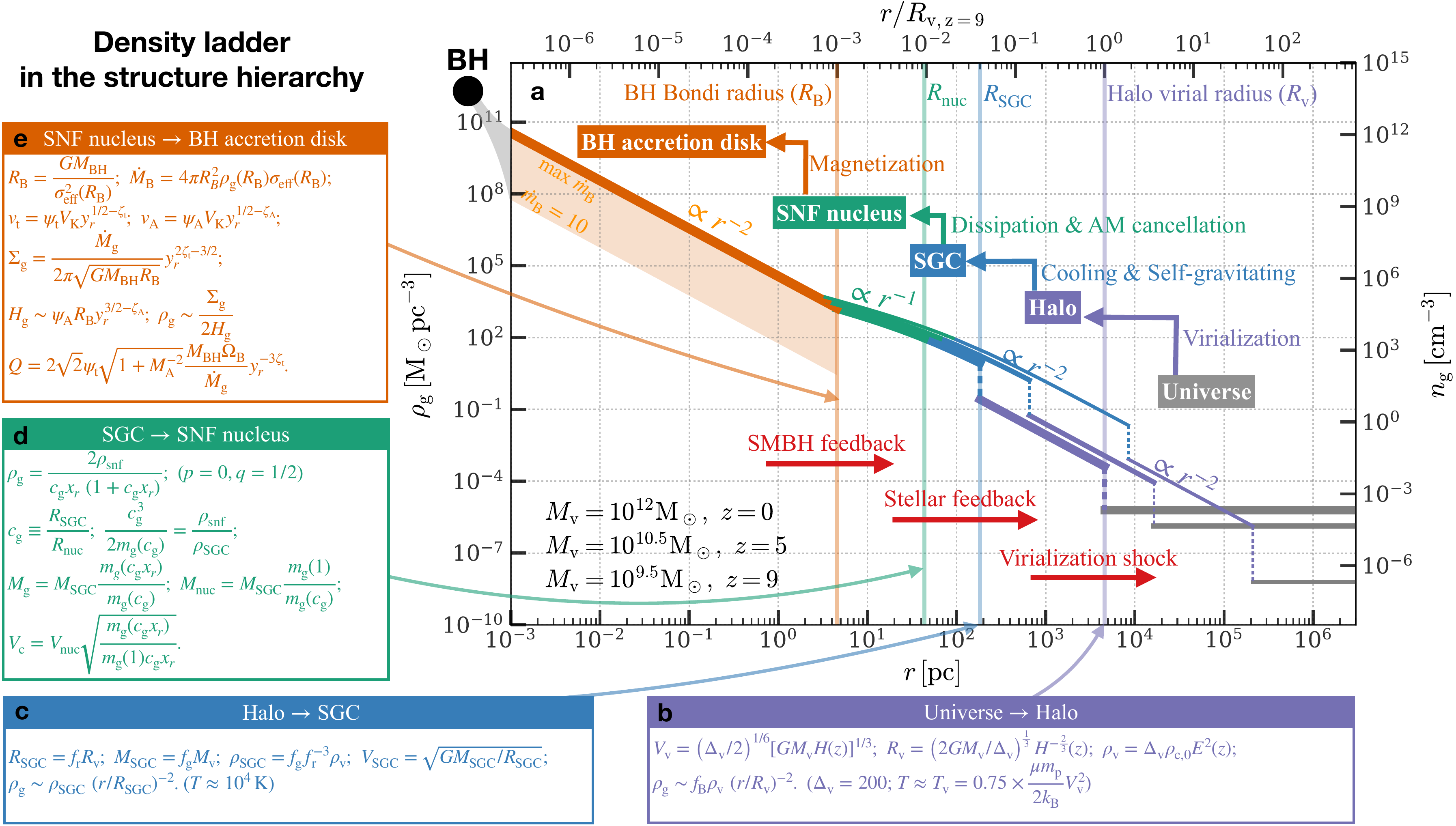}
    \caption{
    {\figem Density ladder of gas in the cosmic structure hierarchy}.
    {\figem a},
    gas density profiles predicted by our model at three redshifts, 
    $z = 0$, $5$ and $9$ (from thin to thick), 
    in and around an MW-size galaxy and its progenitors 
    under global disturbance.
    At each redshift, an SMBH with $M_{\sscBH} = 10^6 \Msun$ is put at the 
    center of the galaxy.
    The profile exhibits a ladder-like structure,
    separated into {\figem five segments}, each corresponding to a level 
    in the structure hierarchy discussed in this paper.
    The innermost segment ({\figem orange}) 
    describes the middle-plane density of the accretion disk around the 
    central BH, with a solid line indicating the
    profile expected in the beginning of the nuclear burst 
    when the turbulence-modified Bondi accretion rate, $\dot{m}_{\sscBH}$ 
    (in the unit of Eddington rate), is the maximum 
    (Eq.~\ref{eq:dot-m-Bondi-numerical-q05}),
    and a shaded area covering the range of the profile during the burst 
    until $\dot{m}_{\sscBH}$ is reduced to $10$. 
    Each of the four {\figem vertical lines} indicates the transition scale of 
    the interface bridging two adjacent levels at $z=9$. 
    {\figem b}--{\figem d}, cards listing the sets of 
    equations bridging the structure properties of any outer level
    to those of the adjacent inner level.
    Processes driving the formation of the structure hierarchy, 
    and feedback resulting from the structure formation,
    are indicated by the texts.
    See Appendix~\ref{app:ssec:hierarchy} for details.
    }
    \label{fig:profile-ladder}
\end{figure*}

The early assembly of dark matter halos at high $z$
is crucial input for the seeding (Appendix~\ref{app:sec:seeding}) 
and growing procedures (Appendix~\ref{app:sec:growing}).
Below we summarize the properties and assembly histories of dark matter halos 
that are relevant to the description of our model.

In the adopted flat $\Lambda$CDM cosmology (see the end of \S\ref{sec:intro}), 
the cosmic baryon fraction is $f_{\rm b} \equiv \Omega_{\rm b,0}/\Omega_{\rm m,0}\approx 0.157$, while the cosmic baryon density at $z \geqslant 0$ is 
\begin{equation}
    \rho_{\rm b}(z) = \rho_{\rm c,0} \Omega_{\rm b,0} (1+z)^3
    = \left[ 4.19 \times 10^{-28}\gperccm \right] (1+z)_{10}^3
    \,,\label{eq:cosmic_rho_b_z}
\end{equation}
where $(1+z)_{10}\equiv (1+z)/10$, and 
$\rho_{\rm c,0}$ is the critical density at $z=0$,

Fig.~\ref{fig:profile-ladder}a shows the gas density as a function of 
the halo-centric distance at three epochs in the progenitor of an MW-size halo. 
The density profile exhibits a ladder-like structure, consisting of five steps that 
correspond to the five levels of the cosmic hierarchy.
The Universe itself is the outermost step of the density ladder,
with a mean baryon density varying from about $10^{-4}\perccm$ at 
$z = 9$ to $10^{-7}\perccm$ at $z = 0$ (grey segments), 
reflecting the factor $(1+z)^3$ that encodes the expansion of the Universe.

In our definition of a virialized halo (\S\ref{ssec:halo-assembly}), 
the virial density, virial radius, virial velocity, and virial temperature 
are given respectively by 
\begin{align} 
    \rho_{\rm v}  
        &\equiv \Delta_{\rm v} \rho_{\rm c,0}E^2(z)
        \nonumber \\
        & \approx \left[ 7.91 \times 10^{-3} \Msunpercpc \right] (1+z)_{10}^{3}
        \,; \label{eq:rho-vir}
\end{align}
\begin{align}      
    R_{\rm v} 
        &\equiv \left(\frac{2 G M_{\rm v}}{\Delta_{\rm v}}\right)^{1/3} H^{-2/3}(z)
        \nonumber \\
        &\approx \left[ 4.57 \Kpc \right] M_{\rm v, 9.5}^{1/3}(1+z)_{10}^{-1}    
        \,; \label{eq:R-vir}
\end{align}
\begin{align} 
    V_{\rm v}  
        &\equiv \left( \Delta_{\rm v}/2 \right)^{1/6}\left[G M_{\rm v} H(z)\right]^{1/3}
        \nonumber \\
        &\approx \left[ 54.5 \kms \right] M_{\rm v,9.5}^{1/3} (1+z)_{10}^{1/2}
        \,; \label{eq:V-vir}
\end{align}
\begin{align} 
    T_{\rm v} 
        &= 0.75 \times \frac{\mu m_{\rm p} }{2 k_{\sscBondi}} V_{\rm v}^2
        \nonumber \\
        &\approx \left[1.62\times 10^5 \Kelvin\right] 
            M_{\rm v,9.5}^{2/3} (1+z)_{10}
        \,, \label{eq:T-vir}
\end{align}
where the first line of each expression is the definition of the quantity;
the approximation in the second line is valid at $z\gg 1$;
$M_{\rm v,9.5}\equiv M_{\rm v}/\left( 10^{9.5}{\rm M}_\odot \right)$. 
The expression of virial temperature is given by e.g.
\citet{bryanStatisticalPropertiesXRay1998} and \citet{wiseSuppressionH2Cooling2007},
with an additional factor 0.75 calibrated by \citet{fernandezH2SuppressionShocking2014}
using their hydro simulations, $m_{\rm p}$ is the proton mass,
and we assume the mean molecular weight $\mu=1.2$.
We model the density profiles of halos using the NFW form: 
\begin{equation}
    \rho(r) \propto \frac{1}{x(1+x)^2} 
    \,,\label{eq:nfw-profile}
\end{equation}
where $x \equiv r/ R_{\rm s}$, 
with $R_{\rm s}$ being a scale radius related to $R_{\rm v}$ by the 
concentration parameter, $c_{\rm v}=R_{\rm v}/R_{\rm s}$ 
\citep[][]{navarroUniversalDensityProfile1997}. 
As shown by e.g. \cite{zhaoAccurateUniversalModels2009} and 
\citet{wangCharacterizingAssemblyDark2024}, 
the value of $c_{\rm v}$ of a halo 
is closely related to the mass assembly history of the halo. 
Finally, we define the dynamical timescale of a halo as 
\begin{equation} \label{eq:t-vir}
    t_{\rm v} \equiv \frac{R_{\rm v}}{V_{\rm v}} = \frac{1}{10H(z)}
    \approx 81.9 \Myr\, (1+z)_{10}^{-3/2} \,.
\end{equation}
As shown by calibrations using numerical 
simulations \citep[e.g.][]{dekelToyModelsGalaxy2013}, the average accretion rate of a virialized 
halo of mass $M_{\rm v}$ at $z \gg 1$ can be approximated as 
\begin{align} 
    \dot{M}_{\rm v}
        &\approx \left[ 10.3 \msunperyr \right] M_{\rm v,9.5}^{1.14}(1+z)_{10}^{2.5}
        \,. \label{eq:dot-M-vir}
\end{align}
This can be integrated to obtain the average assembly histories and provides guidance 
for the growth of individual halos. For reference, 
Fig.~\ref{fig:halo-mahs} shows the median assembly histories of halos 
of given masses at $z=0$, in terms of 
$M_{\rm v}$ (panel a), 
$c_{\rm v}$ (panel b), 
$R_{\rm v}$ (panel c),
$V_{\rm v}$ and $T_{\rm v}$ (panel d),
and $\dot{M}_{\rm v}$ (panel e),
as functions of $z$ (or $\rho_{\rm v}$).   
These curves are obtained by generating a large sample of 
halo merger trees with any given root mass at $z=0$ by a Monte Carlo
algorithm based on the extended Press-Schechter (EPS) formalism
(see Appendix~\ref{app:ssec:extension-merger-tree} for details),
extracting the main branches from the trees, 
and computing the median of each property at each redshift.
These curves show how the properties of a halo evolve along 
its assembly history to the present day. 

In addition to the assembly of dark matter halos, environments of halos can also 
affect the formation of BH seeds through radiation exposure 
(see Appendix~\ref{app:sssec:delay-by-uv-radiation})
and metal contamination (see Appendix~\ref{app:ssec:mass-func-pop3}).
Environmental variations are not well captured by the assembly of 
individual halos, as halo assembly and structure properties correlate only weakly with the 
environment \citep[see fig.~9 of][]{chenRelatingStructureDark2020,
gaoAssemblyBiasClustering2007,wangInternalPropertiesEnvironments2011,wangCharacterizingAssemblyDark2024}.
To model environmental variations more accurately, we thus use halos 
and their merger trees from a cosmological N-body simulation, TNG100-1-Dark (see \S\ref{ssec:halo-assembly}).
The main progenitor of a subhalo is defined as the one with the most massive 
history among all progenitors \citep{deluciaHierarchicalFormationBrightest2007,
rodriguez-gomezMergerRateGalaxies2015}. 
The central subhalo of an FoF halo 
is defined as the one with the most massive history among all subhalos within 
the halo, and the main branch of an FoF halo corresponds to the main branch of 
its central subhalo. 
The concentration $c_{\rm v}$ of a halo is obtained by the 
$V_{\rm max}$-estimator \citep[e.g.][]{klypinResolvingStructureCold2001,wangEfficientRobustMethod2024}.
The specific run, TNG100-1-Dark, used here ensures a balance between statistical
robustness and resolution in halo assembly histories. 

In Fig.~\ref{fig:profile-ladder}a, each purple segment shows the gas density 
at virialization of the halo, assuming an isothermal profile with temperature 
equal to  $T_{\rm v}$ (Eq.~\ref{eq:T-vir}; see also \citealt{crotonManyLivesActive2006,
guoDwarfSpheroidalsCD2011,laceyUnifiedMultiwavelengthModel2016}). 
The card in Fig.~\ref{fig:profile-ladder}b summarizes  
the above equations governing halo properties.
The formation of virialized, dark-matter-dominated halos causes a 
jump in the density to a level about $\Delta_{\rm v} \sim 200$ times the cosmic 
mean density.
The gas density associated with the dark matter halo is also lifted to be
about $\Delta_{\rm v}$ times the cosmic mean baryon density, $\rho_{\rm b}(z)$
(Eq.~\ref{eq:cosmic_rho_b_z}).
At such density, the cooling of the gas is significantly enhanced 
due to the $n^2$ dependence of the two-body cooling processes,
bringing the time needed for the condensation of the gas well below the 
Hubble timescale, $H^{-1}(z)$ (see e.g. the cooling diagram in fig.~2 
of \citetalias{moTwophaseModelGalaxy2024}).
The formation of halos thus forms a necessary step, and sets  
the boundary condition for the subsequent evolution of gas,
star formation and BH growth in galaxies, as we will discuss in the following.




\section{Breeding black-hole seeds}
\label{app:sec:seeding}

\subsection{Producing black-hole seeds in mini-halos}
\label{app:ssec:seed-in-minihalo}

\subsubsection{Formation of self-gravitating clouds}
\label{app:sssec:sgc}

If the angular-momentum support is negligible, as is the case when the 
fraction of the cooled gas in a halo is not much lower than the universal baryon fraction, 
the collapse of gas towards the halo center is expected to lead to the formation of 
an SGC supported by supersonic turbulence, 
as shown in \citetalias{moTwophaseModelGalaxy2024}. 
For a halo with a gas fraction $f_{\rm g} \equiv M_\sscSGC / M_{\rm v}$, 
the gas becomes self-gravitating after contracting by a factor $\sim 1/f_{\rm g}$, 
and the SGC eventually becomes virialized at a radius of $f_{\rm r} R_{\rm v}$, 
where $f_{\rm r}$ is a coefficient to be determined.   
According to \citetalias{chenTwophaseModelGalaxy2025} (see their \S2.2), the radius, density, 
surface density and circular velocity of such an SGC can be written as 
\begin{equation}
    R_\sscSGC 
        = f_{\rm r} R_{\rm v} 
        \approx \left[ 183 \pc \right] 
        f_{\rm r,0.04} M_{\rm v, 9.5}^{1/3}(1+z)_{10}^{-1} 
    \,;\label{eq:r-sgc}
\end{equation}
\begin{equation}
    n_\sscSGC 
        = \frac{f_{\rm g}}{\mu m_{\rm p} f_{\rm r}^3} \rho_{\rm v}
        \approx \left[656 \perccm\right] 
        f_{\rm g, 0.157} f_{\rm r,0.04}^{-3} (1+z)_{10}^3 
    \,; \label{eq:n-sgc}
\end{equation}
\begin{align}
    \Sigma_\sscSGC 
        & \sim \frac{M_\sscSGC }{R_\sscSGC^2}
    \nonumber\\
        & \approx \left[1.49 \times 10^4 \Msunperpcsq\right] 
        f_{\rm g,0.157} f_{\rm r,0.04}^{-2} M_{\rm v,9.5}^{1/3} (1+z)_{10}^{2}
    \,; \label{eq:sigma-sgc}
\end{align}
\begin{align}
    V_\sscSGC 
        &= \sqrt{\frac{GM_\sscSGC}{R_\sscSGC}}
    \nonumber\\
        &\approx \left[ 108 \kms \right] f_{\rm g,0.157}^{1/2} 
        f_{\rm r,0.04}^{-1/2} M_{\rm v,9.5}^{1/3} (1+z)_{10}^{1/2}
    \,,\label{eq:v-sgc}
\end{align}
respectively, and the free-fall timescale ($t_{\rm ff, \sscSGC}$) and  
dynamical timescale ($t_{\rm dyn, \sscSGC}$) are
\begin{align}
    t_{\rm ff, \sscSGC} 
        &= \frac{\pi}{\sqrt{8}} t_{\rm dyn, \sscSGC}
        = \sqrt{ \frac{3\pi}{32 G \rho_\sscSGC}  }
        = \left[1.48\Myr\right] n_{\rm \sscSGC,3}^{-1/2}
    \nonumber\\ 
        &= \left[1.84\Myr\right] f_{\rm g,0.157}^{-1/2} f_{\rm r,0.04}^{3/2} (1+z)_{10}^{-3/2}
    \,,\label{eq:t-ff-sgc}
\end{align}
where the SGC density $\rho_\sscSGC = \mu m_{\rm p} n_\sscSGC$,
and we write $n_\sscSGC = 10^3\perccm\,n_{\sscSGC, 3}$, $f_{\rm g} = 0.157\times f_{\rm g,0.157}$
and $f_{\rm r}= 0.04\times f_{\rm r,0.04}$.

The formation of the SGC produces the jump in density from the halo density  
produced by halo formation. Blue segments in Fig.~\ref{fig:profile-ladder}a 
show the resulting density profiles at this step, where we have assumed an 
isothermal density profile \citep{henriquesLGALAXIES2020Spatially2020,hopkinsFORGEdFIREResolving2024} 
with a temperature, $T \approx 10^4\Kelvin$, at which a cooling barrier is present 
(see e.g. the cooling functions in fig.~4 of \citealt{maioMetalMoleculeCooling2007}
and fig.~1 of \citealt{smithMetalCoolingSimulations2008}).
The blue card in Fig.~\ref{fig:profile-ladder}c summarizes the above equations 
relevant to this step.
Depending on its internal properties and environment, an SGC can fragment to form 
sub-clouds which, in turn, can collapse further to form compact objects of different 
masses, such as dense gas cores and stars. Of particular interest are the 
formation of massive dense cores and the first massive stars at high $z$, and the 
possibility for them to produce BH seeds. In the rest of this subsection, 
we describe in more detail the formation of SGCs in low-mass halos at high $z$ 
and discuss potential channels of BH seed production in SGCs.

\subsubsection{Gas collapse in mini-halos at high $z$}
\label{app:sssec:gas-collapse-in-mini-halos}

\begin{figure} \centering
    \includegraphics[width=\columnwidth]{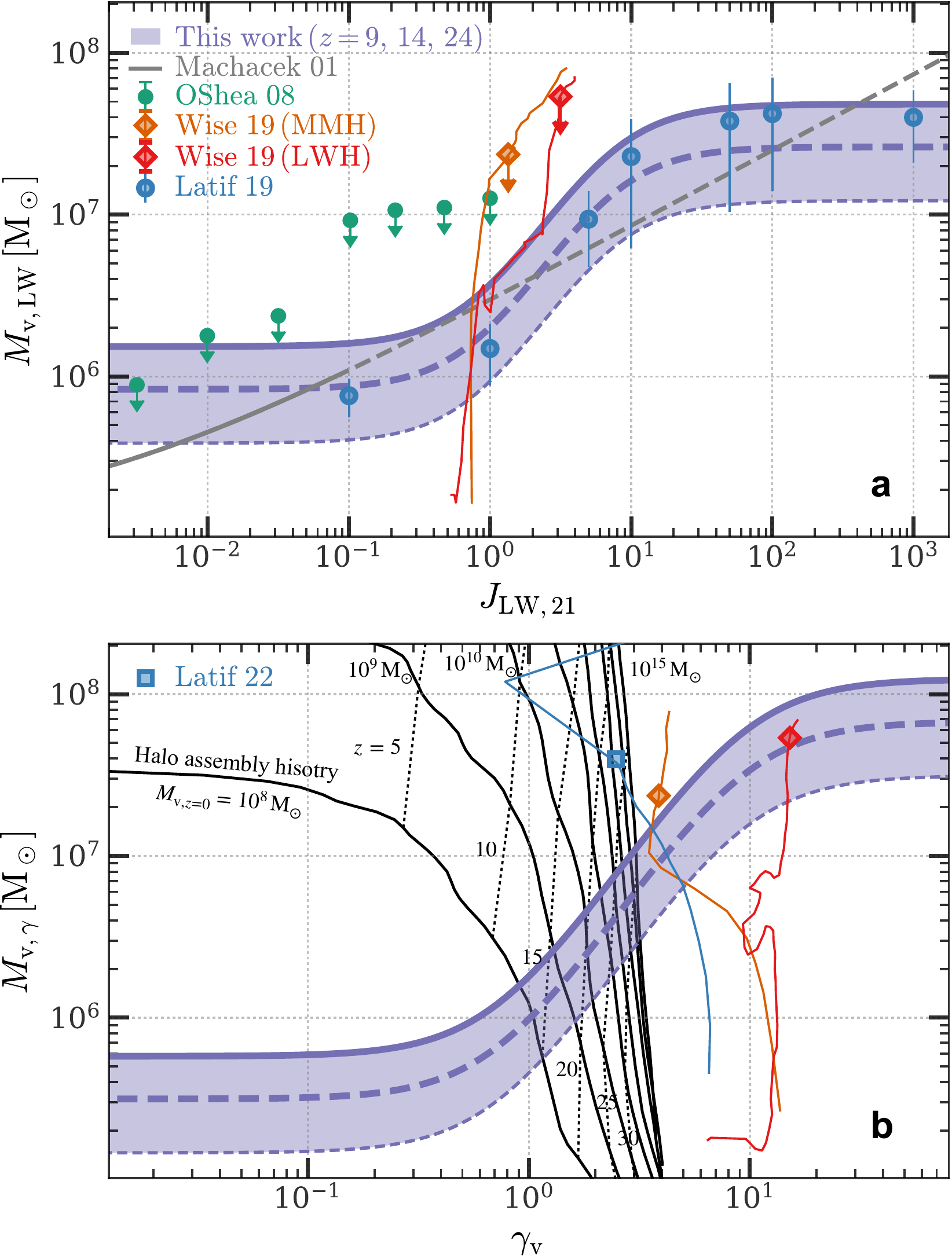}
    \caption{
        {\figem Delay of gas collapse in mini-halos.}
        Here we show the threshold mass of halo expected for the gas to collapse,
        under {\figem a}, different strength of UV background ($J_{\sscLW \rm,21}$,
        measured in the LW band) and {\figem b}, different specific rate 
        of halo accretion ($\gamma_{\rm v}$). The former measures the reduction of
        gas cooling due to the reduction of $\Hmol$ production, while the latter
        measures the dynamical heating due to fast accretion.
        In both panels, solid, dashed and dotted {\figem purple curves}
        show our parameterization of the threshold mass
        at $z=9$, $14$ and $24$, respectively.
        {\figem Black curves} in {\figem b} show the assembly histories of
        halos in the $M_{\rm v}$-$\gamma_{\rm v}$ plane,
        solid for halos with given masses at $z=0$ 
        (from $10^{9}$ to $10^{15}\Msun$, each separated by $1\dex$), 
        and dashed for contours at given $z$ (from 5 to 30), obtained by averaging 
        a large set of Monte Carlo merger trees (see Appendix \ref{app:ssec:extension-merger-tree}).
        The SGC in a pristine halo with dynamical heating thus collapses and forms 
        its first generation of stars once the 
        trajectory of the halo passes the purple band.
        {\figem Colored markers} show simulation results obtained by
        \citet{osheaPopulationIIIStar2008}, 
        \citet{wiseFormationMassiveBlack2019},
        \citet{latifUVRegulatedStar2019} and \citet{latifTurbulentColdFlows2022}.
        The halos used by \citet{osheaPopulationIIIStar2008} and 
        \citet{wiseFormationMassiveBlack2019} are extremely massive ones, 
        dominated by the effect of dynamical heating, and thus we 
        show their results as upper limits in {\figem a}.
        {\figem Colored curves} associated with the markers for \citet{wiseFormationMassiveBlack2019}
        and \citet{latifTurbulentColdFlows2022} are the histories 
        of LW background ({\figem a}) and halo assembly ({\figem b}).
        Grey curve in {\figem a} shows the fitting result,
        $M_{\rm v,cool}/{\rm M}_\odot = 1.25\times 10^5 + 2.86 \times 10^6 J_{\sscLW \rm,21}^{0.47}$,
        obtained by \citet{machacekSimulationsPregalacticStructure2001}
        using their simulations with $J_{\sscLW \rm,21} \lesssim 0.1$,
        with dashed segment indicating the extrapolation.
        See Appendix~\ref{app:ssec:seed-in-minihalo} for details. 
    }
    \label{fig:m-v-cool-fit}
\end{figure}

As discussed in \S\ref{ssec:seeding-main}, the collapse of pristine halo gas  
can occur when a low-mass mini-halo reaches 
$T_{\rm v} \approx T_{\Hmol} = 10^3\Kelvin$ \citep{yoshidaSimulationsEarlyStructure2003,gloverFirstStars2013}.
Using $T_{\rm v}$ given by Eq.~\eqref{eq:T-vir}, we can 
get the critical halo mass for the gas to contract (Eq.~\ref{eq:m-v-cool-H2-main}).
According to Eqs.~\eqref{eq:r-sgc}--\eqref{eq:t-ff-sgc}, the 
mass, radius, density and free-fall timescale are derived 
(see the text below Eq.~\ref{eq:m-v-cool-H2-main}).
The circular velocity of the SGC at $1+z=25$ is $V_\sscSGC = 8.49 \kms$,
a small value that implies gas depletion after seeding.
Purple dotted lines in Fig.~\ref{fig:halo-mahs} show
$M_{\rm v,\Hmol}$ as a function of redshift (panel a) assuming 
$T_{\Hmol,3} = 1$, and the corresponding $R_{\rm v}$ (panel c), 
$V_{\rm v}$ and $T_{\rm v}$ (panel d) obtained 
from $M_{\rm v,\Hmol}$ using Eqs.~\eqref{eq:R-vir},
\eqref{eq:V-vir} and \eqref{eq:T-vir}.
The temperature, $T = T_{\Hmol}$, corresponds to a thermal sound speed of 
$c_{\rm s} \approx 3.4 \kms$.
The implied Jeans mass of the SGC,  
\begin{align} 
    M_{\rm J,\Hmol}
    & = \frac{\pi^{5/2}}{6} \frac{c_{\rm s}^3}{ \sqrt{G^3\rho_\sscSGC} }
    \nonumber\\
    & = \left[ 2.30 \times 10^4 \Msun \right]
        f_{\rm g,0.157}^{-1/2} 
        f_{\rm r,0.04}^{3/2} 
        (1+z)_{25}^{-3/2} T_{\Hmol,3}^{3/2}
    \,, \label{eq:m-jeans-sgc-H2}
\end{align}
sets the lower mass limit for the gas to collapse and fragment. 
The fact that this Jeans mass is comparable to the mass of the SGC
and that the cooling by $\Hmol$ is not rapid ($t_{\rm cool}\sim t_{\rm dyn}$)
implies that the contraction of the SGC proceeds globally without leading to fragmentation 
until other cooling channels are open. For gas with a primordial composition, 
where cooling from metals and dust grains is insignificant, the whole SGC cools 
and contracts gradually to a high density before 
fragmentation and star formation can occur.

The above discussion only takes into account radiative cooling, ignoring 
potential heating sources that may be present. In the presence of heating, the
contraction of the halo gas to form an SGC, and the fragmentation and collapse
of the SGC to form stars, are delayed so that the mass of the Jeans core 
is increased. The central part of the Jeans core may then collapse directly 
to form a BH with mass greater than a few $100\Msun$, 
producing a `heavy-weight' seed to grow into an SMBH. Two sources, one from UV radiation in the environment 
and the other from the energy of the turbulence associated with the fast-accretion regime 
of the host halo, are particularly relevant to our problem, and their effects are described below. 

\subsubsection{Delay of collapse by UV radiation}
\label{app:sssec:delay-by-uv-radiation}

For a given halo, UV radiation produced by young stars in nearby galaxies can 
provide external feedback that suppresses gas cooling and collapse in the halo.
Specifically, UV photons within the window of $11.2$--$13.6\,{\rm eV}$ can 
escape the IGM attenuation \citep[see e.g. \S2.1 of][]{haimanRadiativeFeedbackFirst2000}
and suppress $\Hmol$ cooling \citep[see e.g. \S2 of][]{machacekSimulationsPregalacticStructure2001}, 
and are thus considered a main modulator of seeding.


The threshold mass for a $\Hmol$-cooling mini-halo to collapse is expected to 
increase with increasing $J_{\sscLW}$. 
If the radiation is strong enough ($J_{\rm \sscLW,21} \gg 1$), the collapse of the halo gas 
to form an SGC can be delayed until atomic cooling kicks in, namely when 
$T_{\rm v} \approx T_{\Hatom} \approx 10^4\Kelvin$. 
This gives another critical halo mass, $M_{\rm v, \Hatom}$, for the gas under strong 
LW radiation to collapse (Eq.~\ref{eq:m-v-cool-H-main}),
which is about two orders of magnitude higher than the $\Hmol$-cooling mass.
According to Eqs.~\eqref{eq:r-sgc}--\eqref{eq:t-ff-sgc}, 
the SGC produced has mass $M_\sscSGC = f_{\rm b} M_{\rm v,\Hatom} = 1.92\times 10^6\Msun$,
radius $R_\sscSGC = 11.5\pc$, 
circular velocity $V_\sscSGC = 26.9 \kms$,
and density and free-fall timescale the same as 
those of the $\Hmol$-cooling SGC at a given redshift
(see the text below Eq.~\ref{eq:m-v-cool-H2-main}).
Purple dashed lines in Fig.~\ref{fig:halo-mahs}a, c and d show $M_{\rm v,\Hatom}$
and the corresponding $R_{\rm v}$ and $V_{\rm v}$ ($T_{\rm v}$), respectively,
as a function of redshift. They are parallel to but higher than the corresponding  
$\Hmol$-cooling lines.
More generally, we can parameterize the halo mass for effective cooling to 
drive the formation of SGC as a function of $J_{\sscLW}$ as
\begin{equation}
    M_{\rm v, LW} = \frac{ 
        M_{\rm v, \Hmol} + f_{\rm J} M_{\rm v, \Hatom}
    }{ 
        1 + f_{\rm J} 
    }\,, \label{eq:m-v-LW}
\end{equation}
where $f_{\rm J}$ depends on the intensity of the LW radiation irradiated on the halo, 
\begin{equation}
    f_{\rm J} =
        \left( 
            \frac{J_{\rm \sscLW, 21}}{J_{\rm crit, 21}}
        \right)^{\beta_{\rm J}}
    \,. \label{eq:LW-f-J}
\end{equation}
We set $\beta_{\rm J} = 1.5$ and $J_{\rm crit, 21} = 7.5$ by default to match the results 
from zoom-in hydrodynamic simulations (see below).
So modeled, $M_{\rm v, \sscLW}$ smoothly transits from the $\Hmol$-cooling limit 
to the $\Hatom$-cooling limit at any given redshift.
Fig.~\ref{fig:m-v-cool-fit}a shows our default 
results of $M_{\rm v, \sscLW}$ as a function of $J_{\sscLW}$ at different redshifts, 
in comparison to results obtained by simulations
\citep{machacekSimulationsPregalacticStructure2001,osheaPopulationIIIStar2008,
latifUVRegulatedStar2019,wiseFormationMassiveBlack2019}.
Given the wide range of $J_{\sscLW}$ covered by the set of simulations
of \citet{latifUVRegulatedStar2019}, we calibrate our parameterization 
in Eq.~\eqref{eq:LW-f-J} with it,
yielding the default parameters of $\beta_{\rm J}$ and $J_{\rm crit, 21}$.
The halos of \citet{osheaPopulationIIIStar2008}
and \citep{wiseFormationMassiveBlack2019}
are selected to be extremely massive ones in cosmological volumes,
thus experiencing fast accretion and expected to have strong dynamical heating
(see Appendix~\ref{app:sssec:delay-by-fast-accretion}).
Thus, we show their results for the collapsing threshold as upper limits.

Effects of the UV radiation outside of the LW band, such as the detachment of 
$\Hatom^{-}$ by photons of $> 0.76\,{\rm eV}$, can also affect the formation of 
$\Hmol$.  
Unfortunately, a reliable model of such effects requires detailed modeling of the  
stellar spectral energy distribution (SED), which depends on the star formation and chemical enrichment 
history of the emitting galaxy, and radiative transfer, which depends on the 
properties of the interstellar medium (ISM), circumgalactic medium (CGM)
and IGM, both beyond the scope of this paper.
A less sophisticated model of the UV radiation, as adopted in the hydrodynamic
simulations of \citet{latifUVRegulatedStar2019}, is to assume that the
emitted radiation has a black-body SED 
with some characteristic temperature in the range of $10^4$--$10^5\Kelvin$. 
Our calibration using the results of 
\citet{latifUVRegulatedStar2019} thus partially captures the $\Hatom^{-}$ 
photodetachment.
\citet{agarwalOptimalNeighbourhoodNurture2019} presented a more detailed 
model of the UV radiation by post-processing hydrodynamic simulations,
synthesizing stellar SEDs, and adopting a critical curve for the DCBH
in the 2-D plane of $\Hmol$ photodissociation and $\Hatom^{-}$ photodetachment.
Their results suggest that the net effect of the UV radiation on the delay of 
the SGC collapse is tightly correlated with the star formation rate (SFR) 
of emitting galaxies. This motivates us to use the SFR (or young stars) as a proxy of 
$J_{\sscLW}$ in the threshold variable $f_{\rm J}$
(Eq.~\ref{eq:LW-f-J}) as an effective way to account for the net effect of
the UV radiation. Since our model can predict star formation histories of 
individual galaxies in a cosmological volume, the information needed for the 
modeling of UV radiation is self-consistently included in the model
(see Appendix~\ref{app:ssec:LW-radiation} for details).
The UV radiation produced by young stars, and its effects on the delay of 
SGC formation, are the earliest `radiation feedback' of galaxy 
formation to its environment. 


\subsubsection{Delay of collapse by turbulence associated with fast accretion} 
\label{app:sssec:delay-by-fast-accretion}

An SGC is expected to be a turbulence supported system \citepalias{moTwophaseModelGalaxy2024}.
Depending on the details of the gas collapse 
and the turbulence cascading, the partition between the kinetic and thermal energy 
can be different for different halos (see the different
results obtained by \citealt{wiseFormationMassiveBlack2019}
and \citealt{latifTurbulentColdFlows2022}). 
Since the energy in both forms comes from gravity, we refer to the channel of 
energy injection associated with fast accretion of halos 
collectively as `dynamical heating', regardless of whether the injected 
energy is stored in kinetic or thermal form.

As briefly mentioned in \S\ref{ssec:seeding-main}, 
we first consider an extreme case where energy injection is so efficient that 
radiative cooling is inefficient to dissipate it. If the cooling timescale 
$t_{\rm cool}$ becomes 
comparable or longer than $V_{\rm v}/\dot{V}_{\rm v}$, the timescale for the 
potential energy of a halo to be totally flushed by infalling material,
the gravitational energy associated with the mass accretion of the halo  
can generate a significant impact on the SGC over the cooling timescale. 
Thus, gas in a halo can collapse rapidly only when 
$t_{\rm cool} \lesssim V_{\rm v}/\dot{V}_{\rm v}$. Equating the two timescales, 
we obtain a critical condition for the turbulence to affect the state of 
the SGC: $V_{\rm v} / \dot{V}_{\rm v} \approx t_{\rm cool}$.
Noticing that at the collapse of SGC, $t_{\rm cool} \sim t_{\rm v}$, and 
using Eq.~\eqref{eq:t-vir} for $t_{\rm v}$, we have
\begin{equation}
    \gamma_{\rm v}
    \equiv \frac{\dot{V}_{\rm v}}{V_{\rm v}H(z)} 
    = -\dv{\ln V_{\rm v}}{\ln (1+z)}
    \approx 10
    \,.  \label{eq:gamma-v-crit}
\end{equation}

To incorporate the additional support to SGC by turbulence, we replace the sound 
speed in the Jeans mass (Eq.~\ref{eq:m-jeans-sgc-H2}) with an effective velocity 
dispersion, $\sigma_{\rm eff} = (c_{\rm s}^2 + v_{\rm t}^2)^{1/2}$.
The exact value of the turbulent velocity dispersion
$v_{\rm t}$ is unknown. As a conservative estimate,
we take the value, $v_{\rm t} \approx 20 \kms$, 
obtained by \citet{latifTurbulentColdFlows2022} from their simulation 
of a fast-accreting halo with $\Hmol$ cooling unable to dissipate the turbulence.
This value of $v_{\rm t}$, together with $c_{\rm s}$ given by $T = T_{\Hmol}$ 
and the SGC density given by Eq.~\eqref{eq:n-sgc}, leads to the following Jeans mass: 
\begin{equation} 
    M_{\rm J,max} = \left[4.94\times 10^6 \Msun\right] 
    f_{\rm g,0.157}^{-1/2} 
    f_{\rm r,0.04}^{3/2} 
    (1+z)_{25}^{-3/2}
    \,. \label{eq:m-jeans-sgc-max}
\end{equation}
The SGC thus starts to collapse rapidly only when it reaches this mass scale. 
The corresponding radius and circular velocity are 
$R_\sscSGC=15.7\pc$ and $V_\sscSGC=36.8\kms$, respectively. 
Purple solid lines in Fig.~\ref{fig:halo-mahs}a, c and d show
$M_{\rm v} = M_{\rm J,max}/f_{\rm b}$
and the corresponding $R_{\rm v}$ and $V_{\rm v}$ ($T_{\rm v}$), respectively,
as a function of redshift. They are parallel to but higher than the
corresponding $\Hmol$-cooling lines, and close to the corresponding $\Hatom$-cooling lines.

More generally, a halo can be somewhere between the two extremes, the one without 
turbulence ($v_{\rm t} \ll c_{\rm s}$) and the other with strong turbulence 
($v_{\rm t} \gg c_{\rm s}$). The Jeans mass is thus between 
those given by Eq.~\eqref{eq:m-jeans-sgc-H2} and 
Eq.~\eqref{eq:m-jeans-sgc-max}, with a transition at $\gamma_{\rm v} \approx 10$.
We parameterize the halo mass at which the thermal and turbulent motion is 
no longer able to support the SGC against gravitational collapse 
as a function of $\gamma_{\rm v}$:
\begin{equation}
    M_{\rm v, \gamma} = f_{\rm b}^{-1} \frac{ 
        M_{\rm J,\Hmol} + f_{\gamma} M_{\rm J,max}
        }{ 
            1 + f_{\gamma} 
        }
    \,,\label{eq:m-v-gamma}
\end{equation}
where
\begin{equation}
    f_{\gamma} = 
    \left( 
        \frac{\gamma_{\rm v}}{\gamma_{\rm crit}} 
    \right)^{\beta_{\gamma}} 
    \,, \label{eq:f-gamma-dyn-heat}
\end{equation}
and we adopt $\gamma_{\rm crit} = 10$. The exact value for $\beta_{\gamma}$ 
is not known, and we adopt $\beta_{\gamma} = 2$ as a demonstration of the effect of 
dynamical heating. Future calibrations using detailed hydrodynamic simulations are 
needed to provide a more reliable estimate of $\gamma_{\rm crit}$ 
and $\beta_{\gamma}$, and to account for the synergy of dynamical heating
with other heating processes (e.g. the UV radiation discussed in 
Appendix~\ref{app:sssec:delay-by-uv-radiation}).

Purple curves in 
Fig.~\ref{fig:m-v-cool-fit}b show our model of the collapsing threshold, 
$M_{\rm v, \gamma}$, at three different redshifts, as a function of 
halo specific growth rate, $\gamma_{\rm v}$. The threshold from
$z = 24$ to $9$ covers a narrow band of $\approx 0.3\dex$ due to the dependence of 
the Jeans mass on SGC density (thus redshift). 
The delay of SGC collapse by dynamical heating can lead to an increase of up to 
$\approx 2 \dex$ in $M_{\rm v}$ when $\gamma_{\rm v} \gtrsim 10$.
Since halos with higher $M_{\rm v}$ accrete faster than those with lower $M_{\rm v}$
at a given $z$, formation of BH seeds in the progenitors of massive 
halos at the present day is most likely affected significantly
by dynamical heating.
To see this, we take the EPS-based halo assembly histories used to
produce Fig.~\ref{fig:halo-mahs}, compute the specific growth rate $\gamma_{\rm v}$,
and average over halos with given $M_{{\rm v},z=0}$. 
Solid black curves in Fig.~\ref{fig:m-v-cool-fit}b
show the $M_{\rm v}$-$\gamma_{\rm v}$ trajectories of halos with different 
$M_{{\rm v},z=0}$, while each dashed curve links a snapshot of these 
trajectories at a given redshift. Halos with $M_{{\rm v}, z=0} \geqslant 10^{12}\Msun$
have very similar $M_{\rm v}$ and $\gamma_{\rm v}$ at $z \gtrsim 15$,
consistent with the assembly histories shown in Fig.~\ref{fig:halo-mahs}. 
$\gamma_{\rm v}$ exhibits slow reduction with redshift for all halos 
at high $z$. The significant reduction of $\gamma_{\rm v}$ occurs earlier 
for halos with lower mass, consistent with the findings 
in \citetalias{moTwophaseModelGalaxy2024} (see their fig.~B1) 
that $z_{\rm f}$ shows a negative correlation with $M_{{\rm v}, z=0}$.

The expected epoch of collapse, as well as host-halo properties at this epoch, 
of an SGC being dynamically heated by fast accretion (without taking into account 
other heating sources) can be identified by 
the intersection point of the $M_{\rm v}$-$\gamma_{\rm v}$ trajectory with the purple 
band in Fig.~\ref{fig:m-v-cool-fit}b.
Since the slopes of equal-time contours (dashed black) are higher than 
that of the purple band in the $M_{\rm v}$-$\gamma_{\rm v}$ plane,
SGCs in massive halos are expected to collapse earlier than those in 
less massive halos. Note that this is only a conclusion averaged over 
ensembles of halos, and halo-to-halo variations can be significant.
Halos with $M_{{\rm v}, z=0} = 10^{11}\Msun$ (hosting dwarf galaxies 
with $M_* \approx 10^9\Msun$ at $z=0$) reach the collapsing threshold 
at $z \approx 25$ when $M_{\rm v}$ is a few times $10^6\Msun$
while those with $M_{{\rm v}, z=0} \gtrsim 10^{13}\Msun$ (hosting BCGs
at $z=0$) are expected to collapse at $z \approx 30$ with a 
comparable $M_{\rm v}$.
The values of $M_{\rm v}$ at collapse for these halos are thus $\approx 1\dex$
higher than those of $\Hmol$-cooling halos ($M_{\rm v,\Hmol}$, Eq.~\ref{eq:m-v-cool-H2-main}), 
implying that much more baryonic gas is available for the formation of stars and BHs in the SGCs.
Meanwhile, such a delay of collapse also gives the SGCs more chance to 
be irradiated by the UV photons from nearby star-forming galaxies, causing
further delay, as we will discuss and model in 
Appendix~\ref{app:sssec:seed-in-atomic-cooling-halo}.

For comparison, we show by colored curves in Fig.~\ref{fig:m-v-cool-fit}b
the assembly histories of three pristine halos whose baryonic processes were 
resolved by hydrodynamic simulations. Two of them are the same halos 
as those in panel a, studied by 
\citet{wiseFormationMassiveBlack2019}, and one by 
\citet{latifTurbulentColdFlows2022}. 
All of them are progenitors of extremely massive halos, and were predicted 
to seed DCBHs and then evolve to quasars at $z \approx 6$.
$\gamma_{\rm v}$ of these halos are extremely large, reflecting an accretion 
faster than that expected from progenitors of halos 
with $M_{{\rm v},z=0} \approx 10^{15}\Msun$. Dynamical heating is 
thus expected to be very strong in these halos, causing significant
delay of the SGC collapse. 
These are indeed seen in these simulations,
as shown by the colored markers in Fig.~\ref{fig:m-v-cool-fit}b
that denote the collapses they identified, close to the prediction of 
our model.

\subsubsection{Consequences of the delayed collapse}
\label{app:sssec:seed-in-atomic-cooling-halo}

As shown in Fig.~\ref{fig:m-v-cool-fit}, both UV radiation and fast accretion can delay the collapse of SGCs, as well as the formation of the first generation of stars and the 
breeding of BH seeds. The effects 
are described by the two halo mass thresholds, 
$M_{\rm v, \sscLW}$ and $M_{\rm v, \gamma}$, 
respectively, and the higher one is used as the final threshold for the 
collapse of each SGC (Eq.~\ref{eq:m-v-collapse}).
If either $J_{\sscLW}$ or $\gamma_{\rm v}$ is sufficiently large, the collapse is delayed 
to the regime of atomic cooling (Eq.~\ref{eq:m-v-cool-H-main}).
Consequently, the total available baryon mass, 
$M_{\rm g,th} \approx f_{\rm b} M_{\rm v, th}$, and the halo 
virial velocity, $V_{\rm v}$, are much larger than those for a $\Hmol$-cooling halo.

The purple band in Fig.~\ref{fig:halo-mahs}a summarizes the 
regime in the $z$-$M_{\rm v}$ plane
where the first collapse of an SGC in
a halo can take place. The dotted, dashed and solid curves represent 
$M_{\rm v, \Hmol}$ (Eq.~\ref{eq:m-v-cool-H2-main}),
$M_{\rm v, \Hatom}$ (Eq.~\ref{eq:m-v-cool-H-main}), 
and $f^{-1}_{\rm b} M_{\rm J,max}$ (Eq.~\ref{eq:m-jeans-sgc-max}), respectively.
These masses are plotted on top of the median assembly histories of halos 
of different masses at $z=0$ (black curves). Pop-III stars are expected to form in a
primordial halo when the halo mass, $M_{\rm v}(z)$, crosses $M_{\rm v,\Hmol}$ (dotted line) and
enters the Pop-III band limited by $J_{\sscLW}$ (dashed line)  
or $\gamma_{\rm v}$ (solid line). Note that the region for Pop-III stars in 
the $z$-$M_{\rm v}$ plane is almost fully (partly) covered by that marked Pop-II (Pop-I) GC, 
the region in which Pop-II (Pop-I) globular clusters are expected to form 
\citepalias[see][]{chenTwophaseModelGalaxy2025}. Thus, only halos 
that have not been self-enriched by previous generations of stars or externally enriched 
by star formation in nearby galaxies can host Pop-III stars, 
while the others are likely to host globular clusters.
In the progenitors of halos with $M_{{\rm v}, z=0} \approx 10^9$--$10^{10}\Msun$, 
Pop-III stars can form at $z \lesssim 10$ through the first collapse of the SGC 
provided that the IGM around these halos has not been enriched
and the UV radiation and/or dynamical heating are sufficiently strong
to delay the collapse to that low redshift.
These late-forming Pop-III stars may thus be observable in JWST surveys, such as 
those conducted recently by
\citet{wangStrongHeII2024}, \citet{hsiaoSAPPHIRESExtremelyMetalPoor2025}, 
\citet{suessMediumBandsMega2024},
\citet{naiduAllLittleThings2024}
and \citet{fujimotoGLIMPSEUltrafaint1052025}.
The enrichment history of both the ISM and IGM turns out to be critical in 
disentangling the two populations of stars in the $z$-$M_{\rm v}$ plane, as will be 
discussed in Appendix~\ref{app:ssec:IGM-enrichment}.
For reference, purple bands in Fig.~\ref{fig:halo-mahs}c and d 
show the first-collapse regime in the $z$-$R_{\rm v}$ and $z$-$V_{\rm v}$ ($z$-$T_{\rm v}$)
planes, respectively.

The free-fall timescale of an SGC at its first collapse is about $1 {\rm Myr}$ 
(see the text below Eq.~\ref{eq:m-v-cool-H-main}),
independent of halo mass (Eq.~\ref{eq:t-ff-sgc}).
This timescale is shorter than the `supernova-free' timescale, 
$t_{\rm snf} \approx 1\Myr$ \citep{chenTwophaseModelGalaxy2025,dekelEfficientFormationMassive2023}. 
Thus, supernova (SN) feedback is expected to be unimportant
for the first collapse of the SGC. 
This is particularly true for the Jeans core
that has $n \gtrsim 10^6\perccm$, corresponding to an even shorter $t_{\rm ff}$
of $\lesssim 0.05\Myr$. Without other types of feedback, the upper limit 
of the gas accretion rate onto the central star can then be estimated as 
\begin{equation} 
    \dot{M}_{\rm g}     \leqslant \frac{f_{\rm b} M_{\rm v,th}}{t_{\rm ff,\sscSGC}} 
    \leqslant \frac{M_{\rm J,max}}{t_{\rm ff, \sscSGC}}
    \approx 10.7 \msunperyr
    \,, \label{eq:m-g-accr-upper-lim}
\end{equation}
independent of redshift due to the cancellation of the redshift dependence in
$M_{\rm J,max}$ (Eq.~\ref{eq:m-jeans-sgc-max}) and $t_{\rm ff, \sscSGC}$ (Eq.~\ref{eq:t-ff-sgc}).
As suggested by numerical simulations of the formation of supermassive stars 
\citep[e.g.][]{hosokawaRapidlyAccretingSupergiant2012,hosokawaFormationPrimordialSupermassive2013}, a high accretion 
rate at $\gtrsim 10^{-2} \msunperyr$ causes the outer envelope of the 
star to bloat to a large radius,
\begin{equation} \label{eq:r-sms}
    R_* \approx 2.6\times 10^4 R_\odot \left( \frac{M_*}{100 M_\odot} \right)^{1/2}
    = 12 \au \left( \frac{M_*}{100 M_\odot} \right)^{1/2}\,,
\end{equation}
where $M_*$ is the mass of the star.
The stellar object, with its effective temperature locked at a low value of 
$T_{\rm eff} \approx 5000\Kelvin$ due to the strong temperature dependence of 
the $\Hatom^-$ absorption, is inefficient at producing UV radiation, so that
accretion onto the star continues and the stellar mass can grow
up to $\sim 10^3$--$10^5\Msun$, the regime of supermassive stars (SMSs), 
until the density of the stellar envelope drops to a level that significantly 
reduces the opacity of $\Hatom^-$.
An SMS is expected to collapse when the gas pressure is no longer able to 
counterbalance its weight, leading to the formation of an intermediate-mass, 
direct-collapse black hole (DCBH), with a BH mass comparable to the SMS
\citep[e.g.][]{choiSupermassiveBlackHole2013,latifBlackHoleFormation2013,
wiseFormationMassiveBlack2019,latifRadiationHydrodynamicalSimulations2021,
latifTurbulentColdFlows2022}.

The above estimate ignores potential suppression of the accretion rate by 
feedback, angular momentum support and fragmentation of the gas to form low-mass 
stars. It thus only represents an upper limit of the mass that a DCBH can reach
in a mini-halo. In the following, we consider a more realistic model that
takes into account the effects of radiative feedback from the central star,
and use it to derive the mass spectra of Pop-III stars and BH seeds
expected to form during the first collapse of an SGC in a mini-halo.

\subsection{Radiation-regulated mass function of Pop-III stars and the mass spectrum of black-hole seeds}
\label{app:ssec:mass-func-pop3}

\begin{figure} \centering
    \includegraphics[width=\columnwidth]{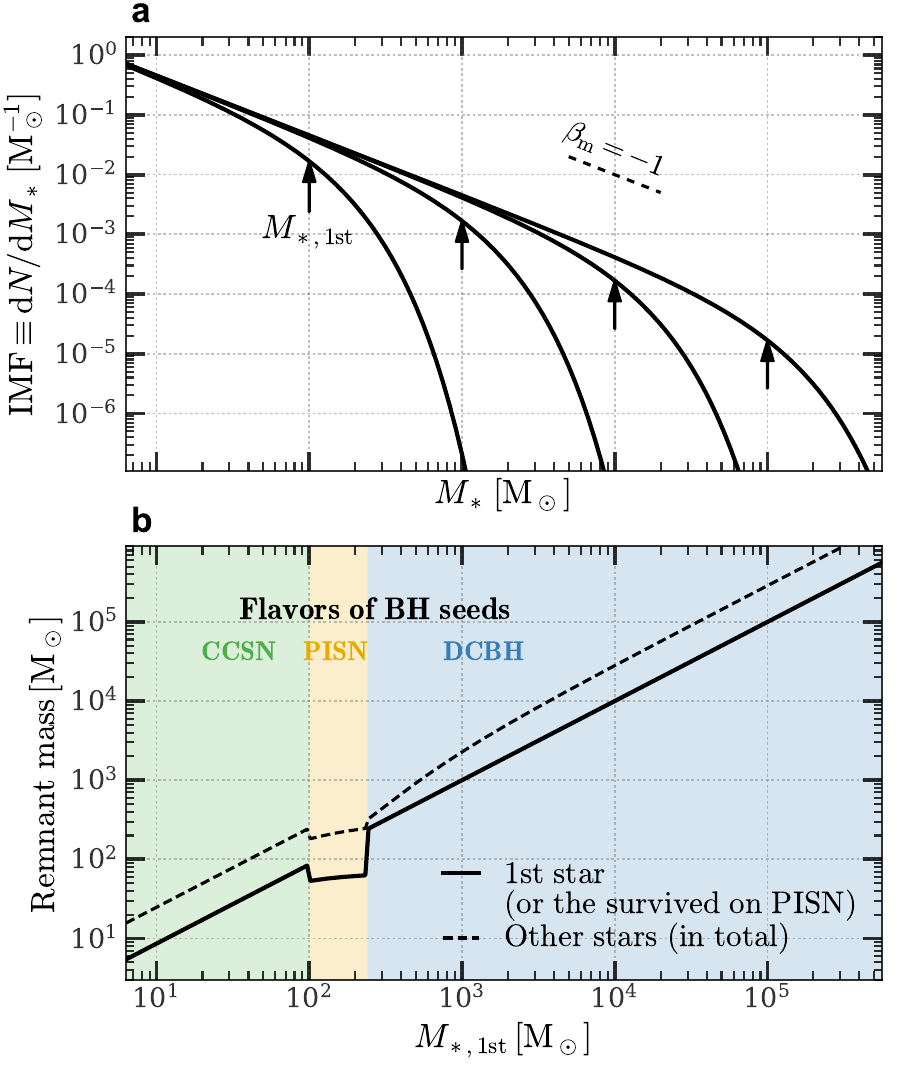}
    \caption{{\figem  Masses of Pop-III stars and BH remnants. }
        {\figem a}, initial mass function (IMF) of stars in a Pop-III 
        star cluster, each with a given mass of the dominant (1st) star 
        ($M_{\rm *,1st}$, indicated by the arrow).
        {\figem b}, remnant mass of stars in a Pop-III star cluster
        as a function of $M_{\rm *,1st}$ of the cluster, 
        {\figem solid} for the dominant star (i.e. the mass of BH seed) 
        and {\figem dashed} for the other stars of the cluster 
        in total. 
        The evolution (CCSN, PISN or DCBH) of the dominant star to end its 
        life is determined by $M_{\rm *,1st}$.
        If the dominant star triggers PISN, we take the 
        remnant mass of the most massive star surviving PISN as 
        a replacement.
        See Appendix~\ref{app:ssec:mass-func-pop3} for the details.
    }
    \label{fig:pop3-imf}
\end{figure}

The description presented above provides the avenue to model the mass 
of individual Pop-III stars expected to form in halos where the 
gas state is regulated by different cooling and heating processes.  
In this subsection, we first use the results to construct a model for 
the initial mass function (IMF) of Pop-III stars. We then combine
the results with assumptions of stellar evolution to obtain the mass spectrum of BH seeds. 

As seen above, the slow cooling of primordial gas has important implications to the star 
formation process. For the radiative cooling to effectively dissipate the energy generated by
the gravitational collapse, the SGC has to contract to achieve a sufficiently high density. 
Gas fragmentation and star formation, which are possible only when the SGC can cool 
effectively, are thus delayed until the gas density is sufficiently 
high \citep[see e.g. fig.~9 of][]{omukaiCanSupermassiveBlack2008}.
The situation is similar to that in the formation of Pop-II GCs in metal-poor 
($Z \sim 10^{-2} \Zsun$) sub-clouds with $n \gtrsim n_{\rm snf} = 10^{3.5}\perccm$ 
\citep{chenTwophaseModelGalaxy2025}, except the more extreme condition produced 
by the fact that the temperature does not drop significantly even when the density reaches 
$n \gtrsim n_{\rm core} = 10^6 \perccm$ in a Pop-III gas 
\citep[see e.g. fig.~1 of][]{latifBirthMassFunction2022}. 
The dense environment favors the continuous accretion by the dominating 
central SMS, which suppresses the growth of off-center stars 
\citep{bonnellCompetitiveAccretionEmbedded2001,chonSupermassiveStarFormation2020}. 
Merger into the SMS is frequent due to its central position and large size \citep[e.g.][]{stacyBuildingPopulationIII2016,chonSupermassiveStarFormation2020},
which can also increase its mass. The IMF of the stars formed in the SGC is 
thus expected to be significantly affected by these processes and to evolve to be top-heavy.
If additional `heating' sources, such as the LW radiation and    
the turbulence generated by the fast accretion, are present, the formation of 
small-scale gas clumps can be further suppressed, making the IMF 
even more top-heavy.

As discussed in Appendix~\ref{app:sssec:seed-in-atomic-cooling-halo}, 
although a gas cloud hosting Pop-III stars is expected to be free 
of SN feedback, radiative feedback from young Pop-III stars can still be present to 
regulate the gas accretion on to the SGC,
`quench' the star formation, and limit the total stellar mass
of the star cluster formed there
\citep[e.g.][]{latifRadiationHydrodynamicalSimulations2021,latifBirthMassFunction2022}. 
Following the same strategy as \citetalias{moTwophaseModelGalaxy2024} 
in modeling the AGN-feedback effects on SMBH formation, and \citetalias{chenTwophaseModelGalaxy2025}
in modeling the stellar-feedback effects on globular-cluster formation, we relate the stellar mass 
of a Pop-III star cluster to the properties of its host SGC by a balance
between the feedback energy/momentum and the SGC binding energy/force,
with model parameters appropriate for the Pop-III regime.
The top-heavy IMF and $L\propto M_*^2$ for massive stars 
imply that the stellar feedback is dominated by the most massive star, 
referred to as the first (1st) star or dominant star in the following.

As one extreme case, we first consider a halo at the $\Hmol$-cooling
threshold, $M_{\rm v,\Hmol}$ (Eq.~\ref{eq:m-v-cool-H-main}).
The 1st star in such a halo cannot grow too massive, and can approach the 
main sequence before the cloud is destroyed 
\citep[see e.g. fig.~6 of][]{stacyBuildingPopulationIII2016}.
The zero-age main sequence (ZAMS) luminosity is
\begin{equation} \label{eq:L-ZAMS}
    L_* = L_{\ssc ZAMS} \approx 1.4 \times 10^4 \Lsun \left(
        \frac{M_*}{10 \Msun}
    \right)^2
\end{equation}
\citep[e.g.][]{schallerNewGridsStellar1992,hosokawaEvolutionMassiveProtostars2010}.
The upper limit of the stellar mass, above which the gas accretion is stopped,
can be estimated as 
\begin{equation} \label{eq:momentum-balance}
    \frac{\eta_{\rm r} L_*}{c} = \frac{G M_\sscSGC^2}{R_\sscSGC^2}\,,
\end{equation}
where the free-parameter $\eta_{\rm r}$ stands for the coupling efficiency
of the stellar radiation to the surrounding gas (and can be $>1$ due to multiple 
scattering and the optically-thick nature of dense SGC).
Taking $M_\sscSGC = f_{\rm b} M_{\rm v, \Hmol}$, 
and using Eq.~\eqref{eq:r-sgc} for $R_\sscSGC$, we can estimate the star formation efficiency as
\begin{align} \label{eq:sfe-ZAMS}
    \epsilon_{\ssc ZAMS} 
    &\equiv \frac{M_*}{M_\sscSGC }
    = \left[
        \frac{ G (10 \Msun)^2 c}{ (1.4\times 10^4 \Lsun) \eta_{\rm r}}
    \right]^{1/2} R_\sscSGC^{-1} 
    \nonumber \\
    &= 2.17 \times 10^{-3} \eta_{\rm r,25}^{-1/2} \left( \frac{R_\sscSGC}{3.63\pc} \right)^{-1}
    \nonumber \\
    &= 2.17 \times 10^{-3} \eta_{\rm r,25}^{-1/2} (1+z)_{25}^{3/2}
    \,,
\end{align}
where our default value, $\eta_{\rm r} = 25$, is chosen to be consistent with
those obtained by the simulations of \citet{latifRadiationHydrodynamicalSimulations2021},
\citet{latifBirthMassFunction2022} and \citet{latifTurbulentColdFlows2022}
for halos of $10^5$--$10^7 \Msun$ at $z \approx 15$--$30$.
The corresponding stellar mass is
\begin{equation}
    M_* = \epsilon_{\ssc ZAMS} M_\sscSGC = 132 \Msun \eta_{\rm r,25}^{-1/2}\,.
\end{equation}
This value does not depend on redshift, implying that the SGC 
collapse in a halo with mass at the $\Hmol$-cooling threshold forms a 
fixed amount of stars. This is a consequence of the fact that 
the redshift dependencies of $\epsilon_{\ssc ZAMS}$ and $M_{\rm v,\Hmol}$
cancel each other. 

As the other extreme, we consider an SGC that is supported by strong turbulence 
in a halo with mass $f_{\rm b}^{-1} M_{\rm J,max}$ (Eq.~\ref{eq:m-jeans-sgc-max}).
In this case, the 1st star has sufficient gas supply to grow massive, 
with its luminosity limited by the Eddington luminosity,
\begin{equation} \label{eq:L-Edd-SMS}
    L_* = L_{\ssc Edd} \approx 3.284 \times 10^6 \Lsun \left(
        \frac{M_*}{100 \Msun}
    \right)
\end{equation}
\citep{hosokawaRapidlyAccretingSupergiant2012}.
Substituting this luminosity into the momentum balance 
(Eq.~\ref{eq:momentum-balance}),
taking $M_\sscSGC = M_{\rm J,max}$, and using 
Eq.~\eqref{eq:sigma-sgc} for $\Sigma_\sscSGC$,
we can estimate the star formation efficiency:
\begin{align}  
    \epsilon_{\ssc Edd}
    & = \frac{ G (100 \Msun) c}{(3.284\times 10^6 \Lsun) \eta_{\rm r}}
        \Sigma_\sscSGC
    \nonumber\\
    &= 5.29 \times 10^{-3}  \eta_{\rm r,25}^{-1} 
        \left(\frac{\Sigma_\sscSGC}{2.00\times 10^4 \Msunperpcsq}\right)
    \nonumber\\
    & = 5.29 \times 10^{-3}  \eta_{\rm r,25}^{-1} (1+z)_{25}^{3/2}
    \,.\label{eq:sfe-Edd}
\end{align}
This efficiency is only slightly higher than that for a $\Hmol$-cooling halo, 
and is a result of the balance between the higher binding force of the more massive SGC 
and the stronger radiative feedback of massive stars. The redshift dependence remains the 
same as that of $\epsilon_{\ssc ZAMS}$, and the stellar mass of the 1st star, 
\begin{equation} 
    M_* = \epsilon_{\ssc Edd} M_\sscSGC 
        = 2.62\times 10^4 \Msun \eta_{\rm r,25}^{-1/2}
    \,,\label{eq:m-star-Edd}
\end{equation}
is again independent of redshift.

Thus, the 1st star in a Pop-III cluster accounts for a few times $10^{-3}$ of the mass of an SGC 
at $z \approx 25$, and scales as
\begin{equation}
    M_* = \epsilon M_\sscSGC \propto M_\sscSGC^{\beta} 
\end{equation}
at a given SGC density, where the star formation efficiency $\epsilon$ varies from 
$\epsilon_{\ssc ZAMS} \propto R_\sscSGC^{-1}$ (Eq.~\ref{eq:sfe-ZAMS}) in pure $\Hmol$-cooling halos
to $\epsilon_{\ssc Edd} \propto \Sigma_\sscSGC$ (Eq.~\ref{eq:sfe-Edd}) in extremely turbulent
$\Hatom$-cooling halos. 
The index, $\beta$, thus varies from
$2/3$ to $4/3$ between the two extremes.
In both cases, the redshift-dependent factors in $\epsilon$ and $M_\sscSGC$ cancel out, 
leading to a time-independent $M_*$-$M_\sscSGC$ relation. 
The delay of the SGC collapse by either the UV radiation or the fast accretion
thus affects the mass of the 1st star by elevating $M_\sscSGC$ ($\approx f_{\rm b}M_{\rm v}$), 
with the star formation efficiency constrained in a narrow range. 
Note that we assumed $\eta_{\rm r}$ to be the same for the two cases, 
which is only a crude approximation. 
In reality, an SMS may be unstable against pulsation, and its radius and 
temperature can experience episodic oscillations due to a strongly varying 
inflow rate, making the coupling efficiency, $\eta_{\rm r}$, mass dependent
\citep[e.g.][see their Extended Data fig. 3]{latifTurbulentColdFlows2022}.
The asymmetry in the internal structure of stars and in the distribution of
gas around the stars can also affect the coupling efficiency
\citep[e.g.][]{susaMassFirstStars2013,latifRadiationHydrodynamicalSimulations2021}
and cause halo-to-halo variations in the mass spectrum of Pop-III stars.
Such details need to be modeled with numerical simulations.

More general cases are expected to fall between the two extremes described above, 
and we model them as follows. Once the halo mass for an SGC about to 
collapse is determined (Eq.~\ref{eq:m-v-collapse}), we obtain the properties
of the SGC using Eqs.~\eqref{eq:r-sgc}--\eqref{eq:v-sgc}, the star formation 
efficiency $\epsilon = \epsilon_{\ssc ZAMS}$ using Eq.~\eqref{eq:sfe-ZAMS}, 
and $M_*$ and $L_*=L_{\ssc ZAMS}$ of the 1st star using Eq.~\eqref{eq:L-Edd-SMS}.
If $L_*$ so obtained does not exceed $L_{\ssc Edd}$ given by $M_*$, we 
take $M_*$ as the mass of the 1st star; otherwise, we compute 
$\epsilon = \epsilon_{\ssc Edd}$ from Eq.~\eqref{eq:sfe-Edd} and use it to 
find the mass of the 1st star.

Once the mass of the 1st star, $M_{\rm *,1st}$, 
is determined through the radiation-regulated processes described above, 
we obtain the full IMF using the parametrization:
\begin{equation}
    \dv{N}{M_*} = 
        \phi_{\rm m} \left(\frac{M_*}{M_{\rm *,1st}}\right)^{-\beta_{\rm m}} 
        \exp\left(-\frac{M_*}{M_{\rm *,1st}}\right)
    \,.\label{eq:imf-pop3}
\end{equation}
We use $\beta_{\rm m} = 1$ as our default choice, as suggested by numerical 
simulations \citep[e.g.][]{stacyBuildingPopulationIII2016,latifBirthMassFunction2022,
chonSupermassiveStarFormation2020} and semi-analytical models 
\citep[e.g.][]{liuGravitationalWavesMergers2024}.
However, we note that the IMF of Pop-III stars is still 
poorly understood, and other values of $\beta_{\rm m}$ are possible 
\citep[see e.g.][for a number of alternatives]{costaMassiveBinaryBlack2023}.
The normalization factor $\phi_{\rm m}$ is determined by the constraint
that the expected number of stars above $M_{\rm *,1st}$ is unity:
\begin{equation}
    1 = \int_{M_{\rm *,1st}}^{\infty} \dv{N}{M_*} \dd{M_*} 
    \,,\label{eq:imf-norm}
\end{equation}
which gives $\phi_{\rm m} = 4.56 / M_{\rm *,1st}$ for $\beta_{\rm m}=1$.
Thus, at a given $\beta_{\rm m}$, the only
parameter needed to fully specify the IMF is $M_{\rm *,1st}$, which 
we have already determined above from SGC properties and the
feedback-regulated scenario.
Fig.~\ref{fig:pop3-imf}a shows the IMF of stars in a Pop-III 
star cluster with the default choice of $\beta_{\rm m}$
for a number of different $M_{\rm *,1st}$ values 
in the range of $10^2$--$10^5\Msun$. At $M_* \ll M_{\rm *,1st}$, all 
IMFs converge to the same logarithmic slope of $-1$, specified 
by $\beta_{\rm m}$, and the same normalization. 
A higher value of $M_{\rm *,1st}$ gives a higher number of 
massive stars other than the 1st star. As we will see below, 
massive stars are expected to evolve along pathways 
different from the less massive ones, thus 
leaving BH seeds with different flavors.

The total stellar mass other than the 1st star is determined by 
\begin{equation} 
    M_{\rm *, other} =  \int_{M_{\rm *,min}}^{M_{\rm *,1st}} M_* \dv{N}{M_*} \dd{M_*} 
    \,,\label{eq:m-star-other}
\end{equation}
where $M_{\rm *,min}$ is the lower limit of the IMF.
Despite the uncertainty of $M_{\rm *,min}$ \citep[e.g.][]{stacyBuildingPopulationIII2016,
wiseFormationMassiveBlack2019,chonSupermassiveStarFormation2020,
latifRadiationHydrodynamicalSimulations2021},
the total mass, $M_{\rm *, other}$, in stars 
other than the 1st is very insensitive to $M_{\rm *,min}$: 
using $M_{\rm *,min} = 0$, the integration of Eq.~\eqref{eq:m-star-other} gives 
\begin{equation}
    M_{\rm *, other} = 2.88 M_{\rm *,1st}
    \,,\label{eq:m-star-other-value}
\end{equation}
and the variation is less than $0.01 M_{\rm *,1st}$ for any  
$M_{\rm *,min} < 0.01 M_{\rm *,1st}$.


The 1st star in a Pop-III star cluster is expected to die 
on a timescale of a few ${\rm Myr}$ and collapse into a BH 
\citep[see e.g. \S2.2 of][]{costaMassiveBinaryBlack2023}.
As dynamical interactions among stars tend to sink the more massive object
into the center, the BH originating from the 1st star is expected to become 
the central BH and to grow further by accreting gas and stars from its neighborhood,
as to be detailed in Appendix~\ref{app:sec:growing}. 
The initial mass of the central BH is thus determined by the remaining mass of 
the 1st star after it dies. 
Here we adopt the result of \citet[][see their \S3.1]{costaMassiveBinaryBlack2023} 
to determine the fate of the 1st star.
For a massive star with $M_* \geqslant M_{\rm *, {\ssc DCBH}} \equiv 242 \Msun$, 
the remnant is a DCBH with mass $M_{\sscBH} = M_{\rm *}$. 
The equality holds as 
stellar wind is ineffective due to the low line-driven opacity arising 
from the low metallicity \citep{vinkMasslossPredictionsStars2001,
baraffeStabilityVeryMassive2001}, and the negligible mass loss due to
pulsational instability compared with the large accretion rate 
\citep[see \S3.2][]{hosokawaFormationPrimordialSupermassive2013}.
With $100 \Msun \equiv M_{\rm *, {\ssc PISN}} \leqslant M_* < M_{\rm *, {\ssc DCBH}}$, the pair-instability 
supernova (PISN) is the trigger to end the life of the star, which totally disrupts 
the star and leaves no remnant behind. For $M_* < M_{\rm *, {\ssc PISN}}$, 
core-collapse supernova
(CCSN) is the trigger to end the stellar evolution. 
We crudely approximate the remnant mass as $M_{\sscBH} = f_{\rm rem} M_*$, 
with $f_{\rm rem} = 0.86$ to match the numerical result of 
\citet{costaMassiveBinaryBlack2023}.
A remnant below $3\Msun$ is classified as a neutron star, but such a remnant is 
not expected owing to the large mass of the 1st star.

In the case where the PISN is triggered in the 1st star, we seed a central BH with 
a mass $M_{\sscBH} = f_{\rm rem} M_{\rm *,surv}$, where $M_{\rm *,surv}$ is the 
initial mass of the most massive star that can survive the PISN, obtained by 
solving
\begin{equation}
    1 =  \int_{M_{\rm *,surv}}^{M_{\rm *, {\ssc PISN}}} \dv{N}{M_*} \dd{M_*} 
    \,.\label{eq:m-star-surv}
\end{equation}
The total mass of other stars able to survive is obtained using equations
similar to Eqs.~\eqref{eq:m-star-other} and \eqref{eq:m-star-other-value}, 
but with the upper limit of integration replaced by $M_{\rm *,surv}$.
In our numerical implementation (Appendix~\ref{app:ssec:seeding}),
the mass loss due to stellar evolution for stars other than the 1st is also 
included when deriving the mass of their remnants.  
Note that our treatment of the stellar evolution 
is a simplification. For example, pulsational pair instability (PPI)
can take place at a mass close to $M_{\rm *, {\ssc PISN}}$ and cause 
fluctuation in the remnant mass
\citep[e.g.][]{farmerMindGapLocation2019,mapelliImpactRotationCompactness2020,
costaMassiveBinaryBlack2023}. Star-to-star variation
is also expected, due to the dynamical nature of the stellar structure
and the dense environment of the host star cluster that leads to 
frequent interactions among stars.

Fig.~\ref{fig:pop3-imf}b shows the remnant mass, separately for
the most massive star (1st star at the birth of the star cluster, 
or the most massive star that can survive the PISN) 
and other stars in total, as a function of $M_{\rm *,1st}$
of the Pop-III star cluster. 
The remnant mass of the most massive star, i.e. the mass of the BH seed,
follows closely $M_{\rm *,1st}$, owing to the insignificant mass loss 
(zero for DCBH, and $1-f_{\rm rem} \approx 0.14$ for CCSN).
The exception is for the case of PISN, where the 1st star at birth of 
the cluster is destroyed and the BH seed is bred by the remnant of 
the most massive star surviving the PISN. Total remnant mass of other stars
also closely follows $M_{\rm *,1st}$, except for the case of PISN,
due to the simple relation between $M_{\rm *,other}$ and $M_{\rm *,1st}$
(Eq.~\ref{eq:m-star-other-value}).

The above discussion for the formation of SGCs, Pop-III star clusters and 
BH seeds assumes a pristine gas environment in which the SGCs reside.
Such a condition is expected only if the IGM is not significantly enriched by
previous generations of stars. Since our model can predict star formation 
of individual galaxies in a cosmological volume, the metallicity of the
IGM can also be self-consistently predicted and used to determine when 
and where the formation of Pop-III stars is possible or prohibited. We present 
the details for the modeling of IGM enrichment in Appendix~\ref{app:ssec:IGM-enrichment}.
Briefly, we find that at $z = 10$, most of the cosmic volume 
has already been filled by SN bubbles that carry metals produced by the star formation,
and the median IGM metallicity goes above
$Z_{\ssc IGM} \approx 10^{-4}\Zsun$.  
This means that halos at $z \gtrsim 10$ form their first generation of 
stars mainly in the form of Pop-III, and are expected to 
seed their central BHs according to the model described above.
The rapid star formation in galaxies during their fast phases then quickly 
enriches the IGM, brings the cosmic star formation into the Pop-II regime, 
and ends the formation of Pop-III stars at $z \approx 10$.
In this paper, we adopt a threshold value of $Z_{\ssc IGM,Pop\text{-}III} = 10^{-4}\Zsun$
for the formation of Pop-III stars \citep{brommFormationFirstSupermassive2003,valianteOriginDustHighredshift2011,costaMassiveBinaryBlack2023,spinosoMultiflavourSMBHSeeding2023}, and assume that halos embedded in 
the IGM with metallicity above this value form their first generation of stars
in the form of Pop-II. The mass of BH seeds in these halos is expected to be 
low, assuming that Pop-II stars do not have a top-heavy IMF. We manually set
the mass of BH seeds in these halos to be $M_{\sscBH} = 10 \Msun$, a typical
value of the remnant mass of massive stars in normal star clusters.

The enrichment of the IGM by the outflow of metal produced by star formation,
and its effect on the formation of stars and BH seeds in nearby halos,
are the earliest `matter feedback' of galaxy formation to its 
environment. This, together with the `radiation feedback' described 
in Appendix~\ref{app:sssec:delay-by-uv-radiation}, is represented by the orange 
markers in Fig.~\ref{fig:cosm-context-seeding}.

\bigskip
The branch in Fig.~\ref{fig:flowchart}b summarizes the seeding procedure in the model 
pipeline. This procedure includes a list of key conditions regarding the 
competition between cooling and heating to be checked 
for the SGC in each halo. These include (i) whether a seed 
has already been bred in the halo, and thus whether the 
SGC is self-enriched to cool effectively; (ii) whether the IGM 
metallicity is high enough so that the SGC is externally enriched
to cool effectively; (iii) whether the SGC is exposed to a strong, external
UV background that stalls the cooling and delays the SGC collapse;
(iv) whether dynamical heating is strong enough to prevent the collapse
and fragmentation of the SGC. The fate of the SGC is then determined
by the outcome of these conditions, and the BH seed is bred 
accordingly. Phenomenologically, condition (i) represents the 
internal effect of galaxy evolution on itself, while conditions (ii)--(iv)
are external effects, from either halo assembly
or large-scale environment, on galaxy evolution. 
Such conceptual separation of `nature' and `nurture' has been built for 
galaxy formation at low $z$
\citep{pengMassEnvironmentDrivers2010,
niemiPropertiesGalaxiesGroups2011,
casadoNatureNurtureClues2015,
paulino-afonsoVIS3COSIINature2019,
mcgibbonMultiepochMachineLearning2022,
muceshNatureNurtureGalaxy2024,
shiNatureNurtureRevisiting2024}, while our scenario suggests that all these 
effects have already been present ubiquitously in the early Universe
at $z \gtrsim 20$. Intriguingly, the recent observations of JWST
have pushed the discoveries of galaxies and the measurements of 
environment towards this high-$z$ regime
\citep[e.g.][]{puskasConstrainingMajorMerger2025,
shuntovConstraintsEarlyUniverse2025,linLargescaleEnvironmentsLowluminosity2026,aritaNatureLowluminosityAGNs2025}, 
providing opportunities to test our predictions.

\section{The growth from seeds to supermassive black holes}
\label{app:sec:growing}

During the formation of a BH seed, gas can be temporarily expelled from the host 
SGC by the radiation and potentially also the SN explosion from Pop-III stars. 
This loss of gas fuel may lead to a quiescent phase in the star formation and AGN 
activity. Such a phase is expected to be transient, since rapid replenishment of 
gas is expected as a combined consequence of rapid halo assembly
($\gamma_{\rm v} \gtrsim 1$, see Fig.~\ref{fig:m-v-cool-fit}) 
at the typical redshift of seed breeding ($z \gtrsim 10$) and 
efficient atomic cooling in mini-halos. 

As summarized in \S\ref{ssec:growing-main}, the critical condition for 
the growth of a BH is to break the barrier of angular momentum,
and the contiguous capturing of sub-clouds proposed in \citetalias{moTwophaseModelGalaxy2024}
provides a natural mechanism to achieve this.
The fraction of the gas captured by the BH thus depends on 
the mass and size of the SGC, which determine the width of the $j$-distribution, 
and on the black hole mass ($M_{\sscBH}$), which determines $j_{\rm cap}$.
This, together with the regulation of the total gas mass of the SGC by 
feedback from the active galactic nucleus (AGN), shapes the $M_{\sscBH}$-$M_{\rm *}$ relation for dynamically hot galaxies
observed at $z \approx 0$. 
In contrast, the BH growth and its feedback effects become much weaker after 
the galaxy transits into 
the slow phase where the formation of a rotation-supported disk
prevents gas from sinking towards the center. This produces an offset in the 
$M_{\sscBH}$-$M_{\rm *}$ relation for dynamically-cold galaxies relative to that 
for dynamically hot galaxies (see \citetalias{moTwophaseModelGalaxy2024}). 
Galaxy mergers play an increasingly important role
in the growth of both $M_*$ and $M_{\sscBH}$ for galaxies residing in more 
massive halos, because of the reduced cold-gas fraction for in-situ 
growth and because of the increase in the merging rate with halo mass.  

This paper extends the above continuous mode of BH growth by taking into account 
the small-scale processes in the galactic nucleus.
The key components include
\begin{enumerate}[topsep=0pt,parsep=0pt,itemsep=0pt]
    \item the formation of a gaseous nucleus that is largely free of supernovae owing to the high density 
    and short dynamical timescale;
    \item
    the turbulence-modified Bondi accretion that is highly efficient in driving
     gas inflow in the nucleus towards the BH;
    \item 
    the formation of a turbulent and magnetized accretion disk 
    around the BH due to the rapid inflow; 
    \item 
    the two modes of accretion, one `wet' and the other `dry',
    that can reach a super-Eddington rate and cause the BH to grow fast in short bursts.
\end{enumerate}
The numerical model for the nuclear processes is then built upon 
these considerations and implemented as a refinement engine of the whole growing procedure 
so that BH growth can be resolved self-consistently within the context of 
the cosmic hierarchy. 
Fig.~\ref{fig:am-redistribute} shows the levels of hierarchy and processes 
relevant to BH growth at each level. Fig.~\ref{fig:profile-ladder} shows the
gas density profile at each level of the hierarchy at three redshifts, in the history 
of a MW-size galaxy whose center BH is predicted to experience super-Eddington 
accretion at these epochs.
The profile is built upon the outer levels established by the Universe,
the halo (see Appendix~\ref{app:sec:halo-assembly}), and the SGC 
(see Appendix~\ref{app:sssec:sgc}), with the inner levels established by the 
nuclear processes.
In the following, we first review the entire hierarchy leading to BH growth, 
and then present the details of the key components listed above.

\subsection{The hierarchy leading to black-hole growth}
\label{app:ssec:hierarchy}

A key property of CDM halos, as mentioned in \S\ref{ssec:halo-assembly}, 
is their two-phase assembly. The early evolution of a dark matter halo 
is thus expected to consist of processes distinct from those later on. These
include: 
\begin{enumerate}[topsep=0pt,parsep=0pt,itemsep=0pt]
    \item fast accretion, with $\gamma_{\rm v} \gtrsim \gamma_{\rm f}$, where the 
    threshold value $\gamma_{\rm f} \approx 0$--$3/8$
    (\citealt{zhaoGrowthStructureDark2003}; \citealt{moreSplashbackRadiusPhysical2015}; \citealt{bocoTwOParametersSemi2023}; 
    see also \S2 of \citetalias{moTwophaseModelGalaxy2024} for a summary);
    \item efficient cooling, with $t_{\rm cool} \lesssim t_{\rm v}$, 
    if halo mass has not grown above a redshift-independent threshold of
    $\approx 10^{12}\Msun$ \citepalias[see e.g. fig.~2 of][]{moTwophaseModelGalaxy2024};
    \item cold streams in halo gas, depending on the configuration of 
           cosmic filaments connected to the halo \citep{keresHowGalaxiesGet2005,
           dekelColdStreamsEarly2009,ceverinoHighredshiftClumpyDiscs2010,
           danovichFourPhasesAngularmomentum2015}.
\end{enumerate}
Processes (i) and (ii) in combination are sufficient to maintain rapid inflow and 
keep the halo center gas-rich before feedback can eject much of the gas. 
The high gas fraction then implies that the gas can become self-gravitating when 
the inflow reaches a radius of $f_{\rm g} R_{\rm v} \lesssim f_{\rm b} R_{\rm v} \approx 0.16 R_{\rm v}$
before angular momentum becomes important to resist the collapse (at a scale of 
$\lambda_{\rm g}R_{\rm v} \approx 0.04 R_{\rm v}$, assuming  
that the spin of the gas, $\lambda_{\rm g}$, follows that of the dark matter).
This leads to the formation of a self-gravitating gas cloud in the halo 
center, as shown schematically in Fig.~\ref{fig:am-redistribute}b.
The motion of gas elements in the SGC at the time of self-gravitation is thus 
dominated by random motion instead of rotation, and the dispersion
of $j$ at an arbitrary direction is about $f_{\rm g} R_{\rm v} V_{\rm c}$,
where $V_{\rm c}$ is the circular velocity at the SGC scale 
(see Fig.~\ref{fig:am-redistribute}a).

Process (ii) alone implies that the gas within the halo can fall freely towards 
the halo center,  with a velocity of $\approx V_{\rm v}$,
without being significantly affected by thermal pressure. 
Shock is thus expected to be produced when the high-flux, high-speed inflow 
collides within the SGC. This, together with the frequent dynamical 
disturbance imparted by the rapid change of the gravitational potential 
during the fast-accretion regime, can generate supersonic turbulence, 
with the Mach number $\mathcal{M}_{\rm s} \equiv v_{\rm t} / c_{\rm s} \gg 1$,
over the entire SGC. Fragmentation in the turbulent medium due to Jeans 
instability and shock compression produces sub-clouds with densities 
as high as $n_{\rm sc} \approx \mathcal{M}_{\rm s}^2 n_{\sscSGC}$, 
making both the cloud-cloud collision and the drag force from the surrounding 
gas negligible. The presence of process (iii) can further enhance the turbulence 
by increasing the fraction of kinetic energy that can penetrate the SGC, 
which can accelerate the formation of sub-clouds by promoting early fragmentation of gas 
in the filamentary inflow. As soon as the collision and drag become unimportant, 
the mixing of angular momentum among sub-clouds ceases, leaving a set of ballistically 
moving sub-clouds that support the structure of the SGC. Sub-clouds with highest density 
($n_{\rm sc} \gtrsim 10^{3.5}\perccm$) are predicted to be the 
sites to form globular clusters, as modeled in \citetalias{chenTwophaseModelGalaxy2025}. 
Fig.~\ref{fig:am-redistribute}c is a schematic plot showing  
the structure of an SGC, formed in a halo (panel b), together with its sub-clouds.   
As a consequence of the formation of sub-clouds, the broad distribution
of $j$ on the SGC scale is preserved, in contrast to the peaked, 
off-center distribution that would be expected for a dynamically-cold disk 
formed in a halo in the slow phase, as indicated in Fig.~\ref{fig:am-redistribute}a. 
A fraction of the sub-clouds with low $j$ in the broad distribution 
is thus expected to reach the influence zone of, and to be captured by, the central BH, 
as indicated by the vertical grey band in Fig.~\ref{fig:am-redistribute}a. 

The additional processes that need to be considered towards the galactic nucleus
have been summarized in \S\ref{ssec:growing-main}.
Among them, the regeneration of low-$j$ sub-clouds is speculated
as a consequence of `positive' SN feedback and other sources of turbulence in 
\citetalias{moTwophaseModelGalaxy2024}, and is supported 
by the findings of hydro simulations that nuclear star formation is not 
passive \citep[e.g.][]{bowerDarkNemesisGalaxy2017,
habouzitSupermassiveBlackHoles2021}. 
Stellar feedback in the nucleus with high density may not be able 
to unbind the gas \citep[e.g.][]{dekelEfficientFormationMassive2023,
hopkinsFORGEdFIREResolving2024}, but can drive turbulence 
\citep[e.g.][]{maSelfconsistentProtoglobularCluster2020} and re-create sub-clouds 
to feed the BH. 
Another process in the galactic nucleus, the mixing of angular momentum, 
can be clearly seen in the simulation of 
\citet[see their fig.~15]{gaspariChaoticColdAccretion2015}, in which 
even weak turbulence promotes the redistribution of angular momentum
in the galactic nucleus and the BH feeding is expected to be replenished 
by the low angular-momentum sub-clouds generated by the redistribution. 
The results of all these processes, however, depend on the competition between 
gas heating and cooling. For example, SN feedback, cloud-cloud collision, drag 
force and AGN feedback can inject energy into the gas, which may 
destroy sub-clouds if cooling is ineffective, instead of promoting sub-cloud 
formation. Thus, these additional processes may cause the gas structure and 
dynamics to deviate from those expected from modeling the outer part 
of the SGC, thereby modifying the density ladder, angular momentum distribution and 
BH accretion rate. We therefore model the extension of the hierarchy to 
small scales in the following.

\subsection{The formation of supernova-free nuclei}
\label{app:ssec:snf-nucleus}

The aforementioned processes within a galactic nucleus require an additional 
refinement on top of the SGC to capture their consequences for gas dynamics, 
star formation and BH accretion, and motivate the definition of a new level,
the SNF nucleus (\S\ref{ssec:growing-main}), in the structure hierarchy.
The threshold number density, $n_{\rm snf}$, implies a mass density of
\begin{equation} \label{eq:rho-snf}
    \rho_{\rm snf} = \mu m_{\rm p} n_{\rm snf} \approx 6.35 \times 10^{-21} \gperccm \,.
\end{equation}  
Using Eq.~\eqref{eq:t-ff-sgc}, the free-fall timescale
of the gas at a density $\rho_{\rm g} = \mu m_{\rm p} n_{\rm g}$ is
\begin{equation}
    t_{\rm ff, g} = \sqrt{ \frac{3\pi}{32G\rho_{\rm g}} }
    = \left[ 0.836 \Myr \right] n_{\rm g, 3.5}^{-1/2}
    \,,\label{eq:t-ff-gas-of-snf}
\end{equation}
where $n_{\rm g} \equiv 10^{3.5}\perccm\,n_{\rm g, 3.5} $.
Thus, a density of $n_{\rm g} \geqslant n_{\rm snf}$ corresponds to a
free-fall timescale of $t_{\rm ff, g} \leqslant t_{\rm snf} \equiv 0.836\Myr$, implying that gas 
dynamics within the nucleus proceeds faster than the lifetime 
of massive stars born in it. Such a nucleus is largely free of 
supernovae, and is therefore referred to as the SNF nucleus
in this paper. The argument in \S\ref{ssec:growing-main} suggests 
that a nuclear burst appears inevitable once such a nucleus is formed,
and the consequence of the burst is a short period of rapid growth of the 
compact NSC, or BH, or both, together with gas ejection by BH feedback.

Fig.~\ref{fig:am-redistribute}d shows the sketch of the SNF nucleus as a zoom-in 
view of the central part of the host SGC. The relations of the nucleus to the  
expected nuclear star cluster (NSC) and central BH are also highlighted.
In the following, we describe the formation and structure of the SNF nucleus 
in detail.

\subsubsection{Criteria for the formation of supernova-free nuclei}
\label{sssec:snf-formation}

Questions remain as to how the SNF nucleus can be formed in the first place.
Several processes can present barriers to the formation, mainly:   
\begin{enumerate}[topsep=0pt,parsep=0pt,itemsep=0pt]
    \item 
    the formation of a dynamically cold disk at the SGC scale, which can build an angular-momentum 
    barrier for the gas flow towards the nucleus
    \citep[e.g.][]{powerAccretionDiscParticle2011,
    rosas-guevaraImpactAngularMomentum2015,curtisResolvingFlowsBlack2016};
    \item 
    star formation and stellar feedback, which may consume/eject much of the gas 
    before it can reach the nucleus
    \citep{mengOriginGiantStellar2020,
    renaudGiantClumpsClouds2021,vandonkelaarGiantClumpsClouds2022,
    vandonkelaarStellarClusterFormation2023,garciaStarClusterFormation2023a,
    renaudGiantClumpsClouds2024,mayerSituFormationStar2025};
    \item
    feedback from the accreting BH, which may affect gas 
    accumulation in the nucleus
    \citep{suSelfregulationBlackHole2023,duttaDissipationAGNJets2024,
    sivasankaranAGNFeedbackIsolated2025,suSelfregulationHighredshiftBlack2025} 
    or may even eject/heat the gas in the SGC and halo
    \citep{zingerEjectivePreventativeIllustrisTNG2020,
    shiColdGasMassive2022,ayromlouFeedbackReshapesBaryon2023,
    weinbergerActiveGalacticNucleus2023,
    wangBlackHolesRegulate2024}.
\end{enumerate}
The formation of an SNF nucleus is thus expected only when a set of conditions is met 
to break these barriers.

In our modeling relevant to the early fast accretion regime of the halo assembly, 
the creation of a dynamically hot, globally disturbed SGC is essential to 
remove those barriers. The `hotness' is needed to prevent the formation of 
dynamically cold disks, so as to break the angular-momentum 
barrier (point i above). Global disturbance can make an SGC out of equilibrium, 
which can produce a strong gas inflow to overpower (at least over a short 
period of time) the gas consumption by star formation/BH accretion 
and ejection by the associated  feedback (points ii and iii above). 
Once these conditions are met, an SNF nucleus can be formed as a consequence of 
the boosted inflow, and the formation of a compact stellar system and 
a period of rapid growth of the central BH seem inevitable.

The conditions for the hotness and global disturbance of an SGC arise naturally 
owing to the fluctuations in the halo assembly expected in the fast accretion regime
(see examples in fig.~1 of \citetalias{moTwophaseModelGalaxy2024}).
The temporary speed-up of the halo mass assembly implies an excessive amount of 
matter being added to the halo, which can perturb the SGC significantly 
by changing the gravitational potential and by the ram pressure of the 
infalling gas. Part of such disturbance is produced by mergers with 
other halos, which can perturb the SGC during a close encounter of  
the merging galaxies, a process known to be critical in transforming galactic 
structure \citep[e.g.][]{springelFormationSpiralGalaxy2005,
keselmanPseudobulgeFormationMajor2012,
puechGalaxyDisksNot2012,
yoonEvidenceImpactGalaxy2022,
jacksonExtremelyMassiveDisc2022,
wrightMergerdrivenFormationClassical2026,
sampaioMorphologyCosmicTime2025,
proctorWeakConnectionStellar2025}. 
A halo temporarily taking an excursion to a sufficiently 
high mass assembly rate is thus expected to be able to impart global disturbance and 
create the hotness required for the formation of an SNF nucleus in the SGC.
These considerations motivate us to adopt the following criterion to trigger
the formation of an SNF nucleus:
\begin{equation}
    \gamma_{\rm v} - \mathcal{S}(\gamma_{\rm v}; t_{\rm v})
    \geqslant \delta_\gamma \gamma_{\rm f} 
    \,, \label{eq:snf-formation-crit}
\end{equation}
where $\mathcal{S}(\gamma_{\rm v}; t_{\rm v})$ denotes the value
of $\gamma_{\rm v}$ smoothed over a timescale of $t_{\rm v}$ along 
the main branch of the halo assembly history (see Appendix~\ref{app:ssec:growing}
for details),
$\gamma_{\rm f} \approx 0$--$3/8$ is the threshold that separates the fast and slow
phases of halo assembly \citepalias{moTwophaseModelGalaxy2024}, 
and $\delta_\gamma \sim 1$ is a parameter that
defines the strength of the excursion needed for the formation of an SNF nucleus.
Here, we adopt $\gamma_{\rm f} = 3/16$, the average of the two extremes, and we 
choose $\delta_\gamma = 1$ to identify excursions that have relaxation timescales 
longer than the typical dynamical time of the halo.

The frequency of the formation of SNF nuclei is expected to be high 
in the early Universe. This can be understood as a result of the specific 
merger rate of CDM halos \citep{dongUniversalSpecificMerger2022} 
in combination with their faster assembly rate at higher $z$ 
(see e.g. Fig.~\ref{fig:halo-mahs}). 
As we will model in Appendix~\ref{app:ssec:nuclear-burst}, 
the growth of the BH in an SNF nucleus is expected to be super-Eddington. This, together with the high frequency 
of formation of SNF nuclei, provides an efficient way for BH seeds to grow into SMBHs
(see \S\ref{sssec:growth-channels}).

\subsubsection{Structure of the supernova-free nuclei}
\label{app:sssec:snf-structure}

The violent gas inflow associated with the formation of SNF nuclei
creates a situation similar to that for the formation of an SGC, albeit 
with smaller volume, higher density, and shorter timescale.
The large mass flux ($\sim M_\sscSGC/t_{\rm dyn,\sscSGC}$) and high velocity 
($\sim V_\sscSGC$) are similar to the conditions of the SGC formation provided 
by a halo of fast assembly and efficient cooling, and are also expected to drive 
supersonic turbulence in the nucleus, giving a gaseous nucleus that is dynamically hot 
(with turbulence velocity $v_{\rm t} \gg c_{\rm s}$) and supported by turbulent motion.

However, the gas dynamics within an SNF nucleus is not entirely the same as that 
in an SGC, because the formation of the nucleus (Appendix~\ref{sssec:snf-formation})
requires a temporarily boosted inflow that outpaces the regulation by supernova and AGN 
feedback. The main processes driving energy transfer during formation 
are instead the release of gravitational potential energy carried by the
inflow and the dissipation of turbulent energy by interacting
gas elements, resembling hydrodynamic simulations 
of isothermal turbulence with a single driving source, such as that mimicked 
by a stochastic force term or a shaking box
\citep{eswaranExaminationForcingDirect1988,schmidtNumericalSimulationsCompressively2009,
federrathSonicScaleInterstellar2021}. The real gas profiles may still differ 
from those in the simulations, as SNF nuclei may not fully develop 
an equilibrium before nuclear star formation and BH accretion disperse the gas, 
introducing uncertainty into the equilibrium-based modeling below.

As an approximation, we divide the processes in an SNF nucleus into a 
formation stage, when gas inflow is boosted to replenish the nucleus before 
feedback regulation is established, and a dispersion stage (the nuclear burst 
itself), when star formation and BH accretion overpower the inflow and reduce 
the gas density back to the original level. For the formation stage, we model 
the structure equations of the gas around and within an SNF nucleus by a 
parametric approach below, with a number of free parameters to capture the 
unknown degrees of freedom; for the dispersion stage, we model the star 
formation, BH accretion and gas reduction in the nuclear burst in 
Appendix~\ref{app:ssec:nuclear-burst}.

We parameterize the gas density ($\rho_{\rm g}$) as a function of the galactocentric 
distance ($r$) within the entire SGC, including that within the SNF nucleus
during the formation stage, by the following profile:
\begin{equation} 
    \rho_{\rm g}(r) = \frac{ 2^{p+2q} \rho_{\rm snf} }{ 
        \left( c_{\rm g}x_r \right)^{2-2q} \left( 1+c_{\rm g}x_r \right)^{p + 2q}
    }
    \,, \label{eq:rho-g-in-nuc}
\end{equation}
where the scaled radius $x_r = r / R_\sscSGC$; the concentration parameter 
$c_{\rm g} \equiv R_\sscSGC/R_{\rm nuc}$;
$R_{\rm nuc}$ is the outer radius of the SNF nucleus;
$\rho_{\rm snf}$ is the supernova-free threshold (Eq.~\ref{eq:rho-snf});
$q \leqslant 1$ is the structure parameter of the turbulence (see below);
$p \geqslant 0$ characterizes the density profile at large radius.
It can then be shown that $\rho_{\rm g}(R_{\rm nuc}) = \rho_{\rm snf}$.
Thus, $\rho_{\rm g} \geqslant \rho_{\rm snf}$ for any 
$r \leqslant R_{\rm nuc}$, consistent with the requirement 
that the nucleus is supernova-free.

The double power-law form of the profile divides the gas structure of 
the SGC into two regimes:
one at $r \gg R_{\rm nuc}$, where the density $\rho_{\rm g} \propto r^{-(2+p)}$,
and the other at $r \ll R_{\rm nuc}$, where the density
$\rho_{\rm g} \propto r^{-(2-2q)}$. This separation is plausibly 
produced by distinct processes. Specifically, the parameter $q$ can be 
understood in terms of the velocity structure of a turbulent medium, 
$\sigma_{\rm v} \propto r^{\rm q}$, namely the velocity dispersion
$\sigma_{\rm v}$ of the gas as a function of the scale $r$. 
Assuming virialization, the mass enclosed within $r$ is 
$M_{\rm g}(r) \propto r \sigma_{\rm v}^2 \propto r^{2q+1}$.
The density profile is thus
$\rho_{\rm g} \equiv \dd{M_{\rm g}}/(4\pi r^2 \dd{r}) \propto r^{-(2-2q)}$,
which is the same as our parameterization (Eq.~\ref{eq:rho-g-in-nuc}) at 
small $r$. The mean density within $r$ can then be written as 
$\bar{\rho}_{\rm g}(r) = 3 \rho_{\rm g}(r) / (2q+1)$.
We choose $q = 1/2$ as the default value. This choice is commonly 
used for well-relaxed, isothermal and supersonic turbulence
\citep{federrathSonicScaleInterstellar2021}, and we adopt it 
because of the supersonic nature of the SNF nuclei 
expected from the rapid inflow driven by the global disturbance.
The choice of $q$ thus implies that Eq.~\eqref{eq:rho-g-in-nuc}
applies to a well-relaxed state of the SNF nucleus, where dissipation by 
radiative cooling is balanced by the energy injection into the gas. 
This balance is expected to be transitional, however, as the bursty star formation 
and BH accretion will soon end the formation stage and take the 
nucleus into the dispersion stage.

We adopt $p=0$ as the default value, which corresponds to an isothermal
profile, $\rho_{\rm g} \propto r^{-2} $, for the volume of the SGC well outside 
the nucleus. Such a gas profile is seen in hydro simulations 
\citep[e.g.][]{hopkinsFORGEdFIREResolving2024}, 
and is adopted by semi-analytical models of galaxy formation  
\citep[e.g.][]{henriquesLGALAXIES2020Spatially2020}.
With $p=0$ and $q=1/2$ given above, Eq.~\eqref{eq:rho-g-in-nuc} becomes 
\begin{equation}
    \rho_{\rm g}(r) = \frac{ 2 \rho_{\rm snf} }{ 
        c_{\rm g}x_r \left( 1+c_{\rm g}x_r \right)
    }
    \,,\label{eq:rho-g-in-nuc-q05}
\end{equation}
and the corresponding asymptotic behaviors are 
$\rho_{\rm g} \propto r^{-2}$,
$\bar{\rho}_{\rm g}(r) = 3 \rho_{\rm g}(r)$,
$M_{\rm g} \propto r$,
and $V_{\rm c} = V_{\ssc SGC} = {\rm const.}$ for $r \gg R_{\rm nuc}$; 
$\rho_{\rm g} \propto r^{-1}$,
$\bar{\rho}_{\rm g}(r) = 3 \rho_{\rm g}(r)/2$, 
$M_{\rm g} \propto r^2$, 
and $V_{\rm c} \propto r^{1/2}$ for $r \ll R_{\rm nuc}$.

To determine the structure of an SNF nucleus from the global properties 
of the host SGC (Eqs.~\ref{eq:r-sgc}--\ref{eq:t-ff-sgc}), we can substitute 
the constraint $M(R_\sscSGC) = M_\sscSGC$ into the profile 
Eq.~\eqref{eq:rho-g-in-nuc}, and obtain the concentration parameter 
$c_{\rm g}$ by solving the following implicit equation:
\begin{equation}
    \frac{2}{c_{\rm g}} \frac{m_{\rm g}(c_{\rm g})}{c_{\rm g}^2} 
    = \frac{\rho_\sscSGC}{\rho_{\rm snf}}
    \,,\label{eq:c-g-def}
\end{equation}
where $m_{\rm g}(x) \equiv 3 \left[ x - \ln(1+x) \right] $.
The radius of the SNF nucleus is then given by the definition, 
$R_{\rm nuc} = R_{\ssc SGC}/c_{\rm g}$.
The mass enclosed within $r$ is obtained by integrating 
$4 \pi r'^2 \rho_{\rm g}(r') \dd{r'}$ over $r' \in [0, r]$:
\begin{equation}
    M_{\rm g}(r) = M_\sscSGC \frac{ m_{\rm g}( c_{\rm g} x_r ) }{ m_{\rm g}(c_{\rm g}) }
    \,.\label{eq:M-g-r-nuc}
\end{equation}
The mass of the SNF nucleus, $M_{\rm nuc} \equiv M_{\rm g}(R_{\rm nuc})$, is then
\begin{equation}
    M_{\rm nuc}
    = M_\sscSGC \frac{m_{\rm g}(1)}{m_{\rm g}(c_{\rm g})}
    \,,\label{eq:M-nuc}
\end{equation}
where $m_{\rm g}(1) \approx 0.921$, and the circular velocity profile is 
\begin{equation}
    V_{\rm c}(r) 
    \equiv \sqrt{ \frac{G M_{\rm g}(r)}{r} } 
    = V_{\rm nuc} \sqrt{ \frac{m_{\rm g}(c_{\rm g} x_r)}{ m_{\rm g}(1) c_{\rm g} x_r } }
    \,,\label{eq:V-c-r-nuc}
\end{equation}
with $V_{\rm nuc} \equiv \left( G M_{\rm nuc}/R_{\rm nuc} \right)^{1/2}$.
For $r \ll R_{\rm nuc}$ (i.e. $c_{\rm g} x_r \ll 1$), $V_{\rm c}$ can be approximated as 
\begin{equation} 
    V_{\rm c} \approx \sqrt{1.5 c_{\rm g}x_r / m_{\rm g}(1)} V_{\rm nuc} 
    \approx 1.28 \left( c_{\rm g}x_r \right)^{1/2} V_{\rm nuc}    
    \,.\label{eq:v-c-in-snf-q05}
\end{equation}

\bigskip
To obtain numerical estimates of some quantities in the above equations, 
we consider $c_{\rm g} \gg 1$, which holds for SGC at $z \lesssim 10$ 
(see Fig.~\ref{fig:sgc-snf-props}b).
In this case, $m_{\rm g}(c_{\rm g}) \approx 3 c_{\rm g}$, and
the above equations for the structure of the SNF nucleus reduce to
\begin{equation} 
    \frac{3}{m_{\rm g}(1)} \frac{M_{\rm nuc}}{M_\sscSGC} 
    \approx \frac{R_{\rm nuc}}{R_\sscSGC} 
    =       \frac{1}{c_{\rm g}} 
    \approx \left( \frac{\rho_\sscSGC}{6\rho_{\rm snf}} \right)^{1/2} 
    \,.\label{eq:structure-nuc-approx}
\end{equation}
Taking the estimates of $R_\sscSGC$ and $n_\sscSGC$ in 
Eqs.~\eqref{eq:r-sgc} and \eqref{eq:n-sgc}, we obtain
\begin{equation}
    c_{\rm g}^{-1} \approx 0.239 f_{\rm g,0.157}^{1/2} f_{\rm r, 0.04}^{-3/2} 
        n_{\rm snf, 3.5}^{-1/2} (1+z)_{10}^{3/2}
    \,; \label{eq:c-g-numeric}
\end{equation}
\begin{equation}
    R_{\rm nuc} \approx \left[43.6 \pc\right] 
        f_{\rm g,0.157}^{1/2} f_{\rm r, 0.04}^{-1/2} 
        n_{\rm snf, 3.5}^{-1/2} M_{\rm v, 9.5}^{1/3} (1+z)_{10}^{1/2}
    \,;  \label{eq:r-nuc-numeric}
\end{equation}
\begin{multline}
    M_{\rm nuc} \approx \left[6.00 \times 10^7 \Msun\right] 
        f_{\rm g,0.157}^{3/2} f_{\rm r, 0.04}^{-3/2}  
        \\
        \times n_{\rm snf, 3.5}^{-1/2} M_{\rm v, 9.5} (1+z)_{10}^{3/2}
    \,,  \label{eq:m-nuc-numeric}
\end{multline}
where $M_{\rm v,9.5}$ is the halo mass scaled by $10^{9.5}\Msun$, roughly 
the value of the progenitor at $z=9$ of an MW-size halo at $z=0$
($M_{\rm v, z=0} \approx 10^{12} \Msun$; see Fig.~\ref{fig:halo-mahs}).
The free-fall timescale and angular frequency of the nucleus are
\begin{align}
    t_{\rm ff, nuc}  &= \sqrt{ \frac{3\pi}{32 G \bar{\rho}_{\rm g}(R_{\rm nuc}) } } \approx 0.616 \Myr\, n_{\rm snf, 3.5}^{-1/2}   
    \, \label{eq:t-ff-nuc} \\
    \Omega_{\rm nuc} &\equiv  V_{\rm nuc}/R_{\rm nuc} \approx 1.80 \Myr^{-1} n_{\rm snf,3.5}^{1/2} 
    \,, \label{eq:omega-nuc}
\end{align}
respectively. Using the structure equations of the nucleus
(Eq.~\ref{eq:structure-nuc-approx}) and the numerical estimate
of $\Sigma_{\ssc SGC}$ (Eq.~\ref{eq:sigma-sgc}),
the surface density of the SNF nucleus is 
\begin{align} 
    \Sigma_{\rm nuc} 
    & \sim \frac{M_{\rm nuc}}{R_{\rm nuc}^2} 
    \approx \frac{m_{\rm g}(1)}{3}c_{\rm g}\Sigma_\sscSGC
    \nonumber\\
    & \approx 3.15 \times 10^4 \Msunperpcsq 
        \left[ \chi_{\rm g}(1+z)_{10} \right]^{1/2} 
        M_{\rm v,9.5}^{1/3} 
    \,,\label{eq:sigma-nuc-numeric}
\end{align}
where $\chi_{\rm g} = f_{\rm g,0.157} n_{\rm snf,3.5} f_{\rm r,0.04}^{-1}$.
The circular velocity of the nucleus, $V_{\rm nuc}$, is related to that of 
the SGC, $V_\sscSGC$, by $V_{\rm nuc} = \sqrt{m_{\rm g}(1)/3}\, V_\sscSGC$.

Green segments in Fig.~\ref{fig:profile-ladder}a extend the density profiles
to the nuclear volumes of SGCs in MW-like progenitors by assuming that SNF 
nuclei have formed. For reference, in Appendix~\ref{app:sec:evolution-of-hierarchy}, we 
show the properties of SNF nuclei
in halos with a wide range of mass over an extended range of redshift.
The two pieces of each SGC profile at
$r \ll R_{\rm nuc}$ and $r \gg R_{\rm nuc}$ are joined smoothly
by the function given by Eq.~\eqref{eq:rho-g-in-nuc-q05}. 
The inner and outer slopes (in logarithmic scale) are -1 and -2, respectively,
as an outcome of our assumption on the structure parameters, $p$ and $q$.
The nucleus size, $R_{\rm nuc} \lesssim 100\pc$,
shows only slow evolution from $z = 9$ to $z = 0$, much 
slower than the evolution of $R_{\ssc SGC}$ of the host SGC (and 
$R_{\rm v}$ of the host halo).
This is due to the requirement of a constantly high density, 
$\rho_{\rm g}(R_{\rm nuc}) = \rho_{\rm snf}$, 
and the cancellation of two effects (see Eq.~\ref{eq:structure-nuc-approx}):
the growth of $R_{\ssc SGC}$ (and $R_{\rm v}$) with galaxy (and halo) assembly,
and the decline of $\rho_\sscSGC$ (and $\rho_{\rm v}$) 
with the cosmic expansion.
The nucleus size is roughly $1/4$ of the SGC size at $z = 9$
($1/80$ at $z = 0$), corresponding to $1/100$ ($1/2000$)
of the halo virial radius, respectively. This indicates that the SNF nucleus 
constitutes a significant part of the host SGC at high $z$, but becomes only 
a tiny part at low $z$. This also implies that 
SNF nuclei play the dominant role in supplying BH growth at high $z$, 
and a negligible role at low $z$ (see \S\ref{sssec:growth-channels} and Figs.~\ref{fig:growth-examples}
and \ref{fig:growth-channels}).
The card in Fig.~\ref{fig:profile-ladder}d summarizes equations 
that link the properties of the SNF nucleus with those of the host SGC.

As discussed at the end of \S\ref{ssec:growing-main}, the equations derived
above provide a way to link the assembly of structures at different 
levels of the hierarchy. For example, we consider the time evolution
of structural properties of the SNF nucleus in an SGC. 
This can be quantified by taking the time derivative of, e.g. $\Sigma_{\rm nuc}$
(Eq.~\ref{eq:sigma-nuc-numeric}), along the main branch of a halo:
\begin{equation} 
    \dv{\ln \Sigma_{\rm nuc}}{\ln a} 
    \approx - \frac{1}{2} + \frac{1}{3} \dv{\ln M_{\rm v}}{\ln a}
    = \frac{2 \Gamma_{\rm v} - 3}{6} 
    \,,\label{eq:diff-sigma-nuc}
\end{equation}
Note that this equation does not describe the evolution of a single nucleus; instead it 
describes the difference between nuclei formed at different epochs in the 
main branch of a halo. The derivative of $\Sigma_{\rm nuc}$
changes its sign at a critical value of $\Gamma_{\rm v} = 1.5$, 
coinciding with the epoch found by \citet{bocoTwOParametersSemi2023}
that separates the fast- and slow-accretion phases 
of a halo (equivalent to $\gamma_{\rm v} = 3/8$, our `Low-$z_{\rm f}$' variant
defined in \S2 of \citetalias{moTwophaseModelGalaxy2024}).
The same arguments applied to $R_{\rm nuc}$ and $M_{\rm nuc}$ 
using Eqs.~\eqref{eq:r-nuc-numeric} and \eqref{eq:m-nuc-numeric}, respectively,
yield the same critical value of $\Gamma_{\rm v}$.
Note that the presence of a critical value of $\Gamma_{\rm v}$, 
and its exact value, $1.5$, are owing to the assumption of the 
isothermal profile in the outer SGC 
(and thus the index of $1/2$ in Eq.~\ref{eq:structure-nuc-approx})
and are independent of the assumption made for the profile at 
$r \lesssim R_{\rm nuc}$.
The drag term, $-1/2$, in Eq.~\eqref{eq:diff-sigma-nuc} comes from the decrease 
of the density of a virialized halo with the cosmic time, 
and thus encodes the effect of cosmological expansion on the growth of 
structures deeply embedded in centers of dark matter halos.

As we will show later in Appendix~\ref{app:ssec:modified-bondi}, 
the turbulent motion of gas within the nucleus can induce strong inflow 
toward the central BH, and both the Bondi radius ($R_{\sscBondi}$; Eq.~\ref{eq:R-Bondi-numerical-q05}) 
and the Bondi accretion rate ($\dot{m}_{\sscBondi}$; Eq.~\ref{eq:dot-m-Bondi-numerical-q05}) 
of the BH are determined by $\Sigma_{\rm nuc}$.
The structure of the accretion disk of the BH, as to be quantified 
in Appendix~\ref{app:ssec:bh-disk} by the profiles of the surface density 
($\Sigma_{\rm g}$; Eq.~\ref{eq:Sigma-bh-disk-numerical-q05}), 
the volume density ($\rho_{\rm g}$; Eq.~\ref{eq:rho-bh-disk-numerical-q05}), 
and the Toomre $Q$ parameter (Eq.~\ref{eq:toomreQ-bh-disk-numerical-q05}),
all depend on $\Sigma_{\rm nuc}$ through the Bondi properties.
This leads to the conclusion made at the end of \S\ref{ssec:growing-main}
that the transition of the halo assembly (at a scale of $r \gtrsim 100\Kpc$ for $M_{\rm v} \gtrsim 10^{11}\Msun$ at $z=0$)
also leads to a transition in the BH-accretion properties 
(at a scale of $r \lesssim $ a few$\pc$ for $M_{\ssc BH} \lesssim 10^6\Msun$).

The mass scale of an SNF nucleus, $M_{\rm nuc}$, is typically a few times 
$10^7\Msun$ (Eq.~\ref{eq:m-nuc-numeric}), about one order of magnitude larger 
than that of a sub-cloud at a comparable density (see Eq. 29 of \citetalias{chenTwophaseModelGalaxy2025}).
The mass scale is also much larger than the typical mass of BH seeds, including 
massive DCBHs (see Eq.~\ref{eq:m-star-Edd} and Fig.~\ref{fig:pop3-imf}),
and is comparable to massive star clusters observed in the local 
Universe \citep[e.g.][]{baumgardtCatalogueMassesStructural2018,krumholzStarClustersCosmic2019,
neumayerNuclearStarClusters2020,brownRadiiYoungStar2021}.
The formation of SNF nuclei thus implies that a large amount of dense
gas is available for star formation and BH growth, as to be described 
below.

\subsection{The turbulence-modified Bondi accretion of black holes}
\label{app:ssec:modified-bondi}

\subsubsection{The role of turbulence}
\label{app:sssec:role-turbulence}

As summarized in \S\ref{ssec:growing-main}, the turbulent environment in the SGC 
is critical for the gas to efficiently flow through the SNF nucleus,
cross the radius of influence (i.e. Bondi radius) of the central BH,
and eventually feed the BH.

In the outer volume of an SGC, the formation and condensation of sub-clouds 
make collisions among them infrequent and the drag force from the surrounding gas 
negligible, as shown in \S3.3 of \citetalias{moTwophaseModelGalaxy2024}.  
This leads to the ballistic motion of sub-clouds, so that angular-momentum 
mixing between them is weak. Towards the inner volume of an SGC, however, 
the collision frequency for a sub-cloud, $\nu_{\rm coll}$, increases. 
The number of collisions expected for a sub-cloud per dynamical time is
\begin{equation}
    \mathcal{N}_{\rm coll} = \nu_{\rm coll} t_{\rm dyn} 
    \propto V_{\rm c} t_{\rm dyn} R_{\rm sc}^2 \rho_{\rm g}
    \propto \rho_{\rm g}^{1/2} R_{\rm sc}^2
    \propto r^{-1} R_{\rm sc}^2 \,,
\end{equation}
where $R_{\rm sc}$ is the typical size of a sub-cloud, and an isothermal
profile is assumed for $\rho_{\rm g}$ of the SGC. Thus, 
$\mathcal{N}_{\rm coll}$ increases rapidly with decreasing $r$, unless 
$R_{\rm sc}$ shrinks. Of particular significance is the case when a sub-cloud 
is close to or within the SNF nucleus, $r \lesssim R_{\rm nuc}$, where the 
environmental density is $\rho_{\rm g} \approx \rho_{\rm snf}$. 
Such a density is comparable to that of sub-clouds themselves, including 
even the densest ones that host globular clusters 
(see e.g. fig.~6 of \citetalias{chenTwophaseModelGalaxy2025}).
Thus, sub-clouds in the nuclear volume are expected to interact 
frequently. A similar estimate can be made for the drag force 
by the ambient gas on a sub-cloud. The force is stronger at smaller 
$r$ and becomes significant at $r\lesssim R_{\rm nuc}$.

The dissipation has important consequences for gas dynamics in the inner SGC.
First, the collision between sub-clouds can compress the gas.
Owing to the more effective cooling at higher gas density, 
the compressed gas does not push back, but becomes denser and more massive 
than the original sub-clouds. This leads to further fragmentation due to the 
reduced Jeans mass. Second, the loss of orbital energy and angular momentum 
due to the collision causes the gas to sink to a smaller orbit. Statistically, 
for a large number of sub-clouds that experience repeated collisions, the orbit sinking 
generates a net inflow of gas from the outer part of the SGC to the nucleus, 
and from the nucleus to the central BH.
This can be understood by considering an extreme case where
two sub-clouds have the same mass, move on circular orbits with the same 
speed but opposite directions. The head-on collision between them 
totally cancels the angular momentum, and enables the gas  
to fall freely towards the central BH.
Thus, the net effect of collisions of sub-clouds on the gas dynamics
is to cause 
(i) mergers of sub-clouds to form denser ones; 
(ii) the conversion of kinetic energy to radiation;
(iii)  the loss of angular momentum; 
and (iv) mass inflow through the SNF nucleus towards the central BH. 
These conclusions hold even if the SGC is metal-poor, since 
the shock compression can raise the gas temperature above $10^4 \Kelvin$, 
the regime of efficient cooling by hydrogen and helium atoms. 
A similar conclusion can be reached by considering the interaction between 
sub-clouds and the ambient gas. Sub-clouds can survive the disruption by 
the surrounding gas due to cooling in the turbulence mixing layer,
if the sub-cloud density is significantly higher than that of the ambient 
gas \citep[e.g.][]{gronkeHowColdGas2020}.
The interaction with the surrounding gas thus provides 
ram-pressure that compresses sub-clouds, and a friction force 
that drags sub-clouds inwards.

The essential role of turbulence can thus be summarized as `deferred mixing',
as discussed in \S\ref{ssec:growing-main}.
In the absence of a strong disturbance, the inflow is expected to be
steady, producing a quasi-equilibrium state in which the gas inflow balances 
gas consumption by star formation, BH growth and associated feedback.
This is the continuous mode of BH growth defined in this paper.     
However, in the presence of a strong disturbance, such as that associated 
with a strong merger, the collision of sub-clouds and the mixing of 
angular momentum can occur over a large volume of the SGC. This produces a 
temporary surge in the gas inflow that can overwhelm the 
outflow generated by the feedback from the existing BH and stars, 
leading to the formation of a massive SNF nucleus that can power a `burst' 
of BH growth. In this case, the feedback may become `positive',  
since the outflow generated by it provides an additional source to 
compress and mix the gas (see \citetalias{moTwophaseModelGalaxy2024}).  
This mode of BH growth is referred to as the bursty mode in the 
paper.
The effect of the deferred mixing of sub-clouds on the distribution of $j$ has 
been schematically illustrated in Fig.~\ref{fig:am-redistribute}a:
a sharply peaked, centralized distribution forms via efficient mixing in the 
inner region of an SGC, on top of a broad distribution expected from 
the lack of mixing in the outer SGC.  Note that the net angular momentum of 
the SGC is not necessarily different from that of a dynamically-cold disk. 
Instead, it is the turbulence that continuously redistributes $j$ and 
generates low-$j$ sub-clouds to feed the central region.


\subsubsection{The turbulence-modified Bondi accretion rate}
\label{app:sssec:Bondi-accretion-rate}

The SNF nuclei formed through the rapid collapse of sub-clouds 
into halo centers are expected to be dynamically hot and turbulent. 
This motivates the use of modified Bondi accretion \citep{hoyleEffectInterstellarMatter1939,
bondiMechanismAccretionStars1944,bondiSphericallySymmetricalAccretion1952}
to model the amount of gas to be accreted into the 
zone of influence of the central BH. The isotropic condition of the Bondi 
accretion may be met here due to the dynamical hotness, but the velocity
dispersion of gas elements is now dominated by the turbulent motion 
rather than the thermal motion. Modified by the turbulence, the Bondi accretion
rate becomes:
\begin{equation}
    \dot{M}_{\sscBondi} = 4 \pi R_{\sscBondi}^2 \rho_{\rm g}(R_{\sscBondi}) 
    \sigma_{\rm eff}(R_{\sscBondi})
    \,,\label{eq:dot-M-Bondi-general}
\end{equation}
where $\sigma_{\rm eff}$ is the effective velocity dispersion of the gas
defined by $\sigma_{\rm eff}^2 = c_{\rm s}^2 + v_{\rm t}^2$, 
accounting for both thermal motion (quantified by the sound speed, $c_{\rm s}$) 
and turbulent motion (quantified by turbulent velocity dispersion $v_{\rm t}$).
The Bondi radius, $R_{\sscBondi}$, is the radius of the zone of influence of the BH,
within which the gravity of the BH dominates the total gravity:
\begin{equation}
    R_{\sscBondi} = \frac{G M_{\sscBH}}{\sigma_{\rm eff}^2}
    = \left[ 0.430  \pc\right] M_{\rm {\sscBH}, 4} \sigma_{\rm eff, 10}^{-2}  
    \,,\label{eq:R-Bondi-general}
\end{equation}
where $\sigma_{\rm eff} = 10 \kms\, \sigma_{\rm eff, 10}$.

In the supersonic, dispersion-dominated regime that is expected for an SNF nucleus,
\begin{equation}
V_{\rm c} \approx \sigma_{\rm eff} \approx v_{\rm t} \gg c_{\rm s}\,.
\end{equation}
Using the general density profile of the SGC (Eq.~\ref{eq:rho-g-in-nuc}), we can express 
the circular velocity profile at $r \ll R_{\rm nuc}$ as
\begin{equation}
    V_{\rm c} = \psi_{\rm nuc} V_{\rm nuc} \left( \frac{r}{R_{\rm nuc}} \right)^q\,,
\end{equation}
where $\psi_{\rm nuc}\sim \mathcal{O}(1)$ is a normalization factor.
In the default case of $q = 1/2$, $\psi_{\rm nuc} \approx 1.28$ (Eq.~\ref{eq:v-c-in-snf-q05}).
The density profile is given by
\begin{equation}
    \rho_{\rm g}(r) = \frac{ (2q+1) \psi_{\rm nuc}^2 \Omega_{\rm nuc}^2 }{4\pi G} \left(\frac{r}{R_{\rm nuc}}\right)^{2q-2}.
\end{equation}
The Bondi radius can then be obtained from Eq.~\eqref{eq:R-Bondi-general}:
\begin{equation} \label{eq:r-bondi-vs-q}
    R_{\sscBondi} = R_{\rm nuc} 
    \left(\frac{ M_{\sscBH} }{ \psi_{\rm nuc}^2 M_{\rm nuc} }\right)^{ 1/(2q+1) } \,.
\end{equation}
We define the velocity and angular frequency at $R_{\sscBondi}$ as 
the Bondi velocity and the Bondi angular frequency, respectively: 
\begin{align}
    V_{\sscBondi} & \equiv V_{\rm c}(R_{\sscBondi}) 
    = \psi_{\rm nuc} V_{\rm nuc} \left(\frac{R_{\sscBondi}}{R_{\rm nuc}}\right)^q
    \,;\\
    \begin{split}
        \Omega_{\sscBondi} 
        & \equiv \frac{V_{\sscBondi}}{R_{\sscBondi}} 
        = \psi_{\rm nuc} \Omega_{\rm nuc} \left(\frac{R_{\sscBondi}}{R_{\rm nuc}}\right)^{q-1}
        \\
        &= \psi_{\rm nuc} \Omega_{\rm nuc} \left(
            \frac{M_{\sscBH}}{ \psi_{\rm nuc}^2 M_{\rm nuc}} 
        \right)^{\gamma_{\rm nuc}}
        \,,
    \end{split}
\end{align}
where $\gamma_{\rm nuc} = (q-1)/(2q+1)$ ranges from $-1$
($q = 0$) to $1/2$ ($q \rightarrow +\infty$).
Using Eq.~\eqref{eq:dot-M-Bondi-general} for $\dot{M}_{\sscBondi}$, we obtain the Eddington 
ratio of the Bondi accretion rate:
\begin{equation} 
    \begin{split}
        \dot{m}_{\sscBondi}
        & \equiv \frac{\dot{M}_{\sscBondi}}{\dot{M}_{\ssc Edd}}
        = (2q+1) t_{\ssc Sal} \Omega_{\sscBondi} 
        \\ 
        & = (2q+1) \psi_{\rm nuc} t_{\ssc Sal} \Omega_{\rm nuc} \left(
            \frac{M_{\sscBH}}{\psi_{\rm nuc}^2 M_{\rm nuc}} 
        \right)^{\gamma_{\rm nuc}}
        \,, \label{eq:m-bh-dot-edd-vs-q}
    \end{split}
\end{equation}
where $\dot{M}_{\ssc Edd} \equiv M_{\sscBH} / t_{\ssc Sal} $
is the Eddington accretion rate\footnote{Since we do not impose a constant 
value on the radiative efficiency (see Appendix~\ref{app:ssec:nuclear-burst}), 
the definition of the Eddington rate of BH accretion 
in this paper assumes a radiative efficiency of unity
for simplicity. We will clarify alternative definitions when discussing 
studies using them.}, 
and $t_{\ssc Sal} \equiv \sigma_{\ssc T}c / ( 4\pi G m_{\rm p} ) \approx 450\Myr$  
is the Salpeter timescale.
Since $\Omega_{\rm nuc} \sim 1\Myr^{-1}$, the factor 
$(2q+1) \psi_{\rm nuc} t_{\ssc Sal} \Omega_{\rm nuc}$ in $\dot{m}_{\sscBondi}$ 
is of the order of $10^3$. For the default choice ($q = 1/2$), 
$\dot{m}_\sscBondi \gtrsim 10^3$ for $M_{\sscBH} \lesssim M_{\rm nuc}$
(or $R_{\sscBondi} \lesssim R_{\rm nuc}$; see Eq.~\ref{eq:r-bondi-vs-q}).
This implies a super-Eddington accretion of the BH at the
beginning of the nuclear burst, provided the gas flowing across the Bondi radius 
can reach the BH (see Appendix~\ref{app:ssec:bh-disk}).

In the default case ($q=1/2$, $\gamma_{\rm nuc}=-1/4$), $R_\sscBondi$ and $\dot{m}_{\sscBondi}$ 
(Eqs.~\ref{eq:r-bondi-vs-q} and \ref{eq:m-bh-dot-edd-vs-q}) can be expressed
neatly as functions of $\Sigma_{\rm nuc}$ and $M_\sscBH$. Using 
Eq.~\eqref{eq:sigma-nuc-numeric}, the dependence on $\Sigma_{\rm nuc}$ 
can then be transformed into a dependence on $\Sigma_\sscSGC$ or  
$M_{\rm v}$. This gives the following two chains of equations:
\begin{align} 
    \begin{split}
        R_{\sscBondi} 
        &= \psi_{\rm nuc}^{-1} \left( \frac{M_{\sscBH}}{\Sigma_{\rm nuc}} \right)^{1/2}
        \\
        &= \psi_{\rm nuc}^{-1} \left( \frac{3}{m_{\rm g}(1)} \frac{M_{\sscBH}}{ c_{\rm g}\Sigma_\sscSGC }  \right)^{1/2}
        \\
        &\approx 4.56\pc
            \frac{M_{\rm {\sscBH}, 6}^{1/2}}
                { \left[\chi_{\rm g} (1+z)_{10}\right]^{1/4} M_{\rm v, 9.5}^{1/6}}
        \,; \label{eq:R-Bondi-numerical-q05}
    \end{split} \\
    \begin{split}
        \dot{m}_{\sscBondi} 
        &= 2 \psi_{\rm nuc}^{3/2} t_{\ssc Sal} G^{1/2} \frac{\Sigma_{\rm nuc}^{3/4}}{M_{\sscBH}^{1/4}}
        \\ 
        &= 2 \psi_{\rm nuc}^{3/2} t_{\ssc Sal} G^{1/2} \left[\frac{m_{\rm g}(1)}{3}\right]^{3/4}
            \frac{ \left(c_{\rm g} \Sigma_\sscSGC\right)^{3/4} }{M_{\sscBH}^{1/4}}
        \\ 
        &\approx 6010 \,
            \left[\chi_{\rm g}(1+z)_{10}\right]^{3/8}
            \frac{M_{\rm v, 9.5}^{1/4}}{M_{\rm {\sscBH}, 6}^{1/4} }
        \,, \label{eq:dot-m-Bondi-numerical-q05}
    \end{split}
\end{align}
Each of the chains is composed of three equations that link quantities at
four different radii: the Bondi radius ($R_{\sscBondi}$), the radius of the SNF nucleus 
($R_{\rm nuc}$), the radius of the SGC ($R_\sscSGC$), and the virial radius of the halo 
($R_{\rm v}$). Note that $R_{\sscBondi}$ depends weakly on $M_{\rm v}$ and $z$
due to the weak dependence of $\Sigma_{\rm nuc}$ on $M_{\rm v}$ and $z$ 
(see the texts below Eq.~\ref{eq:sigma-nuc-numeric}). 
This implies that the Bondi radius in an SNF nucleus is mainly controlled by 
the BH mass. The power-law indices on $\Sigma_{\rm nuc}$ in both chains are less than 
$1$. Combined with the nearly flat evolution of $\Sigma_{\rm nuc}$ during 
the assembly history of any halo (see Fig.~\ref{fig:sgc-snf-props}c),
this implies that both $R_{\sscBondi}$ and $\dot{m}_{\sscBH}$ evolve very slowly 
for given $M_{\sscBH}$ and $M_{{\rm v},z=0}$.

The inner ends of the green segments in Fig.~\ref{fig:profile-ladder}a
show the predicted Bondi radii of BHs with $M_\sscBH = 10^6\Msun$ embedded 
in the SNF nuclei that are expected to form in an MW-size halo and 
its progenitors. The Bondi radii are a few pc, and as expected,
evolve very little from $z=9$ to $z=0$.
For reference, we show the Bondi properties for wide ranges of halo masses and redshifts 
in Appendix~\ref{app:sec:evolution-of-hierarchy} and Fig.~\ref{fig:bondi-props}.

\subsection{Gas structure inside the Bondi radius}
\label{app:ssec:bh-disk}


Within the Bondi radius, the central BH dominates the total gravity.
Gas clouds moving across this radius are tidally torn apart, forming 
a new gaseous structure that fuels the BH. As suggested by hydro 
simulations \citep{angles-alcazarCosmologicalSimulationsQuasar2021,
hopkinsFORGEdFIREResolving2024,hopkinsFORGEdFIREII2024,
shiSeedsSupermassiveBlack2024}, the gas dynamics in this region is highly 
asymmetric, eventually leading to the build-up of an accretion disk surrounding the BH. 
In the regime of the high accretion rate expected in an SNF nucleus,
the accretion disk does not resemble an $\alpha$-disk 
\citep{shakuraBlackHolesBinary1973} in the following aspects 
\citep[see][thereafter \citetalias{hopkinsAnalyticModelMagneticallyDominated2024}, 
for the details]{hopkinsAnalyticModelMagneticallyDominated2024}:
\begin{enumerate}[topsep=0pt,parsep=0pt,itemsep=0pt,label=(\roman*)]
    \item The disk is turbulent, with a sonic Mach number 
    $\mathcal{M}_{\rm s} = v_{\rm t} / c_{\rm s} \gg 1$,
    a consequence of the rapid (super-Eddington) inflow.
    In contrast, turbulence in an $\alpha$-disk is subsonic.
    \item The magnetic field can be strong, with an Alfv\'en velocity 
    $v_{\ssc A} \equiv B / \sqrt{4 \pi \rho_{\rm g}} \gg c_{\rm s}$.
    The strong magnetic field appears to be inevitable. 
    Even if the seed field is initially small, it can be amplified by the 
    turbulent dynamo and reach equal partition with the gas 
    kinetic energy ($v_{\ssc A} \sim v_{\rm t}$, with
    the Alfv\'en Mach number $\mathcal{M}_{\ssc A} \equiv v_{\rm t}/v_{\ssc A} \sim 1$), 
    over a number of dynamical timescales, 
    as suggested by MHD simulations 
    \citep{martin-alvarezThreephaseAmplificationCosmic2018,hopkinsFORGEdFIREResolving2024} and analytical studies 
    \citep{parievExtendingShakuraSunyaevApproach2003,
    begelmanAccretionDiscsStrong2007,
    begelmanMagneticFieldsCatalyse2023}.
    \item The disk is thickened by the large $v_{\rm t}$ and $v_{\ssc A}$,
    in contrast to an $\alpha$-disk which is thin
    ($H_{\rm g}/r \ll 1$) and supported by thermal and radiation 
    pressure.
    \item The gas inflow rate in the disk towards the BH, 
    $\dot{M}_{\rm g}$, is large, which is a consequence of 
    the large kinematic viscosity contributed by the large $v_{\rm t}^2$ (Reynolds stress)
    and $v_{\ssc A}^2$ (Maxwell stress).
    \item The disk is stable against fragmentation and star formation, 
    as the Toomre $Q$ parameter is expected to be large 
    (see Eq.~\ref{eq:toomreQ-bh-disk-numerical-q05}).
\end{enumerate}

The transition of the gas from less structured turbulence at $r \gtrsim R_{\sscBondi}$
to an ordered accretion disk at $r \lesssim R_{\sscBondi}$ implies 
an additional level in the cosmic hierarchy. In particular, the properties of 
the accretion disk determine the possibility of star formation 
of the gas around the BH and eventually the growth rate of the central BH.
We thus explicitly include the accretion disk in our modeling, 
with the outer boundary conditions set by the Bondi properties 
at $R_\sscBondi$ obtained in Appendix~\ref{app:ssec:modified-bondi}
and the inner boundary conditions set by the BH itself.
Briefly, the boundary conditions include
$M_{\sscBH}$ of the BH, $\dot{M}_{\rm g}$, $v_{\rm t}/V_{\sscKep}$
and $v_{\ssc A}/V_{\sscKep}$, all evaluated at $R_{\sscBondi}$,
and some normalization factors.
Here, $V_{\sscKep} \approx V_{\rm c}$ is the Keplerian velocity, which approximates 
the circular velocity at $r \leq R_{\sscBondi}$ as the gravity is dominated by the BH in this region.
The output of this model includes $v_{\rm t}/V_{\sscKep}$, $v_{\ssc A}/V_{\sscKep}$ 
and $\Sigma_{\rm g}$, from which other variables, such as the disk height ($H_{\rm g}$) 
and the Toomre $Q$ parameter, can be derived, all as functions of the BH-centric distance $r$
within the domain of $ r_{\ssc S} \ll r \leq R_{\sscBondi}$.
Here $r_{\ssc S}$ is the Schwarzschild radius, defined as
\begin{align}
    R_{\ssc S} 
    & = \frac{2 G M_{\sscBH}}{c^2}  \nonumber \\
    & = 10^{-4}\pc \, M_{\rm {\sscBH},9}
    = 20 \au \, M_{\rm {\sscBH},9} = 0.02 \au\, M_{\rm {\sscBH},6}\,,
\end{align}
at which relativistic effects become significant.
The model we adopt is developed by \citetalias{hopkinsAnalyticModelMagneticallyDominated2024}, 
which was shown to be capable of reproducing results of their 
hyper-refinement zoom-in simulations carried out in a realistic cosmological context. 
We refer the readers to their paper for details.

The basic equation can be derived by considering the energy
conservation that dissipation of the turbulence/magnetic energy within 
the disk is driven by the release of the gravitational energy of the inflow gas. 
Equating the dissipation rate per unit area, 
$\dd{E}_{\rm t+A}/(\dd{A}\dd{t})$,
and the gravitational power per unit area, 
$\dd{E}_{\rm grav}/(\dd{A}\dd{t})$,
we obtain
\begin{equation} \label{eq:Sigma-g-equality}
    \frac{3}{2 \psi_{\rm t}^2} \Sigma_{\rm g} \Omega_{\sscKep} v_{\rm t}^2
    = \frac{3}{4\pi} \Omega_{\sscKep}^2 \dot{M}_{\rm g} \,.
\end{equation}
Here, $\Omega_{\sscKep} = V_{\sscKep}/r$ is the angular frequency of a circular
orbit, and $\dot{M}_{\rm g}$ is the net inflow (inflow minus outflow)
rate of the gas. As we will see below (Eq.~\ref{eq:toomreQ-bh-disk-numerical-q05}), 
the accretion disk does not 
have significant star formation, and thus the Bondi accretion rate is preserved 
across the accretion disk, i.e. $\dot{M}_{\rm g} = \dot{M}_{\sscBondi} = \dot{M}_\sscBH$.
The condition $\psi_{\rm t}^2 \sim \mathcal{O}(1)$ combines the normalization 
factors on both sides of the equation.
The total specific energy to dissipate is about
$(1/2)( v_{\rm t}^2 + v_{\ssc A}^2 ) \sim v_{\rm t}^2$,
where the approximation holds for a magnetic field close to the equal-partition level. 
The dissipation timescale is $2 \psi_{\rm t}^2 / (3\Omega_{\sscKep})$.
More formally, this equation can be derived from the conservation
of angular momentum \citep[see][for details]{parievExtendingShakuraSunyaevApproach2003}.
An equivalent formalism is $\dot{M}_{\rm g} = 3\pi \Sigma \nu_{\rm visc}$, 
where the viscosity is
$\nu_{\rm visc} = 2 v_{\rm t}^2 / (3 \psi_{\rm t}^2 \Omega_{\sscKep})$.
The inflow velocity, $\left| v_r \right|$, can be deduced 
by its definition,
\begin{equation} \label{eq:v-r}
    \left| v_r \right| \equiv
    \frac{\dot{M}_{\rm g}}{2 \pi r \Sigma_{\rm g}}
    = \left( \frac{v_{\rm t}}{\psi_{\rm t}V_{\sscKep}} \right)^2 V_{\sscKep}.
\end{equation}

To form a closed set for Eq.~\eqref{eq:Sigma-g-equality}, we need an equation to describe 
the turbulence structure, $v_{\rm t} (r)$.
Here we seek similarity solutions and assume that the radial dependence 
of the turbulent velocity can be parameterized as $v_{\rm t} \propto r^{-\zeta_{\rm t}} $. 
We further assume that the gas falls freely at $R_{\sscBondi}$, 
i.e. $\left| v_r(R_{\sscBondi}) \right| = V_{\sscKep} $,
which is used to fix the normalization.
With these assumptions, the equation for the turbulence structure is
\begin{equation}  \label{eq:v-t-scaled}
    \frac{v_{\rm t}}{V_{\sscKep}} = \psi_{\rm t} y_{r}^{1/2- \zeta_{\rm t}} \,,
\end{equation}
where $y_{r} \equiv r/R_{\sscBondi}$.
Substituting Eq.~\eqref{eq:v-t-scaled} into Eq.~\eqref{eq:Sigma-g-equality}, we obtain
\begin{equation} \label{eq:Sigma-g-equality-scaled}
    \Sigma_{\rm g} = 
        \frac{\dot{M}_{\rm g}}{2\pi \sqrt{G M_{\sscBH} R_{\sscBondi}} } 
        y_{r}^{2\zeta_{\rm t}-3/2} \,.
\end{equation}
Interestingly, $\Sigma_{\rm g}$ does not depend on $\psi_{\rm t}$.

The vertical structure of the disk, as suggested by 
\citetalias{hopkinsAnalyticModelMagneticallyDominated2024}, is supported by both turbulence
and the magnetic field, with the latter being even more dominant at smaller radius.
The vertical structure of the disk is thus best parameterized through 
the magnetic field, as $H_{\rm g}/r \sim v_{\ssc A}/V_{\sscKep}$. 
Based on their simulations, \citet{hopkinsFORGEdFIREResolving2024,hopkinsFORGEdFIREII2024} 
found that the magnetic field within $R_\sscBondi$ can be described by 
the flux-frozen/advection scenario extrapolated from the outer boundary
conditions set at $R_\sscBondi$. They thus proposed an effective
equation of state (eEOS) for the magnetic field, as 
$B^2 \propto \rho_{\rm g}^{\gamma_{\sscBondi}} $, or 
$v_{\ssc A} \propto \rho_{\rm g}^{(\gamma_{\sscBondi}-1)/2}$, where
$\rho_{g} \sim \Sigma_{\rm g} / (2H_{\rm g}) \sim \Sigma_{\rm g}\Omega_{\sscKep} / (2v_{\ssc A})$
is the density at the mid-plane of the accretion disk.
Reorganizing $v_{\ssc A}$ to one side of the equation, and using
Eqs.~\eqref{eq:Sigma-g-equality} and \eqref{eq:v-t-scaled} for $\Sigma_{\rm g}$,
we obtain
\begin{equation} \label{eq:v-A-scaled}
    \frac{v_{\ssc A}}{V_{\sscKep}} = \psi_{\ssc A} y_{r}^{1/2- \zeta_{\ssc A}} \,,
\end{equation}
where $\zeta_{\ssc A} = (3-2 \zeta_{\rm t}) (\gamma_{\sscBondi}-1)/(\gamma_{\sscBondi}+1)$,
and $\psi_{\ssc A}$ is a normalization factor. With this equation, 
the height of the disk is $ H_{\rm g} \sim \psi_{\ssc A} R_{\sscBondi} y_{r}^{3/2 - \zeta_{\ssc A}} $.

Eqs.~\eqref{eq:Sigma-g-equality}, \eqref{eq:v-t-scaled} and \eqref{eq:v-A-scaled} 
form a closed set describing the density, turbulence and magnetic field
structure of the accretion disk. The set is fixed once
the structure index $\zeta_{\rm t}$ and $\zeta_{\ssc A}$ (or $\gamma_{\sscBondi}$)
are given/calibrated, and the outer boundary $R_{\sscBondi}$ and 
the boundary conditions on $M_{\sscBH}$, $\dot{M}_{\sscBondi}$, 
$\psi_{\rm t}$ and $\psi_{\ssc A}$ are obtained from the modeling described earlier. 
Other by-products can be derived using these equations, such as the
Toomre $Q$ parameter of the disk:
\begin{equation} \label{eq:toomre-scaled}
    Q \equiv \frac{\sigma_{\rm eff} \kappa}{\pi G \Sigma_{\rm g}} 
    \approx  2\sqrt{2} \psi_{\rm t} \sqrt{1+\mathcal{M}_{\ssc A}^{-2}} 
    \frac{M_{\sscBH}\Omega_{\sscBondi}}{\dot{M}_{\rm g}} 
    y_{r}^{-3\zeta_{\rm t}} \,,
\end{equation}
where we have used $\kappa = \sqrt{2} \Omega_{\sscKep}$, 
$\Omega_{\sscBondi} = \Omega_{\sscKep}(R_{\sscBondi})$,
and $\sigma_{\rm eff}^2 = v_{\rm t}^2+v_{\ssc A}^2$, with the sound speed 
neglected. 

A special case of this model 
\citepalias[the default model of][]{hopkinsAnalyticModelMagneticallyDominated2024}, 
which we adopt in this paper, is given by a flux-frozen magnetic field 
($\gamma_{\sscBondi} = 4/3$) and a trans-Alfv\'enic, supersonic turbulence 
($\mathcal{M}_{\ssc A} \sim 1$, $\mathcal{M}_{\rm s} \gg 1$). 
These imply that  $\zeta \equiv \zeta_{\rm t} = \zeta_{\ssc A} = 1/3$, 
$\psi \equiv \psi_{\rm t} \sim \psi_{\ssc A}$,
and thus
\begin{align}
    \frac{H_{\rm g}}{r} & \sim \frac{v_{\ssc A}}{v_{\sscKep}} 
        \sim \frac{v_{\rm t}}{v_{\sscKep}} = \psi y_{r}^{1/6} 
    \,;\label{eq:v-scaled-q-05} \\
    \Sigma_{\rm g} & = \frac{\dot{M}_{\rm g}}{2\pi \sqrt{G M_{\sscBH} R_{\sscBondi}} y_{r}^{5/6} } 
    \,;\label{eq:Sigma-g-equality-scaled-q-05}\\
    \rho_{\rm g} & \sim \frac{ \dot{M}_{\rm g} }{4\pi\psi \Omega_{\sscBondi} R_{\sscBondi}^3 y_{r}^2}
    \,;\label{eq:rho-scaled-q-05}\\
    Q & \sim \frac{ 4 \psi M_{\sscBH}\Omega_{\sscBondi} }{\dot{M}_{\rm g} y_{r}} 
    \,.\label{eq:toomre-scaled-q-05}
\end{align}
Taking $R_{\sscBondi}$ from Eq.~\eqref{eq:R-Bondi-numerical-q05},
we obtain
\begin{align}
    \begin{split}
        \Sigma_{\rm g} 
            & = 24.7 \Msunperpcsq
            \frac{M_{\rm {\sscBH},6}^{1/2}\dot{m}_{\sscBondi ,10}}
                {R_{\sscBondi , 4.56}^{1/2}y_{r}^{5/6}}        
        \\
            & = 24.7 \Msunperpcsq \left[\chi_{\rm g} (1+z)_{10}\right]^{1/8}
            \frac{M_{\rm v,9.5}^{1/12}M_{\rm {\sscBH},6}^{1/4}\dot{m}_{\sscBondi ,10}}
                {y_{r}^{5/6}}   
        \,; \label{eq:Sigma-bh-disk-numerical-q05}
    \end{split} \\
    \begin{split}
        \rho_{\rm g}
            &\sim 2.70 \Msunpercpc\psi^{-1} \frac{M_{\rm {\sscBH}, 6}^{1/2} \dot{m}_{\rm {\sscBH}, 10} }
                {  R_{\sscBondi , 4.56}^{3/2} y_{r}^2 }
        \\
            &= 2.70 \Msunpercpc\psi^{-1} \left[\chi_{\rm g} (1+z)_{10}\right]^{3/8}
                \frac{ M_{\rm v,9.5}^{1/4}\dot{m}_{\sscBondi ,10}}{ M_{\rm {\sscBH},6}^{1/4} y_{r}^2 }
        \,;\label{eq:rho-bh-disk-numerical-q05}
    \end{split} \\
    \begin{split}
        Q 
            & \sim 1240 \psi 
            \frac{M_{\rm {\sscBH},6}^{1/2}}
                {R_{\sscBondi , 4.56}^{3/2} \dot{m}_{\sscBondi ,10} y_{r}} 
        \\
            & = 1240 \psi
            \left[\chi_{\rm g} (1+z)_{10}\right]^{3/8}
            \frac{ M_{\rm v,9.5}^{1/4} }
                { M_{\rm {\sscBH},6}^{1/4} \dot{m}_{\sscBondi ,10} y_{r} }
        \,, \label{eq:toomreQ-bh-disk-numerical-q05} 
    \end{split}
\end{align}
where $\dot{m}_{\sscBondi, 10} \equiv \dot{m}/10$ is the Eddington ratio 
of the Bondi accretion rate assuming a radiative efficiency of $\chi_\sscAGN = 0.1$ 
\citep{crainEAGLESimulationsGalaxy2015,rosas-guevaraImpactAngularMomentum2015}.
The properties of the accretion disk depend weakly on $M_{\rm v}$ and $M_{\sscBH}$, 
but strongly on $\dot{m}_{\sscBondi}$ and $y_{r}$.
$Q \gg 1$ is implied by 
Eq.~\eqref{eq:toomreQ-bh-disk-numerical-q05} at any $y_{r} \leqslant 1$,
consistent with the assumption that the disk is stable against 
fragmentation (see the text below Eq.~\ref{eq:Sigma-g-equality}).
This is true even when we consider the extremely high accretion rate of 
$\dot{m}_{\sscBondi} \sim 6000$ at the beginning of the nuclear burst 
(Eq.~\ref{eq:dot-m-Bondi-numerical-q05}).

Fig.~\ref{fig:am-redistribute}e schematically shows the structures within the 
Bondi radius -- the last level in the cosmic hierarchy. 
The accretion disk is turbulent, magnetized and thickened. Star formation is 
suppressed in the inner volume of this regime, 
considering the $r^{-1}$ dependence of the Toomre $Q$ 
parameter (Eq.~\ref{eq:toomreQ-bh-disk-numerical-q05}).
Stars might be present in the outer volume, which is not properly 
captured by the approximation of the accretion-disk model.
These stars
might also redistribute their energy and angular momentum via dynamical processes, 
and contribute to the BH growth. Such contribution is referred
to as the `dry' channel, as a complement to the `wet' channel of
gas accretion. Each orange segment in Fig.~\ref{fig:profile-ladder}a
shows the mid-plane density of the accretion disk formed during the 
nuclear burst in an MW-size halo or its progenitors, assuming
a central BH with $M_{\sscBH} = 10^6 \Msun$ and
a maximal accretion rate given by Eq.~\eqref{eq:dot-m-Bondi-numerical-q05}
at the beginning of the burst.
The orange shade covers the range of $\rho_{\rm g}$ as $\dot{m}_\sscBH$
varies from the Eddington level ($\dot{m}_\sscBH = 10$) to the maximum.
The profile of $\rho_{\rm g}$ varies very little 
from $z = 0$ to $z = 10$, due to the weak evolution of 
the Bondi radius and Bondi accretion rate 
(see the texts below Eqs.~\ref{eq:R-Bondi-numerical-q05} and 
\ref{eq:dot-m-Bondi-numerical-q05}). For reference, the card 
in Fig.~\ref{fig:profile-ladder}e summarizes the equations for the 
accretion disk.

The role of turbulence in BH growth is again highlighted here. It magnifies the
magnetic field, suppresses star formation, and preserves the rapid inflow at $R_\sscBondi$
through the accretion disk to feed the BH. As outlined in the beginning
of Appendix~\ref{app:ssec:snf-nucleus}, such a rapid process is expected to be transient,
associated with bursty star formation, BH accretion and feedback.
The evolution of the SNF nucleus in the dispersion stage is thus a process
expected to be largely asynchronous with the evolution of the host galaxy/halo. 
This suggests that a refinement engine working at much finer spatial scales and   
temporal steps than those for modeling the host galaxy/halo is needed to follow 
the evolution of the hierarchy below the SNF nuclei, as described below.

\subsection{The bursty nature of black-hole growth and star formation in the nuclei}
\label{app:ssec:nuclear-burst}

Here we build a model for the evolution of the SNF nucleus 
in the dispersion stage. The nuclear burst, including bursty star formation 
and BH accretion, together with their feedback, is modeled.
A number of depletion timescales in the nucleus are analytically derived and compared 
with each other to gain insights into the mechanisms driving the evolution.
The equations governing the evolution can be solved numerically, 
and will be incorporated as a refinement engine to complete 
the whole modeling procedure (see Appendix~\ref{app:sec:impl-details} and 
Fig.~\ref{fig:flowchart} for an outline).

\subsubsection{Basic considerations}
\label{app:sssec:nuclear-burst-basic-eq}

Simplifications are necessary for the model to be applicable in a cosmological volume, 
while the essence of the nuclear burst must be retained.
Here we model the distributions of the gas and stars within the SNF nucleus 
using spherically averaged, smoothed profiles. 
This takes into account the isotropy expected, at least statistically, 
in a dynamically hot nucleus. Sub-structures, such as gas and stellar clumps, 
are missed in such a treatment. Thus, star formation, gas inflow and interaction
between feedback and gas in the model are treated as effective processes 
representing the average over diffuse and clumpy structures.

The basic equation is the continuity equation of gas density $\rho_{\rm g}(r, t)$:
\begin{multline} 
    \pdv{\rho_{\rm g}}{ t } = 
    - \pdv{\rho_{\rm *}}{ t } -  \frac{\left(\dot{e}_{\ssc AGN} + \dot{e}_{\ssc SN} + \dot{e}_{\ssc SE} + \dot{e}_{\rm ext}\right)\rho_{\rm g}}{v_{\rm out}^2}
    \\
    - \frac{1}{r^2} \pdv{}{r} \left( r^2 \rho_{\rm g} v_r \right)
    \,,\label{eq:pde-rho-g}
\end{multline}
where $v_{\rm r}$ is the radial velocity of inflow.
This equation describes the reduction of the gas density at a given BH-centric distance 
$r$ due to different sinks, each quantified by one term on the right-hand side.
Here, $\rho_{\rm *}(r, t)$ is the stellar density, and the term 
$-\pdv*{\rho_{\rm *}} {t}$ describes the consumption of the gas by star formation.
The terms, $\dot{e}_{\ssc AGN}$, $\dot{e}_{\ssc SN}$, $\dot{e}_{\ssc SE}$,
are the specific (i.e. per unit mass) energy injection rates
by the feedback from the BH, SN explosions and stellar evolution 
within the nucleus, respectively, while $\dot{e}_{\rm ext}$ is from 
star formation outside the nucleus. The quantity 
$v_{\rm out}$ denotes the velocity of the outflow driven by all the feedback
effects. The total baryon density will be denoted as $\rho = \rho_{\rm g} + \rho_*$.
The time domain of this equation is $t \geqslant t_{\rm burst}$, with $t_{\rm burst}$
being the starting epoch of the dispersion stage.
The radial domain of the equation is $R_{\rm in} \leqslant r  \leqslant R_{\rm nuc}$,
with the inner boundary at $R_{\rm in} \ll R_{\sscBondi}$ in the beginning so that 
the time evolution of the Bondi radius can be self-consistently determined 
by its definition (Eq.~\ref{eq:R-Bondi-general}) and the structure equation at 
$r \leqslant R_\sscBondi$ can be determined by the model of the accretion disk 
(Appendix~\ref{app:ssec:bh-disk}).

To estimate the relative importance of each term in Eq.~\eqref{eq:pde-rho-g}, 
we consider the depletion timescale, $t_{{\rm dep}, x} = M_{\rm nuc} / \dot{M}_{x}$, 
for a process $x$ that changes the gas mass at a rate of $\dot{M}_{x}$.
The last sink term in Eq.~\eqref{eq:pde-rho-g} corresponds to a continuous accretion flow
through the nucleus towards the BH. The corresponding mass inflow rate can be 
estimated using the Bondi accretion rate, $\dot{M}_{\sscBondi}$ (Eq.~\ref{eq:m-bh-dot-edd-vs-q}),
and the depletion timescale is:
\begin{equation}
    t_{\rm dep, bhacc} = \frac{M_{\rm nuc}}{\dot{M}_{\sscBondi}} 
    = \frac{M_{\rm nuc}}{ (2q+1) \Omega_{\sscBondi} M_{\sscBH} } \,.
\end{equation}
The first sink term in Eq.~\eqref{eq:pde-rho-g} describes the gas consumption 
by star formation. The corresponding depletion timescale 
may be approximated by the free-fall time of the gas within the nucleus
(Eq.~\ref{eq:t-ff-nuc}),
\begin{equation}
     t_{\rm dep, {\ssc SF}} = t_{\rm ff, nuc} = \frac{\pi}{2\sqrt{2} \Omega_{\rm nuc}}\,.
\end{equation}
The ratio between the two depletion timescales is
\begin{equation} 
    \frac{ t_{\rm dep, bhacc}}{ t_{\rm dep, {\ssc SF}}} 
    = \frac{2\sqrt{2} \psi_{\rm nuc}^{2\gamma_{\rm nuc}-1}  }{\pi (2q+1)}
    \left(\frac{M_{\rm nuc}}{M_{\sscBH}}\right)^{\gamma_{\rm nuc}+1}
    \,.\label{eq:t-dep-acc2sf}
\end{equation}
The prefactor is of the order of unity, while the mass-dependence 
is $(M_{\rm nuc}/M_\sscBH)^{3/4}$ in the default case ($q=1/2$).
The timescale $t_{\rm dep, {\ssc SF}} \ll  t_{\rm dep, bhacc}$ 
if $M_{\sscBH} \ll M_{\rm nuc}$, 
and $t_{\rm dep, {\ssc SF}} \gg  t_{\rm dep, bhacc}$
if $M_{\sscBH} \gg M_{\rm nuc}$.
This suggests a competitive scenario between BH accretion and 
star formation in the SNF nucleus, with the dominant process determined by 
$M_{\rm nuc}/M_\sscBH$.

The second sink term in Eq.~\eqref{eq:pde-rho-g} groups the effects of
different feedback channels of gas depletion.
The energy injection by the SN feedback, $\dot{e}_{\ssc SN}$, is expected
to be negligible before SN explosions from dying massive stars
($t - t_{\rm burst} \lesssim t_{\rm snf}$; see the text below Eq.~\ref{eq:t-ff-gas-of-snf}), 
but to be sufficiently strong 
afterwards to totally disperse the gas from the nucleus. SNe thus place a hard 
limit on the duration of the nuclear burst. 
The depletion timescale by the SN feedback is thus 
$t_{{\rm dep}, {\ssc SN}} = t_{\rm snf}$,
comparable to that by the star formation ($t_{\rm dep, {\ssc SF}}$).
The energy injection by stellar evolution, $\dot{e}_{\ssc SE}$, 
and by external star formation, $\dot{e}_{\rm ext}$,
is expected to be negligible owing to the high density of the nucleus, 
as discussed in Appendix~\ref{app:ssec:snf-nucleus}.
Thus, during a burst of an SNF nucleus, AGN feedback is the only possible 
channel of feedback that is effective to deplete gas.

The exact effects of AGN feedback on the nuclear gas, however, are 
difficult to quantify from first principles. 
The difficulty comes from uncertainties in, e.g.
the `structure' of the BH
\citep[e.g. spin; see][]{huskoSpindrivenJetFeedback2022},
the state and emissivity of the gas in the accretion disk
\citep{yuanHotAccretionFlows2014,
koudmaniUnifiedAccretionDisc2024,
hopkinsAnalyticModelMagneticallyDominated2024} 
and in the circum-BH medium \citep{duttaDissipationAGNJets2024}, 
the potential launch of ultra-fast outflows \citep{wagnerUltrafastOutflowsGalaxyscale2013,
gaspariUnifyingMicroMacro2017}, 
the coupling of the feedback energy (in radiation and outflow) 
with the nuclear gas \citep{suSelfregulationBlackHole2023,suSelfregulationHighredshiftBlack2025}, and the interplay among them.
To simplify the problem, we do not attempt to differentiate channels of the AGN feedback 
\citep[e.g. jet, wind, radiation and cosmic ray; see][]{shiFeedbackregulatedSeedBlack2024}.
Instead, we parameterize the total mass loss rate of the SNF nucleus produced by 
the AGN feedback as
\begin{equation}
    \dot{M}_\sscAGN = \chi_\sscAGN L_{\ssc Edd} / v_{\rm out}^2
    \,,\label{eq:mdot-agn}
\end{equation}
where $L_{\ssc Edd} \equiv M_\sscBH c^2/t_{\ssc Sal}$ is the Eddington 
luminosity; $\chi_\sscAGN$ is the Eddington ratio of feedback 
energy, taking into account the net effect from both
the energy conversion efficiency of the BH accretion and the 
coupling efficiency of the feedback energy to the nuclear gas.
As suggested by e.g. \citet{inayoshiHyperEddingtonAccretionFlows2016},
a thickened accretion disk in the super-Eddington regime is 
radiatively inefficient due to `photon trapping', and is able to 
radiate at the Eddington level. Given the high density in the SNF regime, 
most of the emitted radiation is expected to be coupled with the nuclear gas. 
This implies that AGN feedback is inefficient in radiating but efficient in 
coupling, very different from the sub-grid recipes commonly implemented 
in unresolved hydrodynamic simulations \citep[e.g.][]{schayeEAGLEProjectSimulating2015,
weinbergerSimulatingGalaxyFormation2017,daveSimbaCosmologicalSimulations2019}. 
We thus adopt $\chi_\sscAGN = \chi_{\sscAGN, \infty} = 1$ as our default value at 
the high accretion rate (see below for the treatment of low accretion rate).
The default outflow velocity is $v_{\rm out} = 10^3 \kms$,
as suggested by observations \citep[e.g.][]{sturmMassiveMolecularOutflows2011,greeneSpectacularOutflowObscured2012,pernaXraySDSSSample2017,
pernaMultiphaseOutflowsMkn2019,deugenioFastrotatorPoststarburstGalaxy2024} and feedback 
models \citep[e.g.][]{costaFeedbackActiveGalactic2014,daveSimbaCosmologicalSimulations2019}.
With the adopted parameterization, the depletion timescale for the AGN feedback is 
\begin{equation}
     t_{{\rm dep}, \sscAGN} = \frac{M_{\rm nuc}}{\dot{M}_\sscAGN}
    = \frac{t_{\ssc Sal}}{\chi_\sscAGN} \frac{M_{\rm nuc}}{M_\sscBH} 
    \left( \frac{v_{\rm out}}{c} \right)^2 
    \,.\label{eq:t-dep-agn}
\end{equation}
The ratio of the depletion timescale between the BH accretion and 
the AGN feedback is then 
\begin{equation} 
    \frac{ t_{\rm dep, bhacc}}{ t_{{\rm dep}, \sscAGN}}
    = \frac{ \psi_{\rm nuc}^{2\gamma_{\rm nuc}-1} \chi_{\sscAGN} }{(2q+1)t_{\ssc Sal} \Omega_{\rm nuc}}
      \left(\frac{c}{v_{\rm out}}\right)^2
      \left(\frac{M_{\rm nuc}}{M_{\sscBH}}\right)^{\gamma_{\rm nuc}}
      \,,\label{eq:t-dep-acc2agn}
\end{equation}
where $(c/v_{\rm out})^2 \approx 300^2$ and 
$t_{\ssc Sal}\Omega_{\rm nuc} \approx 450 \times 1.8 \approx 800$.
In the default case ($q=1/2$, $\gamma_{\rm nuc}=-1/4$),
the factor $( M_{\rm nuc}/M_{\sscBH} )^{\gamma_{\rm nuc}} \geqslant 0.01$
for any $M_{\sscBH}/M_{\rm nuc} \gtrsim 10^{-8}$.
Thus, $t_{\rm dep, bhacc} \gg t_{{\rm dep},\sscAGN}$ almost always holds.

The above estimates yield a number of conclusions: 
(i) the depletion of gas in an SNF nucleus is dominated by the AGN feedback; 
(ii) if the AGN feedback does not disperse all the gas until $t - t_{\rm burst} = t_{\rm snf}$, 
     SN feedback kicks in and clears the remaining gas; 
(iii) a short time window, $t - t_{\rm burst} \in [0, t_{\rm snf}]$, is left  
for intense BH accretion and star formation, which leads to bursty growth in mass 
for both the BH and the stellar component. 
The temporary absence of SN feedback in the time window thus gives 
the BH a chance to grow differently from that expected in the continuous mode
where SNe constantly regulate the gas content around the BH.
Such a bursty mode may be missed in numerical simulations that are incapable of 
resolving dense structures at $n_{\rm g} \gg n_{\rm snf}$.
The gas medium around BHs in these simulations may be numerically over-smoothed, 
so that the SN feedback is always effective to regulate the BH growth 
\citep[see e.g.][]{bowerDarkNemesisGalaxy2017,habouzitBlossomsBlackHole2017,
habouzitSupermassiveBlackHoles2021,liPhysicalProcessesCoevolution2025}.
In the following, we present a detailed numerical treatment based 
on these conclusions.

\begin{figure} \centering
    \includegraphics[width=\columnwidth]{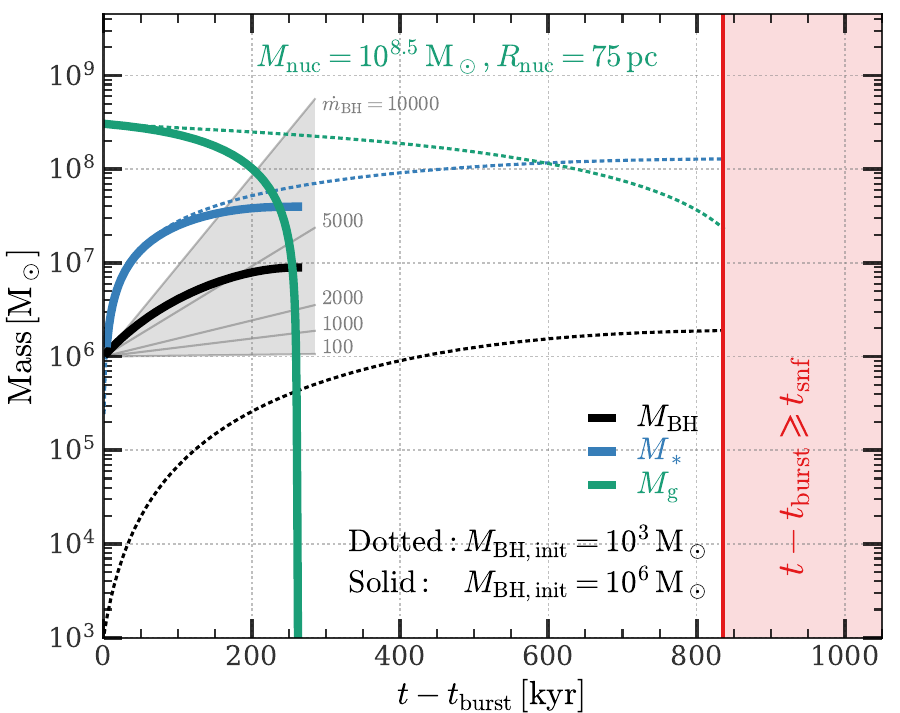}
    \caption{{\figem Evolution of masses in nuclear bursts predicted 
    by the refinement engine.}
    Here we show two events of nuclear bursts as examples, 
    both with a BH placed in the center of the SNF nucleus 
    with $M_{\rm nuc} = 10^{8.5} \Msun$ and $R_{\rm nuc} = 75 \pc$, 
    typical for MW-size galaxies and their progenitors (see e.g. Fig.~\ref{fig:sgc-snf-props}).
    The initial BH mass is $M_{\sscBH} = 10^3 \Msun$ ({\figem dotted}) 
    and $10^6 \Msun$ ({\figem solid}),
    respectively, in the two bursts.
    In each burst, {\figem black}, {\figem blue} and {\figem green} curves 
    show the masses of the BH, stars and gas, respectively, as 
    a function of elapsed time since the beginning of the burst.
    {\figem Red} shaded region indicates the truncation of the burst by 
    SN feedback, which sets a time window with a duration of $t_{\rm snf}$
    for the BH growth and star formation.
    {\figem Grey} lines in grey shaded region represent the evolution
    tracks expected for BHs with constant Eddington ratios, 
    $\dot{m}_\sscBH \equiv \dot{M}_\sscBH / \dot{M}_{\ssc Edd}$.
    Larger initial $M_{\rm BH}$ produces stronger feedback that 
    depletes the gas more quickly, and a self-regulated growth pattern 
    of BH is thus expected (see Fig.~\ref{fig:growth-effect-seeding} and 
    \S\ref{sssec:growth-effect-seeding}).
    A refinement engine is incorporated into the model 
    to numerically obtain the evolution of masses during each nuclear 
    burst.
    See Appendix~\ref{app:ssec:nuclear-burst} for details.
    }
    \label{fig:bh-engine-example}
\end{figure}

\subsubsection{A refined treatment}
\label{app:sssec:nuclear-burst-details}


To solve $\rho_{\rm g}(r,t)$ and $\rho_*(r,t)$ for an SNF nucleus
during the dispersion stage, we need to explicitly formulate 
each sink term in Eq.~\eqref{eq:pde-rho-g}. 
The term of star formation is parameterized as
\begin{equation}
    \pdv{\rho_{\rm *}}{ t } = \frac{\alpha_{\ssc SF}}{f_{\rm Q}} \frac{\rho_{\rm g}}{ t_{\rm ff} } \,,
\end{equation}
where $\alpha_{\ssc SF} \sim 1$ is a free parameter, 
$t_{\rm ff} = \sqrt{ 3\pi/( 32 G \rho ) }$ is the free-fall timescale of gas.
The suppression factor,
\begin{equation}
    f_{\rm Q}
    = \exp\left( \frac{Q - Q_{\sscBondi}}{\delta_{\rm Q}} \right)
    \,,\label{eq:sf-f-q}
\end{equation}
takes into account the suppression of star formation by the
large Toomre $Q$ parameter in the accretion disk, 
which allows a rapid but continuous decrease of the star formation
towards the center of the accretion disk; $\delta_{\rm Q} \sim 1$ controls the sharpness of 
the decrease; $Q$ is obtained from Eq.~\eqref{eq:toomre-scaled-q-05}
at $r \leqslant R_\sscBondi$, and $Q_\sscBondi \equiv Q(R_\sscBondi)$; 
we set $Q = Q_\sscBondi$ for $r > R_\sscBondi$, meaning no suppression
outside the Bondi radius. 

The inflow rate at the Bondi radius is based on the turbulence-modified Bondi 
accretion (Eq.~\ref{eq:dot-M-Bondi-general}), with a free parameter 
$\alpha_{\sscBH} \sim 1$ characterizing the normalization:
\begin{equation} \label{eq:numerical-bh-accretion-rate}
    \dot{M}_{\sscBH} = \alpha_{\sscBH} \dot{M}_{\sscBondi} \,.
\end{equation}
The inflow term $\partial(r^2 \rho_{\rm g} v_r)/\partial r /r^2$
in Eq.~\eqref{eq:pde-rho-g} depends on conditions at the outer boundary, 
$R_{\rm nuc}$,
and the dissipation processes within this boundary. Here we model 
it by a stable solution, $\dot{M}_{\rm g}(r) \equiv \dot{M}_\sscBH$, 
in which the inflow rate $\dot{M}_{\rm g}(r)$ is constant over the 
entire domain of $r \leqslant R_{\rm nuc}$.
This solution implies that the inflow term in Eq.~\eqref{eq:pde-rho-g} vanishes
and the density profile, without being affected by star formation or feedback, 
does not change with time.
This solution relies on the findings in Appendix~\ref{app:sssec:nuclear-burst-basic-eq} 
that the feedback term dominates the gas depletion at $R_\sscBondi < r \leqslant R_{\rm nuc}$
so that an arbitrary choice on the inflow term has negligible effect,
and that star formation is suppressed at $r \leqslant R_\sscBondi$ so that the inflow rate is 
preserved throughout the zone of influence of the BH (see Appendix~\ref{app:ssec:bh-disk}). 

The specific energy injection rate by SNe is modeled by a 
delayed flare, as
\begin{equation}
    \dot{e}_{\ssc SN}(t) = 
    \begin{cases}
        0, \ \ \ \ \ & {\rm if}\ t - t_{\rm burst} \leqslant t_{\rm snf} 
        \,;\\
        \infty,      & \text{otherwise}
        \,,
    \end{cases}
\end{equation}
so that a hard limit of $t_{\rm snf}$ is set for a complete ejection of the gas 
in the SNF nucleus and the end of the dispersion stage.
Other forms of stellar feedback, such as those represented by $\dot{e}_{\ssc SE}$
and $\dot{e}_{\rm ext}$, are expected to be negligible and are ignored.

The specific energy injection rate by AGN is determined implicitly according to 
the parameterization of the total feedback energy (Eq.~\ref{eq:mdot-agn}), as
\begin{equation}
    \chi_{\sscAGN} L_{\ssc Edd} = \int_{R_{\rm in}}^{R_{\rm nuc}} 
    \dot{e}_{\sscAGN} \rho_{\rm g} 4\pi r^2 \dd{r}\,,
\end{equation}
corresponding to a specific mass loss rate of 
$\dot{e}_{\sscAGN} / v_{\rm out}^2$.
The Eddington ratio of the feedback energy, $\chi_{\sscAGN}$, is parameterized as
a function of the Eddington ratio of the accretion rate:
\begin{equation}
    \chi_{\sscAGN} = \chi_{\sscAGN, \infty} 
        \frac{ \alpha_{\chi} \dot{m}_{\sscBondi} }{ 1 + \alpha_{\chi} \dot{m}_{\sscBondi}}
    \,,
\end{equation}
so that $\chi_{\sscAGN}$ is saturated at $\chi_{\sscAGN, \infty}$ for 
$\dot{m}_{\sscBondi} \gg 1$, and $\alpha_{\chi} \sim 0.1$ 
in the sub-Eddington regime where the accretion disk is expected to become
a normal $\alpha$-disk \citep{shakuraBlackHolesBinary1973}
that is often used as a sub-grid model in cosmological hydrodynamic simulations.
The radial dependence of $\dot{e}_\sscAGN$ is determined by the details 
of how the outflow mass and energy are coupled to the nuclear gas. Since these 
are largely unknown, we parameterize $\dot{e}_\sscAGN$ as
\begin{equation}
    \dot{e}_\sscAGN \propto \rho_{\rm g}^{-\beta_{\rm out}} \,,
\end{equation}
which accounts for the expectation that both coupling and cooling (dissipation)
should depend on the gas density. We adopt $\beta_{\rm out} = 0$ as the default,
which implies a radially invariant specific mass loss rate caused by the AGN feedback.
The calibration using hydrodynamic simulations, and the quantification 
for the effects of other relevant state variables, such as gas clumpiness
and metallicity, are worthy of being explored further.
Note that the outflow launched from the SNF nucleus can be very strong
due to the Eddington-level luminosity of the BH.
For $M_{\rm nuc} \sim 10^8\Msun$ and $ t_{\rm dep} \lesssim 1\Myr$, we obtain
$\dot{M}_{\rm w} \gtrsim 100 \msunperyr$.
When propagating outwards, the outflow can continuously couple with the gas in 
the SGC, and potentially launch galaxy-wide outflows.

Fig.~\ref{fig:bh-engine-example} shows two examples of nuclear bursts.
In each, a BH with a given initial mass is placed at the center of the 
SNF nucleus that is typical for MW-size galaxies and their progenitors.
The evolution of the mass components (BH, stars and gas) is shown as functions of
elapsed time since the beginning of the burst.
In both cases, the reduction of gas mass is substantial, much larger than the 
growth of BH and stellar masses. 
The main source of gas depletion is the AGN feedback.
In the case of small initial BH mass, $M_{\sscBH, {\rm init}} = 10^3\Msun$, 
about $1/10$ of the gas remains in the nucleus at $\approx 800\Kyr$ before SNe 
kick in to clear the gas. The central BH grows by more than three 
orders of magnitude in mass, from $10^3\Msun$ to more than $10^6\Msun$.
A nuclear star cluster with $M_* > 10^8\Msun$ also forms in the burst, 
indicating a strong competition with the BH in consuming the gas.
In the case with $M_{\sscBH, {\rm init}} = 10^6\Msun$, the feedback 
efficiency (energy output per unit accreted mass) is much higher,
and the gas depletion terminates the burst at $\approx 200\Kyr$.
Star formation here is less competitive than in the former 
case, and the growth of both $M_\sscBH$ and $M_*$ is about $10^7\Msun$.
The final $M_\sscBH$ is higher than in the
former case, but the final stellar mass is lower.
The competition between the BH growth and star formation thus
depends strongly on $M_{\rm nuc}/M_\sscBH$, consistent with the
expectation based on the depletion timescales (Eq.~\ref{eq:t-dep-acc2sf}).
For comparison, grey lines in Fig.~\ref{fig:bh-engine-example} show the
evolution of the BH mass expected from a constant Eddington ratio ($\dot{m}_\sscBH$) 
between $10^2$ and $10^4$. The average growth rates of the two cases 
during the entire bursts fall in the range of $\dot{m}_\sscBH = 10^3$ to $10^4$, suggesting 
that the bursty growth of the BH is super-Eddington over a wide range in 
the parameter space. Interestingly, observations indeed 
indicate that a small fraction of AGNs are experiencing super-Eddington accretion, 
either based on measurements of the BH mass and accretion rate 
\citep{liuObservationalDifferenceAccretion2021}, 
or based on the weak/variable X-ray emission expected from the obscuration 
by a thick disk and/or clumpy wind surrounding a BH with super-Eddington accretion 
\citep{wangNuSTARObservationsIntrinsically2022,zhangExtremelyXRayVariable2025}.

The diversity in the mass ratio, $M_{\rm nuc}/M_\sscBH$, in 
the efficiency of the AGN feedback and in the BH growth-star formation 
competition is thus expected to produce diversity in the growth paths 
among different galaxy/BH systems. This can be explored further to explain the
observed diversity of high-$z$ galaxies/AGNs 
\citep[e.g.][]{mezcuaOvermassiveBlackHoles2024,onouePoststarburstPathwayFormation2025,
liPrevalentPopulationNormalmass2025},
and our refined treatment for nuclear bursts provides a basis for such 
explorations within the current cosmological framework, as 
we will show in \S\ref{sec:results}.

\subsubsection{Comparison with other growth channels of black holes}
\label{app:sssec:cmp-other-channels}

To see that the nuclear burst is indeed an efficient channel of BH growth, 
we compare it below with other viable channels.

The continuous mode of BH growth, as derived in 
\citetalias{moTwophaseModelGalaxy2024} using the loss-cone argument
\citep{shapiroBlackHolesWhite1986}, predicts an accretion rate of 
$\dot{M}_{\sscBH} \propto (M_{\sscBH} / M_\sscSGC) \dot{M}_{\rm g,cool}$.
Thus, the Eddington ratio of BH accretion rate is
\begin{align} \label{eq:dot-m-bh-continuous}
    \dot{m}_{\sscBH} 
    &\equiv \frac{\dot{M}_{\sscBH}}{\dot{M}_{\ssc Edd}}
    = \alpha_{\rm cap} t_{\ssc Sal} \frac{F_{\rm cool} f_{\rm b}}{f_{\rm gas}} \frac{\dot{M}_{\rm v}}{M_{\rm v}}
    \nonumber \\
    &= 3.67\,\alpha_{\rm cap, 2.5} \frac{F_{\rm cool} f_{\rm b}}{f_{\rm gas}} M_{\rm v,9.5}^{0.14} (1+z)_{10}^{2.5}\,,
\end{align}
where we have adopted Eq.~\eqref{eq:dot-M-vir} for an approximation of 
the halo assembly rate, $\dot{M}_{\rm v}$; 
$F_{\rm cool}$ is the fraction of gas associated with the halo assembly that can 
cool and flow into the SGC; 
$\alpha_{\rm cap} \equiv 2.5 \alpha_{\rm cap, 2.5}$ is a free parameter, adopted 
to be $2.5$ by default.
Most halos at $z \approx 10$ have mass 
$M_{\rm v} < 10^{12}\Msun$ (see Fig.~\ref{fig:halo-mahs}), 
implying effective cooling ($t_{\rm cool} \lesssim t_{\rm v}$; 
see fig.~2 of \citetalias{moTwophaseModelGalaxy2024})
that leads to $F_{\rm cool} \approx 1$ and $f_{\rm gas} \approx f_{\rm b}$.
The dependence of $\dot{m}_\sscBH$ on halo mass is quite weak,
while the factor $(1+z)^{2.5}$ implies a rapid decline of 
$\dot{m}_{\sscBH}$ with cosmic time.
The expected value of $\dot{m}_{\sscBH}$ is $\lesssim 5$ at $z \approx 10$,
marginally touching the Eddington rate, and falls rapidly to 
$\lesssim 0.1$ at $z \lesssim 3$.
This implies a much less efficient accretion for BHs in the continuous mode than 
in the bursty mode. The relation between the BH growth and the halo assembly implied 
by Eq.~\eqref{eq:dot-m-bh-continuous} is $\dot{M}_\sscBH/M_\sscBH \propto \dot{M}_{\rm v}/M_{\rm v}$ 
and that between the BH growth and the star formation is
$\dot{M}_\sscBH/M_\sscBH \propto \dot{M}_*/M_*$ if star 
formation can catch up with the BH accretion in the early Universe.
Given the relation between the BH mass and the galaxy mass, 
$M_\sscBH \propto M_*^\gamma$ with the observed range of $\gamma \gtrsim 1$
in the local Universe \citep[e.g.]{kormendyCoevolutionNotSupermassive2013,
greeneIntermediateMassBlackHoles2020,grahamAppreciatingMergersUnderstanding2023}, 
the continuous mode alone is not sufficient 
to drive the BH seed onto the observed relation, unless the seed is produced above 
the relation. In contrast, the bursty mode may be able to drive the BH seed 
onto or above the relation, as we will discuss in 
Appendix~\ref{app:sec:model-variants}.

In the case where turbulence is negligible at the scale of the SGC,
a rotationally supported disk is expected to form as a result of 
angular-momentum mixing (the dashed curve in Fig.~\ref{fig:am-redistribute}a), 
while thermally supported gas processed by the feedback and left in the spheroid 
is too hot and too diffuse to form stars. This corresponds to a classical 
picture of gaseous components of galaxies in the local Universe. The rate at 
which the hot gas is accreted into the zone of influence of the BH 
can be described by the original form of the Bondi accretion 
(Eq.~\ref{eq:dot-M-Bondi-general}, with $\sigma_{\rm eff} \approx c_{\rm s}$):
\begin{equation}\label{eq:bondi-rate-hot}
\dot{M}_{\sscBondi} = \frac{4 \pi G^2 M_{\sscBH}^2 \rho_{\rm g}}{c_{\rm s}^3} \,.   
\end{equation}
With the approximation that the sound speed $c_{\rm s} \approx V_{\rm c}(R_\sscBondi)$,
the factor $4 \pi G^2 M_{\sscBH}^2 / c_{\rm s}^3$ is similar to 
that of the bursty mode, but the low density $\rho_{\rm g}$
makes the accretion of hot gas much less efficient than the 
bursty-mode accretion. 
On the other hand, the rate at which cold gas in the disk 
is accreted into the zone of influence of the BH 
can be formulated with an additional factor 
accounting for the viscosity, as:
\begin{equation} \label{eq:bondi-rate-cold}
    \dot{M}_{\sscBondi} = 
    \frac{
        4 \pi G^2 M_{\sscBH}^2 \rho_{\rm g}(R_{\sscBondi})
    }{
        V_{\rm c}^3(R_{\sscBondi}) C_{\rm visc}
    } \,,
\end{equation}
where
$C_{\rm visc} = 2\pi \left[\alpha_{\rm visc} (H/R)^2\right]^{-1} $;
$\alpha_{\rm visc}$ is the viscosity number;
$H/R \ll 1$ is the height-to-radius ratio that describes 
disk geometry. This formulation was derived by 
\citet{rosas-guevaraImpactAngularMomentum2015}  by 
comparing the original Bondi timescale and the viscosity timescale,
and was adopted as a sub-grid model in the EAGLE simulation 
\citep{schayeEAGLEProjectSimulating2015,crainEAGLESimulationsGalaxy2015}.
The density $\rho_{\rm g}$ of the cold gas can be large, 
but the viscosity suggested by them is quite small.
This gives a value of $C_{\rm visc} \sim 10^6$ that largely suppresses 
the accretion rate of the BH, supporting the assumption made
in \citetalias{moTwophaseModelGalaxy2024} that BH accretion is negligible 
in a thin galactic disk expected in the slow phase of the halo assembly. 

\section{Detailed implementation of the model}
\label{app:sec:impl-details}

\begin{figure*} \centering
    \includegraphics[width=\textwidth]{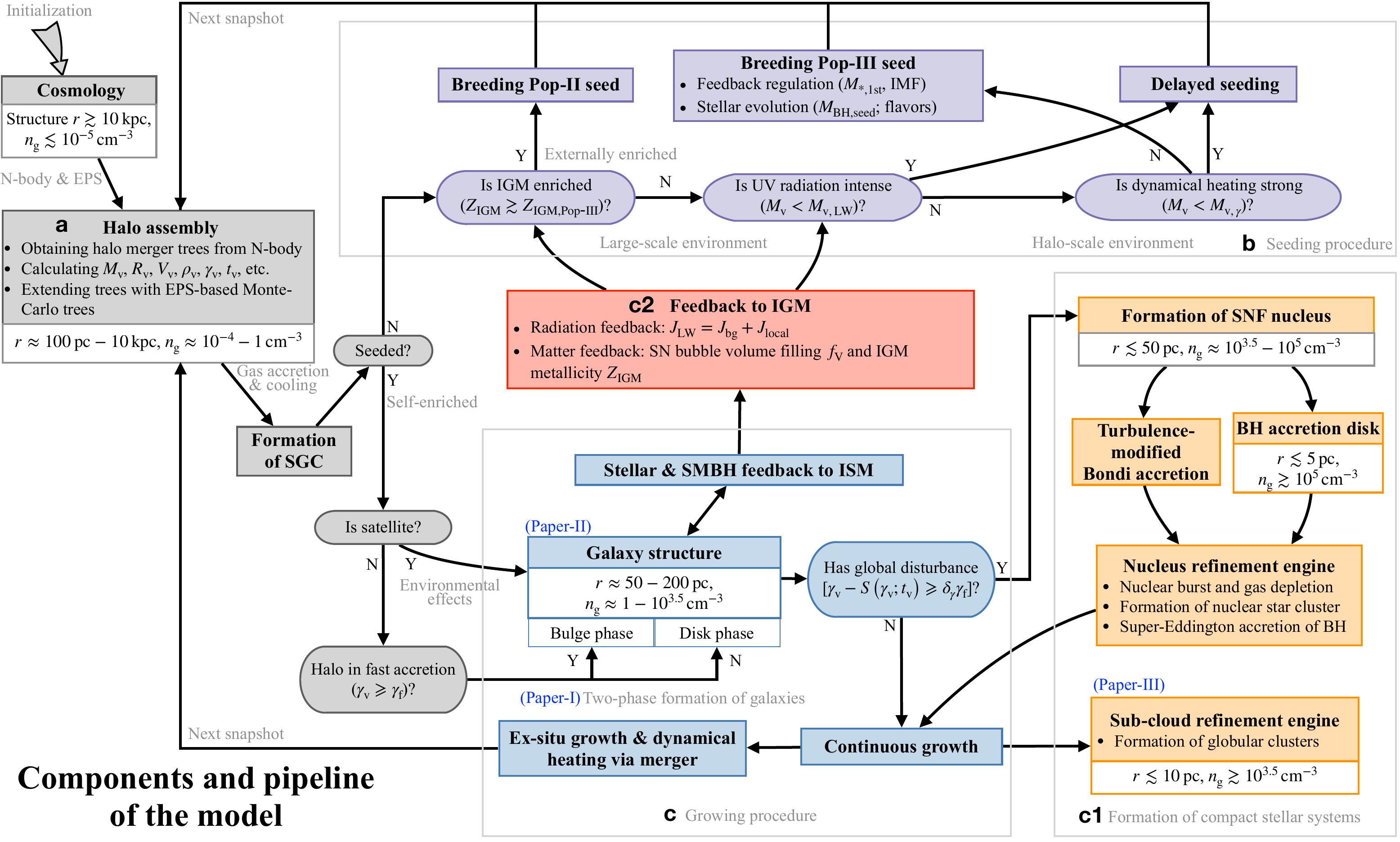}
    \caption{
        {\figem Components and pipeline of the model. }
        The model is built upon the $\Lambda$CDM paradigm, takes
        halos and their assembly histories as input
        ({\figem a}), and uses two main procedures, 
        the {\figem seeding procedure} ({\figem b}; 
        see Fig.~\ref{fig:cosm-context-seeding} for an outline of 
        the relevant physical processes),
        and the {\figem growing procedure} ({\figem c}; 
        see Fig.~\ref{fig:am-redistribute} for the relevant
        physical processes), 
        respectively, to seed and grow SMBHs with their host galaxies and halos.
        Inset {\figem c1}, sub-procedure for modeling
        compact stellar systems, including the refinement engines
        for nuclear bursts and for the formation of GCs.
        Inset {\figem c2}, sub-procedure for modeling radiation and matter feedback of galaxy formation 
        to the large-scale environment, which provides environmental 
        conditions for the seeding procedure and serves as
        an interface connecting the two main procedures.
        Gas densities ($n_{\rm g}$) of structures at different scales are listed, 
        taking the progenitors of MW-size galaxies at $z = 9$ 
        as an example (see Fig.~\ref{fig:profile-ladder} for the profile).
        See Fig.~\ref{fig:flowchart-brief} for a brief version of 
        this figure, and Appendix~\ref{app:sec:impl-details} for the 
        implementation details.
    }
    \label{fig:flowchart}
\end{figure*}

\begin{center}
\begin{table*} 
\renewcommand{\arraystretch}{1.1}
\caption{{\figem List of model components and default
parameters for each component}. See Fig.~\ref{fig:flowchart} 
for a flowchart outlining the components and pipeline,
and Appendix~\ref{app:sec:impl-details} for 
the implementation details and the choices of parameters in the 
{\textem Default} model.
}
\begin{tabular}{ >{\centering\arraybackslash}m{3.75cm} | m{9cm} | m{3.5cm} }
    \hline
    \rowcolor{Gainsboro!60}
        \makecell{Model Component} 		 
        &
        \makecell{Default Parameters}
        &
        \makecell{Outline and Implementation}
        \\
    \hline
        Extension of subhalo merger trees
        &
        $N_{\ssc MC} = 512$;
        $M_{\rm v, anc} = 5\times 10^{8}\msun$
        &
        Fig.~\ref{fig:flowchart}a and 
        Appendix~\ref{app:ssec:extension-merger-tree}
        \\
    \hline
        The global integrator
        &
        $\sigma_{\ssc M} = 0.2\dex$
        &
        Appendix~\ref{app:ssec:integrator}
        \\
    \hline 
            \multirow{3}{*}{
                \makecell{The seeding procedure \\ (Appendix~\ref{app:sec:seeding})}
            }
            &
            {\bf Pop-III threshold:} $Z_{\ssc IGM,Pop\text{-}III} = 10^{-4}\Zsun$;
            &
            \multirow{3}{*}{
                Fig.~\ref{fig:flowchart}b and Appendix~\ref{app:ssec:seeding}
            }
        \\
            &
            {\bf Delay of seeding:}
            $J_{\rm crit,21} = 7.5$;
            $\beta_{\ssc J} = 1.5$;
            $\gamma_{\rm crit} = 10$;
            $\beta_\gamma = 2$;
            &
        \\
            &
            {\bf  Masses of seeds:}
            $\eta_{\rm r} = 25$;
            $\beta_{\rm m} = 1$;
            $M_{\ssc BH,Pop\text{-}II} = 10\Msun$;
            $M_{*,{\ssc Pop\text{-}II}} = 100\Msun$;
            $M_{\rm g,seed} = 10\Msun$
            &
        \\
    \hline
            \multirow{9}{*}{
                \makecell{The growing procedure \\ (Appendix~\ref{app:sec:growing})}
            }
            &
            {\bf Halo assembly:}
            $\gamma_{\rm f}=3/16$;
            &
            \multirow{4}{*}{
                Fig.~\ref{fig:flowchart}c and Appendix~\ref{app:ssec:growing}
            }
        \\
            &
            {\bf Gas cooling:}
            $M_{\rm cool,f} = 10^{13} \msun$; 
            $\beta_{\rm cool,f} = 4$;
            $M_{\rm cool,s} = 10^{11.9} \msun$;
            $\beta_{\rm cool,s,0} = 3.6$;
            $\gamma_{\rm cool,s} = -0.72$;
            &
            
        \\
            &
            {\bf Feedback (continuous mode):}
            $\alpha_{\rm {\ssc SN},f} = 0$; $\beta_{\rm {\ssc SN},f} = 2.5$;
            $V_{\rm w}=250 \kms$;
            $R_{\rm s} = 10^{-0.96}$;
            $\beta_{\rm {\ssc SN}, s} = 1.9$;
            $\alpha_{\rm \sscAGN,f,0} = 5 \times 10^{-3}$;
            $\beta_{\rm \sscAGN,f}=2$;
            $\beta_{\rm \sscAGN,s}=1.8$;
            $f_{\rm ej,f} =0.75$;
            $f_{\rm ej,s} = 1$;
            &
            
        \\
            &
            {\bf Star formation and BH accretion (continuous mode):}
            $\epsilon_{\rm *,f} = 0.75$;
            $\epsilon_{\rm *,s} = 0.32$;
            $R = 0.4$;
            $\alpha_{\rm cap,f} = 2.5$;
            $\alpha_{\rm cap,s} = 0$;
            &
        \\
            &
            {\bf Satellite evolution:}
            $\tau_{\rm sat} = 4\Gyr/\ln 10$;
            &
        \\
            &
            {\bf SNF nucleus:}
            $\delta_\gamma = 1$;
            $n_{\rm snf} = 10^{3.5}\perccm$;
            $p = 0$; $q = 0.5$;
            &
        \\
            &
            {\bf BH accretion disk:}
            $\zeta_{\rm t} = \zeta_{\ssc A} = 1/3$;
            $\psi_{\rm t}/\psi_{\ssc A} = 1$;
            &
        \\
            &
            {\bf Nuclear burst:}
            $\chi_{{\ssc AGN},\infty} = 1$;
            $v_{\rm out} = 10^3\kms$;
            $\beta_{\rm out}$ = 0;
            $\alpha_{\ssc SF}=1$;
            $\alpha_{\ssc BH}=1$;
            $\delta_{\ssc Q}=1$;
            $f_{\rm dry} = 6.8 \times 10^{-3}$;
            &
        \\
            &
            {\bf Merger and dynamical heating:}
            $f_{\rm merge} = 0.5$; $f_{\rm heat} = 0.5$;
            &
        \\
            &
            {\bf Metallicity:}
            $V_{\rm esc} = 75 \kms$; 
            $\beta_{\rm esc}=2$;
            $z_{\rm esc} = 5$; 
            $\gamma_0 = 1$; 
            $\beta_{\rm mix}=2$
            &
        \\
        \hline
        The model of LW radiation field
        &
        $f_{\rm esc}=1$; $\eta_{\ssc LW}=4\times 10^3$;
        $Q_{\ssc LW} = 10^{36} \,{\rm erg} \cdot {\rm s}^{-1} {\rm M}_\odot^{-1}$;
        $J_{\rm local,th,21}=0.01$
        &
        Fig.~\ref{fig:flowchart}c2 and
        Appendix~\ref{app:ssec:LW-radiation}
        \\
    \hline
        The model of IGM enrichment
        &
        $A_{\rm sh}=1$;
        $\epsilon_{\ssc SN}=10^{51}\,{\rm erg} \cdot {\rm M}_{\odot}^{-1}$;
        $l_{\rm sm} = 250\kpc$
        &
        Fig.~\ref{fig:flowchart}c2 and
        Appendix~\ref{app:ssec:IGM-enrichment}
        \\
    \hline
\end{tabular}
\label{tab:parameters}
\end{table*}
\end{center}

In Fig.~\ref{fig:flowchart-brief}, we have briefly outlined the steps to 
implement the model. Here we describe the implementation in detail
following the logical order shown in a more detailed flowchart in
Fig.~\ref{fig:flowchart}. 
For the physical contexts for the seeding and growing procedures,
refer to Figs.~\ref{fig:cosm-context-seeding} and
\ref{fig:am-redistribute}, respectively.
The code for the implementation is publicly available 
(see \nameref{sec:data-availability}).
For reference, we also list the components of the model and
the default parameters adopted in this paper
in Table~\ref{tab:parameters}.

The model starts with a set of subhalo merger trees constructed from an N-body 
simulation (TNG100-1-Dark in this paper; see \S\ref{ssec:halo-assembly}).
A cosmological simulation covering such a volume often has limited resolution, 
and is thus insufficient to resolve the assembly histories of halos down to  
the masses required by the seeding procedure. 
To tackle this issue, we extend the simulated merger trees by joining them 
with those generated by a Monte Carlo algorithm based on the 
extended Press-Schechter (EPS) formalism.
The Monte Carlo trees need to be generated with sufficiently high resolution,
so that the assembly history of any simulated halo, after the extension, 
can be reliably traced back to a mass below the $\Hmol$-cooling limit 
at which the first generation of stars and a BH seed may form
(Eq.~\ref{eq:m-v-cool-H2-main}; see also the purple dotted lines in 
Fig.~\ref{fig:halo-mahs}a,c,d).

\subsection{Extension of subhalo merger trees}
\label{app:ssec:extension-merger-tree}

\begin{figure} \centering
    \includegraphics[width=\columnwidth]{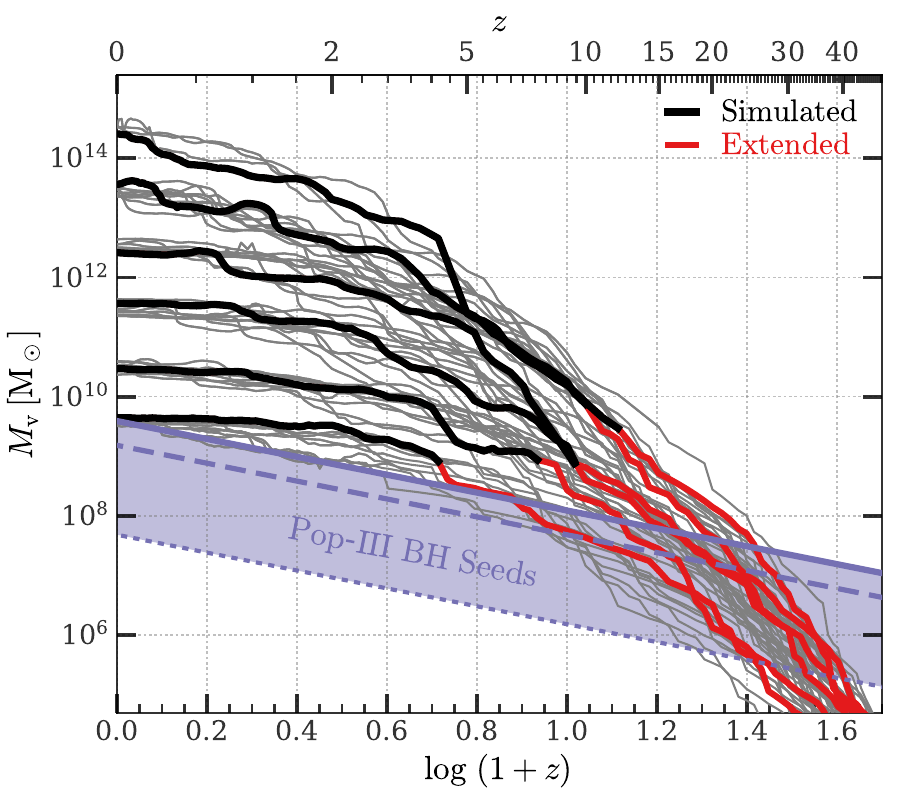}
    \caption{{\figem Extension of subhalo merger trees.}
    Here we show the main-branch mass assembly histories (MAHs)
    for examples of central subhalos at $z=0$ with masses separated by $1\dex$
    in the range of $10^{9.5}$--$10^{14.5}\Msun$.
    For each curve, {\figem black} part is obtained directly from the
    simulation, which is truncated at the anchor point due to limited 
    resolution; 
    {\figem red} part shows the extension by our EPS-based Monte Carlo algorithm 
    that smoothly joins the simulated part at the anchor point.
    Each bundle of {\figem grey} curves shows the MAHs of up to 10 
    additional halos with masses at $z=0$ similar to the example.
    The extension ensures that the assembly histories can 
    be traced back to sufficiently high $z$
    and sufficiently low $M_{\rm v}$ so that the formation of Pop-III stars 
    and BH seeds can be captured, as indicated by the 
    {\figem purple} band (the same as that in Fig.~\ref{fig:halo-mahs}a).
    See Appendix~\ref{app:ssec:extension-merger-tree} for details.
    }
    \label{fig:halo-mahs-ext}
\end{figure}

We follow the algorithm detailed in \citet[see their \S3.2, the `Central-Stage Completion' part]{chenConditionalAbundanceMatching2023}
to perform the extension. We list the steps below.
\begin{enumerate}[topsep=0pt,parsep=0pt,itemsep=0pt]
    \item 
    We decompose each simulated subhalo merger tree into a set of disjoint 
    branches by a recursive method. Starting from the forest $F$ that consists of all the simulated 
    subhalo merger trees, we arbitrarily select a tree $T$ from $F$. 
    We find the root subhalo, $h_{\rm root}$, of $T$, and 
    trace back the main branch, $B$, of $h_{\rm root}$
    (defined as the branch with `most massive' history in TNG; 
    see e.g. \citealt{rodriguez-gomezMergerRateGalaxies2015}).
    We remove $B$ from $T$, which leaves $T$ with a set of disjoint 
    subtrees, $\{T_i\}$, each rooted in a subhalo $h_i$
    whose descendant is $h_{\rm root}$. We put all $T_i$ back into $F$, and 
    repeat the above process until $F$ is empty.
    \item 
    For each branch $B$ obtained above, we start from the leaf subhalo (i.e. that
    with the highest redshift), walk along the branch towards lower redshift to find 
    the `anchor' redshift $z_{\rm anc}$ at which the halo mass $M_{\rm v}$ becomes 
    larger than a threshold $M_{\rm v, anc}$ for the first time. 
    The subhalo at $z_{\rm anc}$ so obtained is referred to as the 
    `anchor subhalo' of $B$, and is denoted as $h_{\rm anc}$.
    \item 
    We start from $h_{\rm anc}$, walk along the branch $B$ towards higher 
    redshift to find the `half-mass' redshift, $z_{\rm form}$, of $h_{\rm anc}$, 
    at which the halo mass drops below $M_{\rm v}(z_{\rm anc})/2$ for the 
    first time.
    \item 
    We follow \cite{parkinsonGeneratingDarkMatter2008} to generate a set of
    $N_{\ssc MC}$ Monte Carlo merger trees with a sufficiently high resolution, 
    each having a root halo $h'_{\rm root}$ with a mass and redshift equal to that of 
    $h_{\rm anc}$. We compute the half-mass formation time, 
    $z'_{\rm form}$, for the root halo of each Monte Carlo tree along the main
    branch. 
    \item 
    We search for the Monte Carlo tree whose root halo has 
    a $z'_{\rm form}$ that is the closest to $z_{\rm form}$ among all the Monte Carlo trees, 
    and we denote the main branch of this root halo as $B'$.
    Physical quantities directly related to $M_{\rm v}$ 
    (e.g. $R_{\rm v}$, $V_{\rm v}$ and $T_{\rm v}$; see Eq.~\ref{eq:R-vir}--\ref{eq:T-vir}) are assigned to the halos 
    of $B'$ according to their definitions. 
    The profile of any halo in $B'$ is assumed to be an NFW profile,
    with a concentration parameter ($c_{\rm v}$; see Eq.~\ref{eq:nfw-profile}) computed from the mass assembly
    history by the analytical fitting of 
    \citet[with slight adaptation to account for the difference
    in the definition of halo boundary]{zhaoAccurateUniversalModels2009},
    and the maximal circular velocity ($V_{\rm max}$) is computed from 
    the profile. 
    \item We prune all halos at $z > z_{\rm anc}$ from $B$, and join $B'$ 
    to $B$ at $z_{\rm anc}$.
\end{enumerate}

\bigskip
The above steps ensure that the Monte Carlo branch $B'$ not only 
has exactly the same $M_{\rm v}$ as $B$ at $z_{\rm anc}$, but also a similar
first-order derivative, since matching $z_{\rm form}$ approximately
provides such a constraint. 
We adopt the Monte Carlo algorithm of \citet{parkinsonGeneratingDarkMatter2008}, 
re-calibrated by N-body simulations to closely match simulation results
(\citealt{jiangGeneratingMergerTrees2014}), which is 
highly efficient due to its carefully chosen bound function for the 
rejection sampling.
The number of Monte Carlo trees, $N_{\ssc MC}$, should be sufficiently large 
so that the sampling of $z'_{\rm form}$ is dense enough to give a close match to
$z_{\rm form}$; we adopt $N_{\ssc MC} = 512$, which produces a smooth extension. 
The threshold mass $M_{\rm v, anc}$ 
that defines the anchor point should be at least twice 
the minimal mass that can be resolved by the simulation used, so that $z_{\rm form}$
of $h_{\rm anc}$ can be reliably determined. Based on the resolution of
TNG100-1-Dark, we set $M_{\rm v, anc} = 5 \times 10^8 \msun$, which is
the mass of about 100 dark-matter particles in the simulation.
Since the seeding procedure requires modeling the
radiation and matter feedback from any galaxy to its large-scale 
environment (see the flowchart Fig.~\ref{fig:flowchart}c2), 
the locations of halos in each $B'$  have to be assigned. 
Here we assign the location of $h_{\rm anc}$ in $B$ to all 
halos in the matched $B'$. This means that the peculiar motion of
a halo at $z > z_{\rm anc}$ is ignored and that the accuracy of
small-scale radiation and matter feedback is compromised.

With the extension, the assembly history of each simulated subhalo can be 
smoothly extended down to the $\Hmol$-cooling limit, and the 
extended merger trees will be used as the input of subsequent modeling
for galaxy formation.
The steps for obtaining the simulated merger trees and extending them are
outlined in the flowchart of Fig.~\ref{fig:flowchart}a.

In Fig.~\ref{fig:halo-mahs-ext}, we show examples of mass 
assembly histories (MAHs, i.e. $M_{\rm v}$ at different redshifts) of simulated 
subhalos before (black) and after 
(black plus red) the extension, over-plotted with the regime in which 
Pop-III stars are expected to form (purple). As expected, the MAHs now cross 
the Pop-III band, thus allowing the formation of the first generation of
stars and BH seeds to be directly modeled.
Note that the Monte Carlo trees themselves can be used as the input 
of the model, if LW radiation and IGM enrichment are available through other 
means. This is the case for the idealized and controlled experiments
presented in \S\ref{sssec:growth-effect-seeding}, where the environment
is set to the extreme so that $M_{\rm {\sscBH}, seed}$ is
minimized or maximized. A similar approach
was used by \citet[][see their \S2.4 and fig.~1]{spinosoMultiflavourSMBHSeeding2023}
to marginally extend the simulated merger trees to the $\Hmol$-cooling
limit at $z \gtrsim 15$, thereby allowing some light-weight seeds to be reared.
The semi-analytical model of \citet{venturaSemianalyticModellingPop2024} for 
Pop-III stars was built on merger trees from an N-body simulation that can resolve 
halos down to $\approx 1.5\times 10^5 \msun$, but with a compromised volume 
of $(10\mpc)^3$. Our mixed approach of using a large-volume simulation
and a Monte Carlo extension thus allows us not only to trace the
formation of Pop-III stars, but also to have a large sample that 
can reduce the effects of cosmic variance \citep[e.g.][]{somervilleCosmicVarianceGreat2004,mosterCOSMICVARIANCECOOKBOOK2011,
chenELUCIDVICosmic2019,chenMassiveDarkMatter2023,jespersenSignificanceRareObjects2025}.

\subsection{The global integrator}
\label{app:ssec:integrator}

The seeding procedure in our model is built on two factors in the large-scale environment of 
subhalos: the LW radiation and the IGM metallicity. Both factors 
are self-consistently modeled by 
tracing the radiation and matter feedback from all other 
subhalos that have already formed stars by the time step in question.
Entanglement among subhalo merger trees is thus inevitable and 
the integration of the formation histories of galaxies must be performed
globally, rather than individually on each subhalo merger tree.
We refer to this integration method as the `global integrator',
implemented as follows.
\begin{enumerate}[topsep=0pt,parsep=0pt,itemsep=0pt]
    \item 
    Starting from the snapshot with the highest $z$, we 
    iterate over all snapshots towards the lower $z$ until $z=0$.
    \item 
    For each snapshot $s$ (at redshift $z_s$),
    we take all the $N_s$ subhalos, $H_s \equiv \left\{ h_i \right\}_{i=1}^{N_s}$, 
    belonging to $s$ from the extended merger trees constructed 
    in Appendix~\ref{app:ssec:extension-merger-tree}. 
    For each subhalo $h \in H_s$, we follow
    the algorithms to be detailed in Appendices~\ref{app:ssec:LW-radiation} 
    and \ref{app:ssec:IGM-enrichment}, respectively, to 
    find the LW intensity ($J_{\sscLW}$) and the IGM metallicity
    ($Z_{\ssc IGM}$) at $\bm{x}_{h}$, the location of $h$. 
    This step is schematically shown in Fig.~\ref{fig:flowchart}c2 as a 
    feedback loop between the two main procedures (seeding and growing).
    \item 
    We check whether $h$ has already formed stars at $z > z_s$. 
    If so, we assume the galaxy in $h$ has been self-enriched, and branch to 
    the growing procedure to be detailed in Appendix~\ref{app:ssec:growing} (see 
    also Fig.~\ref{fig:flowchart}c, including the inset c1, for an outline) to
    model the post-seeding growth of the galaxy.
    If not, we follow the seeding procedure to be detailed in 
    Appendix~\ref{app:ssec:seeding} (see also Fig.~\ref{fig:flowchart}b
    for an outline) 
    to check whether the condition 
    for the collapse of gas in $h$ is satisfied, and breed a BH seed if the
    condition is satisfied.
    Otherwise, we assume that $h$ remains `dark', without any star or BH formed 
    at $z_s$.
\end{enumerate}

\bigskip
With this global integrator, each subhalo is assigned a seeding epoch at which
the first generation of stars and the BH seed are bred, and
the post-seeding growth is driven by different channels that gradually build up 
the BH, stellar and gas contents of the subhalo.
As demonstrated by \citet{chenMassiveDarkMatter2023}, both physical and 
observational uncertainties in the predicted quantities can introduce biases 
in statistical properties of galaxy populations, especially for rare objects.
Accurately accounting for such uncertainties is almost impossible, so
we adopt a tentative solution by injecting stochasticity into 
the predicted quantities.
As shown in e.g. \citet{chenHowEmpiricallyModel2021} and \citet{chenMAHGICModelAdapter2021}
using hydro simulations and regressors with successively refined predictors, many nuanced factors other than halo properties can produce
galaxy-to-galaxy variance in the growth, and such variance may thus be 
missed by a halo-based model. To account for such missed variance, we
introduce stochastic noise, with a log-normal distribution that has
a zero mean and a standard deviation of $\sigma_{\ssc M} = 0.2\dex$,
to the modeled amount of growth, $\Delta M_\sscBH$ and $\Delta M_*$, 
in every time step, at every growth stage
(fast or slow phases, or satellite stage; see below)
and through every growth channel (seeding, or bursty or continuous modes).
In comparison with observations, when appropriate, we additionally add 
stochastic noise to mimic observational uncertainties, which will be clearly 
stated when the comparison is made (see Fig.~\ref{fig:scaling-evol} and 
\S\ref{ssec:mbh-ms-relation}). 

\subsection{The seeding procedure}
\label{app:ssec:seeding}

The seeding procedure takes the internal and environmental properties of a subhalo 
$h \in H_s$ that has not yet formed any star as input, 
determines whether or not the first generation of 
stars and the BH seed can form in $h$ at $z_s$, and breeds the stars and the BH seed. 
The procedure follows prescriptions presented in Appendix~\ref{app:sec:seeding}, 
and is implemented through the following steps (see 
Fig.~\ref{fig:flowchart}b for an outline). 
\begin{enumerate}[topsep=0pt,parsep=0pt,itemsep=0pt]
    \item 
    We check the IGM metallicity, $Z_{\ssc IGM}$, at $\bm{x}_h$.
    If $Z_{\ssc IGM} \geqslant Z_{\ssc IGM,Pop\text{-}III}$, we populate $h$ with a Pop-II star 
    cluster, breed a BH seed from the cluster, and end 
    the seeding procedure for $h$. 
    The mass of the BH seed is set to be $M_{\ssc BH,Pop\text{-}II}$, and the 
    mass of the remaining stars in total is set to be $M_{*,{\ssc Pop\text{-}II}}$.
    \item 
    Otherwise ($Z_{\ssc IGM} < Z_{\ssc IGM,Pop\text{-}III}$), the subhalo is in the 
    Pop-III regime. We take the LW intensity ($J_{\sscLW}$) at $\bm{x}_h$
    and the specific accretion rate ($\gamma_{\rm v}$) of $h$,
    and use Eq.~\eqref{eq:m-v-collapse} to find the threshold of halo mass 
    ($M_{\rm v,th}$) for the gas in $h$ to collapse.
    If the mass of $h$ is below the threshold ($M_{\rm v} < M_{\rm v,th}$), 
    we assume that $h$ forms neither a star nor a BH at $z_s$, and we end the seeding
    procedure for $h$. Otherwise ($M_{\rm v} \geqslant M_{\rm v,th}$), we assume that a 
    Pop-III star cluster forms in $h$, and proceed to the next step
    to breed a BH seed from the cluster.
    \item 
    To determine the properties of the Pop-III star cluster, 
    we first evaluate the properties of the SGC in $h$ according to 
    Eqs.~\eqref{eq:r-sgc}--\eqref{eq:t-ff-sgc},
    assuming that a pristine halo has a baryon fraction equal to the cosmic
    value, $f_{\rm b}$ (see the text around Eq.~\ref{eq:cosmic_rho_b_z}).
    We then use the radiation-regulated scenario (Appendix~\ref{app:ssec:mass-func-pop3}) 
    to determine the mass of the dominant star in the cluster. Specifically, we use 
    Eq.~\eqref{eq:sfe-ZAMS} to compute 
    the zero-age main sequence star formation efficiency
    ($\epsilon_{\ssc ZAMS}$) and luminosity ($L_{\ssc ZAMS}$) 
    of the dominant star. If this luminosity does not exceed the Eddington 
    luminosity ($L_{\ssc ZAMS} < L_{\ssc Edd}$), the mass of the dominant star 
    is set to be $M_{\rm *,1st} = \epsilon_{\ssc ZAMS} M_{\sscSGC}$.
    Otherwise ($L_{\ssc ZAMS} \geqslant L_{\ssc Edd}$), the star reaches the 
    Eddington-level accretion, 
    and we turn to Eq.~\eqref{eq:sfe-Edd} to find the corresponding 
    star formation efficiency ($\epsilon_{\ssc Edd}$) and set 
    $M_{\rm *,1st} = \epsilon_{\ssc Edd} M_{\sscSGC}$.
    \item
    With the mass of the dominant star, the IMF of the entire Pop-III cluster 
    is obtained by Eq.~\eqref{eq:imf-pop3}, which is a power-law
    function with an index of $-\beta_{\rm m}$ and an exponential cutoff
    at $M_{\rm *,1st}$. The normalization of the IMF is obtained 
    by solving Eq.~\eqref{eq:imf-norm} that equates the integral of the IMF
    over $\left[ M_{\rm *,1st},+\infty \right]$ to $1$. This implies that 
    the instantiation of stars from the IMF is a process of 
    `optimal sampling', which is supported by observations of star 
    clusters \citep{yanMostMassiveStars2023}, and is also 
    consistent with theoretical expectations under our assumption
    of self-regulation \citep{kroupaStellarSubStellarInitial2013,schulzMassDistributionsStar2015}.
    \item 
    The evolution of the dominant star is treated by a simplified
    piecewise model (see Appendix~\ref{app:ssec:mass-func-pop3}), 
    with two mass thresholds, 
    $M_{\ssc *,PISN}$ and 
    $M_{\ssc *,DCBH}$. If $M_{\rm *,1st} \geqslant M_{\ssc *,DCBH}$, the 
    star ends its life as a DCBH, which we use as the BH seed.
    The mass of the seed is set to be $M_{\rm {\sscBH},seed} = M_{\rm *,1st}$, 
    assuming no significant mass loss during the stellar evolution 
    and the direct collapse. 
    If $M_{\ssc *,PISN} \leqslant M_{\rm *,1st} < M_{\ssc *,DCBH}$, the star
    ends its life by PISN which totally disrupts the star and leaves no
    remnant. In this case, we seek the most massive star (with a mass $M_{\rm *,surv}$) 
    that can survive PISN by requiring the integral of the IMF 
    over $\left[M_{\rm *, surv}, M_{\ssc *,PISN}\right]$ to be equal to $1$ 
    (Eq.~\ref{eq:m-star-surv}), again implying optimal sampling.
    The remnant of this star is treated as the BH seed, and the mass of 
    the seed is set to be $M_{\rm {\sscBH},seed} = f_{\rm rem} M_{\rm *,surv}$,
    accounting for a fraction of $1-f_{\rm rem}$ of mass loss during the
    stellar evolution.
    If $M_{\rm *,1st} < M_{\ssc *,PISN}$, the dominant star ends its life by CCSN,
    and leaves a BH remnant with a mass of $M_{\rm {\sscBH},seed} = 
    f_{\rm rem} M_{\rm *, 1st}$, which is adopted as the BH seed.
    Once the BH seed is determined, the evolution and remnant mass of any other 
    star in the cluster is obtained similarly by the piecewise model, 
    weighted by the IMF, and integrated over 
    $\left[M_{\rm *,min}, M_{\rm *,1st}\right]$
    (for DCBH or CCSN)
    or $\left[M_{\rm *,min}, M_{\rm *,surv}\right]$ (for PISN) to obtain $M_{\rm *,seed}$, 
    the stellar mass bred together with the BH seed.
\end{enumerate}

\bigskip
The default values of parameters are chosen as follows.
The critical metallicity is set to be $Z_{\ssc IGM,Pop\text{-}III} = 10^{-4}\Zsun$ (see Appendix~\ref{app:ssec:mass-func-pop3}).
The parameters describing the dependence of the collapse mass on LW radiation
are set to be $J_{\rm crit,21} = 7.5$ and $\beta_{\rm J}=1.5$ 
(see Appendix~\ref{app:sssec:delay-by-uv-radiation}),
and on dynamical heating are set to be $\gamma_{\rm crit} = 10$ and
$\beta_\gamma=2$ (see Appendix~\ref{app:sssec:delay-by-fast-accretion}).
The coupling efficiency in the radiation-regulated scenario is set to be
$\eta_{\rm r} = 25$ (see Appendix~\ref{app:ssec:mass-func-pop3}).
The index of the IMF of Pop-III stars is set to be $\beta_{\rm m} = 1$ 
(see Appendix~\ref{app:ssec:mass-func-pop3}).
The mass of a Pop-II BH seed is set to be 
$M_{\ssc BH,Pop\text{-}II} = 10 \Msun$, a typical value for the
remnant of a massive star in normal star clusters.
Unfortunately, the stellar mass associated with a Pop-II BH seed is unknown,
and so an assumption has to be made. We notice that the bursty mode
dominates the growth of galaxies after the seeding,
erases the difference produced by the seeding method, and
sustains a ratio of $M_\sscBH/M_* \approx 1/10$ (see 
\S\ref{sssec:growth-channels}, Figs.~\ref{fig:growth-channels} and 
\ref{fig:scaling-evol}). Thus, we manually set 
$M_{*,{\ssc Pop\text{-}II}} = 10 M_{\ssc BH,Pop\text{-}II}$, so that 
a Pop-II seed appears in an $M_\sscBH$-$M_*$ relation
similar to that shaped by the post-seeding growth without 
introducing an artificial discontinuity.
In both Pop-II and Pop-III cases, the gas density at fragmentation 
is expected to be high due to the lack of effective metal cooling,
and thus the seeding process may be bursty. This implies 
a strong gas depletion during seeding, and a short 
quiescent phase after seeding, as discussed in the beginning of 
\S\ref{ssec:growing-main}. For this reason, we set the remaining gas mass 
after seeding to be negligibly small,
$M_{\rm g,seed} = 10\Msun$.

With the seeding procedure, the halo $h$ either remains `dark', or 
is populated with initial values of $M_{\rm {\sscBH}, seed}$, 
$M_{\rm *, seed}$ and $M_{\rm g, seed}$
so that the post-seeding processes modeled by the 
growing procedure are provided with proper initial conditions.

\subsection{The growing procedure}
\label{app:ssec:growing}

The growing procedure takes a subhalo $h \in H_s$ that has already been seeded 
as the input, and updates the properties of the galaxy in $h$ by modeling 
the evolution within the current time step (i.e. the duration since the previous 
snapshot, $s-1$, to the current snapshot $s$). The growing procedure 
is based on the last version of our model presented in \S2 of
\citetalias{chenTwophaseModelGalaxy2025}, with a major extension to include 
the growth in the bursty mode proposed in this paper and a few updates 
on the other components of the model to accommodate the extension.
Unless otherwise specified, the default parameters adopted in this paper 
are taken from \citetalias{chenTwophaseModelGalaxy2025}.

The assembly of dark matter halos
(quantified by $\gamma_{\rm v}$; see Eq.~\ref{eq:gamma-v-crit}) 
is expected to be the key factor that determines the gas 
dynamics of the central galaxy hosted by the halo, 
and thus also determines how the central BH hosted by the galaxy 
accretes the gas and reacts via AGN feedback 
(see the discussion in Appendix~\ref{app:ssec:hierarchy} and summary in 
Fig.~\ref{fig:am-redistribute}).
Specifically, the BH in a halo during the fast 
phase grows efficiently by
capturing the sub-clouds formed out of the dynamically hot (turbulent) 
SGC, while the BH in a halo during the slow phase grows inefficiently due to the 
barrier of angular momentum ($j$) associated with the formation of 
dynamically cold disk. In addition, a global disturbance, quantified
in this paper by a temporary excursion of the halo assembly to high $\gamma_{\rm v}$, 
induces violent gas inflow that causes the formation of an SNF nucleus,
followed by a nuclear burst that increases the mass of the BH.

The above physical considerations motivate us to treat galaxy formation as a multichannel process 
(see \S\ref{sssec:growth-channels}, Figs.~\ref{fig:growth-examples} 
and \ref{fig:growth-channels}). 
To aid the implementation of the model, we separate each branch $B$ 
of subhalo merger trees constructed in 
Appendix~\ref{app:ssec:extension-merger-tree} into several stages/phases, 
so that numerical prescriptions for modeling galaxies 
hosted by subhalos during each stage can be applied separately.
For this, we first identify the `infall snapshot' ($s_{\rm infall}$) of $B$, defined 
as the last snapshot in which the subhalo in $B$ is still a central subhalo.
Subhalos at and before $s_{\rm infall}$ are defined to be in the `central stage',
and those after $s_{\rm infall}$ are defined 
to be in the `satellite stage'. 
For numerical convenience, we set the value of $M_{\rm v}$ for any subhalo in the 
satellite stage to be that at $s_{\rm infall}$, effectively 
assuming that there is no growth or excursion of halo assembly in the satellite stage.
For the entire branch, we smooth the MAH (i.e. values of $M_{\rm v}$)
by a running Gaussian kernel with a standard deviation
equal to the dynamical timescale of the halo ($t_{\rm v}$; see
Eq.~\ref{eq:t-vir}), and denote the smoothed MAH as 
$\mathcal{S}(M_{\rm v}; t_{\rm v})$.
We then perform a numerical differentiation to obtain the smoothed 
version of the specific accretion rate from $\mathcal{S}(M_{\rm v}; t_{\rm v})$, 
and denote it as $\mathcal{S}(\gamma_{\rm v}; t_{\rm v})$.
If $\mathcal{S}(\gamma_{\rm v}; t_{\rm v}) \geqslant \gamma_{\rm f}$, 
where the default value of the threshold is $\gamma_{\rm f} = 3/16$,
the subhalo is defined to be in the fast phase; otherwise, 
it is defined to be in the slow phase. Note that this separation
of fast and slow phases is slightly different from the one with 
a single transition redshift $z_{\rm f}$ used in
previous papers of this series: the excursion of the halo assembly
to the slow (fast) phase on a timescale of $t_{\rm v}$ can now take 
place at higher (lower) $z$, thus allowing the model to capture
some recurrent transition of galaxy dynamics that has been evident in
observations (\citealt{parlantiALMAHintsPresence2023,
rowlandREBELS25DiscoveryDynamically2024,scholtzTentativeRotationGalaxy2025,
danhaiveDawnDiscsUnveiling2025}; see also the discussion in 
Appendix C of \citetalias{moTwophaseModelGalaxy2024}). 
For convenience, we use $B_{\rm fast}$, $B_{\rm slow}$ and 
$B_{\rm sat}$ to denote the sets of subhalos
in $B$ during the fast phase, slow phase and satellite stage, respectively.

During the current time step, we determine whether the subhalo $h$ has a temporary 
excursion to high $\gamma_{\rm v}$  
by examining the condition $\gamma_{\rm v} - \mathcal{S}({\gamma_{\rm v}; t_{\rm v}})
\geqslant \delta_\gamma \gamma_{\rm f} $, with the default value of $\delta_\gamma = 1$
(see Eq.~\ref{eq:snf-formation-crit}). If the condition is not satisfied, we assume 
that the growth in the bursty mode is not triggered during this time step.
Otherwise, we assume that a global disturbance
has been imparted to the SGC, leading to the formation of an SNF nucleus
in the SGC center, followed by a nuclear burst that depletes the gas 
from the nucleus. Note that such a burst takes place in a much shorter 
timescale ($\lesssim {\rm Myr}$; see Fig.~\ref{fig:bh-engine-example}) 
and within a much smaller volume (see Fig.~\ref{fig:profile-ladder})
than those modeled using the global properties for the host galaxy. 
To properly treat this small-scale process, a refinement engine
is designed and applied on top of the large-scale modeling 
(see Appendix~\ref{app:ssec:nuclear-burst}).
Another example of small-scale processes is the formation of individual sub-clouds 
and star clusters, which is modeled using a similar refined treatment,
as detailed in \citetalias{chenTwophaseModelGalaxy2025}.
Since these small-scale processes are critical in determining the 
growth of compact objects, such as GCs, NSCs and SMBHs, 
which are of particular interest in recent observations at high $z$,
we represent them as a separate inset c1 in Fig.~\ref{fig:flowchart}.

Other unresolved processes, such as stellar evolution and baryon cycling
in individual galaxies, are treated by an instantaneous approximation given 
that their timescales are short compared with the assembly of halos. Hence,
the stellar mass formed in any galaxy during any time step, $\Delta M_*$, 
should be understood as the remaining mass of stars after stellar 
evolution. On the other hand, if the instantaneous (unevolved) star formation rate 
(SFR) is required, we obtain it by
\begin{equation}
    {\rm SFR} = \frac{1}{1-R} \frac{\Delta M_*}{\Delta t}
    \,,\label{eq:mass-return}
\end{equation}
where $R = 0.4$ is the returned fraction of mass assuming a Chabrier IMF 
\citep{bruzualStellarPopulationSynthesis2003} for stellar populations 
other than Pop-III. Note that in Fig.~\ref{fig:growth-cosmic-history}, we have used 
this unevolved SFR to produce the cosmic star formation history.    

With the above definitions, the steps of the growing procedure for a subhalo 
$h \in H_s$ are as follows (see also Fig.~\ref{fig:flowchart}c for an outline).
\begin{enumerate}[topsep=0pt,parsep=0pt,itemsep=0pt]
    \item We find the branch $B$ that contains $h$, and determine which 
    stage/phase $h$ is in ($h \in B_{\rm fast}$, $h \in B_{\rm slow}$ or 
    $h \in B_{\rm sat}$). If $h \in B_{\rm sat}$, we 
    jump to step (ix); otherwise, we continue with the next step
    to model the growth of the BH with its host galaxy and subhalo 
    in the continuous mode.

    \item 
    We compute the total amount of gas in $h$ that is available 
    for subsequent processes:
    \begin{equation}
        \Delta M_{\rm g, avail} = 
        \begin{dcases}
            f_{\rm b} \Delta M_{\rm v}      
            \,, \ \ \ \ \ &\text{if } h\in B_{\rm fast} \,; \\
            \frac{f_{\rm b} M_{\rm v}}{\tau(z)} \Delta t
            \,, &\text{if } h \in B_{\rm slow} \,,
        \end{dcases}
        \label{eq:dM-g-avail}
    \end{equation}
    where $\tau(z) = (1+z)^{-3/2} / (10 H_0)$ is an approximation to the dynamical 
    timescale of the halo (Eq.~\ref{eq:t-vir}), $\Delta t$ is the duration of the 
    current time step,
    and $\Delta M_{\rm v}$ is the change of halo mass in this duration.
    Note that the equation in the second line is expressed by $M_{\rm v}$, 
    instead of $\Delta M_{\rm v}$, which takes into account that
    halo growth ceases in the slow phase, and the gas processed earlier 
    and retained in the halo is recycled back to fuel star formation and
    BH growth (see e.g. \S\ref{sssec:cosmic-bh-sf-history} 
    and Fig.~\ref{fig:growth-cosmic-history} for a discussion).
    
    \item 
    The fraction of available gas in $h$ that can cool down is determined by a 
    cooling factor, $F_{\rm cool}$, which is
    \begin{equation} 
        F_{\rm cool} = \begin{dcases}
            \left[
                1+  \left(\frac{M_{\rm v}}{M_{\rm cool,f}}\right) ^{\beta_{\rm cool, f}}
            \right]^{-1}
            \,, \ \ \ \ \ &\text{if } h\in B_{\rm fast} \,; \\
            \left( 1 + \frac{M_{\rm v}}{M_{\rm cool,s}} \right)^{-\beta_{\rm cool, s}(z)} 
            \,, &\text{if } h \in B_{\rm slow} \,.
        \end{dcases} 
        \label{eq:F-cool}
    \end{equation}
    The factor takes into account the fact that cooling is effective
    only in halos with mass smaller than a threshold of $\sim 10^{12}\Msun$, 
    and that only a fraction of gas in halos with masses larger than the 
    threshold can cool down and flow into the galaxy
    (see e.g. the cooling curves in fig.~2 of \citetalias{moTwophaseModelGalaxy2024}).
    The default parameters are calibrated to be 
    $M_{\rm cool,f} = 10^{13} \msun$, $\beta_{\rm cool,f} = 4$,
    $M_{\rm cool,s} = 10^{11.9} \msun$, 
    $\beta_{\rm cool,s} = \beta_{\rm cool,s,0} (1+z)^\gamma_{\rm cool,s}$,
    $\beta_{\rm cool,s,0} = 3.6$
    and $\gamma_{\rm cool,s} = -0.72$.
    Note that the redshift dependence of both
    $M_{\rm v}$ and $\beta_{\rm cool,s}$ makes $F_{\rm cool}$ 
    a function of redshift.
    With the cooling factor, the amount of gas that can cool down and flow into 
    the galaxy is 
    \begin{equation}
        \Delta M_{\rm g,cool} = F_{\rm cool} \Delta M_{\rm g, avail}
        \,,\label{eq:dM-g-cool}
    \end{equation}
    and the remaining fraction, $1 - F_{\rm cool}$, is referred to as 
    the `hot' gas. Note that the reduction of the specific growth rate of gas mass 
    in the massive example at low $z$
    in Fig.~\ref{fig:growth-channels}c8 is partly due to the
    reduction of $F_{\rm cool}$.

    \item Owing to feedback effects from star formation and AGN, 
    only a fraction of the cooled gas can cool further and become star-forming gas. 
    We model the amount of star-forming gas by modulating the amount of cooled gas
    $\Delta M_{\rm g,cool}$ with two additional factors describing the 
    two sources of feedback, as:
    \begin{equation}
        \Delta M_{\rm g, {\ssc SF}} = F_{\ssc SN} F_{\ssc AGN} \Delta M_{\rm g,cool}
        \,. \label{eq:dM-g-SF}
    \end{equation}
    Here, $F_{\ssc SN}$ describes the effect of supernova feedback and is modeled as
    \begin{equation} 
        F_{\ssc SN} = 
        \begin{dcases}
            \frac{
                \alpha_{\rm {\ssc SN},f}  + (V_{\rm g}/V_{\rm w})^{\beta_{\rm {\ssc SN},f}}
            }{
                1 +(V_{\rm g}/V_{\rm w})^{\beta_{\rm {\ssc SN},f}}
            }
            \,, \ \ \ \ \ &\text{if } h\in B_{\rm fast} \,, \\
            \left(
                \frac{ M_{\rm v} / M_{\rm cool,s} }{R_{\rm s} + M_{\rm v} / M_{\rm cool,s}}
            \right)^{\beta_{\rm {\ssc SN}, s}}
            \,, &\text{if } h \in B_{\rm slow} \,,
        \end{dcases}
        \label{eq:f-SN}
    \end{equation}
    where the default parameters are set to be
    $\alpha_{\rm {\ssc SN},f} = 0$, $\beta_{\rm {\ssc SN},f} = 2.5$, 
    $V_{\rm w}=250 \kms$,
    $R_{\rm s} = 10^{-0.96}$, $\beta_{\rm {\ssc SN}, s} = 1.9$, 
    and $V_{\rm g} = V_{\rm max}$ is the typical velocity dispersion of the 
    galaxy.
    $F_{\sscAGN}$ describes the effect of AGN feedback, and is given by
    \begin{equation} 
        F_{\sscAGN} = 
        \begin{dcases}
            \exp\left[
                - {\alpha_{\rm \sscAGN,f}(z) M_\sscBH c^2 \over M_{\rm b} V_{\rm g}^2}
            \right]
            \,, \ \ \ \ \ &\text{if } h\in B_{\rm fast} \,; \\
            \left(
                \frac{ R_{\rm s} + M_{\rm v} / M_{\rm cool,s} }{ 1 + M_{\rm v} / M_{\rm cool,s} }
            \right)^{\beta_{\rm \sscAGN,s}}
            \,, &\text{if } h \in B_{\rm slow} \,.
        \end{dcases}
        \label{eq:f-AGN}
    \end{equation}
    Here, $M_{\rm b}$ is the total baryon mass (gas, stars and BH) of the 
    bulge of the galaxy, and the default parameter $\beta_{\rm \sscAGN,s}=1.8$.
    The exponential function for the fast phase is similar to the one 
    adopted in \citetalias{moTwophaseModelGalaxy2024} at small $M_\sscBH$,
    but approaches zero more smoothly at large $M_\sscBH$.
    Different from previous papers of this series, 
    the efficiency of AGN feedback in the fast phase, as quantified by $\alpha_{\rm \sscAGN,f}(z)$, 
    is now modeled to be redshift dependent. This is partly
    motivated by the theoretical consideration that the
    fraction of feedback energy that can be dissipated depends on the cooling 
    timescale of the gas affected by the feedback, and thus also depends on
    $z$, owing to the cosmic evolution of the SGC density (Eq.~\ref{eq:n-sgc}).
    This is also supported by empirical evidence that star formation 
    efficiency in high-mass halos is redshift dependent 
    \citep[e.g.][]{luEmpiricalModelStar2014,behrooziUniverse10Predictions2020}.
    Hence, we parameterize the factor $\alpha_{\rm \sscAGN,f}(z)$ as
    \begin{equation} 
        \alpha_{\rm \sscAGN,f}(z) = \alpha_{\rm \sscAGN,f,0} (1+z)^{-\beta_{\rm \sscAGN,f}} 
        \,.
        \label{eq:alpha-ANG-f}
    \end{equation}
    The value of $\beta_{\rm \sscAGN,f}$ is not known, but the 
    bound of it may be obtainable from a timescale argument that 
    takes into account the competition of energy injection and dissipation.
    If we consider an extreme case that the accretion of BH smoothly tracks 
    the growth of the host SGC and thus the timescales of BH growth and injection 
    of feedback energy follow the free-fall timescale of the SGC 
    ($t_{\rm ff, {\sscSGC}}$; Eq.~\ref{eq:t-ff-sgc}), 
    the feedback efficiency may be approximated as 
    $\alpha_{\rm \sscAGN,f} \sim t_{\rm cool, \sscSGC}/t_{\rm ff, {\sscSGC}} \propto n_{\sscSGC}^{-1/2}$,
    where the cooling timescale $t_{\rm cool, \sscSGC} \propto n_\sscSGC^{-1}$
    holds for any two-body radiative process \citep[see e.g. Appendix~B1.4 of][]{moGalaxyFormationEvolution2010}.
    In the other extreme, if the accretion of BH is a highly localized process
    with a microscopic timescale irrespective of the 
    SGC density, the feedback efficiency may be approximated as
    $\alpha_{\rm \sscAGN,f} \propto t_{\rm cool, \sscSGC} \propto n_{\sscSGC}^{-1}$.
    The above timescale argument suggests a bound for the 
    feedback efficiency as $\alpha_{\rm \sscAGN,f} \propto n_{\sscSGC}^{[-1, -1/2]}$,
    and a more realistic value may be somewhere in between.
    In this paper, we adopt the default as $\alpha_{\rm \sscAGN,f} \propto n_{\sscSGC}^{-2/3}$
    (i.e. $\beta_{\rm \sscAGN,f}=2$; see Eq.~\ref{eq:n-sgc} for $n_\sscSGC$),
    which falls between the two extremes. The default value of
    the normalization factor is set to be
    $\alpha_{\rm \sscAGN,f,0} = 5 \times 10^{-3}$, 
    so that $\alpha_{\rm \sscAGN,f}$ at $z \approx 2$ (the redshift at which 
    massive halos with $M_{\rm v} \gtrsim 10^{13}\Msun$ end their fast phase; 
    see Fig.~\ref{fig:halo-mahs}b) is comparable to the value 
    calibrated in \citetalias{moTwophaseModelGalaxy2024}.
    
    \item
    Part of the feedback-affected gas is expected to be expelled from 
    the galaxy (referred to as `ejected gas'), 
    while the rest remains hot in the galaxy and is prevented from 
    forming stars (referred to as `prevented gas'). To separate the 
    two components, we introduce a parameter $f_{\rm ej}$, which is the
    fraction of the feedback-affected gas that is ejected from the galaxy.
    Thus, the amounts of ejected and prevented gas are
    \begin{equation}
        \Delta M_{\rm g,ej} = f_{\rm ej} (1 - F_{\ssc SN} F_{\sscAGN}) \Delta M_{\rm g,cool} 
        \,,\label{eq:dM-g-ej}
    \end{equation}
    and
    \begin{equation}
        \Delta M_{\rm g,prev} = (1-f_{\rm ej}) (1 - F_{\ssc SN} F_{\sscAGN}) \Delta M_{\rm g,cool} 
        \,,\label{eq:dM-g-prev}
    \end{equation}
    respectively. 
    Following \citetalias{chenTwophaseModelGalaxy2025}, 
    we set the default value to be $f_{\rm ej}= f_{\rm ej,f} =0.75$ 
    for the fast phase, so that a galaxy well regulated by feedback has a gas fraction 
    comparable to the typical spin parameter, $0.04$, of dark matter 
    halos, and thus can transit to a stable disk in the slow phase.
    For the slow phase, we set $f_{\rm ej} = f_{\rm ej,s} = 1$,
    meaning that all feedback-affected gas is ejected from the 
    disk, with some recycled back later on a dynamical timescale of the 
    host halo (see Eq.~\ref{eq:dM-g-avail}).
    \item
    The amount of stars formed in the galaxy is related to the amount of
    star-forming gas, $\Delta M_{\rm g, {\ssc SF}}$, by a star formation efficiency, 
    $\epsilon_*$, as
    \begin{equation} 
        \Delta M_* = \epsilon_* \Delta M_{\rm g, {\ssc SF}}
        \,,\label{eq:dM-star}
    \end{equation}
    where $\epsilon_* = \epsilon_{\rm *,f} = 0.75$ and 
    $ \epsilon_* = \epsilon_{\rm *,s} = 0.32$ for the fast and slow phases, 
    respectively, as calibrated in \citetalias{moTwophaseModelGalaxy2024}.
    The stellar mass so obtained is added to the bulge component 
    of the galaxy if $h \in B_{\rm fast}$, or to the disk component if 
    $h \in B_{\rm slow}$.

    \item The growth of the mass of the central BH is obtained by the fraction of low-$j$
    sub-clouds that can approach the center of the galaxy and be 
    captured by the BH, which is derived in \citetalias{moTwophaseModelGalaxy2024} as 
    \begin{equation}
        \Delta M_\sscBH = \alpha_{\rm cap} \frac{M_\sscBH}{M_{\rm b}}
        \Delta M_{\rm g, {\ssc SF}} 
        \,\label{eq:dM-BH}
    \end{equation}
    where $\alpha_{\rm cap} = \alpha_{\rm cap,f} = 2.5$ in the fast phase, 
    and $\alpha_{\rm cap} = \alpha_{\rm cap,s} = 0$ 
    in the slow phase due to the full mixing of gas. 
    Note that the `enhancement factor' is now removed from the equation,
    as it is replaced explicitly by the nuclear burst (see steps x and xi below).

    \item 
    The change of gas mass in the galaxy can be obtained by mass 
    conservation, taking into account the contributions by 
    cooling, consumption by star formation and BH accretion, 
    and ejection by feedback, as    
    \begin{equation}
        \Delta M_{\rm g} = \Delta M_{\rm g, {\ssc SF}} - \Delta M_{\rm *} 
        - \Delta M_\sscBH + \Delta M_{\rm g, prev}
        \,.\label{eq:dM-g}
    \end{equation}
    Similar to the treatment for $M_*$, the update of gas mass is performed 
    on the bulge and disk components, if $h \in B_{\rm fast}$ and $h \in B_{\rm slow}$, 
    respectively.
    Note that the mass and specific rate of gas in Fig.~\ref{fig:growth-channels} (rows 7 and 8)
    are obtained with this equation.

    \item 
    If the subhalo is in the satellite stage ($h \in B_{\rm sat}$),
    the star formation efficiency is expected to be low due to the
    environmental effects, such as ram-pressure stripping
    \citep{gunnInfallMatterClusters1972,duFormationCompactElliptical2019,
    kulierRamPressureStripping2023,carnallMassiveQuiescentGalaxy2023} 
    and tidal stripping \citep{binneyGalacticDynamicsSecond2008,
    readTidalStrippingSatellites2006,
    drakosUniversalModelEvolution2022}. Motivated by the observational results of e.g.
    \citet{pengStrangulationPrimaryMechanism2015}, 
    we model the SFR in this stage by an exponential decay, as
    \begin{equation}
        \dot{M}(t) = \dot{M}(t_{\rm infall}) \exp\left(
            - \frac{t - t_{\rm infall}}{\tau_{\rm sat}}
        \right)
        \,,\label{eq:SFR-sat}
    \end{equation}
    where $t_{\rm infall} \equiv t(s_{\rm infall})$ is the cosmic time 
    at the infall snapshot, and $\tau_{\rm sat} = 4\Gyr/\ln 10$ is the 
    timescale of the decay. The growth of the BH is assumed to be 
    stopped in this stage.
    The amount of gas is updated according to the `close-box' model with 
    instantaneous recycling: the supply of gas is stopped by the 
    environmental effects, and the feedback-affected gas is instantaneously 
    recycled into the ISM. 
    With these assumptions, the only sink term for the gas
    is consumption by star formation.

    \item
    If a nuclear burst is triggered within the current time step,
    we take $M_{\rm g}$ of bulge (obtained from Eq.~\ref{eq:dM-g}) 
    as the mass of SGC ($M_\sscSGC$), 
    derive the radius fraction, 
    $f_{\rm r} \equiv R_\sscSGC/R_{\rm v}$, by the self-gravitating condition,  
    as $f_{\rm r} = M_\sscSGC/M_{\rm v}$ 
    \citepalias[see Eq. 26 of][]{moTwophaseModelGalaxy2024},
    and use Eqs.~\eqref{eq:r-sgc}--\eqref{eq:t-ff-sgc} to compute 
    the properties of the SGC. 
    Using the properties of the SGC, we move a step inward to the 
    galactic nucleus according to the double power-law density profile
    of the entire SGC (Eq.~\ref{eq:rho-g-in-nuc-q05}; see also 
    Fig.~\ref{fig:profile-ladder} for examples),
    and compute the properties of the SNF nucleus 
    using Eqs.~\eqref{eq:c-g-def}--\eqref{eq:V-c-r-nuc}.
    The default parameters to define the SNF nucleus include the 
    supernova-free density, $n_{\rm snf} = 10^{3.5}\perccm$, and
    the structural parameters, $p = 0$ and $q = 0.5$.
    Moving a step further into the zone of influence of the BH ($r \leqslant R_{\sscBondi}$),
    we follow Eqs.~\eqref{eq:v-scaled-q-05}--\eqref{eq:toomre-scaled-q-05}
    to derive the structural properties of the accretion disk.
    The default parameters we adopt for the accretion disk 
    correspond to those describing a turbulent disk with 
    flux-frozen magnetic fields: 
    $\zeta_{\rm t} = \zeta_{\ssc A} = 1/3$ and
    $\psi_{\rm t}/\psi_{\ssc A} = 1$.
    
    \item
    Using the properties of the SNF nucleus, we numerically solve the 
    continuity equation, 
    Eq.~\eqref{eq:pde-rho-g}, that describes the evolution of the gas profile 
    in the SNF nucleus during the nuclear burst, with 
    specific choices of functional forms detailed in 
    Appendix~\ref{app:sssec:nuclear-burst-details}.
    The default parameters controlling the burst
    include those describing the AGN feedback to the nucleus,
    $\chi_{{\ssc AGN},\infty} = 1$,
    $v_{\rm out} = 10^3\kms$
    and $\beta_{\rm out}$ = 0;
    star formation efficiency in the nucleus,
    $\alpha_{\ssc SF}=1$
    and $\delta_{\ssc Q}=1$;
    and BH accretion efficiency,
    $\alpha_{\ssc BH}=1$.
    The numerical solver, referred to as a `refinement engine' in this paper, 
    works with a time step of the order of a year, 
    much finer than the global integrator so that 
    the $\lesssim {\rm Myr}$ timescale of the nuclear burst can be resolved.
    Examples for the evolution of BH mass, 
    stellar mass, and gas mass during individual nuclear bursts are shown in
    Fig.~\ref{fig:bh-engine-example}. Here we take the changes in 
    these masses, $\Delta M_{\rm {\sscBH},burst}$,
    $\Delta M_{\rm *,burst}$ and $\Delta M_{\rm g, burst}$
    during the entire burst as outputs of the refinement engine,
    and add them to the BH mass, stellar mass and gas mass
    of the bulge of the galaxy, respectively.
    BH growth through gas accretion in this way is referred to as 
    the `wet' channel in the bursty mode.
    
    \item 
    The energy of AGN feedback released in the nuclear burst
    could have long-lasting effects on the gas on larger scales,
    which is also modeled here by a formulation similar to 
    the AGN feedback adopted for the continuous growth of the BH.
    Specifically, we remove a fraction
    \begin{equation}
        F_{\rm ej, burst} = 
        f_{\rm ej} \left\{1 - \exp \left[
            - \frac{\alpha_{\rm \sscAGN,f}(z) \Delta M_{\rm {\sscBH}, burst} c^2}{M_{\rm b } V_{\rm g}^2}
        \right] \right\} 
        \label{eq:f-ej-burst}
    \end{equation}
    of gas from that remaining in the bulge, with 
    the parameters $\alpha_{\rm \sscAGN,f}(z)$ and $f_{\rm ej}$
    exactly the same as those used in Eqs.~\eqref{eq:f-AGN}
    and \eqref{eq:dM-g-ej}, respectively.
    This implies that the coupling of AGN feedback with the SGC 
    due to a nuclear burst is assumed to be similar to that due to the 
    continuous growth of the BH, which is certainly a simplified 
    treatment and aims to avoid an over-complication of the 
    model.

    \item
    After a nuclear burst, stars formed in the galactic center
    may continuously adjust their orbits via dynamical relaxation,
    causing some stars to migrate to the center and be captured by 
    the BH. The fraction of stellar mass that can sink and 
    be captured depends on the detailed mass and phase-space 
    distribution of the stars, which has not yet been known.
    To account for this process in a simple way, we move a fraction 
    $f_{\rm dry}$ of stellar mass formed 
    in each nuclear burst ($\Delta M_{\rm *, burst}$) to the BH.
    The default value is set to be $f_{\rm dry} = 6.8 \times 10^{-3}$, 
    derived from the analytical approximation by 
    \citet[see their \S3.1.1]{dekelGrowthMassiveBlack2025}.
    We refer to this process of BH growth as a `dry' channel in the 
    bursty mode, as opposed to the `wet' channel modeled in 
    the above step (xi).
    There should be an order-unity uncertainty in $f_{\rm dry}$.
    However, as we have shown in Fig.~\ref{fig:growth-channels},
    the dry channel is unimportant compared to the wet channel,
    and the exact value of $f_{\rm dry}$ thus does not significantly
    affect our results.

    \item 
    If the subhalo $h$ merges into another subhalo $h_{\rm mp}$
    to form a descendant subhalo (i.e. $h_{\rm mp}$ is a main-branch progenitor 
    of the descendant, while $h$ is not), we add a fraction, $f_{\rm merge}$,
    of $M_\sscBH$ to that of $h_{\rm mp}$. Similarly, we add a fraction, 
    $f_{\rm merge}$, of
    $M_{\rm *}$ and $M_{\rm g}$ of $h$ to those in the bulge of $h_{\rm mp}$.
    The remaining stars and gas are assumed to be tidally stripped or 
    scattered into the halo to become intracluster baryons.
    Since the merger may also dynamically change the orbits of baryons in the disk 
    of $h_{\rm mp}$ and transform its morphology,
    we move some stellar and gas masses in the disk of $h_{\rm mp}$ 
    to those in the bulge. The moved masses
    are assumed to be a fraction, $f_{\rm heat}$, of $M_*$ and $M_{\rm g}$ 
    of the infalling galaxy hosted by $h$, since a more massive 
    infalling galaxy is expected to have a larger dynamical effect 
    on the disk.
    The exact values of $f_{\rm merge}$ and $f_{\rm heat}$ are very 
    uncertain. The reported values in the literature 
    differ significantly \citep[see e.g.][]{luEmpiricalModelStar2014,
    behrooziUNIVERSEMACHINECorrelationGalaxy2019}, 
    and  depend on how the boundaries of galaxies are defined
    \citep{hePhotometricPropertiesLuminosity2013,bernardiMassiveEndLuminosity2013}.
    In this paper, we set $f_{\rm merge} = f_{\rm heat} = 0.5$ 
    as the default only for demonstrating 
    the effects of mergers. However, even with such a
    large value, the contribution of mergers to the growth of
    $M_\sscBH$ and $M_*$ is found to be only significant
    in massive halos at low $z$ (see \S\ref{sssec:growth-channels}, 
    Figs.~\ref{fig:growth-examples} and 
    \ref{fig:growth-channels}) -- a conclusion also reached by
    hydro simulations \citep[see e.g. figs. 2 and 10 of][]{liPhysicalProcessesCoevolution2025}.

    \item 
    Gas-phase metallicity of galaxies is modeled according to the `gas-regulator' 
    scenario \citep[e.g.][]{lillyGasRegulationGalaxies2013,wangGasphaseMetallicityDiagnostic2021,
    wangOriginGalaxySizestellar2026}, with certain modifications
    accounting for the high-$z$ environment (e.g. cold-mode accretion,
    high clumpiness) that produces different metal transport 
    from that in the local Universe
    (see \S2.4 of \citetalias{chenTwophaseModelGalaxy2025}).
    Note that the model of metallicity remains highly uncertain
    due to large uncertainties in both observations and theories.
    For example, for the galaxy in the subhalo $h$,
    the change of the oxygen mass, $\Delta M_{\ssc Z}$, during the current
    time step is given by the continuity equation that combines the effects
    of stellar yield, star formation 
    consumption, and feedback ejection, as
    \begin{align}
        \Delta M_{\ssc Z} &= y_{\rm eff} \frac{\Delta M_*}{1-R} - 
            (\Delta M_* + \Delta M_\sscBH) Z 
        \,;\label{eq:dM-Z} \\ 
        y_{\rm eff} &= y f_{\rm mix} (1-f_{\rm esc})
        \,;\\
        f_{\rm mix} &= \frac{1}{1+ x_{\rm \gamma}}
        \,;\\
        f_{\rm esc} &= \frac{1+\alpha_{\rm esc} x_{\rm v}}{1+ x_{\rm v}}
        \,.
    \end{align}
    Here $\Delta M_*$ and $\Delta M_\sscBH$ are the 
    changes of stellar mass and BH mass, respectively, during the current 
    time step, as modeled in the previous steps;
    $Z \equiv M_{\ssc Z} / M_{\rm g}$ is the metallicity of the galaxy;
    $y_{\rm eff}$ is the effective oxygen yield, which incorporates
    the intrinsic yield of stars, the escape of the decoupled fraction, the 
    ejection by stellar feedback, and the recycling due to cooling,
    all approximated to be instantaneous;
    $y = 0.0163$ is the intrinsic oxygen yield of star formation;
    $f_{\rm mix}$ is the efficiency of metal mixing
    between the inflow gas and the existing ISM, and
    $x_{\rm \gamma} = \left|\gamma / \gamma_0\right|^{\beta_{\rm mix}}$
    characterizes its dependence on $\gamma_{\rm v}$;
    $f_{\rm esc}$ is the escaped fraction of metal from the galaxy,
    $x_{\rm v} = (V_{\rm max}/V_{\rm esc})^{\beta_{\rm esc}} $
    quantifies its dependence on the gravitational potential well $h$,
    and $\alpha_{\rm esc} = z / (z_{\rm esc} + z) $ characterizes its 
    evolution with redshift.
    With the above parameterization, the metal mass ejected from the galaxy
    is
    \begin{equation}
        \Delta M_{\rm {\ssc Z}, ej} = f_{\rm esc} y \frac{\Delta M_*}{1-R}
        \,,\label{eq:dM-Z-ej}
    \end{equation}
    Note that the metal mass brought in by inflow gas from the IGM is ignored 
    in Eq.~\eqref{eq:dM-Z}
    given that the metallicity in the IGM is much lower than that in the ISM.
    The default parameters adopted here are 
    $V_{\rm esc} = 75 \kms$, $\beta_{\rm esc}=2$, 
    $z_{\rm esc} = 5$, $\gamma_0 = 1$, and $\beta_{\rm mix}=2$.
\end{enumerate}

\bigskip
With the above steps, we can update the global properties -- such as the masses of the 
BH, stars and gas -- of the galaxy in $h$ during the current time step. 
The growth of these mass components contributed by different channels 
can be separated, as shown in Figs.~\ref{fig:growth-examples} and \ref{fig:growth-channels} 
for examples of halos. 
The evolution of galaxies also provides the basis for modeling 
their radiation and matter feedback to the large-scale environment,
as detailed below.

\subsection{The model of Lyman-Werner radiation field}
\label{app:ssec:LW-radiation}

\begin{figure} \centering
    \includegraphics[width=\columnwidth]{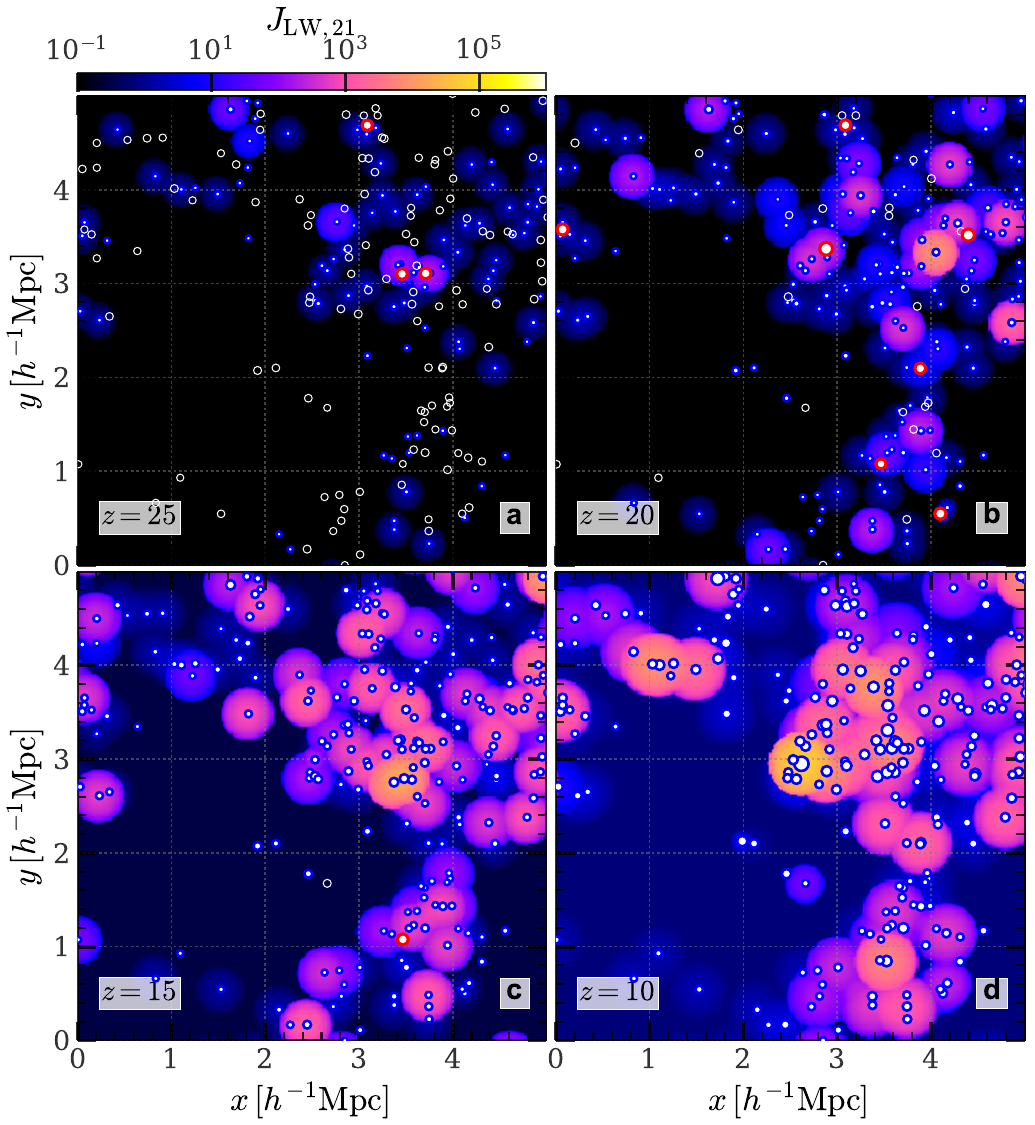}
    \caption{
        {\figem Lyman-Werner radiation field predicted by the model.}
        {\figem a}--{\figem d}, 
        LW intensity, $J_{\sscLW} = J_{\rm bg} + J_{\rm local}$,
        color coded according to the color bar on the top left, 
        in a $5 \times 5\times 2\,(h^{-1}{\rm cMpc})^3$ slice
        of the simulation box (the lower-left corner or that shown in 
        Fig.~\ref{fig:metal-field}) at four different redshifts.
        Each {\figem dot} represents a subhalo, 
        sized proportionally to $R_{\rm v}$, and
        styled according to the status of the galaxy it hosts:
        seeded ({\figem blue edge}),
        unseeded and unilluminated ($J_{\sscLW, 21} < 1$, {\figem white edge}),
        unseeded and illuminated ($J_{\sscLW, 21} \geqslant 1$, {\figem red edge}).
        LW photons represent a type of radiation feedback 
        from galaxy formation to the large-scale environment,
        and play a critical role in delaying the formation of
        the first generations of stars and BH seeds in
        subhalos illuminated by the radiation.
        See Appendix~\ref{app:ssec:LW-radiation} for the detailed steps 
        to model the LW radiation.
    }
    \label{fig:uvb-zoom}
\end{figure}

As shown in \S\ref{sssec:seeding-atlas} and Fig.~\ref{fig:seeding-atlas},
the radiation feedback in the LW band from galaxies to their large-scale 
environment is a critical factor that determines when and how the 
first generations of stars and BHs in halos illuminated by the radiation 
can form.
An accurate modeling of the radiation field requires full spectral 
information from galaxies and a precise algorithm for radiative transfer,
both of which are computationally prohibitive in a cosmological volume.
Below we describe a simplified prescription suitable for 
our purpose.

We first note that the modeling of radiation feedback can take advantage
of the characteristics of the LW photons:
these photons have a large mean free path, up to $\approx 100\mpc$ at $z = 20$ 
(see Eq.~\ref{eq:f-mod-LW} below and \citealt{ahnInhomogeneousBackgroundH2Dissociating2009}),
comparable to the spatial scale of the box of the N-body simulation 
used in this paper.
The intensity of LW radiation can exhibit strong variation (both spatially and temporally),
due to the clustering (i.e. biased distribution) of galaxies in 
the cosmic volume, the strong temporary variation in the star formation 
rate of high-$z$ galaxies (especially when our bursty mode is included;
see e.g. Fig.~\ref{fig:growth-channels}),
and the sharp ($r^{-2}$; see below) radial decay from individual sources. 

To account for both the large mean free path and the strong variation
in spacetime of LW radiation, 
we follow \citet{agarwalUbiquitousSeedingSupermassive2012} 
to decompose $J_{\sscLW}(\bm{x},t)$, the total LW intensity illuminating 
an arbitrary location $\bm{x}$ and a cosmic time $t$, 
into a global term of radiation background, $J_{\rm bg}$, 
and a local term, $J_{\rm local}$, mainly contributed by nearby star-forming
sources, as:
\begin{equation}
    J_{\sscLW}(\bm{x},t) = J_{\rm bg}(t) + J_{\rm local}(\bm{x},t) \,.
\end{equation}
The background term includes the contribution from all radiation sources
in the Universe, and can be obtained by using the total stellar mass density 
$\rho_*(t)$ of all surviving stars in the simulated volume, as
\begin{equation}
    J_{\rm bg}(t) = f_{\rm esc} \frac{h_{\rm P}c}{4\pi m_{\rm p}}
    \eta_{\sscLW} \rho_*(t) (1+z)^3\,.
\end{equation}
Here, $f_{\rm esc} \leqslant 1$ is the fraction of LW photons that escape
from galaxies; $h_{\rm P}$ is the Planck constant;
$c$ is the speed of light; $m_{\rm p}$ is the proton mass;
$\eta_{\sscLW}$ is the number of LW photons emitted per proton mass;
$z = z(t)$ is the redshift at time $t$;
the factor $(1+z)^3$ accounts for the expansion of the universe.
The background is uniform and isotropic, as implied by the cosmological principle.
This formulation implies that $\rho_*$ obtained in the simulated volume
is a fair representation of the entire Universe, which is satisfied
as long as we are using a simulation with a large volume 
and a fairly sampled initial condition
(\S\ref{ssec:halo-assembly}).
The default parameters we use are $f_{\rm esc} = 1$ 
\citep{dijkstraFeedbackregulatedSupermassiveBlack2014,
spinosoMultiflavourSMBHSeeding2023}, assuming that the LW emitters have been 
depleted of $\Hmol$ by recent star formation,  
and $\eta_{\sscLW}=4\times 10^3$
\citep{agarwalUbiquitousSeedingSupermassive2012}
for Pop-II stars. We have not incorporated the contribution from 
Pop-III stars, since their star formation efficiency (SFE) is low 
(see Appendix~\ref{app:ssec:mass-func-pop3}), and the post-seeding 
growth in the bursty mode dominates the growth of galaxies
soon after the seeding (see \S\ref{sssec:growth-channels}
and Fig.~\ref{fig:growth-channels}).
We have also not incorporated the contribution from AGNs, but we 
note that the recently observed high number density of faint AGNs at 
high $z$ suggests that they should be seriously taken into account
(see e.g. Fig.~\ref{sssec:cosmic-bh-sf-history} for the 
cosmic BHAR density).

The local term is modeled by summing up the contribution from individual
source galaxies with a spherically symmetric approximation for the 
emission, as 
\begin{equation}
    J_{\rm local}(\bm{x},t) = \sum_{i} 
    \frac{f_{\rm esc}}{4 \pi \Delta \nu_{\sscLW}} 
    \frac{Q_{\sscLW}M_{{\rm *,alive},i}}{4\pi d_{{\ssc L},i}^2} f_{{\rm mod},i} 
    \,.\label{eq:J-local}
\end{equation}
Here, $Q_{\sscLW}$ is the LW luminosity per unit alive stellar mass;
$\Delta \nu_{\sscLW}$ is the frequency range of LW radiation 
($11.2$--$13.6\,{\rm eV}$); 
$M_{{\rm *,alive},i}$ is the mass of alive stars in the $i$-th source galaxy;
$d_{{\ssc L},i}$ is the luminosity distance between the source galaxy and
the target location $\bm{x}$;
$f_{{\rm mod},i}$ is a modulation factor that accounts for the attenuation
of LW radiation by the IGM.
To avoid over-complication of radiative transfer,
we use the fitting formula obtained by \citet{ahnInhomogeneousBackgroundH2Dissociating2009} 
for $f_{{\rm mod},i}$. Adapting it to our cosmology, we obtain
\begin{equation}
    f_{{\rm mod},i} = 
        1.7 \exp \left[ 
            -\left( \frac{ d_{{\rm c}, i} }{ 110 \alpha_{{\rm mod}, i} \mpc }\right)^{0.68}  
        \right] - 0.7
    \label{eq:f-mod-LW}
\end{equation}
for $d_{{\rm c}, i} \leqslant 92.4 \alpha_{{\rm mod}, i} \mpc$, the LW horizon, and 
$f_{{\rm mod}, i} = 0$ otherwise. 
Here, $d_{{\rm c}, i}$ is the comoving distance between the source galaxy and 
$\bm{x}$, the scaling factor $\alpha_{{\rm mod}, i}$ is 
\begin{equation}
    \alpha_{{\rm mod}, i} =
    \left( \frac{\Omega_{\rm m,0}}{0.3089} \right)^{-1/2}
    \left( \frac{1+z_i}{10} \right)^{-1/2}\,,
\end{equation}
and $z_{\rm i}$ is the redshift of the source galaxy. 
The default value we adopt is 
$Q_{\sscLW} = 10^{36}\,{\rm erg}\cdot {\rm s}^{-1}{\rm M}_\odot^{-1}$,
obtained by averaging the emission of massive O/B stars over their lifetime 
of $\approx 10 \Myr$
\citep[][see their Eq.~6]{dijkstraFeedbackregulatedSupermassiveBlack2014}.
The corresponding mass, $M_{{\rm *,alive}, i}$, is computed by including 
only these young massive stars.
The summation should be performed 
over all galaxies in the past light cone of $(\bm{x},\,t)$, leading to
a computational complexity of $\mathcal{O}(N^2)$ that is
prohibitive for a large number ($N$) of galaxies. 
To proceed, we first note that the light speed is large enough so that 
the emission and reception of LW radiation between two subhalos 
in the simulated volume can be simplified as synchronous events.
Thus, the summation can be performed by using sources within 
the current snapshot. This also leads to an approximation
$d_{{\ssc L}, i} \approx d_{{\rm c}, i}$. 
To reduce the number of summations, we limit the zone of influence
of each source galaxy to a sphere with a radius at which the 
LW intensity emitted by this galaxy drops to a negligible threshold,
$J_{\rm local, th}$.
The default value we adopt is $J_{\rm local, th, 21} = 0.01$,
which is about a factor of $10^{-2}$--$10^{-3}$ smaller than $J_{\rm crit,21}$,
the value that produces a significant delay in gas collapse 
(see Appendix~\ref{app:sssec:delay-by-uv-radiation};
see also Appendix A of \citealt{spinosoMultiflavourSMBHSeeding2023} for 
a similar approach). With these simplifications, 
the computational time for the LW intensity in a single time step
is reduced to a few seconds for our sample of subhalos (\S\ref{ssec:halo-assembly}).

Fig.~\ref{fig:uvb-zoom} shows the LW intensity, $J_{\sscLW}$, in a slice of 
the simulation box at four different redshifts, $z = 25$, $20$, $15$
and $10$. Subhalos with seeds having been bred are represented by dots with 
a blue edge, which are the sources of LW radiation.
Subhalos without seeds are represented by dots with a white ($J_{\sscLW, 21} < 1$)
or a red edge ($J_{\sscLW, 21} \geqslant 1$).
At $z \approx 25$, the universe is in the seeding era (see \S\ref{sssec:cosmic-bh-sf-history}), 
and most subhalos are still on their way to form the first generation of stars 
and BH seeds. 
The large number of unseeded subhalos in panel a reflects this infancy era 
of galaxy formation.
The seeding era ends at $z \approx 20$ when most subhalos have been seeded,
which can be seen from the reduction in the number of unseeded subhalos
from panel a to panel b.
The rapid post-seeding star formation through the bursty mode 
produces intense but localized LW radiation from the seeded galaxies, 
manifested as individual lamp-like illuminated zones in panel b.
A small fraction of unseeded subhalos located in these zones
are illuminated by the LW radiation, and are delayed in forming their first
generation of stars and BH seeds. At $z \lesssim 15$, the rise in the cosmic 
star formation rate density (see also Fig.~\ref{sssec:cosmic-bh-sf-history}) lifts 
the global background.
However, by this time, most subhalos have already been seeded 
(see also Fig.~\ref{sssec:cosmic-seeding-history}), 
providing harbors of star formation and BH growth that are well 
shielded from external radiation.


\subsection{The model of intergalactic-medium enrichment}
\label{app:ssec:IGM-enrichment}

\begin{figure*} \centering
    \includegraphics[width=\textwidth]{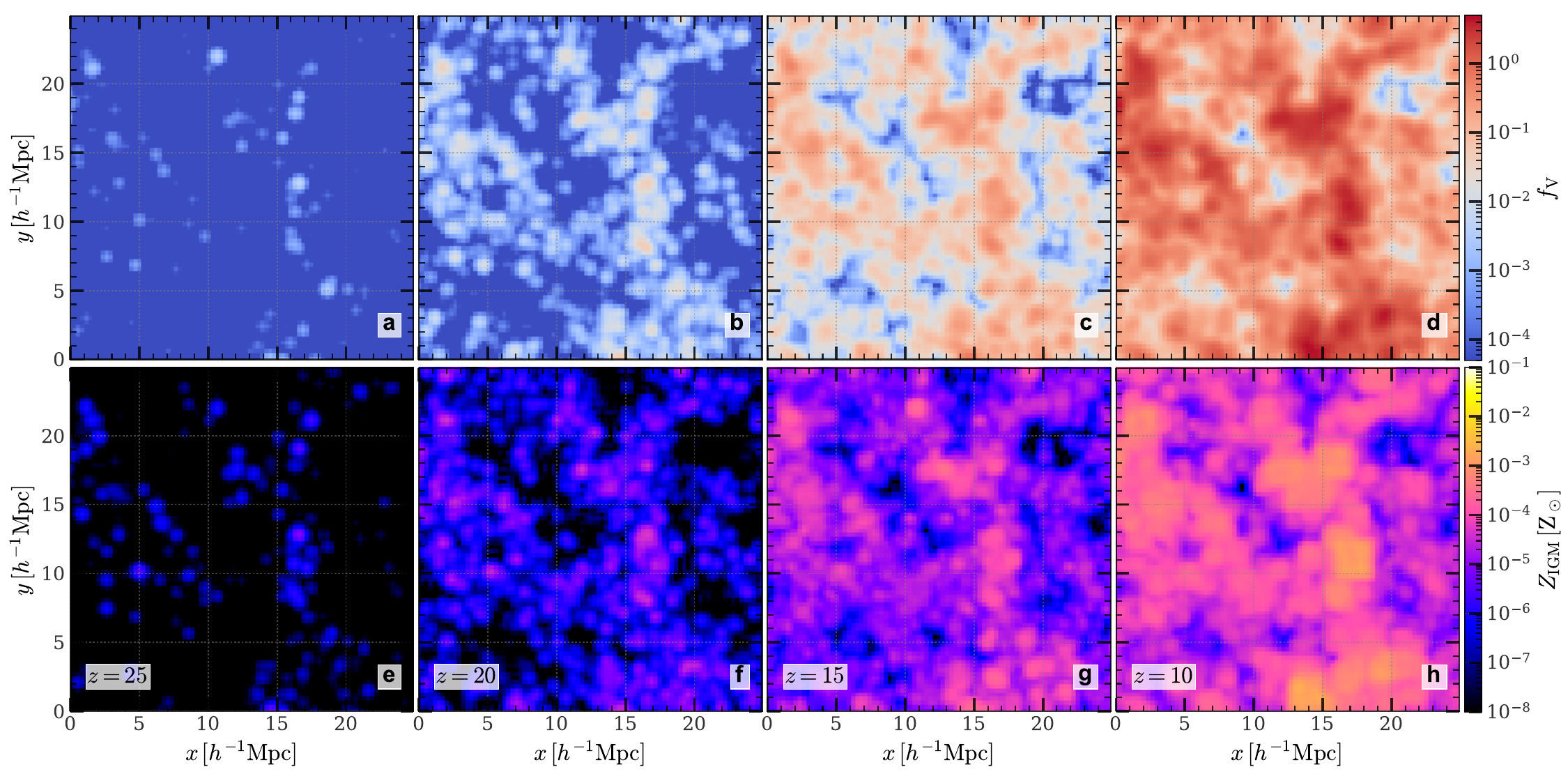}
    \caption{{\figem Enrichment of intergalactic medium 
        predicted by the model.}
        {\figem a}--{\figem d}, distribution of 
        volume filling factor ($f_{\rm V}$) of SN bubbles 
        in a $25 \times 25 \times 2\,(h^{-1}{\rm cMpc})^3$ 
        slice of the simulation box at four different redshifts.
        {\figem e}--{\figem h}, similar but for $Z_{\ssc IGM}$, the 
        metallicity of IGM. 
        Note that $Z_{\ssc IGM}$ shown here is the average value within 
        each entire cell (i.e. $Z_{\ssc IGM} = \Delta M_{\ssc Z}/(f_{\rm b} \rho_{\rm m} \Delta V_{\ssc C})$,
        without an $f_{\rm V}$ in the denominator), 
        not just within the bubble-filled fraction 
        that is used to sample the IGM metallicity for target galaxies 
        (steps iv and v in the enrichment modeling).
        For both $f_{\rm V}$ and $Z_{\ssc IGM}$, mean values along 
        $z$-axis are taken, and are color coded according to the color bar 
        on the right.
        At $z \approx 20$, a noticeable fraction of the cosmic volume
        has been enriched by metals ejected from early-formed 
        galaxies, leading to the significant formation activity of 
        Pop-II seeds; by the time of $z \approx 10$, 
        most cosmic volume has already been enriched to some level,
        marking the prevalence of Pop-II seeds 
        (see e.g. Fig.~\ref{fig:seed-z-func}a for the cosmic seeding history
        and Fig.~\ref{fig:seeding-atlas} for the seeding atlas).
        See Appendix~\ref{app:ssec:IGM-enrichment} for the detailed steps
        to model the IGM enrichment.
    }
    \label{fig:metal-field}
\end{figure*}

The enrichment of the IGM is a type of matter feedback of galaxy formation to the large-scale 
environment. Compared with the radiation feedback described above,
matter feedback also plays a role in determining the formation of seeds in subhalos 
surrounded by the enriched IGM; a major difference is that the diffusion of metals 
is much slower than that of radiation, so that the expansion of 
the zone of enrichment produced by the matter feedback from any galaxy
has to be tracked over a long timescale.

To simplify the model of IGM enrichment so that it can be efficiently
implemented into a large volume, we make the following assumptions:
the intergalactic diffusion of metals is assumed to be a spherically symmetric process
around each source galaxy, and the enriched volume is described by a spherical 
bubble; the energy powering the expansion of the bubble is assumed 
to come from the blast wave generated by SNe, while 
the contribution from Pop-III stars and AGNs is ignored. 
All these assumptions are consistent with
and explained in our treatment of the radiation feedback 
(Appendix~\ref{app:ssec:LW-radiation}).

To estimate the sizes of SN bubbles produced by galaxies,
we adopt a self-similar solution of blast wave \citep[see e.g. \S8.6 of][]{moGalaxyFormationEvolution2010}:
\begin{equation}
    R_{\rm sh} = A_{\rm sh} \left(\frac{E_{\ssc SN}}{\rho_{\rm g}} \right)^{1/5}  \left(t - t_{\rm 0}\right)^{2/5}
    \,, \label{eq:R-sh}
\end{equation}
where $R_{\rm sh}$ is the physical radius of the shock wave,
$A_{\rm sh} \approx 1$ is a normalization constant whose value is determined
by energy conservation,
$E_{\ssc SN}$ is the energy released by SNe during a burst of star formation,
$\rho_{\rm g}$ is the density of the IGM swept by the shock wave, 
and $t - t_0$ is the elapsed time since bubble generation at time
$t_0$. We express the energy released by SNe as
$E_{\ssc SN} = \epsilon_{\ssc SN} R \Delta M_{\rm *,new}$, where
the energy efficiency $\epsilon_{\ssc SN}=10^{51}\,{\rm erg} \cdot {\rm M}_{\odot}^{-1}$
\citep{dijkstraFeedbackregulatedSupermassiveBlack2014,
spinosoMultiflavourSMBHSeeding2023},
$R$ is the fraction of returned mass (Eq.~\ref{eq:mass-return}), and
$\Delta M_{\rm *,new}$ counts all stars newly formed in the past 
duration of $t_{\rm v}$
(Eq.~\ref{eq:t-vir}) in which the replenishment of the gas may be accomplished to trigger 
the next episode of star formation. If there are multiple bubbles generated
in the main branch of a galaxy, we only consider the largest
\citep{venturaSemianalyticModellingPop2024}. However, since SFRs of high-$z$ 
galaxies often increase rapidly with time
(see examples in Fig.~\ref{fig:growth-channels}), 
the largest bubble is most likely produced in the recent
star-forming episode.

Before we implement the IGM enrichment model in our pipeline,
we provide an analytical estimate for the bubble sizes.
We take $\rho_{\rm g} = \epsilon_{\rho} \rho_{\rm b}(z)$, 
where $\rho_{\rm b}$ is the cosmic baryon density (Eq.~\ref{eq:cosmic_rho_b_z}),
and $\epsilon_{\rho} = 60$ \citep{dijkstraFeedbackregulatedSupermassiveBlack2014}, 
roughly the density contrast at $R_{\rm v}$ of any halo at $z$.
The stellar mass formed in the recent $t_{\rm v}$ is:
\begin{align} 
    \Delta M_{\rm *,new}
    &= \epsilon_{\ssc SF} f_{\rm b} \dot{M}_{\rm v} t_{\rm v}
    \nonumber\\
    &= \left[1.73 \times 10^7 \Msun\right] \epsilon_{{\ssc SF},0.1}
        M_{\rm v,9.5}^{1.14}(1+z)_{10}
    \,,\label{eq:Delta-M-star-over-t-vir}
\end{align}
where $\epsilon_{\ssc SF} = 0.1\epsilon_{\ssc SF, 0.1}$ 
is the star formation efficiency, with its typical value 
of $0.1$ taken from observations \citep[e.g.][]{wangJWSTMIRIReveals2025} 
and empirical models for high-$z$ galaxies \citep[e.g.][]{nusserHighredshiftHaloGalaxy2024}, 
and $\dot{M}_{\rm v}$ is the halo accretion rate, approximated using 
Eq.~\eqref{eq:dot-M-vir}. Substituting the above equation into
Eq.~\eqref{eq:R-sh} and taking $t - t_0 = t_{\rm v}$, 
we can obtain the bubble radius in comoving units as
\begin{equation}
    R_{\rm sh,c}
    = (1+z) R_{\rm sh} = 231\Kpc \left[ \chi_{\rm sh} M_{\rm v,9.5}^{1.14} \right]^{1/5}
    \,,\label{eq:R-sh-numeric}
\end{equation} 
where we have defined $\chi_{\rm sh} = \epsilon_{{\ssc SN}, 51} \epsilon_{{\ssc SF},0.1} R_{\rm 0.4}/\epsilon_{\rho,60} $,
$\epsilon_{{\ssc SN}, 51} = \epsilon_{\ssc SN}/\left(10^{51}\,{\rm erg}\cdot {\rm M}_{\odot}^{-1}\right)$,
$R_{\rm 0.4} = R/0.4$ and $\epsilon_{\rho, 60} = \epsilon_{\rho}/60$.
Thus, the comoving size of an SN bubble is only a weak function 
of halo mass, and is redshift insensitive, if $\chi_{\rm sh}$ does 
not vary significantly.
Another useful quantity is the ratio between $R_{\rm sh}$ and 
$R_{\rm v}$ (Eq.~\ref{eq:R-vir}) of the host halo,
\begin{equation}
    R_{\rm sh} / R_{\rm v}
    = 5.05 \, \chi_{\rm sh}^{1/5} M_{\rm v,9.5}^{-0.105}
    \,,\label{eq:R-sh-to-R-vir-numeric}
\end{equation}
which again depends weakly on the halo mass and is largely insensitive to redshift,
suggesting that the radius of an SN bubble is nearly proportional to the 
size of its host halo.

We can also estimate the global volume filling factor, $f_{\ssc V}$,
of SN bubbles (i.e. the fraction of cosmic volume filled by the bubbles).
For this, we multiply the volume of bubble (in comoving unit) with $\Phi(M_{\rm v})$,
the halo mass function (defined similarly to the
mass function of BH seed in Eq.~\ref{eq:def-seed-mass-func}), 
and integrate it over the full range of halo mass:
\begin{equation}
    f_{\ssc V} = \int
        \frac{4}{3} \pi R_{\rm sh,c}^3 \Phi (M_{\rm v}) \dd{\log M_{\rm v}}
    \,.\label{eq:f-V-analytical}
\end{equation}
Since the power index on $M_{\rm v}$ in the halo mass function is smaller 
than $-0.8$ \citep[see \S3.1 of][]{tinkerHaloMassFunction2008},
the integrand scales with halo mass as $R_{\rm sh,c}^3  \Phi \sim M_{\rm v}^\beta$,
with $\beta < 1.14 \times 3/5 - 0.8 \approx -0.12 < 0$.
Thus, the volume filling factor is dominated by low-mass halos.
This signifies the need for modeling the full population of galaxies
in a cosmological volume, particularly in the early Universe where halo
masses are typically small, so that the environment that affects the seeding 
can be fully captured.

In our numerical implementation, we follow \citet{venturaSemianalyticModellingPop2024}
to model the spatial distribution of metal on coarse grids,
and ignore the sub-grid distribution.
The reason for adopting such an approach is to take into account the limitations
arising from the simplifications we have made: (i) the assumptions of 
spherical symmetry and blast-wave expansion are not able 
to capture the sub-structure of bubbles; (ii) interactions among bubbles 
generated by different galaxies or by one galaxy at different episodes
of star formation cannot be resolved; (iii) the locations of 
subhalos extended by our Monte Carlo algorithm 
(Appendix~\ref{app:ssec:extension-merger-tree})
are not exactly known.
Thus, a grid-based approach with spatial resolution comparable to these 
physical limitations is sufficient for our purpose.
The steps of our numerical implementation are as follows:
\begin{enumerate}[topsep=0pt,parsep=0pt,itemsep=0pt]
    \item We partition the simulation box into cubic cells
    with a fixed side length (in comoving unit)
    that is sufficiently fine compared with the spatial scale to be used 
    to smooth the spatial distributions (see step iv below).  
    Any field in the simulation box is then defined by a set of discrete values
    in the cells.
    The first field to compute is $\rho_{\rm m}$, the mass density field 
    of the current snapshot, whose value in each cell $C$ is obtained
    by summing up the masses of all the simulated 
    dark matter particles located in $C$, and dividing it by 
    $\Delta V_{\ssc C}$, the volume of the cell.

    \item For each subhalo, we track the evolution of $R_{\rm sh}$
    of the SN bubble generated by the galaxy hosted by the subhalo,
    and update it over time steps. In the current snapshot $s$,
    we compute $\Delta M_{\rm *,new}$, the 
    mass of newly formed stars over the past duration of $t_{\rm v}$
    in the subhalo $h \in H_s$.
    We compare $\Delta M_{\rm *,new}$ with those evaluated in the previous 
    time steps along the main branch of $h$, and take the largest value
    among them to power the expansion of the SN bubble.
    We find the derivative of $R_{\rm sh}$ with respect to $t$ 
    using Eq.~\eqref{eq:R-sh}, as 
    \begin{equation}
        \dot{R}_{\rm sh} = 
        \frac{2}{5} A_{\rm sh} \left(\frac{E_{\ssc SN}}{\rho_{\rm g}} \right)^{1/5} 
        (t - t_0)^{-3/5}
        \,.\label{eq:dR-sh-dt}
    \end{equation}
    Here $E_{\ssc SN}$ is computed using the largest value of 
    $\Delta M_{\rm *,new}$ and $t_0$ is the time when this value is achieved.
    The gas density $\rho_{\rm g}$ is evaluated at halo-centric distance 
    $R_{\rm sh}$ by assuming an isothermal profile,
    and a total mass of 
    $f_{\rm b} M_{\rm v} - M_{\rm g} - M_* - M_{\ssc BH}$ 
    (i.e. the available baryon mass subtracted by the gas, stellar and BH masses
    in the galaxy hosted by $h$) for the gas within $h$ 
    (Appendix~\ref{app:sec:halo-assembly} and Fig.~\ref{fig:profile-ladder}), if 
    $R_{\rm sh}$ is smaller than $R_{\rm v}$ of $h$; otherwise, we 
    use $\rho_{\rm m}$ at the location of $h$ on a scale of $R_{\rm sh}$
    to find the gas density as $\rho_{\rm g} = f_{\rm b}\rho_{\rm m}$.
    We then use $\dot{R}_{\rm sh}$ to update $R_{\rm sh}$ for one 
    time step.
    
    \item We define the volume filling factor $f_{\ssc V}$ of any cell
    as the fraction of the cell volume that has been filled by SN bubbles,
    and we define the metal mass $\Delta M_{\ssc Z}$ in any cell 
    as the total mass of metal ejected by galaxies into the cell.
    To compute the contributions from a source subhalo $h \in H_s$ to these fields, we identify the intersection of its SN 
    bubble with any cell, and accumulate
    the contribution from $h$ to $f_{\ssc V}$ of the cell
    by using the volume of the intersection;
    similarly, we compute the total mass of metal ejected from $h$ and all 
    of its progenitors, assuming this mass is uniformly distributed within the SN 
    bubble of $h$, and accumulate the contribution
    from $h$ to $\Delta M_{\ssc Z}$ in any cell
    intersected with the bubble.

    \item We smooth the fields, $\rho_{\rm m}$, 
    $f_{\ssc V}$ and $\Delta M_{\ssc Z}$ 
    by a Gaussian kernel with a comoving 
    size of $l_{\rm sm} = 250\kpc$,
    a scale comparable to the typical size of SN bubbles 
    estimated in Eq.~\eqref{eq:R-sh-numeric}.
    We compute the metallicity of the IGM in the bubble-filled volume of any cell as 
    $Z_{\ssc IGM} = \Delta M_{\ssc Z}/(f_{\rm b} f_{\ssc V} \rho_{\rm m} \Delta V_{\ssc C})$,
    and as $Z_{\ssc IGM} = 0$ outside the bubble-filled volume.

    \item For any target subhalo $h \in H_s$, the metallicity of the IGM 
    around the galaxy hosted by $h$ is determined by a stochastic
    algorithm. Specifically, 
    we draw $u_{\ssc Z} \sim U[0,1]$, a random number  
    with a uniform distribution over $[0,1]$, and compare it with the value of
    $f_{\ssc V}$ in the cell containing $h$. 
    If $u_{\ssc Z} \leqslant f_{\ssc V}$, the galaxy is 
    considered to be located in the bubble-filled volume of the cell
    and the value of $Z_{\ssc IGM}$ is assigned accordingly;
    otherwise, it is located outside the bubble-filled volume 
    and $Z_{\ssc IGM} = 0$ is assigned.

\end{enumerate}

Fig.~\ref{fig:metal-field} shows the volume filling factor and IGM metallicity 
of the cells in a slice with a thickness of $2\mpc$ 
(in comoving units) in the simulation box
at four different redshifts, $z = 25$, $20$, $15$ and $10$.
At $z = 25$ (panels a and e), a few SN bubbles can be seen around 
galaxies whose post-seeding star formation has just started owing to the onset of 
nuclear bursts (see Fig.~\ref{fig:growth-channels} for example galaxies).
The sizes of these bubbles are small, as expected from the 
low halo masses (see e.g. Fig.~\ref{fig:seeding-atlas}) and the 
proportional relation between $R_{\rm sh}$ and $R_{\rm v}$ 
(Eq.~\ref{eq:R-sh-to-R-vir-numeric}).
The metallicity of the enriched volume is as low 
as $\sim 10^{-6}\Zsun$, which is a consequence of the low star formation 
efficiency in low-mass halos (see
e.g. Fig.~\ref{fig:growth-cosmic-history}b for a comparison 
between $\dot{\rho}_*$ and virialized $\dot{\rho}_{\rm b}$).
From $z = 25$ to $z = 20$ (panels b and f), 
the bubble-filled volume expands significantly from the sites densely 
populated with galaxies that have been newly seeded and 
experiencing intense star formation through the bursty mode. 
The metallicity in some bubble-filled zones increases beyond
$10^{-4}\Zsun$, and the unseeded subhalos swept by the bubbles 
have increased chances to be fueled with enriched gas and to form 
Pop-II stars and BH seeds. The rapid increase of cosmic seeding activity contributed by Pop-II seeds
(Fig.~\ref{fig:seed-z-func}a, green shade)
at $z \gtrsim 20$ can thus be explained. At $z \lesssim 20$, most subhalos 
had already been seeded, while their SFRs through the bursty mode continue to increase
due to the fast halo assembly (Fig.~\ref{fig:halo-mahs}a).
The SN bubbles continue to expand and overlap with each other
to form extended structures, leaving only a small fraction of
the cosmic volume pristine (panels c and g).
By the time of $z = 10$ (panels d and h), most cosmic volume has been
filled by SN bubbles, with some hot spots having $Z_{\ssc IGM} \gtrsim 10^{-3}\Zsun$.
In this era, Pop-II seeds are bred at a rate comparable to late-formed 
Pop-III seeds, as can be seen by a comparison among cosmic BH seeding 
histories contributed by different flavors in Fig.~\ref{fig:seed-z-func}a.

\section{Evolution of the structure hierarchy}
\label{app:sec:evolution-of-hierarchy}

\begin{figure*} \centering
    \includegraphics[width=0.825\textwidth]{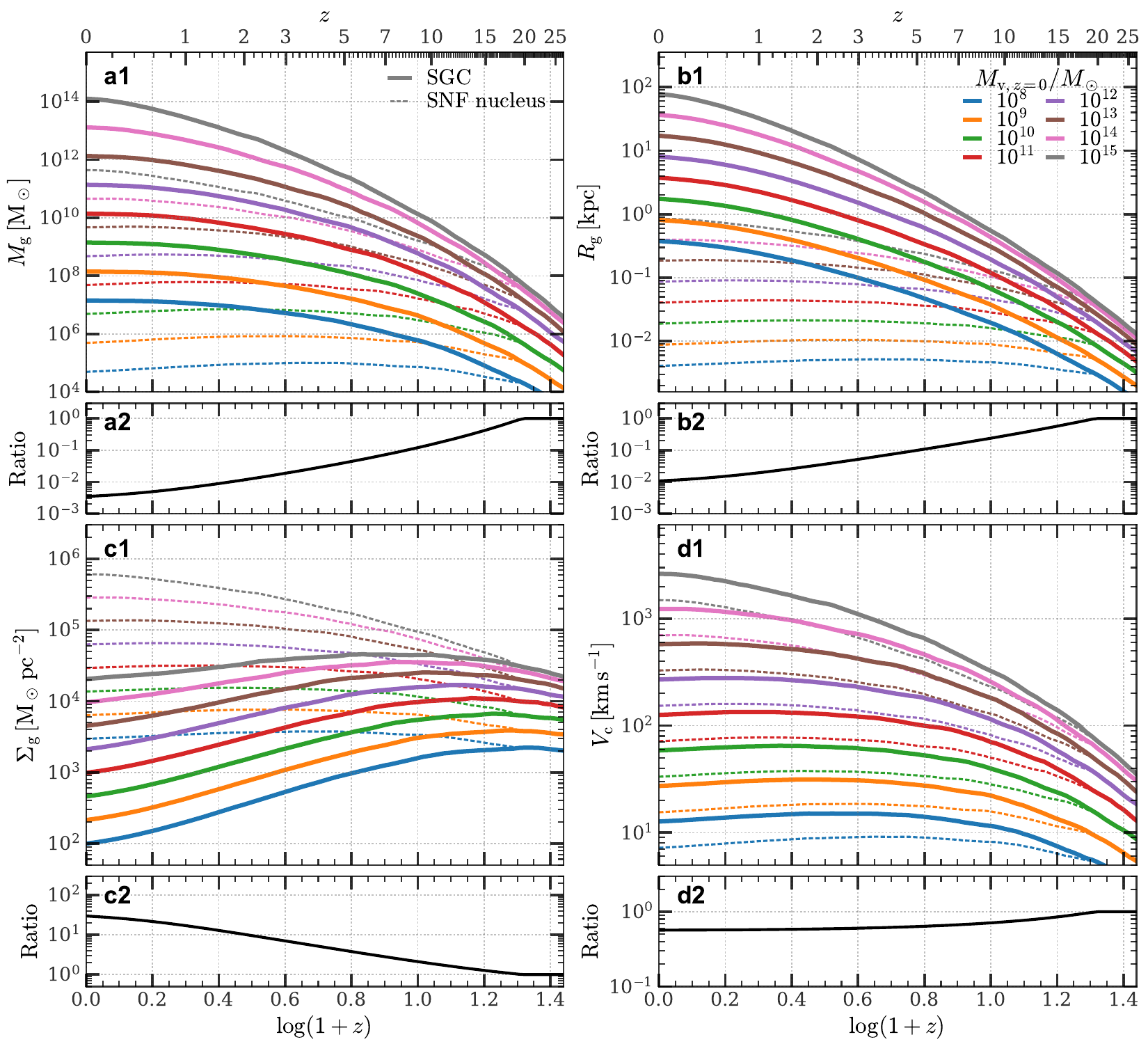}
    \caption{
    {\figem Properties of self-gravitating gas cloud and supernova-free nucleus in the histories of halos 
    with different masses.} 
    Four properties for SGC ({\figem solid}) and SNF nucleus ({\figem dashed})
    are shown: gas mass ($M_{\rm g}$; {\figem a1}),
    gas radius ($R_{\rm g}$; {\figem b1}),
    gas surface density ($\Sigma_{\rm g}$; {\figem c1})
    and circular velocity ($V_{\rm c}$; {\figem d1}).
    {\figem a2}--{\figem d2} show the ratios of the corresponding properties
    between the SNF nucleus and the SGC.
    Each property is shown as a function of $z$ evaluated based on median  
    assembly history of halos with a given mass $M_{\rm v}$ at $z=0$ ($M_{{\rm v},z=0}$).
    The assembly histories of halos used in this figure 
    are exactly the same as those shown in Fig.~\ref{fig:halo-mahs}.
    Analytical approximations for the properties of SGC can be found in 
    Eqs.~\eqref{eq:r-sgc}--\eqref{eq:t-ff-sgc}, and 
    of SNF nucleus can be found in
    Eqs.~\eqref{eq:c-g-numeric}--\eqref{eq:sigma-nuc-numeric}.
    See Appendix~\ref{app:sec:evolution-of-hierarchy} for details.
    }
    \label{fig:sgc-snf-props}
\end{figure*}

\begin{figure*} \centering
    \includegraphics[width=0.825\textwidth]{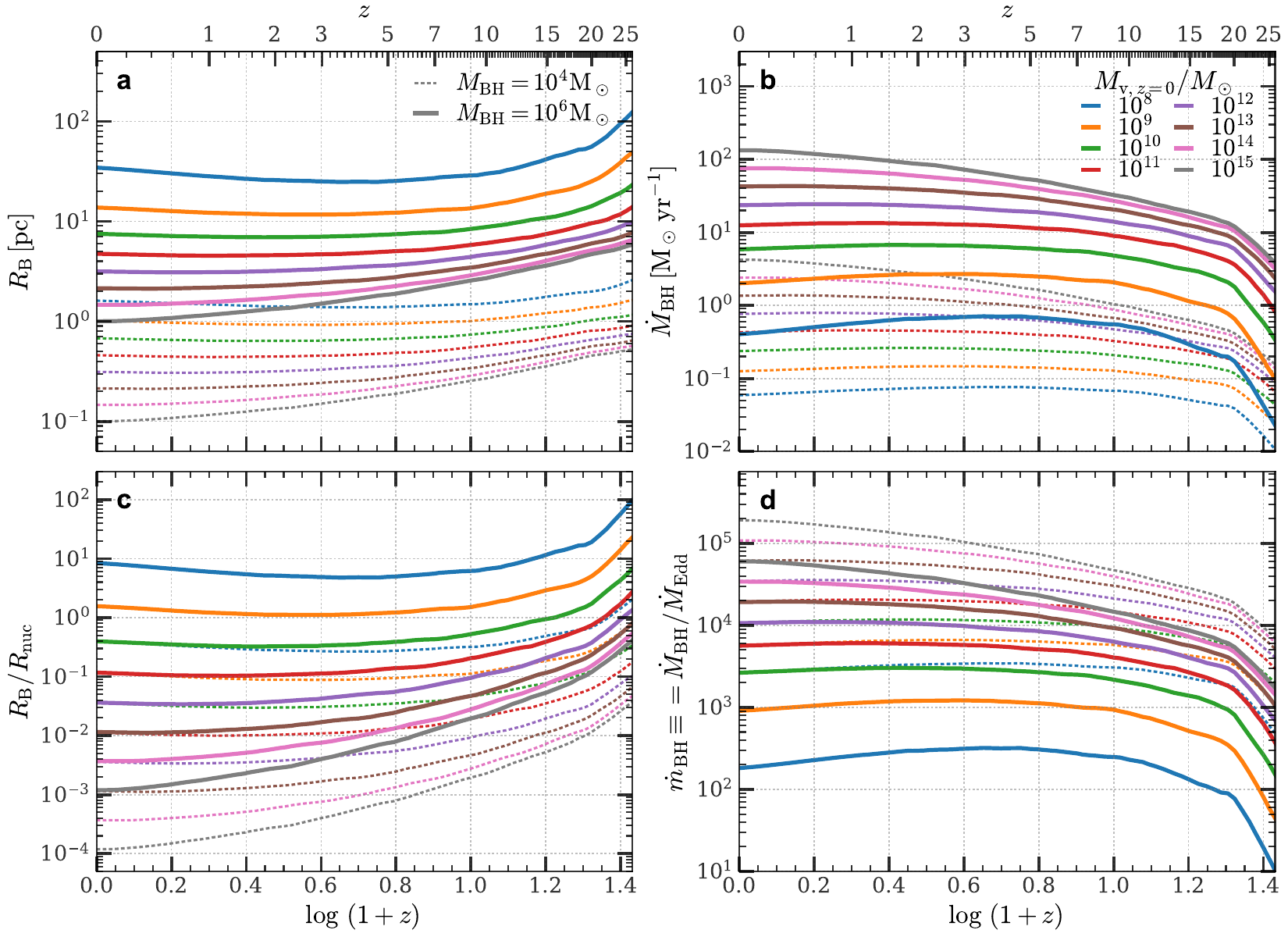}
    \caption{
    {\figem Bondi properties of central black holes in the histories of halos 
    with different masses.} Here we show four properties:
    Bondi radius ($R_{\sscBondi}$; {\figem a}),
    Bondi accretion rate ($\dot{M}_{\rm B}$; {\figem b}),
    the ratio between $R_{\sscBondi}$ and the radius of the 
    SNF nucleus hosting the BH ({\figem c}),
    and the Eddington ratio of Bondi accretion rate
    ($\dot{m}_\sscBH \equiv \dot{M}_{\rm B}/\dot{M}_{\ssc Edd}$; {\figem d}).
    Each property is shown as a function of redshift evaluated based on 
    properties of the host SGC and SNF nucleus (Fig.~\ref{fig:sgc-snf-props})
    in halos with a given mass $M_{\rm v}$ at $z=0$ ($M_{{\rm v},z=0}$), assuming
    a BH with a mass of $M_{\sscBH} = 10^4\Msun$ ({\figem dashed}) or  
    $10^6\Msun$ ({\figem solid}) is placed at galactic center.
    See Appendix~\ref{app:sec:evolution-of-hierarchy} for details.
    }
    \label{fig:bondi-props}
\end{figure*}

In Appendix~\ref{app:ssec:hierarchy}, we described the structure hierarchy 
(Fig.~\ref{fig:am-redistribute})
that leads to BH growth and showed the ladder-like gas profile 
for a halo of MW size and its progenitors at different redshifts
(Fig.~\ref{fig:profile-ladder}). Here we present the properties 
of individual levels on the ladder, and the evolution of these properties,
in the histories of halos.
In applications, one can start from any level on the ladder, 
use it as a large-scale boundary condition for the processes that lift gas to 
higher levels, and resolve the formation and growth of 
structures on smaller scales.

Fig.~\ref{fig:sgc-snf-props} shows the evolution of the properties of the 
SGC and the SNF nucleus in the assembly histories of halos with 
different masses at $z=0$ ($M_{{\rm v},z=0}$).
To produce these properties, we first take the median assembly history 
of halos with a given $M_{{\rm v},z=0}$ in the same way as that used 
to obtain Fig.~\ref{fig:halo-mahs}.
We then assume that a global disturbance has been induced in the SGC hosted by a halo
at a redshift $z$, and that the disturbance generates an SNF nucleus in the center of the SGC.
We compute the properties of the SGC according to Eqs.~\eqref{eq:r-sgc}--\eqref{eq:t-ff-sgc} 
(the first equality in each of them) and the properties of the SNF nucleus according
to Eqs.~\eqref{eq:rho-snf}--\eqref{eq:v-c-in-snf-q05}.
We perform the calculation for halos of any $M_{{\rm v},z=0}$ and at
any $z$, and show the gas mass ($M_{\rm g}$), radius ($R_{\rm g}$), 
surface density ($\Sigma_{\rm g}$) and circular velocity ($V_{\rm c}$)
in panels a1--d1, respectively, for both SGC (solid) and SNF nucleus (dashed).
For reference, we also show the ratio between the properties of the SNF nucleus 
and the SGC in panels a2--d2.
The gas fraction assumed here is $f_{\rm g,0.157} = 1$,
which describes a gas-rich SGC that has not been depleted by feedback;
the radius fraction is $f_{\rm r,0.04} = 1$, roughly the critical 
value at which rotation is still unable to provide enough support;
the feedback-free threshold is $n_{\rm snf, 3.5} = 1$, which corresponds to a free-fall
timescale of about $1\Myr$. Halos with values of $f_{\rm g,0.157}$, $f_{\rm r,0.04}$
or $n_{\rm snf, 3.5}$ other than 1 can also be analyzed by the equations listed
above. Note that at $z \gg 1$ (when the universe is approximately an Einstein-de Sitter universe),
the dependence of SGC properties on $M_{\rm v}$ and $z$ can be 
analytically approximated using Eqs.~\eqref{eq:r-sgc}--\eqref{eq:t-ff-sgc};
if, in addition, $z \lesssim 10$ (i.e. the concentration
$c_{\rm g} \equiv R_{\sscSGC}/R_{\rm nuc} \gg 1$),
the dependence of the properties of the SNF nucleus on $M_{\rm v}$ and $z$ can also be
analytically approximated using Eqs.~\eqref{eq:c-g-numeric}--\eqref{eq:sigma-nuc-numeric}.

The most obvious feature of the SNF nucleus compared to the host SGC is
the redshift evolution. At $z \gtrsim 20$ (when the age of the universe is less than $200\Myr$), 
an SGC is entirely supernova-free,
as seen from the ratio of $M_{\rm g}$ (or $R_{\rm g}$) between
the SNF nucleus and the SGC. 
At lower redshifts, the SNF nucleus occupies a smaller fraction of the SGC:
at $z \approx 5$, the SNF nucleus takes about $10\%$ of the SGC
mass, and at $z = 0$, the fraction drops to about $1\%$ (panel a2);
a similar trend is seen for the radius (panel b2).
The surface density of the SNF nucleus, however, remains higher 
than $10^{3}\Msunperpcsq$ at any redshift for halos with any mass
shown in Fig.~\ref{fig:sgc-snf-props}, and the ratio of $\Sigma_{\rm g}$ between the
SNF nucleus and the SGC continuously increases with time.
The evolution of the properties of the SNF nucleus and the SGC reflects
that the importance of the SNF nucleus in supplying gas to the central BH of 
a galaxy declines with time.
This, together with the less frequent global disturbance at lower 
redshift, explains the decrease of BH growth through the bursty mode
with time (see \S\ref{sssec:growth-channels} and Fig.~\ref{fig:growth-channels}).

Climbing one level higher on the ladder, one can deduce the Bondi properties
of the central BH hosted by an SNF nucleus. The properties most relevant to the
modeling of BH growth are the Bondi radius ($R_{\sscBondi}$; Eq.~\ref{eq:R-Bondi-general}) 
and the Bondi accretion rate ($\dot{M}_{\sscBondi}$, or
$\dot{m}_\sscBondi \equiv \dot{M}_\sscBondi / \dot{M}_{\ssc Edd}$;
Eq.~\ref{eq:dot-M-Bondi-general}), which form the boundary condition
to determine the structure of the accretion disk around the BH that 
experiences the bursty mode accretion.
In Fig.~\ref{fig:bondi-props}, we show the evolution of $R_{\sscBondi}$
and $\dot{M}_{\sscBondi}$ in panels a and b, respectively,
for a BH with a mass of $M_{\sscBH} = 10^4\Msun$ (dashed) or 
$10^6\Msun$ (solid) placed in the center of an SNF nucleus
in the history of a halo with a given $M_{{\rm v},z=0}$.
At $1 \ll z \lesssim 10$, $R_{\sscBondi}$ and $\dot{m}_{\sscBH}$ can also 
be approximated analytically by linking them to the properties of
the SNF nucleus, the SGC, or the halo, using the chained 
equalities
in Eqs.~\eqref{eq:R-Bondi-numerical-q05} and
\eqref{eq:dot-m-Bondi-numerical-q05}.
As expected from these approximations, $R_{\sscBondi}$ 
increases with $M_{\sscBH}$, reflecting the larger zone of influence
of more massive BHs, and decreases with $M_{\rm v}$, due to 
the deeper gravitational potential well in the centers of more massive 
halos. On the other hand, $\dot{M}_{\sscBondi}$ increases with both
$M_{\sscBH}$ and $M_{\rm v}$.
Note that in some cases where a too massive BH is placed in a too small SNF nucleus, 
$R_{\sscBondi}$ is obtained by extrapolating the SGC density profile (Eq.~\ref{eq:rho-g-in-nuc-q05}) 
outward beyond $R_{\rm nuc}$, which can lead to $R_{\sscBondi} / R_{\rm nuc} > 1$. 
In realistic applications where BHs grow from well-defined initial and boundary conditions, 
such cases should not occur.

\section{Variants of the model}
\label{app:sec:model-variants}

\begin{figure*} \centering
    \includegraphics[width=\textwidth]{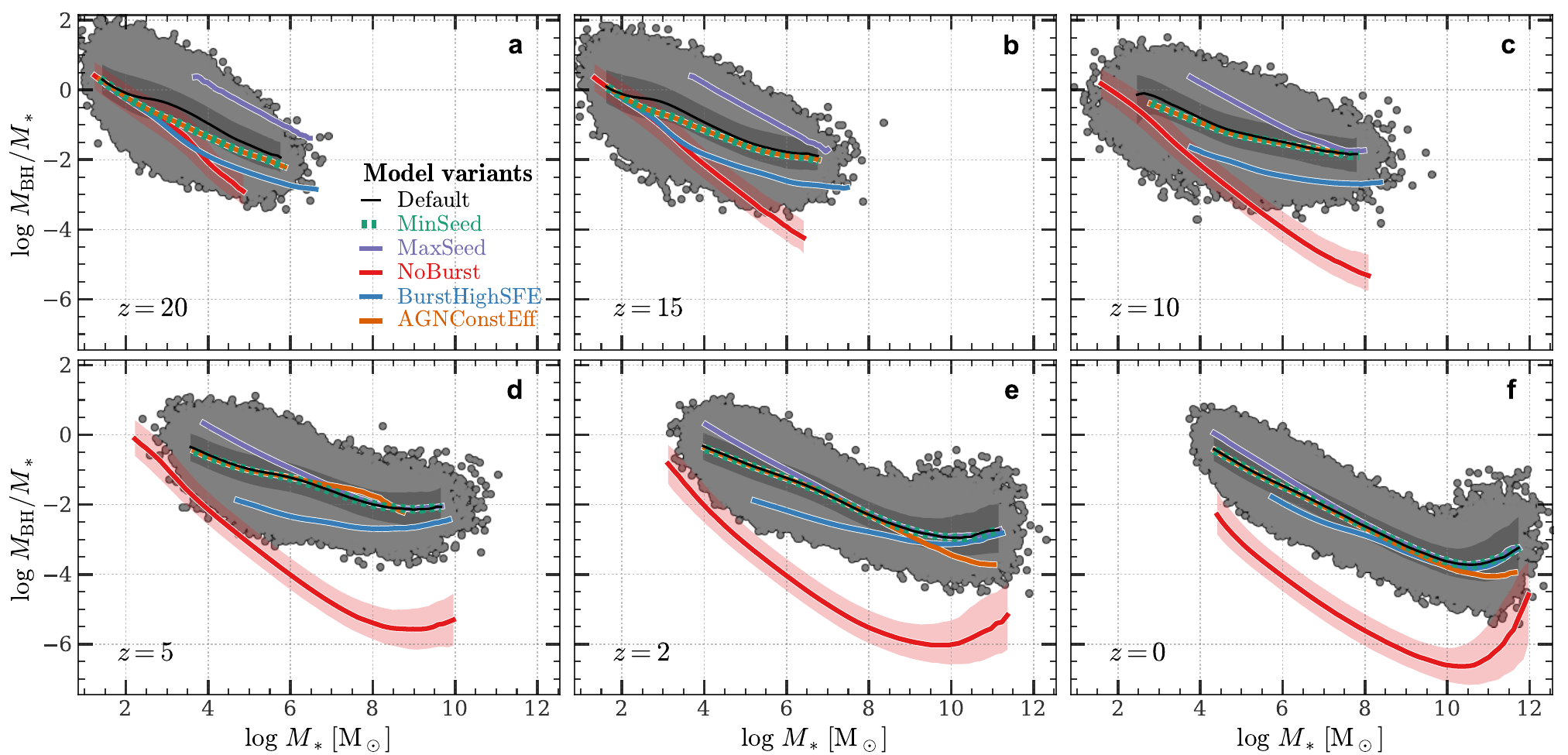}
    \caption{
    {\figem $M_\sscBH/M_*$-$M_*$ relations predicted by the variants of the model.}
    {\figem a}--{\figem f}, $M_\sscBH/M_*$ versus $M_*$ predicted 
    at different $z$.
    In each panel, curves show the median values of $M_\sscBH/M_*$
    as a function of $M_*$ predicted by
    six variants.
    {\figem Scatter} points show the predictions for individual galaxies
    by the {\textem Default} variant.
    For {\textem Default} and {\textem NoBurst}, we also show the
    1-$\sigma$ range ($16^{\rm th}$--$84^{\rm th}$ percentiles)
    by shaded regions. The variants demonstrate the effects of
    adapting the physical recipes for different growth channels
    in the build-up of BH and stellar contents of galaxies.
    See Appendix~\ref{app:sec:model-variants} for details.
    }
    \label{fig:scaling-evol-variants}
\end{figure*}

The bursty mode requires a global disturbance to generate the hotness
of the gas. This is more likely to happen if the merger rate of the host 
halo is high, and is thus expected to be more common in the early Universe
and in a halo with higher mass. 
The accretion rate of a BH in the bursty mode is stronger if 
$M_\sscBH$ is smaller, so that the Eddington ratio is initially higher 
and the feedback regulation is less effective, or if 
$M_{\rm v}$ of the host halo is larger, so that the SNF nucleus 
is more massive to supply more gas to the BH (see texts below 
Eq.~\ref{eq:dot-m-Bondi-numerical-q05} for an estimate). 
Thus, the growth of a BH through the bursty mode is expected to be 
more significant in the early Universe (before the BH has grown massive)
or in a halo with higher mass.
A similar conclusion applies to star formation in the bursty mode,
since a less massive BH is less competitive compared to
star formation (see Appendix~\ref{app:sssec:nuclear-burst-basic-eq}), and a more massive halo can supply
a more massive nucleus to form stars. However,
the continuous mode depends on the fast accretion of the host halo
and the global properties of the SGC. Such a mode is thus expected to be 
efficient during the entire fast phase of the halo, even if 
the SNF nucleus is not formed or is negligibly small.
Finally, the contribution from mergers dominates if other 
channels are not active. This is more likely in later times
and/or in a halo with higher mass where galaxy quenching 
suppresses the growth in the other modes.

As described in \S\ref{sssec:growth-channels}, the growth of stellar and 
BH masses in galaxies is modeled as a multichannel process, with 
each channel governing a certain stage of the growth.
To cover the broad scales of time, space and mass involved in the growth,
we have incorporated certain assumptions for 
each channel, and used the parameters listed in Table~\ref{tab:parameters}
to characterize the degree of freedom (DoF) in the model.
Some parameters have already been assigned values or ranges based
on either analytical estimates, numerical simulations or
observational/empirical constraints, while others remain uncertain.
Here, we perform a set of controlled experiments by introducing a number
of variants of the model. The purpose is neither to perform an exhaustive
exploration of the full parameter space nor to find a best-fit calibration,
but to demonstrate the effects of different physical recipes on model
predictions for the stellar and BH contents of galaxies and to
explore implications for future observations.

To maximize the coverage of the parameter space, the variants to be considered
include adaptations for each of the growth channels.
To minimize the entanglement of the effects of different processes,
we limit ourselves to changing the parameters of only one process at 
a time to produce a new variant.
Note that the modifications made on one process may affect the
consequences of other processes. For example, a change in
the star formation efficiency in any channel also modifies
the LW radiation field and IGM enrichment, thus affecting
the seeding in other halos affected by the radiation or matter
feedback. To avoid such complication, new variants introduced here
are based on the {\textem MinSeed} variant (see \S\ref{sssec:growth-effect-seeding}),
which pins down the seeding process by always producing minimal masses of
BH seeds in mini-halos. All variants are applied to the same set of
merger trees (obtained from TNG100-1-Dark; see \S\ref{ssec:halo-assembly})
as used for the main text. 

The variants considered here, including those defined in the main text, are listed below.
\begin{enumerate}[topsep=0pt,parsep=0pt,itemsep=0pt]
    \item {\textem Default}: 
    this is the default case
    used to produce the results in the main texts (except those shown
    in Fig.~\ref{fig:growth-effect-seeding}),
    with the values of parameters listed in Table~\ref{tab:parameters}.
    \item {\textem MinSeed}: 
    this variant is the same as {\textem Default}, except that
    Pop-III seeds with minimal masses are forced to form in all 
    mini-halos. To achieve this, we set the parameter
    $\gamma_{\rm crit} \rightarrow + \infty$ and force
    $J_{\rm LW} = 0$ and $Z_{\rm IGM} = 0$ for all halos.
    This ensures that the factors $f_{\rm J}$ (Eq.~\ref{eq:LW-f-J}) 
    and $f_{\gamma}$ (Eq.~\ref{eq:f-gamma-dyn-heat}) are both zero,
    so that the delay of gas collapse in mini-halos is completely
    disabled.
    \item {\textem MaxSeed}: 
    this variant is the same as {\textem Default}, except that
    Pop-III seeds with maximal masses are forced to form in all 
    mini-halos. This is achieved by setting the 
    parameter $\gamma_{\rm crit} \rightarrow 0$ and forcing
    $J_{\rm LW} = +\infty$ and $Z_{\rm IGM} = 0$ 
    for all halos, so that the delay of seeding and the 
    masses of seeds are maximized.
    \item {\textem NoBurst}: 
    this variant is the same as {\textem MinSeed}, except that
    nuclear burst is completely turned off. This is achieved by setting
    $\delta_\gamma = +\infty$ (see Eq.~\ref{eq:snf-formation-crit}), so that the formation 
    of SNF nucleus never occurs.
    \item {\textem BurstHighSFE}: 
    this variant is the same as {\textem MinSeed}, except that the star formation
    efficiency in nuclear burst is boosted, 
    so that the competition between star formation and BH accretion is tilted towards the former
    and thus BH cannot grow by the wet channel in the bursty mode. 
    This is achieved by setting $\alpha_{\sscBH} = 0$
    and $\delta_{\ssc Q} \rightarrow +\infty$ (see Appendix~\ref{app:sssec:nuclear-burst-details}).
    Note that the dry channel in the bursty mode is still open in this variant.
    \item {\textem AGNConstEff}: 
    this variant is the same as {\textem MinSeed}, 
    except that the efficiency of AGN feedback is set to be 
    redshift-independent. This is achieved by setting
    $\beta_{\rm {\sscAGN},f} = 0$ (see Eq.~\ref{eq:f-AGN}).
\end{enumerate}

Among the above variants, {\textem MinSeed} and {\textem MaxSeed} are designed to bound the effects of 
seeding, and are used to demonstrate the degeneracy between seeding and 
growing (see \S\ref{sssec:growth-effect-seeding}). IGM enrichment is disabled
(by forcing $Z_{\ssc IGM} = 0$) in both variants to avoid the entanglement 
introduced by Pop-II seeds (see the seeding atlas, Fig.~\ref{fig:seeding-atlas}).

Fig.~\ref{fig:scaling-evol-variants} shows $M_\sscBH$ and $M_*$ predicted 
by the above model variants at different redshifts. 
To limit the dynamic range of displayed values, we show the ratio,
$M_\sscBH/M_*$, instead of $M_\sscBH$ itself, as a function of $M_*$.
Obvious differences can be seen among the variants with adaptations 
for the seeding methods ({\textem Default}, {\textem MinSeed} and {\textem MaxSeed}).
At $z = 20$ (panel a), about the ending point of the seeding era 
(see \S\ref{sssec:cosmic-bh-sf-history} and 
Fig.~\ref{fig:growth-cosmic-history}), the {\textem MaxSeed} ({\textem MinSeed}) variant
predicts the highest (lowest) $M_\sscBH/M_*$ at any $M_*$, and the {\textem Default} 
variant predicts a value in between the two extremes.
The largest value of $M_*$ produced by the {\textem MaxSeed} is also the largest
among the three variants,
while that produced by the {\textem MinSeed} variant is similar to that 
produced by the {\textem Default} variant. 
The largest values of $M_*$ have already exceeded $10^6\Msun$ at $z=20$, 
implying that significant growth through the bursty mode has occurred 
at $z \gtrsim 20$ (see Fig.~\ref{fig:growth-channels} for some examples). 
Such a growth is expected to be most significant
in the {\textem MinSeed} variant due to the least effective AGN feedback,
thus explaining the similar ranges of $M_*$ produced by the {\textem MinSeed} and
{\textem Default} variants.
The differences among the three variants become less significant when 
galaxies evolve to higher masses, and vanish at $M_* \gtrsim 10^7\Msun$.
This is a direct consequence of the self-regulated growth 
of BHs, as demonstrated by the converged evolution of $M_\sscBH$ 
and $M_*$ in individual galaxies using a series of idealized and 
controlled experiments done in \S\ref{sssec:growth-effect-seeding}.
The differences in $M_\sscBH$ and $M_*$ produced by
the seeding methods are thus observable only
for extremely low-mass galaxies ($M_* \lesssim 10^7\Msun$) or
galaxies at very high redshift ($z \gtrsim 10$), 
for which current observations are still very limited 
(see \S\ref{ssec:mbh-ms-relation} for a data compilation).

In contrast to the seeding methods, adaptations for the bursty mode
can produce larger and longer-lasting effects on the BH and stellar contents 
of galaxies. This can be seen by comparing the predictions from the 
three variants with different treatments of nuclear bursts
({\textem MinSeed}, {\textem NoBurst} and {\textem BurstHighSFE}). As discussed in 
\S\ref{ssec:mbh-ms-relation}, the growth through the bursty 
mode plays an essential role in boosting the masses of low-mass BHs 
to the regime of SMBHs ($M_\sscBH \geqslant 10^6\Msun$).
The complete absence of the bursty mode in the {\textem NoBurst} variant
thus produces the lowest $M_\sscBH$ at any $M_*$ and any $z$, 
and leads to an $M_\sscBH$-$M_*$ relation that is significantly
below the observational results (see \S\ref{ssec:mbh-ms-relation}).
However, disabling only the wet channel in the bursty mode
({\textem BurstHighSFE}) yields a smaller deficiency in $M_\sscBH$, reinforcing
the conclusion reached phenomenologically in \citetalias{moTwophaseModelGalaxy2024} 
(see their \S5.2 and fig.~7) that moderate boosts of BH masses are sufficient to bring 
the BHs into the regime where the continuous mode takes over the responsibility of 
growing BHs to higher masses. Thus, the dry mode alone seems sufficient to grow 
BHs to the $M_\sscBH$-$M_*$ relation observed at $z \approx 0$ (panel f).
However, at $z \approx 5$--$10$, the BH masses predicted by the {\textem BurstHighSFE}
variant are more than $0.5\dex$ lower than those predicted by the 
{\textem MinSeed} (or {\textem Default}) variant at $M_* \lesssim 10^9\Msun$, 
as expected from the inefficiency of the dry channel in growing BHs
compared to the wet channel (see \S\ref{sssec:growth-channels}). 
Another interesting feature that can be expected for
the {\textem BurstHighSFE} variant is the peculiar 
properties of nuclear star clusters (NSCs).
This expectation is built on the consideration that AGN feedback
dominates the depletion of gas in SNF nuclei in the {\textem Default} variant, 
and that little BH accretion during a nuclear burst also implies 
little competition from the AGN feedback (see Appendix~\ref{app:sssec:nuclear-burst-basic-eq}).
Intense nuclear starbursts are thus inevitable in this variant,
which may produce NSCs that are much more massive than those 
produced in other variants. The presence of such massive NSCs may
significantly bias the total light of galaxies towards their centers
and produce dot-like morphologies. Fortunately, 
this regime overlaps with the faint AGN population that has recently
been probed by JWST (see \S\ref{ssec:mbh-ms-relation}), 
thus providing a potential way to distinguish the contributions
to BH growth through the wet and dry channels in nuclear bursts,
and to constrain the coevolution of galaxies, NSCs and BHs. 
We will come back to this topic in an upcoming paper.

The final variant considered here is {\textem AGNConstEff}, in which we impose a 
constant efficiency of AGN feedback at all redshifts (see Eq.~\ref{eq:f-AGN}).
Compared with the {\textem MinSeed} variant, {\textem AGNConstEff} produces significantly 
stronger AGN feedback in the continuous mode at $z \gg 0$, and thus 
prevents the growth of BH and stellar masses in massive galaxies.
Noticeable differences can be found at $z \lesssim 5$ when galaxy formation has 
transited to the continuous era (see \S\ref{sssec:cosmic-bh-sf-history}). At $z\approx 5$, 
the upper limit of $M_*$ produced by the {\textem AGNConstEff} variant 
(panel d, orange curve) is about $1\dex$ lower than that produced by the 
{\textem MinSeed} variant (green curve). 
The self-regulation of BH growth also leads to a reduction in $M_\sscBH$
larger than that in $M_*$, so that the ratio $M_\sscBH/M_*$ at a given $M_*$
is also lower in the {\textem AGNConstEff} variant in the high-mass end at $z \lesssim 2$ (panels e and f).
The deficit in the number of over-massive galaxies and BHs at $z \approx 5$,
and the low values of $M_\sscBH/M_*$ at $z \lesssim 2$ suggest
that {\textem AGNConstEff} may not be able to match observations of massive galaxies and BHs, 
such as high-$z$ quasars and local massive ellipticals. Instead, 
these observations seem to prefer a redshift-dependent efficiency of the AGN feedback, 
such as the one adopted in the {\textem Default} variant.


\bsp	
\label{lastpage}
\end{document}